%% file: Disser2.tex
\documentclass[twoside]{report}

\usepackage{fancyheadings}

\input{ThLech.tex}

\begin{document}
\pagenumbering{roman}

%%%%%%%%%%%%%%%%%%%%%%%%%%%%%%%%%%%%%%%%%%%%%%%%%%%%%%%%%%%%%%%%%%
%%%%%%%%%%%%%%%%%%%%%%%%%%%%%%%%%%%%%%%%%%%%%%%%%%%%%%%%%%%%%%%%%%
%
% Title page
%
%%%%%%%%%%%%%%%%%%%%%%%%%%%%%%%%%%%%%%%%%%%%%%%%%%%%%%%%%%%%%%%%%%
%%%%%%%%%%%%%%%%%%%%%%%%%%%%%%%%%%%%%%%%%%%%%%%%%%%%%%%%%%%%%%%%%%

%\begin{figure}
%\centering
%\includegraphics[width=16cm,height=25cm]{TPREF.ps}
%\end{figure}

%\commentout{
\begin{center}
{\Large \bf Noise Induced Dissipation in Discrete-Time Classical and Quantum Dynamical Systems}
\vskip 0.19in
\centerline{By}
\vskip 0.1in
\centerline{\large  Lech Wo{\l}owski}
\centerline{\large B.S. (Maria Curie-Sk{\l}odowska University in Lublin, Poland; 1999)}
\vskip 0.25in
\centerline{\large  DISSERTATION}
\vskip 0.1in
\centerline{\large Submitted in partial satisfaction of the requirements for the
degree of}
\vskip 0.1in
\centerline{\large  DOCTOR OF PHILOSOPHY}
\vskip 0.1in
\centerline{\large in}
\vskip 0.1in
\centerline{\large APPLIED MATHEMATICS}
\vskip 0.2in
\centerline{\large in the}
\vskip 0.1in
\centerline{\large  OFFICE OF GRADUATE STUDIES}
\vskip 0.1in
\centerline{\large of the}
\vskip 0.1in
\centerline{\large  UNIVERSITY OF CALIFORNIA}
\vskip 0.1in
\centerline{\large  DAVIS}
\end{center}
\vskip 0.09in
\begin{center}
{\large Approved:}
\end{center}
\begin{center}
\vskip 0.06in
\centerline{\underbar{Prof. Albert Fannjiang}}
\vskip 0.17in
\centerline{\underbar{Prof. Bruno Nachtergaele}}
\vskip 0.17in
\centerline{\underbar{Prof. John Hunter}}
\vskip 0.2in
\centerline{\large Committee in Charge}
\vskip 0.2in
\centerline{\large 2004}
\end{center}
%}%commentout

\newpage
\thispagestyle{empty}

$ $

%%%%%%%%%%%%%%%%%%%%%%%%%%%%%%%%%%%%%%%%%%%%%%%%%%%%%%%%%%%%%%%%%%
%%%%%%%%%%%%%%%%%%%%%%%%%%%%%%%%%%%%%%%%%%%%%%%%%%%%%%%%%%%%%%%%%%

%\newpage
\large
\tableofcontents

%%%%%%%%%%%%%%%%%%%%%%%%%%%%%%%%%%%%%%%%%%%%%%%%%%%%%%%%%%%%%%%%%%
%%%%%%%%%%%%%%%%%%%%%%%%%%%%%%%%%%%%%%%%%%%%%%%%%%%%%%%%%%%%%%%%%%

\newpage
\thispagestyle{empty}

$ $

\newpage
\section*{Acknowledgments}

\parindent=1cm

This dissertation was written during the spring quarter of 2004 and is
based on the research results obtained by the author in collaboration 
with Prof. Albert Fannjiang and Prof. St\'{e}phane Nonnenmacher.
All the results were obtained within the period from September 2000 
to April 2004, that is, during the author's enrollment in the 
Graduate Group in Applied Mathematics (GGAM) at UC Davis.

The author would like to express his gratitude to his advisor, Professor
Albert Fannjiang, for helping him to choose the research topic which proved
to be appropriate for the author's mathematical skills and interests as well as for 
his scientific support.

The author also wants to express his gratitude to the Department of Mathematics at UCD and the 
Office of Graduate Studies for providing him with substantial financial support in his graduate
program through various types of fellowships and stipends, including Departmental Block 
Grant Programs, Summer Research Fellowships, G. and D. Zolk Fellowship, travel and conference 
grants as well as some special additional awards, e.g., A. Leung Prize. 
Without this support it would have been impossible to complete this work in the above mentioned 
period of time.

Apart from these awards and fellowships, the author received constant scientific and administrative 
support from faculty members and the staff of the Mathematics Department and the GGAM group. 
In this context the author especially wishes to thank
Prof. Bruno Nachtergaele and Prof. Gerry Puckett (present and former Chairs of the GGAM),
Prof. John Hunter and Prof. Motohico Mulase (present and former Chairs of the Department) 
and Celia Davis (Graduate Coordinator) for their help and friendly encouragement. 

I would like to thank Prof. Bruno Nachtergaele for organizing the informal quantum statistical 
mechanics seminar, during which I had an opportunity to acquire the necessary background in 
statistical physics and to present and develop my own research results. I would like to thank him
for spending his time to attend my talks and discuss them with me.   
I would also like to thank Prof. Nachtergaele for making it possible for me to attend  
XIV ICMP in Lisbon, which had a great impact on my scientific development and also
for introducing me to Prof. Stephan De Bi\`evre and via him to Prof. St\'ephane Nonnenmacher.

Professor Nonnenmacher played a truly crucial role in the development of this work. I would like
to express my gratitude to him for spending his time explaining to me, with great patience, many complicated 
and completely new to me notions from ergodic theory, semiclassical and spectral analysis, as well as some 
important elements of the theory of open quantum systems. 

I want to mention that I benefitted a lot during my stay here in Davis from discussions and friendship with 
some of my peer graduate students: Shannon Starr, Momar Dieng, Jeremy Clark and Rudy Yoshida, 
to mention a few. 

In a special way I want to thank Prof. Tomasz Komorowski who encouraged me to apply to the graduate
program at UCD and provided me with essential help during the whole application process.
I would like to thank my parents Franciszek and Danuta, my brother Witold, his wife Katarzyna and their little daughter 
Maria for staying in touch with me and keeping me in their thoughts and prayers.

As far as spiritual aspect of my life is concerned, which internally cannot be separated from the professional one, 
I want to thank Fr. Daniel A. Looney, the Pastor of the St. James Parish, for providing me with an unconditional help
during some periods of difficulties which I had to face.
And finally, I would like to express my most heartfelt gratitude to the author of \cite{FK},
for her constant presence in, and the impact she made on, my life.

\hspace{1cm} 

Lech Wo{\l}owski     \hspace{6.5cm}                                                            Davis, Spring 2004.

%%%%%%%%%%%%%%%%%%%%%%%%%%%%%%%%%%%%%%%%%%%%%%%%%%%%%%%%%%%%%%%%%%
%%%%%%%%%%%%%%%%%%%%%%%%%%%%%%%%%%%%%%%%%%%%%%%%%%%%%%%%%%%%%%%%%%

\newpage 
\begin{center}
\underline{\bf \Large Abstract} 
\end{center}

In this dissertation, written under the supervision of Prof. Albert Fannjiang, 
we study statistical and ergodic properties of randomly 
perturbed (noisy) classical and quantum dynamical systems. We concentrate 
on the discrete time dynamics generated by Lebesgue measure preserving maps 
defined on $d$-dimensional torus. 
We introduce the notion of the {\em dissipation time} which enables us
to test how the system responds to the noise and in particular to measure the 
speed at which an initially closed, conservative system converges to the 
equilibrium when subjected to noisy interactions with its environment. 

We study the asymptotics of the dissipation time in the limit of vanishing 
noises and prove that it provides a robust criterion of the chaoticity of 
the underlying conservative system. The results formalize in a rigorous 
and quantitative way the idea that the dissipation is {\em fast} 
for chaotic systems and {\em slow} for regular ones. In the classical setting,
we show that chaotic systems, e.g., Anosov diffeomorphisms 
possess logarithmic dissipation time while for non-chaotic maps the 
corresponding asymptotics is of a power-law type. 
In case of diagonalizable ergodic toral automorphisms we compute the exact 
value of the dissipation rate constant and show that it is equal to the
reciprocal of the minimal dimensionally averaged KS entropy among all
irreducible components of the rational block diagonal decomposition of the map.

In case of quantum systems we introduce the notion of the dissipation
time for both finite and infinite dimensional quantizations on the torus. 
We study the simultaneous semiclassical and small noise asymptotics of the
quantum dissipation time and relate it to the notions of the Ehrenfest 
time and the dynamical entropy of the quantum system.
We concentrate on quantum toral symplectomorphisms (generalized cat maps)
for which we compute the exact asymptotics of their quantum dissipation 
time and show that it coincides with a classical one in the semiclassical
regime in which the magnitude of the Planck constant does not exceed the
the size of the noise.

%%%%%%%%%%%%%%%%%%%%%%%%%%%%%%%%%%%%%%%%%%%%%%%%%%%%%%%%%%%%%%%%%%%%%%%%%%%%%%%%%%%%%%%%%%%%%
%%%%%%%%%%%%%%%%%%%%%%%%%%%%%%%%%%%%%%%%%%%%%%%%%%%%%%%%%%%%%%%%%%%%%%%%%%%%%%%%%%%%%%%%%%%%%
%            C H A P T E R 1 : Introduction                                                 % 
%%%%%%%%%%%%%%%%%%%%%%%%%%%%%%%%%%%%%%%%%%%%%%%%%%%%%%%%%%%%%%%%%%%%%%%%%%%%%%%%%%%%%%%%%%%%%
%%%%%%%%%%%%%%%%%%%%%%%%%%%%%%%%%%%%%%%%%%%%%%%%%%%%%%%%%%%%%%%%%%%%%%%%%%%%%%%%%%%%%%%%%%%%%

\newpage

\thispagestyle{empty}

$ $

\newpage
\pagenumbering{arabic}

%mark command must be specified before \chapter to get correct
%header on the chapter title page
\markboth{ \rm \normalsize Chapter 1. \hspace{0.25cm} Introduction}
{ \rm \normalsize Chapter 1. \hspace{0.25cm} Introduction}
\chapter{Introduction}
\label{IntroThesis}
%pagestyles must be specified AFTER \chapter to get custimized headers:
\pagestyle{fancy} %pagestyle for the whole chapter
\thispagestyle{fancy} %pagestyle for the chapter title page
\lhead[\thepage]{\rightmark}
\rhead[\leftmark]{\thepage}
\cfoot{}

\parindent=1cm

The main subject of this dissertation is the study of statistical and ergodic properties of
noisy dynamical systems. We investigate the problems of irreversibility and approach to 
equilibrium for randomly perturbed classical and quantum systems exhibiting various
degrees of chaoticity.

The origin of irreversibility in dynamical systems is usually modeled by
small stochastic perturbations of the otherwise reversible evolution. These
perturbations may be attributed to many different sources:
uncontrolled interactions with the environment, internal stochasticity of
the system or unavoidable simplifications made in theoretical models of 
real-life experiments; e.g., some internal variables neglected in the equations. 
In experimental or numerical investigations, stochasticity or noise is introduced 
respectively by finite precision
of the preparation and measurement apparatus, and by rounding-off errors
due to finite precision of numerical computations. 
The important common feature is that noises, intrinsic (internal stochasticity)
or extrinsic (random influence from the environment), can, on appropriately long time scales, 
induce or emphasize effects that would be absent or difficult to discern without noise.  

In this work we concentrate mainly on one such effect, the effect of {\em dissipation}. 
The term dissipation refers in our study to the loss of the
energy of fluctuations of densities, or equivalently, observables of the system during
the course of noisy evolution.
The strength of dissipation can be determined by measuring the speed at which an
initially closed, conservative 
system converges to the equilibrium when subjected to noisy interactions with the
environment. The latter task can be accomplished by studying an appropriate
time scale on which the influence of the noise becomes noticeable, i.e.,
affects the dynamics on characteristic spatial scales of the whole system. 
Such time scale will be called the {\em dissipation time} and will
constitute the main object of our study. 

Intuitively speaking, the dissipation time is a time scale on which the
magnitude of initial density fluctuations is brought below a certain fixed threshold 
and hence the system finds itself in an intermediate state, roughly
speaking, 'half-way' from its final equilibrium.
From a physical point of view, the dissipation time captures
the time scale on which the system, due to the action of
environmental noise, achieves a certain fixed level of Boltzmann-Gibbs entropy (cf. Section \ref{BGEnt}). 

The main goal of this work is to determine the relation between {\em ergodic}, and 
in particular {\em chaotic}, properties of unperturbed, conservative systems and 
{\em dissipative} properties of their noisy counterparts. 
The main method is to study the asymptotics of the dissipation time in
the limit of vanishing noises. 
Our main task in the first part of the work is to characterize in a rigorous and
quantitative way the rate of the dissipation for classical systems. The results will support and formalize 
an intuitive understanding that dissipation should be {\em fast} for chaotic dynamics and {\em slow} for 
a regular one, the difference being more and more visible as the magnitude of the noise decreases.
The fact that we are mainly interested in the small noise limit has a direct 
physical interpretation.

Indeed, in the experimental setting considerable effort is usually made to eliminate the 
influence of the noise on the system by isolating it
from its environment (at least to some reasonable degree). It is, however, impossible to 
eliminate the noise completely. 
In the theoretical approach such situations are usually modeled by limiting procedures 
(the magnitude of the noise is positive but assumed to be arbitrary small). 

The notion of the dissipation time, as described above and defined in Section \ref{DTdef}, is relatively new. 
It has been introduced in the context of classical, continuous-time systems in \cite{F0,F} (cf. Section \ref{DTmot})
and later developed and extended to discrete-time classical and quantum systems in the following series 
of works \cite{FNW,FNWq,FW}. 

One of the most important findings is that the asymptotics of 
the dissipation time provides clear and robust characteristics of
chaoticity of underlying conservative (noiseless) systems. To explain the connection
between the dissipation time and chaoticity we need to review briefly some basic 
notions from the theory of chaotic systems and relate them to our results. 

Chaotic behavior of classical dynamical systems has been studied with an exceptional 
intensity over the period of the last fifty years. Many equivalent ways of defining or
characterizing chaos were developed over that time.
Among the most important and well-known approaches to chaoticity one has to mention at least
the following 

\begin{itemize}
\item {\em Algebraic approach}: K-systems. In this approach chaoticity is characterized through the Kolmogorov property 
(see Definition \ref{Ksystem}), which encodes in an algebraic language the idea that the system, although deterministic,
behaves effectively like a memoryless process. This approach allows for some generalization to quantum systems 
,i.e., to the noncommutative algebraic setting with the classical definition recovered as a particular (commutative) case 
(cf. Section \ref{prelim} and Chapter \ref{Inter}).

\item {\em Geometric approach}: Hyperbolicity. Chaoticity is characterized here by uniform hyperbolicity 
,i.e., strict positiveness of all Lyapunov exponents. The condition expresses the geometric picture of two nearby
trajectories separating from each other at an exponential rate in time. Different, slightly weaker, formulations are 
also allowed in this approach, e.g., almost uniform hyperbolicity or quasihyperbolicity (with a typical example given 
by ergodic but not hyperbolic toral automorphisms, cf. \cite{B,RSO}).

\item {\em Ergodic approach}: Strong mixing. In this approach, fast, i.e., at least exponential mixing 
is required if the system is to be called chaotic. This characterization is especially useful if the dynamics is modeled on 
the level of densities or observables of the system (not directly on the phase space).
In particular the property is equivalent to fast (at least exponential) decay of correlations and the 
approach is particularly useful in spectral analysis (cf. Sections \ref{SP} and \ref{UB}).
  
\item {\em Entropic approach}: Positiveness of KS Entropy. The system is called chaotic here if it has {\em completely positive} 
Kolmogorov-Sinai (KS) entropy. The term completely positive entropy \cite{P} refers to the property that 
the entropy of {\em any} partition of the phase space is strictly positive. Mere positiveness of KS entropy do not 
guarantee chaoticity of the system, as it is not difficult to construct an example of a nonergodic map with positive 
entropy (an example will be given later in this Introduction). One of the advantages of this approach lies in the fact
that it gives a clear information-theoretical interpretation of the notion of chaoticity. 
\end{itemize}

We note that in the classical setting some of these approaches are equivalent. For example, by Pinsker theorem \cite{P} 
(see also \cite{RoSi}), the system has K-property iff it has completely positive KS entropy. 
However, it is worth noting here that some attempts to
generalize both notions to quantum systems led to nonequivalent counterparts (for more details see Chapter \ref{Inter}
or \cite{BNS}).
Some of these approaches allow one to introduce different degrees of chaoticity. In the geometric approach,
different levels of hyperbolicity can be specified. In the ergodic formulation, one
may require a particular speed of decay of correlations within a particular class of observables 
determined, e.g., by some regularity properties.

In this dissertation we introduce another characteristics of chaoticity, namely {\em the asymptotics of the dissipation time}. 
The difference between the above-mentioned approaches and the present one lies in the fact that 
chaoticity is tested here in an extrinsic way. We test how the system responds to the noise.
Given this information, and not necessarily knowing all the details 
of the underlying conservative dynamics, one can distinguish chaotic from regular behavior. This makes the
dissipation time a robust criterion of chaoticity (cf. remark before Proposition \ref{supexp} in Section \ref{MainT}).  
Moreover, since all real-life experimental systems are inevitably subject to noisy interactions with their
environments the present approach is well suited for practical purposes.

Another reason for considering the asymptotics of the dissipation time as a test of chaoticity
is that the notion can be almost literally and quite successfully adapted from the classical to the quantum setting. 
Similarly as for the classical dynamics, the quantum dissipation time provides here a good criterion  
which enables us to distinguish chaotic from regular behavior in an appropriate semiclassical regime 
(cf. Section \ref{CLTS}). 
The importance of this observation can be understood if one takes into account the fact that 
despite great progress made in the field of quantum chaos in the last two decades, there is still no agreement
on what quantum chaoticity really means. The main problem lies in the fact that one cannot simply
take any particular classical definition of chaoticity and apply it directly to a quantum system.
Indeed, as mentioned above, in classical dynamics chaoticity is usually connected with the notion of a trajectory
of a system and the arbitrary closeness of two nearby trajectories (the geometric approach), or equivalently
with theoretical ability to resolve the details of the phase space to arbitrary small scales (the entropic approach). 
For obvious reasons neither of these notions has any direct counterpart in quantum case. On the other hand, even if 
some quantum generalization can be constructed (e.g., via the algebraic approach), the way to obtain it is 
usually non-unique, and the same classical notion can have many nonequivalent quantum counterparts (for more
detailed discussion and references see Chapter \ref{Inter}). 
The theory of the quantum dissipation time developed in this dissertation can be viewed as one of the many possible ways 
of approaching the problem of chaoticity in quantum systems. The second part of this work will be entirely devoted 
to this subject. 

Now we pass to a more systematic discussion of our main results. We start with the classical setting considered
in the first part of the dissertation. One general comment is appropriate here. Namely, the notion of the {\em classical} 
dissipation time is independent of any particular mathematical model of the noisy dynamical system one chooses to
work with. However, in order to fix the attention and, more importantly, be able to derive concrete, 
rigorous results one needs to choose a certain class of models for which a uniform framework can be 
constructed and the results for different systems can be compared, provided that they belong to 
the selected class. 
In this dissertation we choose to work with discrete time systems (maps) defined on compact 
phase spaces (represented by $d$-dimensional tori) with Lebesgue measure as a natural 
invariant measure for both conservative and noisy dynamics. 

As a matter of illustration and to build up some intuition we start by presenting 
some simple but representative examples of classical maps for which exact results regarding the
asymptotics of their dissipation time are available. 
The simplest examples (toy models) of that kind are as follows
\begin{itemize}
\item[I.] \textbf{Translations on $\IT^{d}$}, defined by $F\bx=\bx+\bv$,
represent the simplest examples of nonergodic, 
if $1,v_{1},..,c_{d}$ are rationally dependent, or otherwise ergodic
but not weakly-mixing maps. 
\item[II.] \textbf{Cat maps on $\IT^{2}$}, i.e., elements $F\in SL(2,\IZ)$ satisfying $|\Tr F|>2$
 projected on the torus (see \cite{Arnold}), provide simple
examples of uniformly hyperbolic, exponentially mixing, fully chaotic systems.
\item[III.] \textbf{Angle doubling map} $Fx= 2x \bmod 1$ provides an example 
of a uniformly expanding, exponentially mixing, noninvertible chaotic map.
\end{itemize}

Let $\ep$ denote the strength of the noise and $T_{\ep}$ the the operator representing the
action of the noisy dynamics associated with the above conservative maps on the observables of the system. 
The classical dissipation time $\tau_{c}$ 
(``c'' stands for ``classical'') is defined as follows (for detailed definition see Section \ref{DTdef}).
\bean
\tau_{c}=\min\{n \in \IZ_{+}:\|T_{\ep}^{n}\|< e^{-1}\},
\eean
where $\|T_{\ep}^{n}\|$ denotes the operator norm of $T_{\ep}^{n}$.

The corresponding asymptotics of $\tau_{c}$ as a function of positive, but vanishing magnitude of 
the noise $\ep$ is summarized in Table \ref{Tab1}

\begin{table}
\centering
\begin{tabular}{|l|c|c|c|c|}
\hline   $\quad$Map           &   Ergodic Properties  &  Dissipation Time        \\
\hline I. Translations    &   Not ergodic or not weakly-mixing       &   $\tau_{c}=\ep^{-2} + \cO(1)$   \\
\hline II. Cat maps        &   Exponentially mixing   &  $\tau_{c}=\frac{2}{h(F)}\ln(\ep^{-1}) + \cO(1)$ \\
\hline III. Angle doubling  &   Exponentially mixing   &  $\tau_{c}=\frac{1}{\ln 2}\ln(\ep^{-1}) + \cO(1)$\\
\hline
\end{tabular}
\caption{Asymptotics of dissipation time for typical maps}
%label must be put AFTER caption, sice the later generates the number for the float.
\label{Tab1}
\end{table}

Two observations emerge from the above picture. 
The first general observation is that two qualitatively different behaviors of the 
dissipation time can immediately be noticed:
\begin{itemize}
\item If the dynamics is {\em regular} then $\tau_{c}(\ep^{-1})$ diverges (as $\ep$ vanishes) in
 a power-law fashion. Here we speak of 
{\em slow} or {\em simple} dissipation (long dissipation time). 
\item If the dynamics is chaotic then $\tau_{c}(\ep^{-1})$ has logarithmic 
asymptotics. In this case we speak of {\em fast} dissipation (short dissipation time). 
\end{itemize}

We note that when the rate of divergence of $\tau_{c}$ as a function of $\ep^{-1}$, with $\ep\rightarrow 0$,
is fast then the dissipation is slow (dissipation time is long), and vice versa.

\begin{figure}
\centering
\includegraphics[width=14cm,height=10cm]{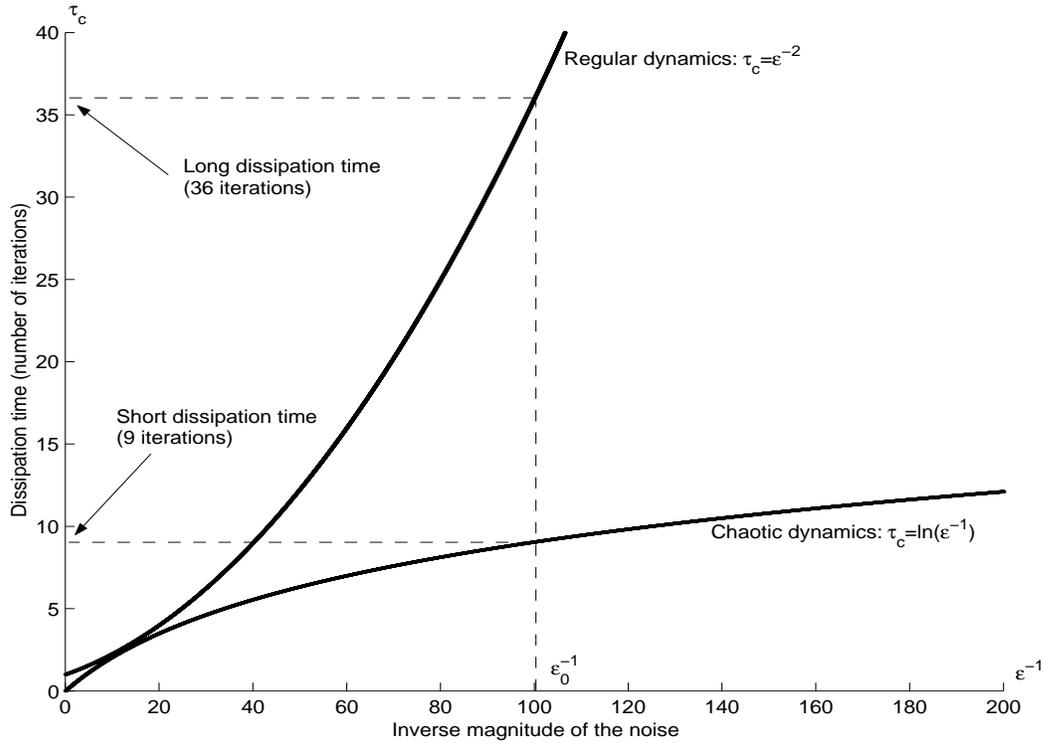}
\caption{Asymptotics of dissipation time}
\label{Fig1}
\end{figure}

Figure \ref{Fig1} illustrates both behaviors and explains this terminology.
Indeed, the number of iterations required to keep $\|T_{\ep}^{n}\|$
at a constant level (here $e^{-1}$), i.e., the dissipation time is plotted here against 
the inverse magnitude of the noise for typical regular (upper curve) and chaotic (lower curve) 
systems. If we fix some small amount of noise, say $\ep_{0} = 10^{-2}$, the reduction of the
norm of $T_{\ep}^{n}$ to the prescribed level is achieved much faster (9 iterates) in
a chaotic (logarithmic) regime than in a regular (power-law) one (36 iterates).

The second observation is of a rather particular nature. Namely, we 
note that in the case of these simple chaotic systems,
the constant of the logarithmic asymptotics is reciprocally proportional to their
Kolmogorov-Sinai entropy $h(F)$. As we will see later this observation does
not generalize in any obvious way to higher dimensions (cf. Theorem \ref{thm3}) and
to more complicated maps.

In view of the above observations a natural question arises whether the power-law and the 
logarithmic are the only possible asymptotic behaviors of the dissipation time.

To address this question, we first note that it is possible to provide quite sharp general upper and lower 
bounds for the dissipation time within a vast class of dynamical systems. 
Indeed, using methods from the spectral theory of non-normal operators and in particular 
the notion of the pseudospectrum (see Section \ref{SP}), and investigating basic geometric 
properties of conservative maps (hyperbolicity and local expansion rates studied in Section \ref{LB}), 
we arrive at the following results :

\begin{itemize}
\item
The dissipation time of an arbitrary dynamical system generated by
a measure-preserving map is never finite 
(i.e. $\tau_{c}\rightarrow \infty$, as $\ep\rightarrow 0$).
\item
The rate of divergence of $\tau_{c}$ is never faster than power-law in $\ep$,
\bean 
\tau_{c}\leq \ep^{-\alpha}, \alpha \in (0,2].
\eean
\item
If the map $F$ is $C^{1}$ then the divergence of $\tau_{c}$ is never slower than logarithmic: 
\bean
 \frac{1}{ \ln\|DF\|_\infty} \ln(\ep^{-1}) \leq {\tau}_{c},
\eean
where $\|DF\|_{\infty}=\sup_{\bx\in\IT^d} \|(DF)(\bx)\|$ denotes the highest expansion rate of $F$
(if $\|DF\|_{\infty}=1$ then $\tau_{c}\sim \ep^{-\alpha}$).
\item
Almost all  non weakly-mixing systems (a modicum of regularity is required, cf. Theorem \ref{weakmix}) 
undergo power-law (i.e., the slowest possible) dissipation: $\tau_{c} \sim \ep^{-\alpha}, \alpha \in (0,2]$. 
\end{itemize}

The question which arose from the first observation is now reduced to the problem of 
establishing an logarithmic upper bound for the dissipation time of a largest possible class of 
systems exhibiting some chaotic properties, or equivalently deciding whether there exists a map
for which an intermediate (i.e., contained strictly between power-law and logarithmic) asymptotics hold.

The second observation suggests that for systems with logarithmic dissipation time, 
the value of the dissipation rate constant (i.e., the prefactor of the asymptotics) should provide
valuable information about the underlying conservative dynamics.

Chapter \ref{DTCCS} is entirely devoted to the study of these two problems.

As to the first problem, we developed two different methods which allowed us to establish logarithmic
asymptotics respectively for linear (toral automorphisms - Section \ref{MainT}) and 
nonlinear ($C^{3}$ Anosov diffeomorphisms - Section \ref{Anosov}) hyperbolic maps in arbitrary 
phase space dimension. 

Both methods rely eventually on quite advanced number theoretical or respectively spectral analysis 
and it seems that there is no 'short-cut' way to establish logarithmic asymptotics
for any chaotic dynamical system except for simple $1$- or $2$-dimensional toy models (e.g., cat maps).

As far as abstract (i.e., not related to any particular map) results are concerned, we derive in Section \ref{UB} 
a general connection between mixing properties (the rate of decay of correlations) of both conservative and 
noisy dynamics and the rate of the divergence of the dissipation time.
In particular we show that within a large class of maps, strong (exponentially fast) mixing implies logarithmic 
dissipation time. On the other hand we also prove that methods used in the computation of the dissipation can be used to 
determine in certain cases the (precise) rate of decay of correlations
(cf. Proposition \ref{supexp} in Section \ref{MainT}). 

Let us also comment here briefly on the second problem. The exact solution is now only available in the case of 
diagonalizable toral automorphisms. The result is established in Theorem \ref{thm3}, where the dissipation rate 
constant is proved to be equal to the reciprocal of the minimal dimensionally averaged KS entropy among 
irreducible components of the rational block diagonal decomposition of the map. 
At this point another question arises: why does the minimal dimensionally averaged KS-entropy appear in the constant 
instead of KS-entropy itself?

The complete answer to this question is not known. However, the following considerations can shed some light on it.
It is known that the knowledge of KS-entropy itself (e.g., its positiveness) is not sufficient to determine whether the 
system is chaotic or not. Indeed, consider the following toral automorphism in 4-dim represented
in the block-diagonal form 
\bean
F=\begin{bmatrix}
\begin{matrix}
1 & 0 \\
0 & 1 \\
\end{matrix} & \pmb{0} \\
\pmb{0} & 
\begin{matrix}
1 & 1 \\
1 & 2 \\
\end{matrix} \\
\end{bmatrix}
\eean
The first block is simply the identity and the second is a hyperbolic automorphism 
(the famous Arnold's cat \cite{Arnold}).
The entropy of $F$ is positive and equals the entropy of Arnold's cat, but the system is 
not even ergodic (toral automorphisms are ergodic
iff no root of unity lies in their spectrum - see Section \ref{prelim}).
The minimal dimensionally averaged entropy is in this case $0$ and the system undergoes
slow (power-law) dissipation characteristic for non-chaotic systems.

The fact that the dissipation rate constant averages the KS entropy over the
dimension of the irreducible block is of separate importance. We will not explore it fully here. Let us only mention the following 
simple example. Consider two matrices $F_{1}\in SL(d_{1},\IZ)$
and $F_{2}\in SL(d_{2},\IZ)$ and assume that they have the same or almost the same spectra, but with different 
degeneracies. If it happens that $d_{1} \gg d_{2}$ then also $h_{KS}(F_{1}) \gg  h_{KS}(F_{2})$ 
while their dimensionally averaged counterparts are of the same order $\hat{h}(F_{1})\sim \hat{h}(F_{2})$. 
This reflects the natural intuition that if the strengths of Lyapunov exponents of two systems are comparable 
then the degree of their chaoticity should also be comparable (i.e., independent of the dimension).

As far as nonlinear maps in the context of the second problem are concerned, we derive lower and upper bounds for the 
dissipation rate constant for $C^{3}$ Anosov systems (Theorem \ref{TAnosov}) but the exact value 
(and even its existence, not to mention its connection to the KS entropy) remains unknown.

To conclude the description of the first part of the dissertation we want to mention that in Section \ref{GandA} we collect 
some of the many possible generalizations and applications of the above described results, especially the ones 
concerning toral automorphisms. In particular we investigate the
possibility of defining the dissipation time for maps with degenerate noise kernels (Section \ref{degnoise}), and we
study the relations between the asymptotics of the dissipation time and some typical time scales encountered in the
study of the so-called kinematic dynamo problem (Section \ref{dynamo}).  

Now we pass to the description of the second part of the present work, devoted to the
study of the dissipation time in the quantum mechanical setting.

The second part begins in Chapter \ref{Inter}, called the {\em Interludium} as it
constitutes a separate and almost independent part of this work. It is meant as a historical overview 
and in the same time as a quick but comprehensive introduction to the specific area of quantum mechanics 
on the torus. We describe there the two most important and most commonly encountered in the literature quantization 
schemes for toral maps. We put a special emphasis on careful explanation of their origins. We also discuss
similarities and differences between these two approaches. In particular, we concentrate on the role which 
semiclassical analysis of spectral properties of quantum chaotic systems on the one hand, and 
the introduction of several nonequivalent notions of quantum dynamical entropy, on the
other, played in the development of both quantization methods (and vice versa).
   
In the first part of Chapter \ref{QuDiTi} (cf. Sections \ref{QTK} and \ref{QTD}) both quantization methods are 
presented in a systematic and rigorous way. We develop a general framework (based on Weyl quantization), which 
unifies both approaches (i.e., each can be derived as a special case of the general scheme). 
This prepares the ground for an introduction of 
the quantum noise (Section \ref{Qnoise}) and the notion of quantum dissipation time (Section \ref{SAQDT}). 
The final part of the chapter is devoted to our
results on semiclassical analysis of the dissipation time of canonical toral maps
(Sections \ref{DTQL} and \ref{CLTS}).   

Before we discuss these results in more detail a word of caution is necessary here. Namely, as 
mentioned above and explained in the Interludium, the quantization of any classical system and 
in particular a toral canonical map can be performed in many different ways.
We would like our results to depend as little as possible on the particular 
quantization scheme and this is why certain effort was made in this work to present the
results in most unified way possible. 
Nevertheless, complete independence is not possible and
it is necessary to describe precisely the quantization principles and methods 
adopted in a given approach before the results can be stated 
(this non-uniqueness of the setting and dependence on quantization procedures constitutes one of 
the most fundamental differences between classical and quantum descriptions).
We sketch some of the quantization principles briefly here and refer to Chapters \ref{Inter} and \ref{QuDiTi} 
for a detailed presentation. 

Quantization on the torus is usually approached from the following two, non-equivalent points of view:
\begin{itemize}
\item {\em Finite dimensional approach}. This method (sometimes referred to as canonical quantization) 
was originally introduced in \cite{HB}, and later generalized and developed in a number of works 
\cite{E,EGI,KMR,BozDB,RSO}.
\item {\em Infinite dimensional approach}. Usually referred to as $^*$-algebraic noncommutative 
deformation of the torus, see, e.g., \cite{BNS,KL}. 
\end{itemize}

In this work we do not distinguish between these two approaches in terms of
whether they are 'canonical' or 'algebraic' (cf. \cite{DB}). 
The reason is that the real difference between these two quantizations lies in the 
choice of the fundamental Hilbert space of pure quantum states of the system whose 
classical counterpart has the $2d$-dimensional torus as a phase space. After the choice 
is made (i.e., the space is chosen to be either finite or infinite dimensional), both 
quantizations can be studied in any, including in particular 'abstract' $^*$-algebraic, 
framework.   

It turns out, however, that the finite dimensional approach is much more suitable for 
semiclassical analysis of the dissipation time, and most of
the results of Chapter \ref{QuDiTi} are stated in this setting. For completeness and to
illustrate the difference we also consider the infinite dimensional case. 
In particular we prove that in this case, regardless of the value of the Planck constant, 
quantum and classical evolutions of toral symplectomorphisms coincide.

In the finite dimensional setting, the geometry of the torus (the phase space)
and standard requirements of quantum mechanics (conjugacy between the
position and momentum representations) restrict
the space of admissible wave functions (pure states) to quasiperiodic 
Dirac delta combs. 
Planck's constant $h=2\pi\hbar$ is then restricted to reciprocals of integers
$N\in \IZ_{+}$, and the resulting quantum Hilbert space of pure
states is $N^{d}$-dimensional (for detailed explanations see Section \ref{CanQuant} in 
Appendix \ref{A}). The unitary and noisy quantum dynamics can 
be implemented either on the set of all quantum states, i.e., density operators
(the Schr\"odinger picture corresponding to the Frobenius-Perron
approach in classical case) or on the $N^{2d}$-dimensional algebra of quantum 
observables (the Heisenberg picture - quantum counterpart of classical the Koopman 
formalism). 

For a fixed finite dimensional quantum system (i.e., fixed Planck constant) 
and vanishing noise the presence of a pure point, unitary 
spectrum of the quantum propagator forces the dissipation time to have the same, trivial 
(i.e., power-law) asymptotics regardless whether the underlying conservative classical map 
is chaotic or not (cf. Proposition \ref{Nfixed}). 
To recover useful information about the dynamics
one needs to perform simultaneously the small noise and the semiclassical limit. It is thus clear
that the notion of the quantum dissipation time is intrinsically of a semiclassical nature.
It is, however, not enough to consider just 'a semiclassical limit', since the way the 
limits are taken here matters considerably.
 
The reason for this is the following: the classical dissipation time becomes 
larger and larger in the small-noise limit, but the correspondence
between classical and
quantum evolutions holds only up to a certain ''breaking time'', which diverges only in
the semiclassical limit (see Chapter \ref{Inter} for detailed discussion). 
Therefore, when seeking traces of chaoticity
in quantum systems with classical chaotic counterparts one needs to consider sufficiently
fast semiclassical limit.

On the other hand, one has to avoid falling into triviality waiting on another extreme of the problem. 
Namely, if the Planck constant is sent to zero too fast w.r.t. the size of the noise, the
dissipation time can be shorter or even much shorter than the ''breaking time'' and the quantization 
effects disappear too quickly to be noticeable in the asymptotic behavior of the system. 

The above situation is common to all approaches to quantum chaoticity in finite dimensional
systems (see, e.g., results regarding the spectrum of noisy quantum propagators \cite{Br,N}, 
or the study of the decoherence rate and quantum dynamical entropy \cite{AF94,ALPZ,BPS,PZ,GSS,BCCFV}).

The problem one really needs to solve here is to determine the relation between, 
on the one hand, spatial scales represented by the 
the size of the noise and the magnitude of the Planck constant, and on the other hand, temporal
scales, such as dissipation and ''breaking'' times, on which some traces of original 
classical chaoticity are still present in quantized systems.

The difficulty of the problem lies in the fact that in order to solve the problem one needs to control 
simultaneously the behavior of four different asymptotic parameters 
(noise, the Planck constant, dissipation and breaking times).

It may be of some interest to remark that the problem resembles (to some extent) some problems
occurring in numerical simulations of chaotic or turbulent dynamical systems. 
Any chaotic system develops, in a relatively short time, extremely complicated
structures on smaller and smaller spatial scales.
The quantization as well as numerical discretization inevitably imposes a finite 
resolution of the details of the phase space, the quantum 'mesh spacing' being constrained 
by the Heisenberg uncertainty principle $\Delta q \Delta p \gtrsim \hbar$. 
The introduction of a small amount of noise corresponds to
numerical instabilities due to finite precision of any numerical computation (rounding errors). 

Just as numerical approximations break down on sufficiently long time scales, one
expects a similar breakdown of the quantum-classical correspondence. In the
quantum framework, the corresponding time scale (the above ''breaking time'') is usually referred
to as the {\em Ehrenfest} time $\tau_E$ (see Chapter \ref{Inter}). For chaotic systems, the first signs of 
discrepancy may appear around $\tau_E\approx \lambda^{-1} \ln(\hbar^{-1})$, where $\lambda$ is
the largest Lyapunov exponent (this is the earliest time scale on which the chaotic dynamics might develop
structures beyond the reach of quantum phase-space resolution). 
Numerous works, both theoretical and numerical, have been devoted to
study  this phenomenon. Recently some rigorous results \cite{BonDB1}
have been obtained, describing the breakdown of the classical-quantum correspondence on
such a time scale.  However, this breakdown effectively occurs for
a very particular class of maps and values of $\hbar$, 
and hence does not necessarily reflect the generic behavior (cf. Chapter \ref{Inter}). 

For a given noise strength $\ep>0$, a quantum
dynamical system always resembles its
classical counterpart if Planck's constant is small enough.  
This is reflected in Proposition
\ref{AQC}, which states that for any canonical map and noise strength $\ep$, 
the quantum dissipation time
converges to its classical counterpart
(see also Corollary \ref{QTAnosov}).
For a general map, it is more difficult to determine precisely a
regime where both $\ep,\ \hbar\to 0$, and such that
classical and quantum dissipation times have the same asymptotics. 
We address this problem
in the case of quantized toral automorphisms.

For symplectic toral automorphisms in arbitrary dimension, we prove
(Theorem \ref{QDTthm} and Corollary \ref{QC}) this asymptotic correspondence between the classical
and quantum dissipation times $\tau_c\approx\tau_q$ in the regime 
where $1\gg \ep\geq C \hbar$. In this case, one indeed has $\tau_{q} \lesssim \tau_{E}$,
which intuitively justifies the correspondence.

Around the ``boundary'' of this regime, that is for $\ep\sim C\hbar$, the noise strength 
is comparable with the ``quantum mesh'', and the dissipation and the Ehrenfest times are
of the same order.
It is important to stress that our methods allows us to prove that the asymptotics of the quantum and 
classical dissipation times coincide in an exact way, that is, including
the prefactor in front of the logarithmic asymptotics (the ``dissipation rate constant''). 

In case of quantum coarse graining (cf. def. (\ref{qcoarse})) we derive similar semiclassical result but 
under a little bit stronger assumption on the convergence rate for $\hbar$ - the existence
of a positive $\beta$ such that $\ep^{\beta}\hbar^{-1}>1$. 

We also investigate the opposite situation, when the classical dissipation time is 
longer than $\tau_E$. In the situation where $\ep\ll\hbar^{1+\delta}$,
we obtain (for ergodic toral automorphisms):
\bean
\tau_{q}\geq C\left(\frac{\hbar}{\ep}\right)^2 \gg \tau_{c} \sim
\ln(\ep^{-1}) 
\gg |\ln\hbar|\sim \tau_{E},
\eean

In this case, three different time ranges can be distinguished.
Before $\tau_{E}$, classical and quantum evolutions are identical 
(and do not dissipate). In the range
$\tau_{E} \ll t \ll \tau_{c}$, 
the evolutions may differ, but neither dissipates yet. On the
time scale 
$\tau_{c} \ll t \ll \tau_{q}$, the classical system dissipates, while
the quantum one remains dissipationless until $t>\tau_{q}$.

This result shows in particular that after passing the Ehrenfest time
the quantum and classical evolutions do not have
to part from each other in an immediate and complete fashion. For example, the
difference between the systems is still not visible on
that time scale if one restricts the observations
only to dissipative properties of both systems.

To finish the description of our results we want to remark that
a lot of important questions remain still open in this area and that
the whole theory is rather in the initial stage of its development. 
The first problem is to generalize the above sharp estimates to nonlinear quantum 
maps (e.g. Anosov diffeomorphisms). This requires the application of different 
techniques (cf. \cite{FNWq}). In particular appropriate estimates are needed on the constant in 
Egorov theorem for such maps \cite{BR} (the problem is nonexistent in the
linear case due to the fact that the semiclassical approximation is exact).

Also recent results on quantum dynamical entropy for finite 
dimensional systems \cite{ALPZ,BCCFV} seem to indicate that just like
in the case of classical maps, it should be possible to establish a link
between the dissipation rate constant and the quantum mechanical entropy of
the system. It is still too early however to formulate any concrete
conjectures (since the relation is not yet established for a sufficiently general
class of classical systems). We discuss this problem 
briefly in Chapter \ref{Inter}.

We now conclude the Introduction with a few remarks regarding the organization of the material in
this work. The dissertation is divided into five chapters and two appendixes (in the list below the descriptions do
not necessary match the titles)

Chapter 1. Introduction.

Chapter 2. Definition and general properties of dissipation time.

Chapter 3. Dissipation time of classically chaotic systems on $\IT^{d}$.

- Section 3.1 and 3.2 - Linear maps: toral automorphism and generalizations.

- Section 3.3 and 3.4 - Nonlinear maps: Anosov systems. 

Chapter 4. Interludium.

Chapter 5. Dissipation time of quantum systems on $\IT^{d}$.

- Section 5.1 - Weyl quantization on the Torus.

- Section 5.2 - Semiclassical analysis of dissipation time.

Appendix A. The dynamics of Cat maps.

Appendix B. Wigner transform.

Two chapters (1 and 4) are of expository character. The remaining chapters contain the 
main results of the work and constitute the original contribution to the field. Appendixes
contain necessary technical background regarding two specific topics. Throughout the work
attention was paid to ensure mathematical correctness and completeness. All results are 
presented with full proofs. For the convenience of the reader, technical proofs of
secondary importance are collected at the end of each non-expository chapter in a 
separate section.      

%%%%%%%%%%%%%%%%%%%%%%%%%%%%%%%%%%%%%%%%%%%%%%%%%%%%%%%%%%%%%%%%%%%%%%%%%%%%%%%%%%%%%%%%%%%%%%%%%%%%%%%%%%
%%%%%%%%%%%%%%%%%%%%%%%%%%%%%%%%%%%%%%%%%%%%%%%%%%%%%%%%%%%%%%%%%%%%%%%%%%%%%%%%%%%%%%%%%%%%%%%%%%%%%%%%%%%
%%%%%%%%%%%%%%%%%%%%%%%%%%%%%%%%%%%%%%%%%%%%%%%%%%%%%%%%%%%%%%%%%%%%%%%%%%%%%%%%%%%%%%%%%%%%%%%%%%%%%%%%%%%
%%%%%%%%%%%%%%%%%%%%%%%%%%%%%%%%%%%%%%%%%%%%%%%%%%%%%%%%%%%%%%%%%%%%%%%%%%%%%%%%%%%%%%%%%%%%%%%%%%%%%%%%%%
%
%
%                  *****************      P  A  R  T    I         ******************
%
%
%%%%%%%%%%%%%%%%%%%%%%%%%%%%%%%%%%%%%%%%%%%%%%%%%%%%%%%%%%%%%%%%%%%%%%%%%%%%%%%%%%%%%%%%%%%%%%%%%%%%%%%%%%
%%%%%%%%%%%%%%%%%%%%%%%%%%%%%%%%%%%%%%%%%%%%%%%%%%%%%%%%%%%%%%%%%%%%%%%%%%%%%%%%%%%%%%%%%%%%%%%%%%%%%%%%%%%
%%%%%%%%%%%%%%%%%%%%%%%%%%%%%%%%%%%%%%%%%%%%%%%%%%%%%%%%%%%%%%%%%%%%%%%%%%%%%%%%%%%%%%%%%%%%%%%%%%%%%%%%%%%
%%%%%%%%%%%%%%%%%%%%%%%%%%%%%%%%%%%%%%%%%%%%%%%%%%%%%%%%%%%%%%%%%%%%%%%%%%%%%%%%%%%%%%%%%%%%%%%%%%%%%%%%%%

\newpage
\thispagestyle{empty}

$ $

\newpage
\markright{ \rm \normalsize Part I. \hspace{0.25cm} Classical Mechanics}
$ $
\vskip 2.7in
\begin{center}
\begin{huge}
\textbf{Part I}
\vskip 0.5in
\textbf{Classical Mechanics}
\end{huge}
\end{center}
\thispagestyle{fancy}
\lhead[\rightmark]{\rightmark}
\rhead[\thepage]{\thepage}
\cfoot{}

%%%%%%%%%%%%%%%%%%%%%%%%%%%%%%%%%%%%%%%%%%%%%%%%%%%%%%%%%%%%%%%%%%%%%%%%%%%%%%%%%%%%%%%%%%%%%%%%%%%%%%%%%%
%%%%%%%%%%%%%%%%%%%%%%%%%%%%%%%%%%%%%%%%%%%%%%%%%%%%%%%%%%%%%%%%%%%%%%%%%%%%%%%%%%%%%%%%%%%%%%%%%%%%%%%
%%%%%%%%%%%%%%%%%%%%%%%%%%%%%%%%%%%%%%%%%%%%%%%%%%%%%%%%%%%%%%%%%%%%%%%%%%%%%%%%%%%%%%%%%%%%%%%%%%%%%%%
%            C H A P T E R 2 :  Dissipation time - definition, properties and general results         % 
%%%%%%%%%%%%%%%%%%%%%%%%%%%%%%%%%%%%%%%%%%%%%%%%%%%%%%%%%%%%%%%%%%%%%%%%%%%%%%%%%%%%%%%%%%%%%%%%%%%%%%%
%%%%%%%%%%%%%%%%%%%%%%%%%%%%%%%%%%%%%%%%%%%%%%%%%%%%%%%%%%%%%%%%%%%%%%%%%%%%%%%%%%%%%%%%%%%%%%%%%%%%%%%
%%%%%%%%%%%%%%%%%%%%%%%%%%%%%%%%%%%%%%%%%%%%%%%%%%%%%%%%%%%%%%%%%%%%%%%%%%%%%%%%%%%%%%%%%%%%%%%%%%%%%%%%%%

%Setup automatic chapter and section counters and markers for the rest of the document:

\renewcommand{\chaptermark}[1]{\markboth{ \rm \normalsize Chapter \thechapter . \hspace{0.25cm} #1 }{}}

\renewcommand{\sectionmark}[1]{\markright{\rm \normalsize Section \thesection . \hspace{0.25cm} #1}}

\newpage

%take care of chapter title page:

\markright{  \rm \normalsize  Chapter 2. \hspace{0.25cm} Dissipation time - definition, properties and general results}
\chapter{Dissipation time - definition, properties and general results}
\label{GBT}
\thispagestyle{fancy}

%Setup header and footer for the chapter

\lhead[\thepage]{\rightmark}
\rhead[\leftmark]{\thepage}
\cfoot{}

\parindent=0cm

%%%%%%%%%%%%%%%%%%%%%%%%%%%%%%%%%%%%%%%%%%%%%%%%%%%%%%%%%%%%%%%%%%%%%%%%%%%%%%%%%%%%%%%%%%%%%%%%%%%%%%%
%%%%%%%%%%%%%%%%%%%%%%%%%%%%%%%%%%%%%%%%%%%%%%%%%%%%%%%%%%%%%%%%%%%%%%%%%%%%%%%%%%%%%%%%%%%%%%%%%%%%%%%%%%

\section{Setup and notation}
\label{setup}

%%%%%%%%%%%%%%%%%%%%%%%%%%%%%%%%%%%%%%%%%%%%%%%%%%%%%%%%%%%%%%%%%%%%%%%%%%%%%%%%%%%%%%%%%%%%%%%%%%%%%%%
%%%%%%%%%%%%%%%%%%%%%%%%%%%%%%%%%%%%%%%%%%%%%%%%%%%%%%%%%%%%%%%%%%%%%%%%%%%%%%%%%%%%%%%%%%%%%%%%%%%%%%%%%%

%%%%%%%%%%%%%%%%%%%%%%%%%%%%%%%%%%%%%%%%%%%%%%%%%%%%%%%%%%%%%%%%%%%%%%%%%%%%%%%%%%%%%%%%%%%%%%%%%%%%%%%

\subsection{Evolution operators}
\label{evop}

%%%%%%%%%%%%%%%%%%%%%%%%%%%%%%%%%%%%%%%%%%%%%%%%%%%%%%%%%%%%%%%%%%%%%%%%%%%%%%%%%%%%%%%%%%%%%%%%%%%%%%%

Let $(\IT^{d},\EuScript{B}(\IT^{d}),m)$ denote the $d$-dimensional torus, 
equipped with 
its $\sigma-$field of Borel sets and the Lebesgue measure
$m$.  Let $F:\IT^{d}\to \IT^{d}$ be a map on
the torus preserving the Lebesgue measure: for any set $B\in
\EuScript{B}(\IT^{d})$ we have $m(F^{-1}(B))=m(B)$. In general, $F$ is 
not supposed to be invertible. In the following we call such a map
`volume preserving' with implicit reference to the Lebesgue measure.

The map $F$ generates a discrete time dynamics on $\IT^{d}$,
which in terms of  pathwise description can be represented
by the forward trajectory 
$\{F^{n}(\bx_{0})$, $n\in \IN\}$ of any initial point (particle) $\bx_{0}\in \IT^{d}$.
However, instead of looking at the evolution 
of a single particle, one can consider the statistical description of
the dynamics, that is the evolution of a density (more generally 
a measure) describing the initial statistical configuration of the system. 

Let $\ml{M}(\IT^{d})$ denote the set of all Borel measures on $\IT^{d}$. 
For any $\mu\in\ml{M}(\IT^{d})$ and $f\in C^{0}(\IT^{d})$ we write
\bean
\mu(f)=\int_{\IT^{d}}f(\bx)d\mu(\bx).
\eean
The map $F$ induces a map $F^{*}$ on $\ml{M}(\IT^{d})$ given by
\bean
 (F^{*}\mu)(f)=\mu(f \circ F), \qquad \text{for all} \; f\in C^{0}(\IT^{d}).
\eean 
This map can also be defined as follows: 
\bean
 (F^{*}\mu)(B)=\mu(F^{-1}(B)), \qquad \text{for all} \; B\in \EuScript{B}(\IT^{d}).
\eean
In particular if $\mu=\del_{\bx_0}$ then $F^{*}(\mu)=\del_{F(\bx_0)}$ 
and one recovers the pathwise description. 

If $\mu$ is absolutely continuous w.r.t. $m$, then $F^{*}(\mu)$ preserves
this property (since the measure-preserving map $F$ is nonsingular w.r.t. $m$,
see \cite[p.42]{LM}). 
The corresponding densities $g=\frac{d\mu}{dm}\in L^1(\IT^d)$ 
are transformed by the Frobenius-Perron
or transfer operator $P_{F}$ \cite{B}:
\bean
P_F\left(\frac{d\mu}{dm}\right)=\frac{d(F^{*}\mu)}{dm}.
\eean

If the map $F$ is invertible, $P_F$ is given explicitly by:
\bean
%\label{exF-P}
(P_{F}g)(\bx)= (g\circ F^{-1})(\bx)\frac{dF^{*}m}{dm}(\bx)
=g\circ F^{-1}(\bx).
\eean
If the map $F$ is differentiable, and the preimage
set of $\bx$ is finite for all $\bx$, the Perron-Frobenius operator is
given by
\bean
%\label{0exF-P}
(P_{F}g)(\bx)=\sum_{\by|F(\by)=\bx}\frac{g(\by)}{|J_{F}(\by)|},
\eean
where $J_{F}(\by)$ is the Jacobian of $F$ at $\by$. 

%Applying this formula to
%the constant function, the property $F^*m=m$ implies
%\bea\label{preserving}
%\forall \bx,\qquad 1=\sum_{\by|F(\by)=\bx}\frac{1}{|J_{F}(\by)|}.
%\eea

On the other hand one can consider the dual
of the Frobenius-Perron operator, called
the Koopman operator, which governs the evolution of
observables
$f\in L^{\infty}(\IT^{d})$  instead of that of densities
$g\in L^1$. The   Koopman
operator 
$U_{F}$ is defined as
\bea
\label{exK}
U_{F}f=f\circ F.
\eea

Due to the nonseparability of the Banach space $L^{\infty}(\IT^{d})$, it
is often more convenient to consider its closure in some weaker $L^p$ norm, 
which yields larger (but separable) spaces of observables  
$L^{p}(\IT^{d})$. Here we will be mainly concerned with the space $L^2(\IT^d)$ and
its codimension-$1$ subspace of zero-mean functions 
\bea
\label{0-mean}
L^2_0(\IT^d)=\{f\in L^2(\IT^2) : m(f)=0\}.
\eea
This subspace is obviously invariant under $U_F$ and $P_F$, due to the assumption
$F^*m=m$.
Throughout the whole work, $\|\cdot\|$ will always refer to the $L^2$-norm (and corresponding
operator norm) on $L^2_0(\IT^d)$ (any other norm will carry an explicit subscript).

For any measure-preserving map $F$, the operator $U_F$ is an isometry on $L^2(\IT^d)$
and $L^2_0(\IT^d)$. When $F$ is invertible, $U_F$ is unitary on these spaces, 
and satisfies $U_F=P_F^{-1}=P_{F^{-1}}$.

Although just introduced operators will mostly be considered on $L^2_{0}(\IT^d)$, 
we will need from time to time to act on some more specific spaces defined usually by certain
regularity properties of functions belonging to them. Most typically these will be H\"older and 
Sobolev spaces, the definitions of which we now briefly recall and use this opportunity to fix the 
appropriate notation.

For any $m\in\IN$, we denote by $C^m(\IT^d)$ the space of $m$-times continuously
differentiable functions, with the norm
\bean
\|f\|_{C^{m}}=\sum_{|\bmal|_{1}\leq m}\|D^{\bmal}f\|_{\infty}
\eean
(we use the norm $|\bmal|_1=\alpha_1+\ldots+\alpha_d$ for the multiindex $\bmal\in\IN^d$).
For any $s=m+\eta$ with $m=[s]\in\IN$, $\eta\in (0,1)$, let $C^{s}(\IT^d)$ denote the space of
$C^m$ functions for which the $m$-derivatives are $\eta$-H\"{o}lder continuous; this
space is equipped with the norm
\bean
\|f\|_{C^{s}}=\|f\|_{C^{m}}+
\sum_{|\bmal|_{1}= m}\sup_{\bx\not=\by}\frac{|D^{\bmal} f(\bx)-D^{\bmal} f(\by)|}{|\bx-\by|^\eta}.
\eean 

The Fourier transforms of functions $g\in L^1(\IR^d)$ and $f\in L^1(\IT^d)$ 
are defined as follows:
\bea
\label{FT}
\forall \bxi\in \IR^d,\quad
\hat{g}(\bxi)&=& \int_{\IR^{d}}g(\bx)e^{-2\pi i \bx \cdot \bxi} d\bx, \\
\forall \bk\in \IZ^d,\quad
\hat{f}(\bk)&=& \int_{\IT^{d}}f(\bx)e^{-2\pi i \bx \cdot \bk} d\bx=\la\be_{\bk},f \ra.
\eea
Above we used the Fourier modes on the torus 
$\be_{\bk}(\bx):=e^{2 \pi i \bx\cdot\bk}$.
For any $s\geq 0$, we denote by $H^s(\IT^d)$ and $H^s(\IR^d)$ the Sobolev spaces of $s$-times
weakly differentiable $L^2$-functions equipped with the norms $\|\cdot\|_{H^s}$ defined
respectively by 
\bean
\|g\|^2_{H^s(\IR^d)}&=&\int_{\bxi\in \IR^d}(1+|\bxi|^2)^s |\hat{g}(\bxi)|^2d\bxi,\\
\|f\|^2_{H^s(\IT^d)}&=&\sum_{\bk\in \IZ^d}(1+|\bk|^2)^s |\hat{f}(\bk)|^2.
\eean
Finally, for any of these spaces, adding the subscript $0$ will mean that we consider the 
($U_F$-invariant) subspace
of functions with zero average, e.g. $C^j_0(\IT^d)=\{f\in C^j(\IT^d),\ m(f)=0\}$.

%%%%%%%%%%%%%%%%%%%%%%%%%%%%%%%%%%%%%%%%%%%%%%%%%%%%%%%%%%%%%%%%%%%%%%%

\subsection{Noise operator}
\label{Cnoise}

%%%%%%%%%%%%%%%%%%%%%%%%%%%%%%%%%%%%%%%%%%%%%%%%%%%%%%%%%%%%%%%%%%%%%%%%%%%%%%%%%%%%%%%%%%%%%%%%%%%%%%%%%%

To construct the noise operator we first define the {\em noise generating density}
i.e. an arbitrary probability density function $g\in L^1(\IR^d)$ symmetric w.r.t. the origin
$g(\bx)=g(-\bx)$.
The \emph{noise width}
(or \emph{noise level}) will be given by a single nonnegative parameter, which we call 
$\ep$. To each $\ep > 0$ there
corresponds the noise kernel on $\IR^d$:
\bean
g_{\ep}(\bx)=\frac{1}{\ep^{d}}g\left(\frac{\bx}{\ep}\right),
\eean
with the convention that $g_0=\del_0$. 
The noise kernel on the torus is obtained by periodizing $g_{\ep}$, yielding the
periodic kernel
\bea
\label{ank}
\tilde{g}_{\ep}(\bx)=\sum_{\bn\in \IZ^d} g_{\ep}(\bx+\bn).
\eea
We remark that the Fourier transform of $\tilde{g}_{\ep}$ is related to that of $g$ by
the identities $\hat{\tilde{g}}_{\ep}(\bk)=\hat{g}_{\ep}(\bk)=\hat{g}(\ep\bk)$.

The action of the noise operator $G_{\ep}$ on any function $f\in L^2_0(\IT^d)$ is defined
by the convolution:
\bean
G_{\ep}f &=& \tilde{g}_{\ep}*f.
\eean
As a convolution operator with kernel from $L^1$, $G_{\ep}$ is compact on $L^2_0(\IT^d)$ 
(if $g$ is square-integrable, $G_\ep$ is Hilbert-Schmidt). 
The Fourier modes $\{\be_{\bk}\}_{0\not= \bk\in\IZ^d}$ form an orthonormal basis of
eigenvectors of $G_\ep$, yielding the following 
spectral decomposition:
\bea
\label{specdecompo}
\forall f\in L_0^2(\IT^d),\quad G_{\ep}f= \sum_{0\not=\bk\in \IZ^d} \hat{g}(\ep\bk) 
\la\be_{\bk},f \ra \,\be_{\bk}.
\eea
This formula shows that the eigenvalue associated with $\be_{\bk}$ is $\hat{g}(\ep\bk)$.
Since $g$ is a symmetric function, this eigenvalue is real, so that $G_\ep$ 
is a self-adjoint operator. Its spectral radius $r_{sp}(G_{\ep})$ is therefore 
given by
\bea
\label{Gspr}
r_{sp}(G_{\ep})=\|G_{\ep}\|=\sup_{0\not=\bk\in \IZ^d}|\hat{g}(\ep\bk)|.
\eea
Since the density $g$ is positive, $\hat g$ attains its maximum 
$\hat g(0)=1$ at the
origin and nowhere else. Besides, because $g\in L^1(\IR^d)$, 
$\hat g$ is a continuous function vanishing at infinity.  As a result, for small enough $\eps>0$, 
the supremum on the RHS of (\ref{Gspr}) is reached at some point $\ep\bk$ close to the origin, and 
this maximum is strictly smaller than $1$.
This shows that the operator $G_{\ep}$ is strictly contracting on $L_{0}^{2}(\IT^{d})$:
\bea
\label{strcon}
\forall \ep>0,\quad \|G_{\ep}\|=r_{sp}(G_{\ep})<1.
\eea
In the next section we study this noise operator more precisely, starting from appropriate
assumptions on the noise generating density.

%%%%%%%%%%%%%%%%%%%%%%%%%%%%%%%%%%%%%%%%%%%%%%%%%%%%%%%%%%%%%%%%%%%%%%%

\subsection{Noise kernel estimates}

%%%%%%%%%%%%%%%%%%%%%%%%%%%%%%%%%%%%%%%%%%%%%%%%%%%%%%%%%%%%%%%%%%%%%%%%%%%%%%%%%%%%%%%%%%%%%%%%%%%%%%%%%%

In this subsection we present some estimates regarding the noise operator. These estimates
will later play a crucial role in the derivation of the asymptotics of the dissipation time.

We will be interested in the behavior of the system in the limit 
of small noise, that is the limit $\ep\to 0$.
It will hence be useful to introduce the following asymptotic notation.
Given two variables $a_{\ep}$, $b_{\ep}$ depending on $\ep>0$,
we write 
\bea
a_{\ep} &\lesssim& b_{\ep} \text{ if } \limsup_{\ep
\rightarrow
0}\frac{a_{\ep}}{b_{\ep}} < \infty, \\
a_{\ep} &\approx& b_{\ep} \text{ if } \lim_{\ep \rightarrow 0}\frac{a_{\ep}}{ b_{\ep}}=1,\\
a_{\ep} &\sim& b_{\ep} \text{ if }a_{\ep} \lesssim b_{\ep}\text{ and }b_{\ep} \lesssim a_{\ep}.
 \label{asymnot}
\eea

In order to obtain interesting estimates on the noise operator $G_\ep$, it will be necessary to 
impose some additional conditions on its generating density $g$, regarding e.g. its rate of
decay at infinity, or the behavior of its Fourier transform near the
origin.

The weakest condition we are going to impose is the existence of some
positive moment of $g$, by which we mean that for some $\alpha \in (0,2]$,
\bea
\label{Moment}
M_{\alpha}=\int_{\IR^d}|\bx|^{\alpha}g(\bx)d\bx < \infty
\eea
(we take the length $|\bx|=(x_1^2+\ldots+x_d^2)^{1/2}$ on $\IR^d$). 
This condition implies the following properties of the Fourier transform $\hat g$ (proved in Section \ref{Ch2proofs})

\begin{lem}
\label{Mom-Four}
For any $\alpha\in (0,2]$ there exists a universal constant $C_\alpha$ such that, 
if a normalized density $g$ satisfies (\ref{Moment}), then the following inequalities hold:
\bea\label{Moment-Fourier}
\forall \bxi\in\IR^d,\quad 0\leq 1-\hat g(\bxi)\leq C_\alpha M_\alpha |\bxi|^{\alpha}.
\eea
If (\ref{Moment}) holds with $\alpha=2$, we have the more precise information:
$$
1-\hat g(\bxi)\sim |\bxi|^2\quad\mbox{in the limit}\ \bxi\to 0.
$$
\end{lem}

In the case $\alpha<2$, we will sometimes assume a stronger property than \eqref{Moment-Fourier}, namely that
\bea
\label{noise1}
 1-\hat{g}(\bxi) \sim |\bxi|^{\alpha}\quad \mbox{in the limit }\bxi\to 0.
\eea

Note that this behavior 
implies a uniform bound $1-\hat g(\bxi)\leq C|\bxi|^\gamma$ for any $\gamma\leq\alpha$ and $C$ independent of
$\gamma$.

Typical examples of noise kernels satisfying (\ref{noise1})
include the Gaussian kernel and more general symmetric $\alpha-$stable kernels
\cite[p.152]{Stroock} defined for $\alpha\in(0,2]$:
\bea
\label{Akernel}
g_{\ep,\alpha}(\bx):=\sum_{\bk \in \IZ^{d}}e^{-(\bQ(\ep \bk))^{\alpha/2}}\be_{\bk}(\bx),
\eea
where $\bQ$ denotes an arbitrary positive definite quadratic form. For the values
of $\alpha$ indicated, the function $g_{\ep,\alpha}(x)$ is positive on $\IR^d$.

\medskip

In view of Eq.\eqref{Gspr}, the properties \eqref{Moment} or \eqref{noise1}
determine the rate at which  $G_{\ep}$ 
contracts on $L_{0}^{2}(\IT^{d})$. For instance, \eqref{noise1} implies that
in the limit $\ep\to 0$,
\bea
\label{contr}
1-\|G_{\ep}\| \sim \ep^{\alpha}.
\eea

The following proposition describes the effect of the noise on various types of
observables, in the limit of small noise level.
The proof is given in Section \ref{Ch2proofs}. 

\begin{prop}
\label{nke}
i) For any noise generating density $g\in L^1(\IR^d)$ and any observable $f\in L^2_0(\IT^d)$, one has
\begin{equation}\label{gen}
\|G_{\ep}f-f\| \epto 0.
\end{equation}
To obtain information on the speed of convergence, we need to impose constraints on
both the noise kernel and the observable.

ii) If for some $\alpha\in(0,2]$ the kernel $g$ satisfies (\ref{Moment}) or (\ref{noise1}),
then for any $\gamma>0$ there exists a constant $C>0$ such that for any observable
$f\in H^{\gamma}(\IT^d)$,
\bea
\label{H1est}
\|G_{\ep}f-f\| \leq  C \ep^{\gamma \wedge \alpha} \|f\|_{H^{\gamma \wedge \alpha}},
\eea
where $ \gamma\wedge \alpha:=\min\{\gamma,\alpha\}$.
If $f\in C^{1}(\IT^d)$, the above upper bound can be replaced by
\bea
\label{C1est}
\|G_{\ep}f-f\| \leq C \ep^{1\wedge \alpha} \|\nabla f\| 
\leq C \ep^{1\wedge\alpha }\|\nabla f\|_{\infty}.
\eea
\end{prop}

Using the noise operator, we are now in position to define the noisy (resp. the
coarse-grained) dynamics generated by a measure-preserving map $F$.

%%%%%%%%%%%%%%%%%%%%%%%%%%%%%%%%%%%%%%%%%%%%%%%%%%%%%%%%%%%%%%%%%%%%%%%

\subsection{Noisy evolution operators}
\label{Noevop}

%%%%%%%%%%%%%%%%%%%%%%%%%%%%%%%%%%%%%%%%%%%%%%%%%%%%%%%%%%%%%%%%%%%%%%%%%%%%%%%%%%%%%%%%%%%%%%%%%%%%%%%%%%

The noisy evolution through the map $F$ is constructed by successive application of
the Koopman operator $U_{F}$ and the noise operator $G_{\ep}$. The noisy dynamics
is then generated by taking powers of the \emph{noisy propagator}
\bean
T_{\ep}=G_{\ep}U_{F}.
\eean
In general, the operator $T_{\ep}$ is not normal, but satisfies 
$r_{sp}(T_\ep)\leq \|T_\ep\|=\|G_\ep\|$. 
We will also consider a \emph{coarse-grained dynamics} 
defined by the application of the noise kernel only at
the beginning and at the end of the evolution. Hence we define
the following family of operators:
\bean
\tilde{T}^{(n)}_{\ep}=G_{\ep}U^{n}_{F}G_{\ep},\quad n\in\IN.
\eean

In view of the contracting properties of $G_{\ep}$,
the inequalities $\|T_\ep^{n}\|\leq \|G_\ep\|^n$, $\|\tilde{T}_\ep^{(n)}\|\leq \|G_\ep\|^2$ 
imply that 
both noisy and coarse-grained operators are strictly contracting on $L_{0}^2(\IT^d)$.

%%%%%%%%%%%%%%%%%%%%%%%%%%%%%%%%%%%%%%%%%%%%%%%%%%%%%%%%%%%%%%%%%%%%%%%%%%%%%%%%%%%%%%%%%%%%%%%%%%
%%%%%%%%%%%%%%%%%%%%%%%%%%%%%%%%%%%%%%%%%%%%%%%%%%%%%%%%%%%%%%%%%%%%%%%%%%%%%%%%%%%%%%%%%%%%%%%%%%

\section{General definition of dissipation time}
\label{DTgen}

%%%%%%%%%%%%%%%%%%%%%%%%%%%%%%%%%%%%%%%%%%%%%%%%%%%%%%%%%%%%%%%%%%%%%%%%%%%%%%%%%%%%%%%%%%%%%%%%%%
%%%%%%%%%%%%%%%%%%%%%%%%%%%%%%%%%%%%%%%%%%%%%%%%%%%%%%%%%%%%%%%%%%%%%%%%%%%%%%%%%%%%%%%%%%%%%%%%%%

Once the notation has been set up, we can pass to the precise definition of the dissipation time.
We prefer to start, however, with some remarks regarding the motivation. In particular we briefly 
recall the original, continuous-time physical setting considered in \cite{F0}, 
where the notion had been introduced for the first time. 
We then generalize the original definition to an abstract, discrete-time setting 
and after discussing some of the most basic properties
of this notion we conclude the section with another physical interpretation,
this time coming from statistical physics and expressed in terms of Boltzmann-Gibbs Entropy.

%%%%%%%%%%%%%%%%%%%%%%%%%%%%%%%%%%%%%%%%%%%%%%%%%%%%%%%%%%%%%%%%%%%%%%%%%%%%%%%%%%%%%%%%%%%%%%%%%%

\subsection{Motivation}
\label{DTmot}

%%%%%%%%%%%%%%%%%%%%%%%%%%%%%%%%%%%%%%%%%%%%%%%%%%%%%%%%%%%%%%%%%%%%%%%%%%%%%%%%%%%%%%%%%%%%%%%%%%
In order to gain a bit of physical intuition it is useful to consider the original problem
described in \cite{F0}, where the notion of dissipation time was introduced in the context 
of continuous-time dynamical system of a passive tracer in a fluid flow.
The flow is prescribed in terms of given {\em periodic}, {\em incompressible} velocity filed
and the path of the tracer is randomly perturbed by  
collisions with fluid particles. The latter phenomena being modeled by standard Brownian motion.
In pathwise description the evolution of the tracer is given by the following Langevin-type 
equation
\bean
 d\mathbf{x}^{\vep}(t) =\mathbf{u}(\mathbf{x}^{\vep}(t))dt + \sqrt{\vep}d\mathbf{w}(t), \quad
 \nabla \cdot \mathbf{u}(\mathbf{x})=0,
\eean
where $\mathbf{w}$ stands for the standard Brownian motion.
In Statistical description, the dynamics on the densities is
given by the corresponding Fokker-Planck (advection-diffusion) equation
\bea
\label{FP}
\frac{\pa \rho}{\pa t}=\mathbf{u}\cdot \nabla\rho + \frac{\vep}{2}\Delta\rho, \quad
 \nabla \cdot \mathbf{u}=0,
\eea
Let us note that unperturbed skew-symmetric operator $\mathbf{u}\cdot \nabla$
generates an unitary (conservative) group $U^{t}=e^{t\mathbf{u}\cdot \nabla}$, which
corresponds (for $t=1$) to our conservative Koopman operator $U_{F}$ for some $F$.

Now, the perturbed generator $\cL_{\ep}=\mathbf{u}\cdot \nabla + \frac{\vep}{2}\Delta$ gives
rise to a semi-group of contractions $P_{\ep}^t=e^{t\cL_{\ep}}$, which heuristically can be
thought of as continuous time-1 counterpart of the generator $T_{\ep}=G_{\ep}U_{F}$ of our 
discrete-time noisy dynamics in case of Gaussian noise (we say 'heuristically', 
because due to noncommutativity of both terms in $\cL_{\ep}$ there is no obvious way of 
associating $T_{\ep}$ with $P_{\ep}$ for general velocity fields $\mathbf{u}$). 

In order to predict long time behavior of the tracer and in particular the
influence of the noise on its trajectory, one is interested in determining
the speed of contraction of the semigroup $P_{\ep}$.
In \cite{F0} (see also more recent version in \cite{F}) the following time scale $t_{diss}$ 
defined by equation $\|P^{t_{diss}}_{\ep}\|=1/2$ was suggested for consideration regarding 
this problem and termed as {\em dissipation time}. Determining the asymptotics of $t_{diss}$ 
(in continuous-time setting) proved however to be exceedingly difficult and except
for some very special and simple cases (e.g. cellular flow considered in \cite{F}) the problem has not 
been solved up to date. Also there is no known example of non-trivially short dissipation time (i.e. 
logarithmic in $\ep^{-1}$) in continuous setting.    
The main difficulty, as was already observed in \cite{F0}, lies in the fact that to get this result one 
would need to consider fully chaotic system, while it is very difficult to construct one i.e. to
write down a simple differential equation which would exhibit fully chaotic behavior.
On the other hand there is no problem in constructing fully chaotic discrete-time systems and
in fact the ergodic theory literature abounds in such examples.
This situation provided natural motivation for the generalization of the notion of the
dissipation time to discrete-time systems.

%%%%%%%%%%%%%%%%%%%%%%%%%%%%%%%%%%%%%%%%%%%%%%%%%%%%%%%%%%%%%%%%%%%%%%%%%%%%%%%%%%%%%%%%%%%%%%%%%%

\subsection{The definition}
\label{DTdef}

%%%%%%%%%%%%%%%%%%%%%%%%%%%%%%%%%%%%%%%%%%%%%%%%%%%%%%%%%%%%%%%%%%%%%%%%%%%%%%%%%%%%%%%%%%%%%%%%%%

In its general form the classical dissipation time $\tau_{c}(p)$ for discrete-time noisy dynamics $T_{\ep}$
is defined in terms of the norm $\|\cdot\|_{p,0}$ on the space $L^p_0(\IT^d)$ and w.r.t. a threshold $\eta\in(0,1)$.
\begin{defin}
Let $T_{\ep}$ denote discrete-time noisy dynamics. We define
\bea
\label{tdiss2}
\tau_{c}(p,\eta):=\min\{n \in \IZ_{+}:\|T_{\ep}^{n}\|_{p,0}< \eta\}, \quad 1\leq p\leq\infty.
\eea
\end{defin}

\begin{figure}
\centering
\includegraphics[width=15cm,height=15cm]{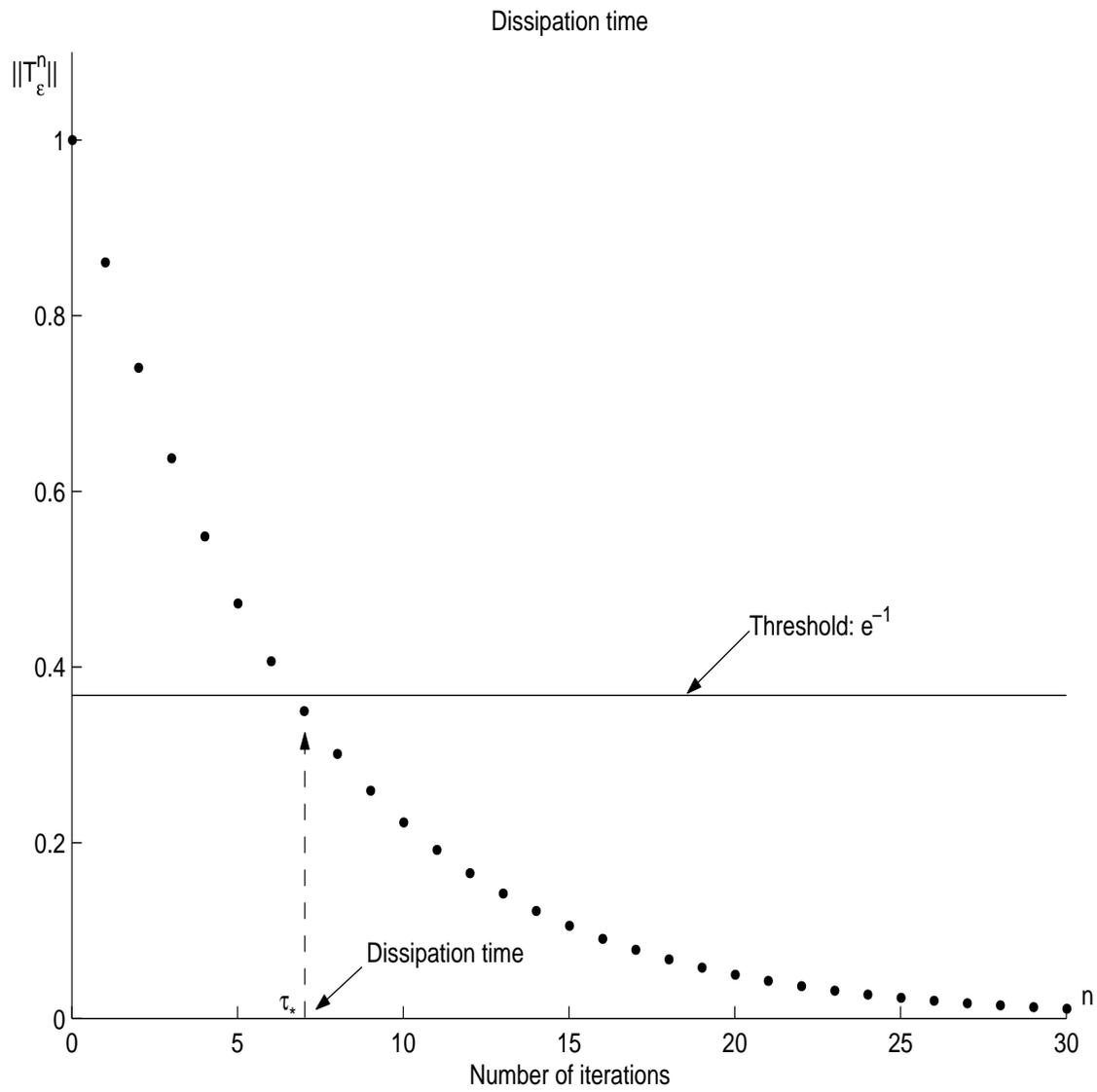}
\caption{Definition of the dissipation time}
\label{Fig2}
\end{figure}

The fact that $T_{\ep}$ acts on every $L^p_0(\IT^d)$ as a strict contraction ensures existence and
uniqueness of $\tau_{c}(p,\eta)$, for each $\eta\in(0,1)$.

Figure \ref{Fig2} illustrates the definition.

We need to show that the value of the threshold $\eta$ in (\ref{tdiss2}) does not affect 
the order of divergence of $\tau_{c}(p,\eta)$, as $\ep$ tends to zero.
\begin{prop}
\label{prop1}
For any $0<\tilde{\eta},\eta<1$,  $\tau_{c}(p,\tilde{\eta}) \sim \tau_{c}(p,\eta)$.
\end{prop}
\textbf{Proof.}
Assume $0<\tilde{\eta}<\eta<1$. 
Obviously $\tau_{c}(p,\tilde{\eta}) \geq \tau_{c}(p,\eta)$. On the other hand let $k$ be a 
positive integer such that $\eta^{k}<\tilde{\eta}$. Then
\bean
 \|T_{\ep}^{\tau_{c}(p,\eta)}\|_{p,0}<\eta \Rightarrow \|T_{\ep}^{k\tau_{c}(p)}\|_{p,0}
<\eta^{k}<\tilde{\eta}.
\eean
Hence $k\tau_{c}(p,\eta) \geq \tau_{c}(p,\tilde{\eta})$, which implies 
$\tau_{c}(p,\eta) \sim \tau_{c}(p,\tilde{\eta})$.$\qquad \blacksquare$

Following the argument of \cite{Ros} one can use the Riesz convexity theorem to establish
also the asymptotic equivalence of the $\tau_{c}(p)$, for all $1<p<\infty$ (to alleviate
the notation we drop $\eta$).
\begin{prop}
\label{pdiss}

i) For any $1<q,p<\infty$,  $\tau_{c}(q) \sim \tau_{c}(p)$.

ii) For any $1<p<\infty$, $\tau_{c}(p) \lesssim \tau_{c}(1)$ and $\tau_{c}(p) \lesssim \tau_{c}(\infty)$.
\end{prop}

We postpone a technical proof of this proposition to Section \ref{Ch2proofs}.

In view of the above results, instead of working in general setting, one can choose some
convenient values of $p$ and $\eta$ and perform, without any loss of generality, all necessary asymptotic calculations in
one notationally simplified setting. We will usually choose $p=2$ and $\eta=e^{-1}$ for computational convenience.
Following this choice we introduce the convention that $\tau_{c}(p):=\tau_{c}(p,e^{-1})$ and $\tau_{c}:=\tau_{c}(2,e^{-1})$.
The obvious dependence of the dissipation time on $\ep$ will always be implicitly assumed but rarely explicitly denoted.

The corresponding dissipation time for a coarse-grained dynamics is defined in fully analogous way and
denoted respectively by $\tilde{n}_{c}(p,\eta)$, $\tilde{n}_{c}(p)$ and $\tilde{n}_{c}$. In particular 
\bea
\label{ttdiss1}
\tilde{\tau}_c:=\min\{n \in \IN:\|\tilde{T}_{\ep}^{(n)}\| < e^{-1}\}.
\eea

We note that the dissipation time 
does not depend on whether the dynamics is applied to
densities (i.e. by the Frobenius-Perron operator) or to
observables (by the Koopman operator). 
Indeed, the norm of
an operator equals the norm of its adjoint \cite[p.195]{Y}, so that
\bean
\|\tilde{T}_{\ep}^{(n)}\|=
\|G_{\ep}U^{n}_{F}G_{\ep}\|=
\|(G_{\ep}U^{n}_{F}G_{\ep})^{*}\|
=\|G_{\ep}P^{n}_{F}G_{\ep}\|,
\eean
and similarly for the noisy operator $T_\ep$.
In particular, for invertible maps the dissipation time does not depend
on the direction of time.

\medskip

As mentioned in the Introduction, we will distinguish two qualitatively different asymptotic behaviors of dissipation time
in the limit $\ep\to 0$.
We say that the operator $T_{\ep}$ (or the map $F$ associated with it) respectively has
\begin{itemize}
\item[I)] {\em simple} or {\em power-law} dissipation time if there exists $\beta>0$ such that
\bean
{\tau}_{c} \sim 1/\ep^{\beta},
\eean
\item[II)] {\em fast} or {\em logarithmic} dissipation time if 
\bean
{\tau}_{c} \sim \ln(1/\ep).
\eean
\end{itemize}
We will also speak about {\em slow} dissipation time whenever there exists 
some $\beta>0$ s.t.
\bean
{\tau}_{c} \gtrsim 1/\ep^{\beta}.
\eean
In case of logarithmic dissipation time, the dissipation rate constant $R_{c}$, when it exists, is 
defined as
\bea
\label{Rdiss}
  R_{c}=\lim_{\ep \rightarrow 0} \frac{{\tau}_{c}}{\ln(1/\ep)}.
\eea    
A similar terminology will be applied to the coarse-grained dissipation time $\tilde{\tau}_{c}$.

%%%%%%%%%%%%%%%%%%%%%%%%%%%%%%%%%%%%%%%%%%%%%%%%%%%%%%%%%%%%%%%%%%%%%%%%%%%%%%%%%%%%%%%%%%%%%%%%%%

\subsection{Physical interpretation via Boltzmann-Gibbs entropy}
\label{BGEnt}

%%%%%%%%%%%%%%%%%%%%%%%%%%%%%%%%%%%%%%%%%%%%%%%%%%%%%%%%%%%%%%%%%%%%%%%%%%%%%%%%%%%%%%%%%%%%%%%%%%

In this section we briefly discuss the connection between
dissipation time and Boltz-mann-Gibbs entropy. The results formalize
an intuitive physical interpretation of the dissipation time
outlined in the Introduction. 

First we note that on the scales exceeding $\tau_{c}$, the Boltzmann-Gibbs entropy
approaches its maximal equilibrium value (i.e. $0$) as can
be seen from the following simple estimate (cf. \cite{LM}).
Let us first restrict considerations to bounded initial states, i.e., 
$f\geq 0, f\in L^\infty$ and $\|f\|_{1}=1$. Let 
\[
\eta(u)=\begin{cases}
-u\ln{u},& u>0\\
0,& u=0
\end{cases}
\]
and let $D_n=\{\bx\in \IT^d: 1\leq T^n_{\ep}f\}$. On one hand, we have
\bea
&&\left|\int_{D_n}\eta(T^n_{\ep}f(\bx))d\bx\right| \label{fi}\nonumber \\
&\leq&\int_{D_n}\left|\int_{1}^{T^n_{\ep}f(\bx)}\frac{d\eta(u)}{du}du\right|
d\bx \notag \\
&\leq&\sup_{1\leq u\leq \|T_{\ep}^n f\|_\infty}(1+\ln{u})\int_{D_n}
|T_{\ep}^n f(\bx)-1|d\bx \notag \\
&\leq& (1+\ln{\|T_{\ep}^n f\|_\infty}) \|T_{\ep}^nf-1\|_1 \notag \\
&\leq&(1+\ln{\| f\|_\infty})\|T_{\ep}^nf-1\|_1 \label{li}.
\eea
On the other hand, we have
\[
0 \geq \int_{\IT^d}\eta(T_{\ep}^n f(\bx))d\bx 
\geq \int_{D_n}\eta(T_{\ep}^n f(\bx))d\bx.
\]
In view of the inclusion relation: $L^\infty(\IT^d)\subset L^2(\IT^d)
\subset L^1(\IT^d)$, we then obtain that for $n\gg \tau_{c}$ 
\[
\sup_{f\geq 0,\|f\|_\infty\leq c}\left|\int_{\IT^d}\eta(T_{\ep}^n f(\bx))d\bx
\right|
\stackrel{\ep \downarrow 0}{\longrightarrow} 0,
\quad \forall c>0.
\]

For unbounded initial states, we note that, by Young's inequality,
\[
 \|T_{\ep}^n f\|_\infty \leq  \|T_{\ep} f\|_\infty \leq 
 \|g_{\ep}\|_\infty \|f\|_{1}=\|g_{\ep}\|_\infty
\] 
from which we have, instead of (\ref{li}), the following estimate 
\[
\left|\int_{D_n}\eta(T^n_{\ep}f(\bx))d\bx\right| \leq
(1+\ln{\|g_{\ep}\|_\infty})\|T_{\ep}^nf-1\|_1.
\] 
where in view of (\ref{contr}) 
\[
 \ln\|g_{\ep}\|_\infty
 \sim \ln(1/\ep).
\]
Therefore for sufficiently fast diverging $n\gg \tau_{c}(1)$ such 
that
\bea
\label{entropy2}
 \ln(1/\ep)\|T_{\ep}^n(f-1)\|_{1,0}
 \stackrel{\ep \downarrow 0}{\longrightarrow} 0
\eea 
one obtains
\[
\sup_{f\geq 0,\|f\|_1=1}\left|\int_{\IT^d}\eta(T_{\ep}^n f(\bx))d\bx
\right|
\stackrel{\ep \downarrow 0}{\longrightarrow} 0.
\]
The condition (\ref{entropy2}) typically results in
a slightly longer time scale than $\tau_{c}(1)$.

On the other hand, we  can bound the $L_1$ distance between
the probability density function $f$ and the Lebesgue measure
by their relative entropy via Csisz\'{a}r's inequality \cite{Cs}
\[
\int_{\IT^d} 
|f(\bx)-g(\bx)|d\bx \leq \sqrt{2\int_{\IT^d} f(\bx)\ln{(f(\bx)/g(\bx))}d\bx}
\]
with $g(\bx)=1$. We see immediately that the decay rate
of 
\[
\sup_{f\geq 0,\|f\|_1=1}\left|\int_{\IT^d}\eta(T_{\ep}^n f(\bx))d\bx
\right|
\]
provides an estimate for $\tau_{c}(1)$ and, consequently, for
$\tau_{c}(p), p\in (1,\infty)$.

%%%%%%%%%%%%%%%%%%%%%%%%%%%%%%%%%%%%%%%%%%%%%%%%%%%%%%%%%%%%%%%%%%%%%%%%%%%%%%%%%%%%%%%%%%%%%%%%%
%%%%%%%%%%%%%%%%%%%%%%%%%%%%%%%%%%%%%%%%%%%%%%%%%%%%%%%%%%%%%%%%%%%%%%%%%%%%%%%%%%%%%%%%%%%%%%%%%%

\section{Dissipation time and spectral analysis}
\label{SP}

%%%%%%%%%%%%%%%%%%%%%%%%%%%%%%%%%%%%%%%%%%%%%%%%%%%%%%%%%%%%%%%%%%%%%%%%%%%%%%%%%%%%%%%%%%%%%%%%%%
%%%%%%%%%%%%%%%%%%%%%%%%%%%%%%%%%%%%%%%%%%%%%%%%%%%%%%%%%%%%%%%%%%%%%%%%%%%%%%%%%%%%%%%%%%%%%%%%%%

In this section we investigate the connection between the dissipation time of the noisy propagator $T_{\ep}$ and 
its pseudospectrum together with some spectral properties of $U_{F}$ and $G_{\ep}$. All the operators considered
in this section are defined on $L^{2}_{0}(\IT^d)$. In the framework of continuous-time 
dynamics, some connections have recently been obtained between, on one side, the
pseudospectrum of the (non-selfadjoint) generator $A$, and on the other side, the norm of the evolution operator 
$e^{t A}$ \cite{Dav}. We consider complementary, discrete-time setting, which allows for generalizations and more 
transparent proofs.
We start with the definition of the pseudospectrum, and then derive general abstract lower and upper bounds
for the dissipation time. In the following sections we will apply these results to determine the asymptotics
of the dissipation time under some dynamical assumptions regarding the underlying conservative maps (e.g. 
lack of weak-mixing).

%%%%%%%%%%%%%%%%%%%%%%%%%%%%%%%%%%%%%%%%%%%%%%%%%%%%%%%%%%%%%%%%%%%%%%%%%%%%%%%%%%%%%%%%%%%%%%%

\subsection{Pseudospectrum}

%%%%%%%%%%%%%%%%%%%%%%%%%%%%%%%%%%%%%%%%%%%%%%%%%%%%%%%%%%%%%%%%%%%%%%%%%%%%%%%%%%%%%%%%%%%%%%%%%%

In this short subsection we define the pseudospectrum of a bounded operator \cite{Var} and state
some of its properties. 

\begin{defin}
\label{psp}
Let $T$ be a bounded linear operator on a Hilbert space $\mathcal{H}$ (we note $T\in \mathcal{L}(\mathcal{H})$).
For any $\del>0$, the $\del$-pseudospectrum 
of $T$ (denoted by $\sigma_{\del}(T)$) can be defined in the following three equivalent ways:
\bean
&(I)& \sigma_{\del}(T)=\{\lam \in \IC: \|(\lam-T)^{-1}\|\geq \del^{-1}\}, \\
&(II)& \sigma_{\del}(T)=\{\lam \in \IC: \exists v\in\mathcal{H}, \ \|v\|=1, \quad \|(T-\lam)v\|\leq \del \}, \\
&(III)& \sigma_{\del}(T)=\{\lam \in \IC: \exists B\in \mathcal{L}(\mathcal{H}),\ 
\|B\|\leq \del, \quad \lam \in \sigma(T+B) \}.
\eean
\end{defin}
We will apply these definitions to the operator $T_\ep$. 
For brevity, the resolvent of this operator will be denoted by 
$R_\ep(\lam)=(\lam - T_\ep)^{-1}$.
We call $S^{r}$ the circle $\{\lam \in \IC: |\lam|=r\}$ in the complex plane, and define the following 
{\em pseudospectrum distance function}:
\bean
d_\ep(r):=\inf \{ \del>0 :  \sigma_{\del}(T_\ep)\cap S^r \not = \emptyset\}.
\eean
From the definition (I) of the pseudospectrum, one easily shows that
this distance is also given by
\bea
\label{psdist}
d^{-1}_\ep(r)=\sup_{|\lam|=r} \|R_\ep(\lam)\|.
\eea

We have the following property (proved in Section \ref{Ch2proofs}):

\begin{prop}
\label{disto0}
For any isometry $U$ and noise generating function $g$, one has
\bea
\label{depto0}
d_\ep(1)\epto 0.
\eea
This means that for any fixed $\delta>0$, the pseudospectrum $\sigma_\delta(T_\ep)$ will intersect
the unit circle for small enough $\ep$.
\end{prop}

\subsection{General bounds for the dissipation time}

In this section we consider both fully noisy and coarse grained dynamics. We start with 'non-finiteness' results.

\begin{prop}\label{nonfinite}
For any measure-preserving map $F$ and any noise generating function $g$,
both fully noisy and coarse-grained dissipation times diverge in the small-noise limit $\ep \rightarrow 0$.
\end{prop}

\textbf{Proof.} We skip the subscript $F$ to alleviate the notation. 
We only use the fact that $U=U_F$ is an isometry. 
We start with the full noisy case and prove by induction the following strong convergence of operators
$$
\forall f\in L^2_0(\IT^d),\quad \forall n\in\IN,\qquad \|T_\ep^n f-U^n f\|\epto 0.
$$
From Proposition~\ref{nke}{\em i)}, this limit holds in the case $n=1$. 
Let us assume it holds at the rank $n-1$.
Then we write
$$
T_\ep^n f=UT_\ep^{n-1} f + (G_\ep -I)UT_\ep^{n-1} f.
$$
From the inductive hypothesis, $T_\ep^{n-1}f\epto U^{n-1}f$, so that the first term on the
RHS converges to $U^nf$. Applying Proposition~\ref{nke}{\em i)} to the
function $U^n f$, we see that the second term vanishes in the limit $\ep\to 0$.
From the isometry of $U$, we obtain that for any $n>0$, $\|T_\ep^n\|\epto 1$, 
so that ${\tau}_{c}\epto\infty$.
In coarse-grained version, similarly as above, we have 
\bean
\|\tilde{T}_{\ep}^n f - U^{n}f\|&=&\|G_{\ep}U_{F}^n(G_{\ep}-I)f+(G_{\ep}-I)U^nf\|\\
&\leq& \|(G_{\ep}-I)f\|+\|(G_{\ep}-I)U^nf\|\rightarrow 0. \qquad \blacksquare
\eean

Now we pass to abstract spectral bounds.

\begin{thm}
\label{gb}
For any isometric operator $U$ on $L^2_0(\IT^d)$ and noise operator $G_\ep$,
the dissipation time of the noisy evolution operator $T_\ep=G_{\ep}U$ satisfies the following estimates:
\begin{align}
\frac{1-e^{-1}}{d_\ep(1)} \leq  {\tau}_{c} &\leq
\frac{1}{\left|\ln(\|G_\ep\|)\right|}+1,\label{firstline}\\
 {\tau}_{c} &\leq\inf_{r_{sp}(T_\ep)<r<1}\frac{1}{|\ln (r)|} \ln\left(\frac{e}{d_{\ep}(r)}\right).
\label{second-upper}
\end{align}
\end{thm}

We notice that the first upper bound does not depend on $U$ at all, but only on the noise.
Using the estimate \eqref{contr}, we obtain the following obvious corollary:

\begin{cor}
\label{noise1->upper}
If the noise generating density satisfies the
estimate \eqref{noise1} for some $\alpha\in (0,2]$, then for any measure-preserving map $F$ the
noisy dissipation time is bounded from above as follows ${\tau}_{c} \lesssim  \ep^{-\alpha}$.
\end{cor}

\textbf{Proof of Theorem \ref{gb}.} 

1. \underline{Lower bound}

We use the following series expansion of the resolvent \cite[p.211]{Y}
valid for any $|\lam|> r_{sp}(T_\ep)$:
\bea
\label{NSR}
R_\ep(\lam)=\sum_{n=0}^{\infty}\lam^{-n-1}T_\ep^{n}.
\eea 
Considering that $r_{sp}(T_{\ep})\leq \|G_{\ep}\|< 1$, we may take  $|\lam|=1$, and cut this sum
into two parts:
\bean
 R_{\ep}(\lam)=
\sum_{n=0}^{{\tau}_{c}-1}\lam^{-n-1}T_{\ep}^{n}+
\lam^{-{\tau}_{c}}T_{\ep}^{{\tau}_{c}}R_{\ep}(\lam).
\eean
Taking norms and applying the triangle inequality, we get
\bean
\|R_{\ep}(\lam)\| &\leq& \|\sum_{n=0}^{{\tau}_{c}-1}\lam^{-n-1}T_{\ep}^{n}\|
+|\lam|^{-{\tau}_{c}}\|T_{\ep}^{{\tau}_{c}}\|\|R_{\ep}(\lam)\|\\
&\leq& {\tau}_{c}+e^{-1}\|R_{\ep}(\lam)\|\\
\Longrightarrow \|R_{\ep}(\lam)\|(1-e^{-1})&\leq& {\tau}_{c}. 
\eean
Taking the supremum over $\lambda\in S_1$ yields the lower bound. 

\bigskip

2. \underline{Upper bounds}

To get both upper bounds, we use the following trivial lemma.

\begin{lem}\label{simple}
Assume that (for some value of $\ep$) the powers of $T_\ep$ satisfy
$$
\forall n\in\IN,\quad \|T_\ep^n\|\leq \Gamma(n),
$$
where the function $\Gamma(n)$ is strictly decreasing, and $\Gamma(n)\nto 0$.
Then the dissipation time is bounded from above by
$$
{\tau}_{c}\leq \Gamma^{(-1)}(e^{-1})+1,
$$
where $\Gamma^{(-1)}$ is the inverse function of $\Gamma$.
In particular, for the geometric decay
$\Gamma(n)=C r^n$ with $r\in(0,1)$,
$C \geq 1$, one obtains  ${\tau}_{c}\leq
\frac{\ln(e C)}{|\ln r|}+1$.
\end{lem}

The upper bound in Eq.~\eqref{firstline} comes from the obvious estimate
\bean
\|T_{\ep}^n\| \leq \|G_{\ep}\|^n,
\eean
on which we apply the lemma with $C=1$, $r=\|G_\ep\|$.

To prove the second upper bound, we use the representation of $T_\ep^{n}$ in terms of the resolvent:
\bean
T_\ep^{n}=\frac{1}{2\pi i}\int_{S^r}\lam^{n}R_{\ep}(\lam)d\lam
\eean
valid for any $r>r_{sp}(T_{\ep})$. Thus for all $r\in(r_{sp}(T_\ep),1)$, one has
\bean
\|T_\ep^{n}\|\leq \frac{1}{2\pi}\int_{S^{r}}|\lam|^{n}\|R_{\ep}(\lam)\||d\lam|
\leq \sup_{|\lam|=r}\|R_{\ep}(\lam)\| r^{n+1}=\frac{1}{d_\ep(r)}\,r^{n+1}.
\eean
We then apply Lemma \ref{simple} on the geometric
decay for any radius $r_{sp}(T_\ep)<r<1$, with $C=\frac{r}{d_\ep(r)}\geq 1$.

\hfill$\blacksquare$

%%%%%%%%%%%%%%%%%%%%%%%%%%%%%%%%%%%%%%%%%%%%%%%%%%%%%%%%%%%%%%%%%%%%%%%%%%%%%%%%%%%%%%%%%%%%%%%%%%%%%%%%%%%%%%
%%%%%%%%%%%%%%%%%%%%%%%%%%%%%%%%%%%%%%%%%%%%%%%%%%%%%%%%%%%%%%%%%%%%%%%%%%%%%%%%%%%%%%%%%%%%%%%%%%%%%%%%%%%%%%

\section{Dissipation time of not weakly-mixing maps}
\label{NWM}

%%%%%%%%%%%%%%%%%%%%%%%%%%%%%%%%%%%%%%%%%%%%%%%%%%%%%%%%%%%%%%%%%%%%%%%%%%%%%%%%%%%%%%%%%%%%%%%%%%%%%%%%%%%%%%
%%%%%%%%%%%%%%%%%%%%%%%%%%%%%%%%%%%%%%%%%%%%%%%%%%%%%%%%%%%%%%%%%%%%%%%%%%%%%%%%%%%%%%%%%%%%%%%%%%%%%%%%%%%%%%

In order to better control the growth of ${\tau}_{c}$, we need more 
precise information on the noise and the dynamics. In the present section, we restrict ourselves to
the dynamical property of weak-mixing.
We recall \cite{CoFoSi} that the map $F$ is ergodic (resp. weakly-mixing) iff $1$ is not an
eigenvalue of $U_{F}$ (resp. iff $U_{F}$ has no eigenvalue) on $L^2_0(\IT^d)$.
We now use Theorem~\ref{gb} in the case where $U=U_F$ is the Koopman operator for some measure-preserving
map $F$ on $\IT^d$ to establish the following important result. 

\begin{thm}
\label{weakmix}
Assume that the noise generating density $g$ satisfies the estimates  
\eqref{Moment} or \eqref{noise1} with exponent $\alpha\in (0,2]$.
If $F$ is not weakly-mixing and at least one eigenfunction of 
$U_{F}$ belongs to $H^{\gamma}(\IT^d)$ for some $\gamma>0$, 
then $T_{\ep}$ has slow dissipation time: 
\bean
\ep^{-(\alpha\wedge\gamma)} \lesssim {\tau}_{c}.
\eean 
\end{thm}
\textbf{Proof.} 
Let $h\in H^\gamma(\IT^d)$ be a normalized eigenfunction of $U_F$ with eigenvalue $\lam$. 
Applying Proposition~\ref{nke}~{\em ii)}, we get
$$
\|(\lam -T_{\ep})h\|= \|(I-G_{\ep})h\|\leq K\ep^{\gamma\wedge\alpha}
$$
for some constant $K>0$ depending on $g$ and $h$. 
This implies that $\|R_\ep(\lam)\|\geq\frac{1}{K\ep^{\gamma\wedge\alpha}}$, therefore
taking the supremum over $|\lam|=1$ yields $d_\ep(1)^{-1}\geq\frac{1}{K\ep^{\gamma\wedge\alpha}}$.
The lower bound in Theorem~\ref{gb} then implies 
\bea
\label{low-gamma}
\frac{1-e^{-1}}{K\ep^{\gamma\wedge \alpha}} \leq  {\tau}_{c}.\qquad \blacksquare
\eea

\begin{rem}
Recall that if $g$ satisfies \eqref{noise1} with exponent
$\alpha$, then  the dissipation time is also bounded from
above, as shown in Corollary~\ref{noise1->upper}. If one eigenfunction $h$ has
regularity $H^\gamma$ with $\gamma\geq\alpha$, then both results
imply that the dissipation is {\em simple}, with exponent
$\alpha$.
\end{rem}

\begin{rem}
The above results can be stated in more general form: $U_{F}$ does not
need to be a Koopman operator associated with a map $F$. The result
holds true for any isometric operator $U$ on $L^2_0$ with an eigenfunction of Sobolev regularity.
\end{rem}

The dependence of the lower bound in (\ref{low-gamma}) on $\gamma$ can be intuitively 
explained as follows. In case of non-weakly-mixing maps the eigenfunctions of $U_F$
are, in general, responsible for slowing down the dissipation. The rate of the dissipation 
is affected by the regularity of the smoothest eigenfunction. In principle, irregular functions
undergo faster dissipation giving rise to slower asymptotics of ${\tau}_{c}$.
It is not clear, however, whether the actual asymptotics of the dissipation time
will be slower than power law in case when all eigenfunctions of $U_F$ on $L^2_0(\IT^d)$ are
outside any space $H^{\gamma}(\IT^d)$ with $\gamma>0$.

The above theorem serves as a source of examples of 'non-chaotic' ergodic dynamical systems. 
A typical example of ergodic but not weakly mixing transformations for which this corollary applies
is the family of 'irrational' shifts on $\IT^{d}$ i.e. maps $F\bx=\bx+\mf{c}$ on $\IT^{d}$, 
where $\mf{c}=(c_{1},..,c_{d})$ is a constant vector such that the numbers $1,c_{1},..,c_{d}$ are 
linearly independent over rationals. 
More general and less trivial examples of ergodic maps giving rise to a slow dissipation time will be 
discussed in Section \ref{aAT} (cf. Remark \ref{remerg}). 

In Corollary~\ref{disto0} we have shown that for any map $F$ and
arbitrary small $\del>0$, 
the pseudospectrum $\sigma_\del(T_\ep)$ 
intersects the unit circle for sufficiently small $\ep>0$.
If $F$ is not weakly-mixing, the spectral radius of $T_\ep$ (that is, the modulus of its largest eigenvalue) 
is believed to converge to $1$ when $\ep\to 0$, and the associated eigenstate $h_\ep$ should converge 
to a ``noiseless eigenstate'' $h$.
This ``spectral stability'' has been discussed for several cases in the continuous-time as well as
for discrete-time maps on $\IT^2$ \cite{Kif,N}.

On the opposite, if $F$ is an Anosov map on $\IT^2$ (see Section~\ref{Anosov}), 
the spectrum of $T_{\ep}$ does not approach the unit circle, but
stays away from it uniformly: $r_{sp}(T_\ep)$
is smaller than some $r_0<1$ for any $\ep>0$ \cite{BKL}. Simultaneously, $\|T_\ep\|\to 1$,
so we have here
a clear manifestation of the {\em nonnormality} of $T_\ep$ for such a map. In some
cases (see \cite{N} and the linear examples of Section~\ref{examples}), the operator $T_{\ep}$ is even 
quasinilpotent, meaning that $r_{sp}(T_{\ep})=0$ for all $\ep>0$. 
For such an Anosov map, the spectral radius of $T_\ep$ is therefore
``unstable'' or ``discontinuous'' in the limit $\ep\to 0$, while in the same limit 
the (radius of its) pseudospectrum
$\sigma_\del(T_\ep)$ (for $\delta>0$ fixed) is  ``stable''.

We end this section by determining the coarse-grained dissipation time for
non weakly-mixing maps. We have

\begin{prop}\label{infinite} 
Let $F$ be a measure-preserving map. If $F$ is not weakly-mixing then 
$\tilde{\tau}_{c}=\infty$ for small enough $\ep>0$.
\end{prop}
\textbf{Proof.}
Let $h\in L^2_0(\IT^d)$ be a normalized eigenfunction of $U_F$, then
\bean
\|\tilde{T}^{(n)}_{\ep}h\| &=& \|G_{\ep}U^n_F(h+(G_\ep -I)h)\|
\geq\|G_{\ep}h\|-\|G_{\ep}U^n_F(G_\ep -I)h\|\\
&\geq& 1-2\|(G_\ep -I)h\|.
\eean
Since the RHS above is independent of $n$, we see that $\|\tilde{T}^{(n)}_{\ep}\|$ is close to $1$ for
all times and sufficiently small $\ep>0$. \hfill$\blacksquare$
Thus as opposed to the noisy case (see Prop.~\ref{noise1->upper}), the coarse-grained evolution through a
non-weakly-mixing map does not dissipate.

%%%%%%%%%%%%%%%%%%%%%%%%%%%%%%%%%%%%%%%%%%%%%%%%%%%%%%%%%%%%%%%%%%%%%%%%%%%%%%%%%%%%%%%%%%%%%%%%%%%%%%%%%%%%%%
%%%%%%%%%%%%%%%%%%%%%%%%%%%%%%%%%%%%%%%%%%%%%%%%%%%%%%%%%%%%%%%%%%%%%%%%%%%%%%%%%%%%%%%%%%%%%%%%%%%%%%%%%%%%%%

\section{Local expansion rate and general lower bound}
\label{LB}

%%%%%%%%%%%%%%%%%%%%%%%%%%%%%%%%%%%%%%%%%%%%%%%%%%%%%%%%%%%%%%%%%%%%%%%%%%%%%%%%%%%%%%%%%%%%%%%%%%%%%%%%%%%%%%
%%%%%%%%%%%%%%%%%%%%%%%%%%%%%%%%%%%%%%%%%%%%%%%%%%%%%%%%%%%%%%%%%%%%%%%%%%%%%%%%%%%%%%%%%%%%%%%%%%%%%%%%%%%%%%

We saw in the previous section that there exists no general upper bound for coarse-grained dynamics $\tilde {\tau}_{c}$.
On the opposite, we will prove below a general {\em lower} bound for both coarse-grained and noisy evolutions,
valid for any measure-preserving map $F$ of regularity $C^1$.
We note that Propositions~\ref{nonfinite} and \ref{infinite}{\em i)} (which are valid
independently of any regularity assumption) do not provide an explicit lower bound.

First we introduce some notation. For any map $F\in C^1$, 
$DF(\bx)$ is the tangent map of $F$ at the point $\bx\in\IT^d$, mapping
a tangent vector at $\bx$ to a tangent vector at $F(\bx)$. Selecting the canonical (i.e. Cartesian) basis 
and metrics on $T(\IT^d)$, this map can be represented as a $d\times d$ matrix. 
The metrics naturally yields a norm $\bv\in T_{\bx}(\IT^d)\mapsto |v|$ on the tangent space, and
therefore a norm on this matrix: $|DF(\bx)|=\max_{|\bv|=1}|DF(\bx) \cdot \bv|$.
We are now in position to define the maximal expansion rate of $F$:
\bean
\mu_{F}=\limsup_{n\rightarrow \infty} \|DF^n\|_{\infty}^{1/n},\quad\mbox{where}\quad
\|DF^n\|_{\infty}=\sup_{\bx\in\IT^d} |(DF^n)(\bx)|.
\eean
Since $F$ preserves the Lebesgue measure, the Jacobian $J_F(\bx)$ satisfies
$|J_F(\bx)|\geq 1$ at all points. 
In the Cartesian basis, $J_F(\bx)=\det(DF(\bx))$, so
that we have $\|DF^n(\bx)\|\geq 1$ for all $\bx\in\IT^d$, $n\geq 0$. One can
actually prove the following:
\begin{rem}
Although $|(DF^n)(\bx)|$ and $\|DF\|_{\infty}$ may depend on the choice of the metrics, $\mu_F$ does not, 
and satisfies $1\leq \mu_F\leq \|DF\|_{\infty}$.
\end{rem}

From the definition of $\mu_{F}$, for any
$\mu > \mu_F$ there exists
a constant $A\geq 1$ such that
\bea
\label{expogrowth}
\forall n\in \IN,\qquad \|DF^n\|_{\infty} \leq A\mu^n.
\eea
In some cases one may take $\mu=\mu_F$ in the RHS.
In case $\mu_F=1$, $\|DF^n\|_{\infty}$ can sometimes grow as a power-law:
\bea
\label{power-law}
\|DF^n\|_{\infty} \leq A n^\beta, \qquad n\in \IN
\eea
for some $\beta>0$, or even be uniformly bounded by a constant ($\beta=0$). 

\medskip

The relationship between, on one side, the local expansion of the map $F$ and on the other side,
the dissipation time, can be intuitively understood as follows. A lack of 
expansion ($\|DF\|_\infty=1$) results in the transformation of ``soft'' or ``long-wavelength'' oscillations
into ``soft oscillations'', both being little affected by the
noise operator $G_\ep$. On the opposite, a locally strictly expansive map ($\|DF\|_\infty>1$) will
quickly transform soft oscillations into ``hard'' or ``short-wavelength'', 
the latter being much more damped by the noise.

The following theorem precisely measures this relationship, in terms of {\em lower bounds}
for the dissipation times.

\begin{thm}
\label{nln}
Let $F$ be a measure-preserving $C^1$ map on $\IT^d$, and assume that the 
noise generating density $g$ satisfies (\ref{Moment}) or
(\ref{noise1}) for some $\alpha\in (0,2]$.

i) If $\|DF\|_{\infty}> 1$, resp. $\mu_{F}>1$, then
there exist a constant $c$, resp. constants $\mu\geq\mu_F$ and $\tilde{c}$, such that for small enough $\ep$,
\bea
\label{nlb}
{\tau}_{c} \geq \frac{\alpha\wedge 1}{\ln(\|DF\|_\infty)} \ln(\ep^{-1})+c, \quad\mbox{resp.}\quad
\tilde{\tau}_{c} \geq \frac{\alpha \wedge 1}{\ln\mu} \ln(\ep^{-1})+\tilde{c}.
\eea
If $F$ is a $C^1$ diffeomorphism, then (\ref{nlb}) holds
with
$\|DF\|_{\infty}$ replaced by
$\|DF\|_{\infty}\wedge\|D(F^{-1})\|_{\infty}$, resp. with some
$\mu\geq\mu_{F}\wedge\mu_{F^{-1}}$.

ii) If $\|DF\|_{\infty}=1$ then $T_{\ep}$ has slow 
dissipation time, ${\tau}_{c}\gtrsim \ep^{-(\alpha\wedge 1)}$. 
If the noise kernel satisfies the condition (\ref{noise1}) for $\alpha\in (0,1]$, then 
the dissipation time is simple, ${\tau}_{c}\sim \ep^{-\alpha}$.

iii) If $\mu_{F}=1$ and $\|DF^n\|_\infty$ grows as a power-law as in Eq.~\eqref{power-law}
with $\beta>0$, then $\tilde{\tau}_{c} \gtrsim \ep^{-(\alpha\wedge 1)/\beta}$.
If $\|DF^n\|_\infty$ is uniformly bounded above by a constant, then $\tilde{\tau}_{c}=\infty$ 
for small enough $\eps$.
\end{thm}

\medskip

\begin{rem}
\label{class}
This theorem shows that classical systems on $\IT^d$ (i.e. $C^1$ diffeomorphisms) cannot have
a dissipation time growing slower than $C\ln(\ep^{-1})$. In view of the results 
for toral automorphisms (cf. Proposition~4), this lower bound
on the dissipation time is sharp and consistent with Kouchnirenko's upper
bound on the entropy of the classical systems, namely
all classical systems have a finite (possibly zero) Kolmogorov-Sinai
entropy (Theorem 12.35. in \cite{Arnold}, see also \cite{AM}, \cite{Kou}).
\end{rem}

\textbf{Proof of the Theorem \ref{nln}.}
The following trivial lemma (similar to Lemma~\ref{simple}) will be crucial in the proof.

\begin{lem}\label{simple2}
Assume that there exists some $\alpha>0$ and a strictly increasing function $\gamma(n)$, $\gamma(0)=0$ such that
\bea
\forall n\geq 1,\qquad \|T_\ep^n\|\geq 1-\ep^\alpha \gamma(n).
\eea
Then the dissipation time is bounded from below as:
\bea
{\tau}_{c}\geq \gamma^{(-1)}\left(\frac{1-e^{-1}}{\ep^\alpha}\right),
\eeq
where $\gamma^{(-1)}$ is the inverse function of $\gamma$.

The same statement  holds for  the coarse-grained version.
\end{lem}

Our task is therefore to bound $\|T_\ep^n\|$ (resp. $\|\tilde T_\ep^{(n)}\|$) from below.
A simple computation shows that for any $f\in C^0(\IT^d)$, 
$\|G_{\ep}f\|_{\infty}\leq \|f\|_{\infty}$. Since convolution commutes
with differentiation, for $f\in C^1$ we also have $\|\nabla(G_\ep f)\|_\infty\leq \|\nabla f\|_\infty$. 
We use this fact to estimate the gradient of $T_{\ep}f$:
\bean
\|\nabla (T_{\ep}f) \|_{\infty}&=&\|\nabla(G_{\ep}U_{F}f)\|_{\infty}\\
&\leq& \|\nabla(f\circ F) \|_{\infty}=\|(\nabla f)\circ F \cdot DF \|_{\infty}\\
&\leq& \|(\nabla f)\circ F\|_{\infty}\|DF\|_{\infty}= \|\nabla f\|_{\infty}\|DF\|_{\infty}.
\eean
Repeating the above procedure $m$ times, we get
\bea
\label{nab}
\|\nabla(T^{m}_{\ep}f) \|_{\infty}
\leq \|\nabla f\|_{\infty}\|DF\|_{\infty}^m,\quad 
\|\nabla(U_FT^{m}_{\ep}f) \|_{\infty}
\leq \|\nabla f\|_{\infty}\|DF\|_{\infty}^{m+1}.
\eea
We now choose some arbitrary $f\in C^1_0(\IT^d)$, with $\|f\|=1$.
We first apply the triangle inequality:
$$
\|T^n_{\ep}f\|= \|G_{\ep}U_{F}T^{n-1}_{\ep}f\|
\geq  \|U_{F} T^{n-1}_{\ep}f\| - \|(G_{\ep}-I)U_{F}T^{n-1}_{\ep}f\|.
$$
To estimate the second term on the RHS we use the bound \eqref{C1est} and
the estimate \eqref{nab} to obtain
$$
\|T^n_{\ep}f\|\geq \|T^{n-1}_{\ep}f\|-C\ep^{\alpha \wedge 1}\|\nabla f\|_{\infty}\|DF\|_{\infty}^n.
$$
Applying the same procedure iteratively to the first term on the RHS, we finally get (remember $\|f\|=1$):
\bea\label{est-noise}
\|T^n_\ep\|\geq\|T^n_{\ep}f\|\geq 1-C\ep^{\alpha \wedge 1} \|\nabla f\|_{\infty}\sum_{m=1}^n\|DF\|_{\infty}^m.
\eea
The computations in the case of the coarse-grained operator are even simpler:
\bea
\|\tilde{T}^{(n)}_{\ep}f\|&=&\|G_{\ep}U_{F}^n G_{\ep}f\|\non\\
&\geq&
1-C\ep^{\alpha\wedge 1} \|\nabla f\|_{\infty} - C\ep^{\alpha\wedge 1} \|\nabla(G_{\ep}f)\|_{\infty}\|DF^n\|_{\infty}
\non\\
&\geq&
1- 2C\ep^{\alpha\wedge 1} \|\nabla
f\|_{\infty}\|DF^n\|_{\infty}.\label{est-coarse}
\eea
Notice that from the assumptions on $f$, 
$\|\nabla f\|_\infty$ cannot be made arbitrary small, but is necessarily larger than some positive constant.
We choose some arbitrary function, say $f=\be_{\bk}$ with $\bk=(1,0)$ which satisfies 
$\|\nabla f\|_\infty=2\pi$.

\medskip

The estimate (\ref{est-noise}) has the form given in Lemma~\ref{simple2}. 
The growth of the function $\gamma(n)$ depends
on whether $\|DF\|_{\infty}$ is equal to or larger than $1$, 
which explains why the lower bounds are qualitatively different in the two cases. 
 
In case $\|DF\|_{\infty}$ is strictly larger than $1$, then the function $\gamma(n)$ 
grows like an exponential, therefore the lower bound is of the type \eqref{nlb}.
For the coarse-grained version, a growth of $\|DF\|_{\infty}$ of the type \eqref{expogrowth} 
yields the lower bound for $\tilde{\tau}_{c}$ in \eqref{nlb}.

In the case $\|DF\|_{\infty}=1$, $\gamma(n)$ is a linear function, so that 
${\tau}_{c}\geq \frac{1-e^{-1}}{C\|\nabla f\|_\infty}\ep^{-(\alpha\wedge 1)}$. 

In the coarse-grained version, if $\mu_F=1$ and $\|DF^n\|_\infty$ grows like in \eqref{power-law} 
with $\beta>0$,  the dissipation is slow:
$\tilde {\tau}_{c}\geq C\ep^{-(\alpha\wedge 1)/\beta}$.
In the case where $\|DF^n\|_\infty$ is uniformly bounded by some constant, 
the norm of the coarse-grained propagator
stays larger than some positive constant for all times, so that for small enough noise
$\tilde {\tau}_{c}$ is infinite. \hfill$\blacksquare$

%%%%%%%%%%%%%%%%%%%%%%%%%%%%%%%%%%%%%%%%%%%%%%%%%%%%%%%%%%%%%%%%%%%%%%%%%%%%%%%%%%%%%%%%%%%%%%%%%%%%%%%%%%%%%%%%%%
%%%%%%%%%%%%%%%%%%%%%%%%%%%%%%%%%%%%%%%%%%%%%%%%%%%%%%%%%%%%%%%%%%%%%%%%%%%%%%%%%%%%%%%%%%%%%%%%%%%%%%%%%%%%%%%%%

\section{Decay of correlations and general upper bound}
\label{UB}

%%%%%%%%%%%%%%%%%%%%%%%%%%%%%%%%%%%%%%%%%%%%%%%%%%%%%%%%%%%%%%%%%%%%%%%%%%%%%%%%%%%%%%%%%%%%%%%%%%%%%%%%%%%%%%%%%%
%%%%%%%%%%%%%%%%%%%%%%%%%%%%%%%%%%%%%%%%%%%%%%%%%%%%%%%%%%%%%%%%%%%%%%%%%%%%%%%%%%%%%%%%%%%%%%%%%%%%%%%%%%%%%%%%%%

For any two functions $f,g \in L^2_0(\IT^d)$ the dynamical correlation function for
the map $F$ is defined as the following function of $n\in\IN$ 
(see e.g. \cite{B}):
\bean
C_{f,g}(n)=C^0_{f,g}(n)=m(fU_{F}^{n}g)=\la \bar f,U_F^n g\ra= \la P_F^n \bar f,g\rangle.
\eean
The same quantity may be defined for the noisy evolution:
\bean
C^{\ep}_{f,g}(n)=m(fT_{\ep}^{n}g).
\eean
We recall that a map $F$ is {\em mixing} iff for any $f,\ g\in L^2_0$, 
\bean
C_{f,g}(n)\rightarrow 0, \quad \text{as}\quad n\to \infty.
\eean
The correlation function can easily be measured in (numerical or real-life) experiments, so it is
often used to characterize the dynamics of a system.

To focus the attention, we will only be concerned with maps 
for which correlations decay in a precise way. We assume that
there exist H\"older exponents $s_*,s\in \IR_+$, $0\leq s_*\leq s$
together with some decreasing
function $\Gamma(n)=\Gamma_{s_*,s}(n)$ with $\Gamma(n)\nto 0$, 
such that for any observables $f\in C_0^{s_*}(\IT^d)$,
$g\in C_0^{s}(\IT^d)$ 
and for sufficiently
small $\eps\geq 0$ (sometimes only for $\ep=0$), 
\bea
\label{dc}
\forall n\in\IN,\quad|C^{\ep}_{f,g}(n)|\leq \|f\|_{C^{s_*}} \|g\|_{C^{s}}\Gamma(n).
\eea
In general, such a bound can be proved only if the map $F$ has regularity 
$C^{s+1}$. The reason why we do not necessarily take the same norm for the functions 
$f$ and $g$ will be clear below.

We will be mainly interested in the following two types of decay

\begin{itemize}
\item[i)] Power-law decay: there exists $C>0$, $\beta>0$ such that,  
\bea
\Gamma(n)=C n^{-\beta}.
\eea
This behavior is characteristic of intermittent maps, e.g. maps possessing one or several 
neutral orbits \cite{Bal-decay}.
\item[ii)] Exponential decay: there exists $C>0$, $0<\sigma<1$ such that, 
\bea
\Gamma(n)=C \sigma^n. 
\eea
Such a behavior was proved in the case of uniformly expanding or hyperbolic maps on the torus 
(see Section~\ref{examples}), as well as many other cases \cite{Bal-decay}.
\end{itemize}

The central result of this section is a relationship between, on one side, the
decay of noisy (resp. noiseless) correlations and on the other side, the small-noise
behavior of the noisy (resp. coarse-graining) dissipation time. The intuitive picture is similar
to the one linking the local expansion rate to the dissipation: namely, a fast decay of 
correlations is generally due to the transition of ``soft'' into ``hard'' fluctuations of the
observable through the evolution, which is itself induced by large expansion rates of the map.
Still, as opposed to what we obtained in last Section, the following theorem and its corollary
yields {\em upper bounds} for the dissipation time.

\begin{thm}
\label{ccln}
Let $F$ be a volume preserving map on $\IT^d$ with correlations
decaying as in Eq.~(\ref{dc}) for some indices $s$, $s_*$ and decreasing function 
$\Gamma(n)$, at least in the noiseless limit $\ep=0$.
Assume that the noise generating function $g$ is $([s]+1)$-differentiable, and that
all its derivatives of order $|\bmal|_1\leq  [s]+1$ satisfy
\[|D^{\bmal}g(\bx)|\lesssim
\frac{1}{|\bx|^M},\quad |\bx|\gg 1,
\] with a power $M>d$.

Then there exist constants $\tilde C>0$, $\ep_o>0$ such that the coarse-grained propagator satisfies
\bea
\forall\ep\leq\ep_o,\quad\forall n\geq 0,\qquad
\|\tilde{T}^{(n)}_{\ep}\| \leq \tilde{C}\; \frac{\Gamma(n)}{\ep^{d+s+s_*}}.
\eea
If the decay of correlations \eqref{dc} also holds for sufficiently small $\ep>0$ 
(and assuming the Perron-Frobenius operator $P_F$ is bounded in $C^{s}(\IT^d)$), then
the noisy operator satisfies (for some constants $C>0$, $\ep_o>0$):
\bea
\forall \ep\leq\ep_o,\quad\forall n\geq 0,\qquad
\|T^{n}_{\ep}\| \leq C\; \frac{\Gamma(n)}{\ep^{d+s+s_*}}.
\eea
\end{thm}

From these estimates, we straightforwardly obtain the following bounds on both
dissipation times (the assumptions on $F$ and the noise generating function $g$
are the same as in the Theorem):

\begin{cor}
\label{correl-upper}

I) If the correlation function satisfies the bound \eqref{dc} for $\ep=0$, then
 the coarse-grained dissipation time is well defined
($\tilde{\tau}_{c}<\infty$). Moreover,
\begin{itemize}
\item[i)] if $\Gamma(n)\sim n^{-\beta}$ then there exists a constant
$\tilde{C}>0$ such that
\bean
\tilde{\tau}_{c} \leq \tilde{C} \ep^{-\frac{d+s+s_*}{\beta}}
\eean
\item[ii)] if  $\Gamma(n)\sim\sigma^n$ then there exists a constant
$\tilde{c}$ such that 
\bean
\tilde{\tau}_{c} \leq \frac{d+s+s_*}{|\ln\sigma|} \ln(\ep^{-1})+\tilde{c}, 
\eean
\end{itemize}

II) If Eq.~\eqref{dc} holds for sufficiently small $\ep>0$, then
\begin{itemize}
\item[i)] if $\Gamma(n)\sim n^{-\beta}$, there exists a constant $C>0$ such that
\bean
{\tau}_{c} \leq C \ep^{-\frac{d+s+s_*}{\beta}}
\eean
\item[ii)] if  $\Gamma(n)\sim\sigma^n$, there exists a constant $c$ such that
\bean
{\tau}_{c} \leq \frac{d+s+s_*}{|\ln\sigma|} \ln(\ep^{-1})+c. 
\eean
\end{itemize}
\end{cor}

\textbf{Proof of Theorem \ref{ccln}.}

\underline{$1^{st}$ step}: We represent the action of $T_{\ep}^n$  
(resp. $\tilde{T}_{\ep}^{(n)}$) on an observable 
$f\in L^2_0(\IT^d)$
in terms of the correlation functions $C^\ep(n)$ (resp. $C(n)$). 
To do this we Fourier decompose both $T_{\ep}^{n+2}f$ and $f_1=U_F f$, and use Eq.~\eqref{specdecompo}:
\bean
T_{\ep}^{n+2}f &=& 
\sum_{0\neq \bj\in \IZ^d}\la \be_{\bj}, G_{\ep}U_{F}T_{\ep}^{n}G_{\ep}f_{1} \ra \be_{\bj}\\
&=&
\sum_{0\neq \bj\in \IZ^{d}}\sum_{0\neq \bk\in \IZ^{n}}\hat{f}_{1}(\bk)
\la G_{\ep}\be_{\bj},U_{F}T_{\ep}^{n}G_{\ep}\be_{\bk} \ra \be_{\bj}\\&=&
\sum_{0\neq 
\bj\in \IZ^{d}}\sum_{0\neq \bk\in \IZ^{d}}\hat{f}_{1}(\bk)
\hat{g}_{\ep}(\bj)  \hat{g}_{\ep}(\bk)
\la P_{F}\be_{\bj},T_{\ep}^{n}\be_{\bk} \ra 
  \be_{\bj}.
\eean
(remember that $\hat g$ is a real function). 
A similar computation for the coarse-grained propagator yields:
\bean
\tilde{T}_{\ep}^{(n)}f=\sum_{0\neq 
\bj\in \IZ^{d}}\sum_{0\neq \bk\in \IZ^{d}}\hat{f}(\bk)
\hat{g}_{\ep}(\bj)  \hat{g}_{\ep}(\bk)
\la \be_{\bj},U_{F}^{n}\be_{\bk} \ra 
  \be_{\bj}.
\eean
Taking the norms on both sides, we get in the noisy case:
\bea
\|T_{\ep}^{n+2}f\|^{2}&=&\sum_{0\neq \bj\in \IZ^{2}}
\bigg| \sum_{0\neq \bk\in \IZ^{d}}\hat{f}_{1}(\bk)
\la P_{F}\be_{\bj},T_{\ep}^{n}\be_{\bk} \ra \hat{g}_{\ep}(\bj) \hat{g}_{\ep}(\bk) \bigg|^{2}\nonumber\\
&\leq&
\sum_{0\neq \bj\in \IZ^{d}} 
\bigg( \sum_{0\neq \bk\in \IZ^{d}}|\hat{f}_{1}(\bk)|^{2}\bigg)
\sum_{0\neq \bk\in \IZ^{d}}|\la P_{F}\be_{\bj},T_{\ep}^{n}\be_{\bk} \ra|^{2}
|\hat{g}_{\ep}(\bj)\hat{g}_{\ep}(\bk)|^2\nonumber\\
\Longrightarrow \|T_{\ep}^{n+2}f\|^{2}&\leq& \|f_1\|^2
\sum_{0\not = \bj,\bk\in \IZ^{d}}|C^{\ep}_{P_{F}\be_{-\bj},\be_{\bk}}(n)|^{2}
  |\hat{g}(\ep\bj)\hat{g}(\ep\bk)|^2, \label{noisy-ineq} 
\eea
and in the coarse-graining case
\bea
\label{coarse-ineq}
\|\tilde{T}_{\ep}^{(n)}f\|^2 \leq \|f\|^2 \sum_{0\neq \bj,\bk\in \IZ^{d}}|C_{\be_{-\bj},\be_{\bk}}(n)|^{2}
  |\hat{g}(\ep\bj)\hat{g}(\ep\bk)|^2. 
\eea
These two expressions explicitly relate the dissipation with the correlation functions.
\medskip

\underline{$2^{nd}$ step:} 
We now apply the estimates (\ref{dc}) on correlations for the observables $\be_{\bk}$, $\be_{-\bj}$, 
$P_{F}\be_{-\bj}$. 
In the coarse-grained case, it yields (using simple bounds of the type of Eq.~\eqref{cosin}):
\bean
\forall\bj,\ \bk\in\IZ^d\setminus \{0\},\qquad |C_{\be_{-\bj},\be_{\bk}}(n)|
\leq C'\;|\bj|^{s}|\bk|^{s_*}\Gamma(n).
\eean
In the noisy case, we need to assume that the Perron-Frobenius operator $P_F$ is bounded in the space
$C^{s}(\IT^d)$. This property is in general a prerequisite in the proof of estimates
of the type \eqref{dc}, so this assumption is quite natural here.
\bea
\forall\bj,\ \bk\in\IZ^d\setminus \{0\},\qquad
|C^{\ep}_{P_{F}\be_{-\bj},\be_{\bk}}(n)| &\leq& C\; \|P_{F}\be_{-\bj}\|_{C^{s}} \|
 \be_{\bk}\|_{C^{s_*}}\Gamma(n)\non\\
&\leq& C\|P_F\|_{C^{s}} |\bj|^{s}|\bk|^{s_*}\Gamma(n).\label{Cep}
\eea
We insert these bounds on the decay of correlations in the expressions (\ref{noisy-ineq}-\ref{coarse-ineq}), 
for instance in the coarse-grained case we get:
\bea
\forall n\geq 0,\qquad
\|\tilde{T}^{(n)}_{\ep}\|^{2} \leq C\; \Gamma(n)^2 \bigg(\ep^{-(s+s_*)}\sum_{0\neq \bk\in \IZ^{d}}|\ep\bk|^{s+s_*}
\hat{g}(\ep\bk)^2 \bigg)^{2}.\label{step2}
\eea

\medskip

\underline{$3^{rd}$ step:}
We finally estimate the $\ep$-dependence of the RHS of the above inequality. 
Up to a factor $\ep^{-d}$, the sum in the brackets is
a Riemann sum for the integral $\int |\bxi|^{s+s_*}\hat g(\bxi)^2 d\bxi<\infty$. 
A precise connection is given in the following lemma, proved in Section \ref{Ch2proofs}:
\begin{lem}
\label{sum-int}
Let $f\in C^0(\IR^d)$ be symmetric w.r.t. the origin and decaying at infinity as $|f(\bx)|\lesssim |\bx|^{-M}$ with $M>d$. 
Then the following small-$\ep$ estimate holds in the limit $\ep\to 0$:
\bea
\label{sum-integ}
\ep^d\sum_{\bk\in\IZ^d}\hat f(\ep\bk)^2
=\int_{\IR^{d}} \hat f(\bxi)^2 d\bxi+\cO(\ep^M).
\eea
\end{lem}
Let $m\in\IN$ satisfy $2m\leq s+s_*\leq 2m+2$ (notice that $m\leq [s]$ since we assumed $s_*\leq s$).
From the obvious inequality 
$$
\forall x>0,\qquad x^{s+s_*}\leq x^{2m} +x^{2m+2},
$$ 
we may replace in the RHS of \eqref{step2} the factor $|\ep\bk|^{s+s_*}$ by $|\ep\bk|^{2m}+|\ep\bk|^{2m+2}$. 
Applying Lemma~\ref{sum-int} to the derivatives of $g$ of order $m$ and $m+1$, we end up with
the following upper bound, which proves the first part of the
theorem: 
\bean
\|\tilde{T}^{(n)}_{\ep}\|^{2} 
&\leq&C\;\Gamma(n)^2 
\left(\frac{1}{\ep^{d+s+s_*}}\int_{\IR^{d}} \big(|\bxi|^{2m}+|\bxi|^{2(m+1)}\big)\hat g(\bxi)^2 d\bxi+\cO(\ep^M)\right)^{2} \\
&\leq& C'\;\frac{\Gamma(n)^2}{\ep^{2(d+s+s_*)}}\;\|g\|^4_{H^{m+1}}.
\eean
The computations follow identically
for the case of the noisy operator, yielding the second part of the theorem. \hfill$\blacksquare$

%%%%%%%%%%%%%%%%%%%%%%%%%%%%%%%%%%%%%%%%%%%%%%%%%%%%%%%%%%%%%%%%%%%%%%%%%%%%%%%%%%%%%%%%%%%%%%%%%%%%%%%%%%%%%%%
%%%%%%%%%%%%%%%%%%%%%%%%%%%%%%%%%%%%%%%%%%%%%%%%%%%%%%%%%%%%%%%%%%%%%%%%%%%%%%%%%%%%%%%%%%%%%%%%%%%%%%%%%%%%%%%

\section{Dissipation time and optimization problems}
\label{Opt}

%%%%%%%%%%%%%%%%%%%%%%%%%%%%%%%%%%%%%%%%%%%%%%%%%%%%%%%%%%%%%%%%%%%%%%%%%%%%%%%%%%%%%%%%%%%%%%%%%%%%%%%%%%%%%%%
%%%%%%%%%%%%%%%%%%%%%%%%%%%%%%%%%%%%%%%%%%%%%%%%%%%%%%%%%%%%%%%%%%%%%%%%%%%%%%%%%%%%%%%%%%%%%%%%%%%%%%%%%%%%%%%

In general the problem of computing the dissipation time is rather complicated. In some cases 
it can be reformulated as an asymptotic optimization problem. 
To see it, one can represent the action of a given unitary operator $U$ in the Fourier basis
\bea
\label{uk}
 U\be_{\mf{\bk}}=\sum_{0 \not = \bk' \in \IZ^{d}}u_{\bk,\bk'}\be_{\bk'},
\eea
where for each $\bk$
\bea
\label{ul2}
 \sum_{0 \not = \bk' \in \IZ^{d}}|u_{\bk,\bk'}|^{2}=1.
\eea
Next we introduce the notation
\bean
\ml{U}_{n}(\bk_0,\bk_{n})&=&\sum_{0 \not = \bk_{1},...,\bk_{n-1} \in \IZ^{d}}
u_{\bk_{0},\bk_{1}}...u_{\bk_{n-1},\bk_{n}} \prod_{l=1}^{n}\hat{g}_{\ep}(\bk_{l})\\
\ml{S}_n(\bk_{n})&=&\{\bk_{0} \in \IZ^{d}\backslash \{0\} : \ml{U}_{n}(\bk_0,\bk_{n})\not = 0\}.
\eean
Then for any $f\in L_{0}^{2}(\IT^{d})$  we have
\bea
\nonumber
\|T_{\ep}^{n}f\|^{2} &=& 
\left\|\sum_{0 \not = \bk_{0} \in \IZ^{d}}\hat{f}(\bk_{0})T_{\ep}^{n} 
\be_{\bk_{0}}\right\|^{2}= 
\left\|\sum_{0 \not = \bk_{0} \in \IZ^{d}}\hat{f}(\bk_{0})\sum_{0 \not = \bk_{n} \in \IZ^{d}} 
\ml{U}_{n}(\bk_0,\bk_{n})\be_{\bk_{n}}\right\|^{2} \\
\label{Tnorm1}
&=&\sum_{0 \not = \bk_{n} \in \IZ^{d}}\left| \sum_{0 \not = \bk_{0} \in \IZ^{d}}\hat{f}(\bk_{0})
 \ml{U}_{n}(\bk_0,\bk_{n})\right|^{2}\\
&=&
\sum_{0 \not = \bk_{n} \in \IZ^{d}}\left| \sum_{\bk_{0} \in \ml{S}_{n}(\bk_{n})}\hat{f}(\bk_{0})
 \ml{U}_{n}(\bk_0, \bk_{n})\right|^{2}.
\eea
The following general upper bound for $\|T_{\ep}^{n}f\|$
holds.
\begin{lem}
\label{gub}
For any $f\in L_{0}^{2}(\IT^{d})$,
\bea
\label{uppernorm}
\|T_{\ep,\alpha}^{n}f\|^{2} &\leq&
\sum_{0 \not = \bk_{n} \in \IZ^{d}}
\sum_{\bk_{0} \in \ml{S}_{n}(\bk_{n})}
|\hat{f}(\bk_{0})|^{2}
\sum_{\bk_{0} \in \ml{S}_{n}(\bk_{n})}
|\ml{U}_n(\bk_{0},\bk_{n})|^{2}.
\eea
\end{lem}
For the proof we refer to Section \ref{Ch2proofs}.

In order to see how this lemma can work in practice let us consider a concrete example.
To this end we focus on a case when $u_{\bk,\bk'}$ is a Kronecker's delta function 
\bea
\label{linear}
u_{\bk,\bk'}=\del_{A\bk,\bk'},
\eea
where $A:\IZ^{d}\mapsto \IZ^{d}$ is a linear surjective map.

Under this assumption the upper bound (\ref{uppernorm}) can be used to obtain
an identity for $\|T_{\ep}^{n}\|$. 
Indeed, first observe that 
\bean
\label{linear1}
 \ml{U}_n(\bk_{0},\bk_{n})= \prod_{l=1}^{n}\hat{g}_{\ep}(A^{l}\bk_{0})\del_{A^{n}\bk_{0},\bk_{n}}
\eean
and hence (\ref{uppernorm}) becomes
\bean
\|T_{\ep}^{n}f\|^{2} &\leq&
\sum_{0 \not = \bk_{0} \in \IZ^{d}}
|\hat{f}(\bk_{0})|^{2}\prod_{l=1}^{n}\hat{g}_{\ep}(A^{l}\bk_{0})\leq
\|f\|^{2}\max_{0 \not=\bk \in \IZ^{d}}\prod_{l=1}^{n}\hat{g}_{\ep}(A^{l}\bk)
\eean
On the other hand for any nonzero $\bk\in \IZ^{d}$, one can take in (\ref{Tnorm1}) $f=\be_{\bk}$   
and get
\bean
\|T_{\ep}^{n}f\|^{2}= \prod_{l=1}^{n}\hat{g}_{\ep}(A^{l}\bk)
\eean
and therefore
\bea
\label{Tnorm3p}
\|T_{\ep}^{n}\| &=& \max_{0 \not=\bk \in \IZ^{d}} \prod_{l=1}^{n}\hat{g}_{\ep}(A^{l}\bk).
\eea
Let us now determine the class of maps $F$ such that the corresponding Koopman operator 
$U_{F}$ satisfies (\ref{linear}). The relation (\ref{linear}) implies 
\bean
U_{F}\be_{\bk}=\be_{A\bk}=e^{2\pi i \la A\bk,\bx\ra}.
\eean 
On the other hand
\bean
U_{F}\be_{\bk}(\bx)=\be_{\bk}(F\bx)=e^{2\pi i \la \bk,F\bx\ra}.
\eean
Thus 
\[
\la \bk,F\bx\ra=\la A\bk,\bx\ra \,\,\hbox{mod}\,\, 1,\quad\forall \bx \in \IR^d,\,\, \bk\in \IZ^d,
\] that is, $A$ is linear and $A^\dagger$ equals the lifting of $F$
from $\IT^{d}$ onto $\IR^{d}$. 
Moreover,  the matrix $A$ has integer entries and  determinant equal to $\pm 1$, i.e., $A$ (and $F$)
is a toral automorphism.
In the next Chapter we will use formula \ref{Tnorm3p} to derive an exact asymptotics of the 
dissipation time for virtually all toral automorphisms.

%%%%%%%%%%%%%%%%%%%%%%%%%%%%%%%%%%%%%%%%%%%%%%%%%%%%%%%%%%%%%%%%%%%%%%%%%%%%%%%%%%%%%%%%%%%%%%%%%%%%%%%%%%
%%%%%%%%%%%%%%%%%%%%%%%%%%%%%%%%%%%%%%%%%%%%%%%%%%%%%%%%%%%%%%%%%%%%%%%%%%%%%%%%%%%%%%%%%%%%%%%%%%%%%%%%%%

\section{Technical proofs}
\label{Ch2proofs}

%%%%%%%%%%%%%%%%%%%%%%%%%%%%%%%%%%%%%%%%%%%%%%%%%%%%%%%%%%%%%%%%%%%%%%%%%%%%%
%%%%%%%%%%%%%%%%%%%%%%%%%%%%%%%%%%%%%%%%%%%%%%%%%%%%%%%%%%%%%%%%%%%%%%%%%%%

%%%%%%%%%%%%%%%%%%%%%%%%%%%%%%%%%%%%%%%%%%%%%%%%%%%%%%%%%%%%%%%%%%%%%%%%%%%%

\subsection*{Proof of Lemma \ref{Mom-Four}}

%%%%%%%%%%%%%%%%%%%%%%%%%%%%%%%%%%%%%%%%%%%%%%%%%%%%%%%%%%%%%%%%%%%%%%%%%%%%

We use the following upper bound: for any $\alpha\in(0,2]$, there is a constant $C_\alpha$ 
such that
\bea
\label{cosin}
\forall x\in\IR,\quad 0\leq 1-\cos(2\pi x)\leq C_\alpha |x|^\alpha.
\eea
Besides, one has the asymptotics $1-\cos(x)\approx \frac{x^2}{2}$ for small $x$. We simply apply these
estimates to the following integral:
\bean
1-\hat g(\bxi)&=&\int_{\IR^d}(1-\cos(2\pi\bx\cdot\bxi))g(\bx)d\bx\\
&\leq&\int_{\IR^d}C_\alpha |\bx\cdot\bxi|^\alpha g(\bx)d\bx\\
&\leq& C_\alpha|\bxi|^\alpha\int_{\IR^d}|\bx|^\alpha g(\bx)d\bx=C_\alpha M_\alpha |\bxi|^\alpha.
\eean
In the case $g$ admits a second moment, we have in the limit $\bxi\to 0$:
\bean
\int_{\IR^d}(1-\cos(2\pi\bx\cdot\bxi))g(\bx)d\bx&\approx& \int_{\IR^d}2\pi^2(\bx\cdot\bxi)^2 g(\bx)d\bx\\
&\approx&2\pi^2|\bxi|^2\int_{\IR^d}(\bx\cdot\hat{\bxi})^2 g(\bx)d\bx,
\eean
where we have used the notation $\hat{\bxi}=\frac{\bxi}{|\bxi|}$ for any $\bxi\neq 0$.\hfill$\blacksquare$

%%%%%%%%%%%%%%%%%%%%%%%%%%%%%%%%%%%%%%%%%%%%%%%%%%%%%%%%%%%%%%%%%%%%%%%%%%%%

\subsection*{Proof of Proposition \ref{nke}}

%%%%%%%%%%%%%%%%%%%%%%%%%%%%%%%%%%%%%%%%%%%%%%%%%%%%%%%%%%%%%%%%%%%%%%%%%%%%

The statement i) is standard in the context of distributions \cite[p.157]{Y}. 
In our case, assume that $f\in L^2$ is normalized to unity and consider an arbitrary small $\del>0$. Since
$f\in L^2(\IT^d)$, there exists $K>0$ s.t.
$\sum_{|\bk|\geq K} |\hat{f}(\bk)|^2<\del$. Since $\hat g$ is continuous and $\hat g(0)=1$, 
there exists $\eta$ such that $(1-\hat{g}(\bxi))^2<\del$ if $|\bxi|<\eta$. 
Thus using spectral decomposition \eqref{specdecompo} of $G_\ep$, we obtain for all $\ep<\frac{\eta}{K}$
\bea\label{decompo2}
\|G_{\ep}f-f\|^2= \sum_{\bk\in \IZ^d}(1-\hat{g}(\ep\bk))^2 |\hat{f}(\bk)|^{2} 
\leq \del \sum_{|\bk|<K} |\hat{f}(\bk)|^{2}+  \sum_{|\bk|>K} |\hat{f}(\bk)|^{2}
\leq 2\del.
\eea

To prove the next statement, first notice that if $g$ satisfies
the estimate \eqref{Moment-Fourier} for the exponent $\alpha$, it also satisfies it for the
exponent $\gamma\wedge\alpha$. Using once again spectral decomposition of $G_\ep$, and applying the
estimate \eqref{Moment-Fourier} with the latter exponent we get
\bea
\|G_{\ep}f-f\|^2 &\leq&\sum_{\bk\in \IZ^d}(C_{\gamma\wedge\alpha}M_{\gamma\wedge\alpha}
|\ep\bk|^{\gamma\wedge\alpha})^2 |\hat{f}(\bk)|^{2}\non\\
&\leq&(C_{\gamma\wedge\alpha}M_{\gamma\wedge\alpha})^2\ep^{2(\gamma\wedge\alpha)}
\sum_{\bk\in \IZ^d}|\bk|^{2(\gamma\wedge\alpha)}|\hat{f}(\bk)|^{2}\label{interm}\\
&\leq&(C_{\gamma\wedge\alpha}M_{\gamma\wedge\alpha})^2\ep^{2(\gamma\wedge\alpha)}
\|f\|_{H^{\gamma\wedge\alpha}}^2.\non
\eea

To obtain the last statement, we notice that any $f\in C^1(\IT^d)$ is automatically in 
$H^1(\IT^d)$, and that its gradient satisfies
\bean
\|\nabla f \|_\infty^2\geq \|\nabla f \|^2 = 4\pi^2\sum_{\bk \in \IZ^d}|\bk|^2|\hat{f}(\bk)|^2 
\geq  4\pi^2\sum_{\bk \in \IZ^d}|\bk|^{2(1\wedge\alpha)}|\hat{f}(\bk)|^2.
\eean
The inequality \eqref{interm} with $\gamma=1$ then yields the desired result.\hfill$\blacksquare$

%%%%%%%%%%%%%%%%%%%%%%%%%%%%%%%%%%%%%%%%%%%%%%%%%%%%%%%%%%%%%%%%%%%%%%%%%%%%

\subsection*{Proof of Proposition \ref{pdiss}}

%%%%%%%%%%%%%%%%%%%%%%%%%%%%%%%%%%%%%%%%%%%%%%%%%%%%%%%%%%%%%%%%%%%%%%%%%%%%%%%%%%%%%%%%%%%%%%%%%%%%%%%%%

The proof will be based on the Riesz convexity theorem (see \cite{Zyg}, pp. 93-100) which
states that for any operator $T$ defined
on $L^p(\IT^d), 1\leq p\leq \infty,$  $\ln{\|T\|_p}$
is a convex function of $p^{-1}$. 
On the space $L^p(\IT^d)$ we consider the operator
 $\widetilde{T}:=T_{\ep,\alpha}-\la\cdot\ra$ and we have 
the relation $\widetilde{T}^n f= T^n_{\ep,\alpha} (f-\la f\ra), \forall f\in L^p(\IT^d), n\geq 1$ 
because $T_{\ep,\alpha}$ is conservative.
Now since $\|f -\la f\ra\|_p\leq 2\|f\|_p$, it follows
that
\bea
\label{14}
\|\widetilde{T}^n\|_p &\leq &2 \|T^n_{\ep,\alpha}\|_{p,0}\leq 2 \\
\label{15}
\|T^n_{\ep,\alpha}\|_{p,0} &\leq &\|\widetilde{T}^n\|_p
\eea
for $1\leq p\leq \infty, n\geq 1$.
The Riesz convexity theorem implies that if $p<q<\infty$
\bea
\label{16}
\ln{\|\widetilde{T}^{n}\|_q}\leq \frac{p}{q}\ln{\|\widetilde{T}^{n}\|_p}+\left(1-\frac{p}{q}
\right)\ln{\|\widetilde{T}^{n}\|_\infty}
\eea
while if $1<q<p$
\bea
\label{17}
\ln{\|\widetilde{T}^{n}\|_q}\leq \left(\frac{1-1/q}{1-1/p}\right)
\ln{\|\widetilde{T}^{n}\|_p}+\left(1-\frac{1-1/q}{1-1/p}
\right)\ln{\|\widetilde{T}^{n}\|_1}.
\eea
From (\ref{16})-(\ref{17}) we have the interpolation relations
\bea
\label{16'}
\|\widetilde{T}^n\|_q &\leq& \|\widetilde{T}^n\|_p^{p/q}\|\widetilde{T}^n\|_\infty^{1-p/q}, \quad p<q<\infty\\
\label{17'}
\|\widetilde{T}^n\|_q &\leq &
\|\widetilde{T}^n\|_p^{(1-q^{-1})/(1-p^{-1})}\|\widetilde{T}^n\|_1^{1-(1-q^{-1})/(1-p^{-1})},\quad
1<q<p
\eea
which, along with (\ref{14})-(\ref{15}), imply
\bean
\|T^n_{\ep,\alpha}\|_{q,0}&\leq& 2\|T^n_{\ep,\alpha}\|_{p,0}^{p/q},\quad p<q<\infty\\
\|T^n_{\ep,\alpha}\|_{q,0}&\leq& 2\|T^n_{\ep,\alpha}\|_{p,0}^{(1-q^{-1})/(1-p^{-1})},
\quad 1<q<p
\eean
This proves that the order of divergence
of $n_{diss}(p)$ are the same for $1<p<\infty$.
Estimates (\ref{16'})-(\ref{17'}) also show
that the order of divergence of
$n_{diss}(1)$ and $n_{diss}(\infty)$ is at least as high as
$n_{diss}(p), 1<p<\infty$. $\qquad \blacksquare$

%%%%%%%%%%%%%%%%%%%%%%%%%%%%%%%%%%%%%%%%%%%%%%%%%%%%%%%%%%%%%%%%%%%%%%%%%%%%%%%%%%%%%%%%%%%%%%%%%%%%%%%%%

\subsection*{Proof of Proposition \ref{disto0}}

%%%%%%%%%%%%%%%%%%%%%%%%%%%%%%%%%%%%%%%%%%%%%%%%%%%%%%%%%%%%%%%%%%%%%%%%%%%%%%%%%%%%%%%%%%%%%%%%%%%%%%%%%

We prove the limit $d_\ep(1)\epto 0$ by contradiction. Assume that there is some constant $a\in(0,1)$ such that
for all $\ep>0$, the distance $d_\ep(1)>a$. We will show that the following triangle inequality holds:
\bea\label{distance}
\forall \ep>0,\quad d_\ep(1-a/2)> a/2.
\eea
First of all, notice that the assumption $d_\ep(1)>a$ means that for any $\lambda\in S^1$, $\|R_\ep(\lam)\|<a^{-1}$.
We apply the following identity \cite{Y}:
$$
R_\ep(\lam')=R_\ep(\lam)\{1+\sum_{n\geq 1}(\lam-\lam')^n R_\ep(\lam)^n\}
$$
with $\lam'=r\lam$, for $1-a<r<1$. Taking norm of both sides yields the bound
$\|R_\ep(\lam')\|\leq \frac{1}{r-(1-a)}$, uniformly w.r.t $\ep$.  
Since this upper bound holds for any $|\lam'|=r$, it shows that the spectral 
radius $r_{sp}(T_\ep)\leq 1-a$, and proves \eqref{distance} by taking $r=1-a/2$.
We can now use \eqref{distance} in the upper bound \eqref{second-upper} of 
Theorem~\ref{gb}:
this $\ep$-independent upper bound shows that ${\tau}_*$ remains finite in the 
limit $\ep\to 0$, which
contradicts Proposition~\ref{nonfinite}.\hfill$\blacksquare$

%%%%%%%%%%%%%%%%%%%%%%%%%%%%%%%%%%%%%%%%%%%%%%%%%%%%%%%%%%%%%%%%%%%%%%%%%%%%%%%%%%%%%%%%%%%%%%%%%%%%%%%%%

\subsection*{Proof of Lemma \ref{sum-int}}

%%%%%%%%%%%%%%%%%%%%%%%%%%%%%%%%%%%%%%%%%%%%%%%%%%%%%%%%%%%%%%%%%%%%%%%%%%%%%%%%%%%%%%%%%%%%%%%%%%%%%%%%%

Considering its decay at infinity, the function $f$ is automatically in $L^2(\IR^d)$. 
The function $\hat f^2$ is the Fourier transform of the self-convolution $f\ast f$. Therefore,
using the parity of $f$ and applying the Poisson summation formula to the LHS of \eqref{sum-integ} 
yields
\bea
\label{poisson}
\ep^d\sum_{\bk\in\IZ^d}\hat f(\ep\bk)^2=\int\hat{f}^{2}(\bxi)d\bxi+ 
\sum_{0\not=\bn\in\IZ^{d}}(f\ast f)\big(\frac{\bn}{\ep}\big).
\eea
A simple computation shows that
$(f\ast f)(\bx)$ also decays as fast as $|\bx|^{-M}$. This 
piece of information is now sufficient to control the RHS of \eqref{poisson}, yielding the result,
Eq.~\eqref{sum-integ}.
\hfill$\blacksquare$

%%%%%%%%%%%%%%%%%%%%%%%%%%%%%%%%%%%%%%%%%%%%%%%%%%%%%%%%%%%%%%%%%%%%%%%%%%%%%%%%%%%%%%%%%%%%%%%%%%%%%%%%%

\subsection*{Proof of Lemma \ref{gub}}

%%%%%%%%%%%%%%%%%%%%%%%%%%%%%%%%%%%%%%%%%%%%%%%%%%%%%%%%%%%%%%%%%%%%%%%%%%%%%%%%%%%%%%%%%%%%%%%%%%%%%%%%%

Using the notation introduced in Section \ref{Opt} one has
\bean
&&T_{\ep,\alpha}^{n}\fb_{\bk_{0}}=
(G_{\ep,\alpha}U)^{n}\fb_{\bk_{0}}
=(G_{\ep,\alpha}U)^{n-1}\sum_{0 \not = \bk_{1} \in \IZ^{d}}u_{\bk_{0},\bk_{1}}
e^{-\ep |\bk_{1}|^{2\alpha}}\fb_{\bk_{1}} \\ 
&=& \sum_{0 \not = \bk_{1},...,\bk_{n} \in \IZ^{d}}u_{\bk_{0},\bk_{1}}
u_{\bk_{1},\bk_{2}}...u_{\bk_{n-1},\bk_{n}} e^{-\ep \sum_{l=1}^{n}|\bk_{l}|^{2\alpha}}
\fb_{\bk_{n}}=\sum_{0 \not = \bk_{n} \in \IZ^{d}} 
\ml{U}_{n}(\bk_0,\bk_{n})\fb_{\bk_{n}}.
\eean

We note that for any $n$ and $\bk_{n} \in \IZ^{d}$, the sequence 
$\ml{U}_n(\bk_{0},\bk_{n})$ (indexed by $\bk_{0}\in \IZ^{d}$) belongs to $l^{2}(\IZ^{d})$.
Indeed, using the Cauchy-Schwarz inequality and identity (\ref{ul2}) one gets for $n=2$,
\bean
&&\sum_{0 \not = \bk_{0} \in \IZ^{d}}|\ml{U}_2(\bk_{0},\bk_{2})|^{2}=
\sum_{0 \not = \bk_{0} \in \IZ^{d}}\left|\sum_{0 \not = \bk_{1} \in \IZ^{d}}
u_{\bk_{0},\bk_{1}}u_{\bk_{1},\bk_{2}} 
e^{-\ep (|\bk_{1}|^{2\alpha}+|\bk_{2}|^{2\alpha})}\right|^{2} \\
&\leq&\sum_{0 \not = \bk_{0} \in \IZ^{d}}\sum_{0 \not = \bk_{1} \in \IZ^{d}}
|u_{\bk_{0},\bk_{1}}|^{2}e^{-\ep |\bk_{1}|^{2\alpha}}
\sum_{0 \not = \bk_{1} \in \IZ^{d}}|u_{\bk_{1},\bk_{2}}|^{2}
e^{-\ep |\bk_{1}|^{2\alpha}}e^{-2\ep |\bk_{2}|^{2\alpha}} \\
&\leq&\sum_{0 \not = \bk_{1} \in \IZ^{d}}e^{-\ep |\bk_{1}|^{2\alpha}}
\sum_{0 \not = \bk_{1} \in \IZ^{d}}e^{-\ep |\bk_{1}|^{2\alpha}}
e^{-2\ep |\bk_{2}|^{2\alpha}} = Ke^{-2\ep |\bk_{2}|^{2\alpha}},
\eean
where K denotes a constant which depends only on $\ep$ and $\alpha$. 
Similar estimates hold for $n>2$.

Now applying the Cauchy-Schwarz inequality in (\ref{Tnorm1}) we get
\bea
\|T_{\ep,\alpha}^{n}f\|^{2} &\leq&
\sum_{0 \not = \bk_{n} \in \IZ^{d}}
\sum_{\bk_{0} \in \ml{S}_{n}(\bk_{n})}
|\hat{f}(\bk_{0})|^{2}
\sum_{\bk_{0} \in \ml{S}_{n}(\bk_{n})}
|\ml{U}_n(\bk_{0},\bk_{n})|^{2}.
\eea $\qquad \blacksquare$

%%%%%%%%%%%%%%%%%%%%%%%%%%%%%%%%%%%%%%%%%%%%%%%%%%%%%%%%%%%%%%%%%%%%%%%%%%%%

%%%%%%%%%%%%%%%%%%%%%%%%%%%%%%%%%%%%%%%%%%%%%%%%%%%%%%%%%%%%%%%%%%%%%%%%%%%%%%%%%%%%%%%%%%%%%
%%%%%%%%%%%%%%%%%%%%%%%%%%%%%%%%%%%%%%%%%%%%%%%%%%%%%%%%%%%%%%%%%%%%%%%%%%%%%%%%%%%%%%%%%%%%%
%%%%%%%%%%%%%%%%%%%%%%%%%%%%%%%%%%%%%%%%%%%%%%%%%%%%%%%%%%%%%%%%%%%%%%%%%%%%%%%%%%%%%%%%%%%%%
%            C H A P T E R 3 : Dissipation Time of Classically Chaotic Systems              % 
%%%%%%%%%%%%%%%%%%%%%%%%%%%%%%%%%%%%%%%%%%%%%%%%%%%%%%%%%%%%%%%%%%%%%%%%%%%%%%%%%%%%%%%%%%%%%
%%%%%%%%%%%%%%%%%%%%%%%%%%%%%%%%%%%%%%%%%%%%%%%%%%%%%%%%%%%%%%%%%%%%%%%%%%%%%%%%%%%%%%%%%%%%%
%%%%%%%%%%%%%%%%%%%%%%%%%%%%%%%%%%%%%%%%%%%%%%%%%%%%%%%%%%%%%%%%%%%%%%%%%%%%%%%%%%%%%%%%%%%%%

\newpage

\markright{  \rm \normalsize Chapter 3. \hspace{0.25cm} Dissipation time of classically chaotic systems}
\chapter{Dissipation time of classically chaotic systems}
\label{DTCCS}
\thispagestyle{fancy}

\lhead[\thepage]{\rightmark}
\rhead[\leftmark]{\thepage}
\cfoot{}

\parindent=0cm

%%%%%%%%%%%%%%%%%%%%%%%%%%%%%%%%%%%%%%%%%%%%%%%%%%%%%%%%%%%%%%%%%%%%%%%%%%%%%%%%%%%%%%%%%%%%%
%%%%%%%%%%%%%%%%%%%%%%%%%%%%%%%%%%%%%%%%%%%%%%%%%%%%%%%%%%%%%%%%%%%%%%%%%%%%%%%%%%%%%%%%%%%%%

\section{Dissipation time of toral automorphisms}

%%%%%%%%%%%%%%%%%%%%%%%%%%%%%%%%%%%%%%%%%%%%%%%%%%%%%%%%%%%%%%%%%%%%%%%%%%%%%%%%%%%%%%%%%%%%%
%%%%%%%%%%%%%%%%%%%%%%%%%%%%%%%%%%%%%%%%%%%%%%%%%%%%%%%%%%%%%%%%%%%%%%%%%%%%%%%%%%%%%%%%%%%%%

%%%%%%%%%%%%%%%%%%%%%%%%%%%%%%%%%%%%%%%%%%%%%%%%%%%%%%%%%%%%%%%%%%%%%%%%%%%%%%%%%%%%%%%%%%%%%

\subsection{Preliminaries}
\label{prelim}

%%%%%%%%%%%%%%%%%%%%%%%%%%%%%%%%%%%%%%%%%%%%%%%%%%%%%%%%%%%%%%%%%%%%%%%%%%%%%%%%%%%%%%%%%%%%%

It is well known (see \cite{AP}) that (the lifting map corresponding to) any 
toral homeomorphism $H:\IT^{d} \mapsto \IT^{d}$ can be decomposed
into three parts $H=L+P+c$, where $L$, the linear part, is an element of $SL(d,\IZ)$ - the set of all 
matrices with integer entries and determinant equal to $\pm 1$, $P$ is periodic i.e. 
$P(\bx+\bv)=P(\bx)$ for any $\bv \in \IZ^{d}$, and $c$ is a constant shift vector.

Every algebraic and measurable automorphism of the torus is continuous.
Each continuous toral automorphism is a homeomorphism with zero periodic 
and constant parts and hence can be identified with an element of $SL(d,\IZ)$. 
And vice versa, each element of $SL(d,\IZ)$ uniquely determines a measurable, algebraic 
toral automorphism.
Thus from now on the term {\em toral automorphism}
will simply be reserved for elements of $SL(d,\IZ)$.
We recall here that all Anosov diffeomorphisms on $\IT^d$ are topologically conjugate 
to the toral automorphisms (\cite{Fr}, \cite{Man}).

Below we summarize a few definitions from ergodic theory along with some well known 
ergodic properties of toral automorphisms 
(cf. \cite{Katok} p. 160, \cite{Katznelson} and \cite{Arnold}).

\begin{defin}
\label{Ksystem}
A dynamical system $(\IT^{d},\mu,F)$ is called a K-system (possesses K-property) if there 
exists subalgebra $\cA$ of the algebra $\cM$ of all $\mu$-measurable sets such that
\begin{itemize}
\item{i)} $ \underset{n\in \IZ^{+}}{\forall}\quad \cA \subset F^{n}\cA $
\item{ii)} $\underset{n \geq 0}{\bigvee} F^{n}\cA = \cM$
\item{ii)} $\underset{n \geq 0}{\bigwedge} F^{-n}\cA = \cC$
\end{itemize}
where $\cC$ denotes the algebra of sets of measure $O$ or $1$. We note that the definition
easily extends to abstract noncomutative version (in this case $\cC$ stands for $c 1\!\!1$).
\end{defin}

\begin{prop}
\label{ergodicTA}
Let $F$ be a toral automorphism. 

The following statements are equivalent

a) no root of unity is an eigenvalue of $F$.

b) $F$ is ergodic.

c) $F$ is mixing.

d) $F$ is a K-system.

\end{prop}

In the sequel we will use the following result (cf. \cite{Yuz}).
\begin{prop} The entropy $h(F)$ of any toral endomorphism $F$ is computed by the formula
\bea
\label{entropy}
 h(F)=\sum_{|\lam_{j}| \geq 1}\ln{|\lam_{j}|},
\eea 
where $\lam_{j}$ denote the eigenvalues of $A$.
\end{prop}
From the formula (\ref{entropy}) one immediately sees that a toral automorphism has 
zero entropy iff all its eigenvalues are of modulus 1. In fact much stronger result holds
(for proof see Section \ref{Ch3proofs}).
\begin{prop}
\label{zeroentropy}
 A toral automorphism has zero entropy iff all its eigenvalues are roots of unity. 
 In particular all ergodic toral automorphisms have positive entropy.
\end{prop}

Given any toral automorphism $F$ we denote by $P$ its characteristic polynomial and by 
$\{P_{1},...,P_{s}\}$ the complete set of its distinct irreducible (over $\IQ$) factors. 
Let $d_{j}$ denote the degree of polynomial $P_{j}$ and $h_{j}$ the   
KS-entropy of any toral automorphism with the characteristic polynomial $P_{j}$. 
For each $P_{j}$ we define its dimensionally averaged KS-entropy as 
\bea
 \hat{h}_{j}=\frac{h_{j}}{d_{j}}.
\eea 

\begin{defin}
\label{MDAE}
Assuming the above notation, we define minimal dimensionally averaged entropy of $F$ (denoted $\hat{h}(F)$) as
\bean
 \hat{h}(F)=\min_{j=1,...,s}\hat{h}_{j} 
\eean
\end{defin}

%%%%%%%%%%%%%%%%%%%%%%%%%%%%%%%%%%%%%%%%%%%%%%%%%%%%%%%%%%%%%%%%%%%%%%%%%%%%%%%%%%%%%%%%%%%%%%

\subsection{Main theorems}
\label{MainT}

%%%%%%%%%%%%%%%%%%%%%%%%%%%%%%%%%%%%%%%%%%%%%%%%%%%%%%%%%%%%%%%%%%%%%%%%%%%%%%%%%%%%%%%%%%%%%%

In this section we state two main theorems of Part I of this work.
Both theorems concern the asymptotics of the dissipation time for toral automorphisms in
arbitrary dimension. Our main task is to derive not only the logarithmic order
of the asymptotics in the case of chaotic maps but also to find the exact value of the constant
and relate it to dynamical properties of the map via the connection with the minimal
dimensionally averaged KS-entropy of its irreducible blocks. 

In order to be able to perform exact calculations we need to specify more concretely the family
of noise kernels we are going to work with. Namely, we assume here that the noise kernel 
is $\alpha$-stable, cf. (\ref{Akernel}), that is
\bea
\label{A1kernel}
g_{\ep,\alpha}(\bx):=\sum_{\bk \in \IZ^{d}}e^{-|\ep \bk|)^{\alpha}}\be_{\bk}(\bx),
\eea
Under this assumption we have the following results regarding the dissipation time of
$\|T_{\ep,\alpha}^{n}\|$ (in order to emphasize that our calculations depend in an
essential way on the choice of the noise kernel we denote noisy operators associated with 
$\alpha$-stable kennels by of $T_{\ep,\alpha}$) 
\begin{thm}
\label{thm2}
Let $F$ be any toral automorphism, $U_{F}$ the Koopman operator
associated with $F$, $G_{\ep,\alpha}$ $\alpha$-stable noise 
operator and $T_{\ep,\alpha}=G_{\ep,\alpha}U_{F}$. Then
 
i) $T_{\ep,\alpha}$ has simple dissipation time iff $F$ is not ergodic.

ii) $T_{\ep,\alpha}$ has logarithmic dissipation time iff $F$ is ergodic. 

iii) If $T_{\ep,\alpha}$ has logarithmic dissipation time then the dissipation rate constant 
     satisfies the following constraint
\bean
\frac{1}{\hat{h}(F)} \leq R_{c} \leq \frac{1}{\tilde{h}(F)},
\eean
where $\tilde{h}(F)$ is a positive constant satisfying $\tilde{h}(F)\leq\hat{h}(F)$. 
\end{thm}

The natural question arises, whether the lower bound for the dissipation rate constant 
given in the above theorem is best possible. 
The next theorem and its corollary provides a strong argument in favor of this conjecture.

\begin{thm}
\label{thm3}
If $F$ is ergodic and diagonalizable then 
\bean
 \tau_{c}\approx \frac{1}{\hat{h}(F)}\ln(1/\ep).
\eean
That is, the dissipation rate constant of 
$T_{\ep,\alpha}$ is given by
\bean
 R_{c}=\frac{1}{\hat{h}(F)}.
\eean
\end{thm}

In this case ergodicity of toral automorphisms is equivalent
to hyperbolicity (in higher dimension it is not true, since there exists
so called quasihyperbolic automorphisms, which although ergodic, possesses
eigenvalues of modulus one \cite{B,RSO}). Two dimensional hyperbolic toral 
automorphisms are usually referred to as the \emph{cat maps}.
  
Using Corollary \ref{23dim} and applying Theorem \ref{thm3} in
two and three dimensions one gets the following
\begin{cor}[Cat maps]
\label{cat}
 Let $F$ be any ergodic, two or three dimensional toral automorphism. Then 
\bean
 \tau_{c}\approx \frac{1}{\hat{h}(F)}\ln(1/\ep),
\eean
\end{cor}

We end this section with a remark that toral automorphisms provide a good example on 
which robustness of the dissipation time (i.e. its independence of unimportant 
details of the underlying conservative dynamics) can be tested. To this end we
compare the dissipation time with another characteristics of chaoticity -
the decay of correlations.
As we will show in Section \ref{Anosov} the dissipation time has the same logarithmic 
asymptotics among general class of Anosov diffeomorphisms while the decorrelation may be 
exponential (generic Anosov case) or super-exponential depending on particular map.

Indeed, we illustrate this fact below by the following result (obtained in the similar way
as the above theorems) on the decay of correlations 
for $d$-dimensional toral automorphisms (for a proof see Section \ref{Ch3proofs}). 

\begin{prop}
\label{supexp}
Let $F$ be a diagonalizable ergodic toral automorphism, $U_{F}$ its Koopman operator 
and $\lam$ any constant such that  $0<\lam<\hat{h}(F)$. 
Then for any $f,h \in L^{2}_{0}(\IT^{2d})$ the correlation function for noisy 
dynamics generated by $F$ and and any $\alpha$-stable noise decays superexponentially i.e.
\bean
C^{\ep}_{f,h}(n) = \la \bar{f}, T_{\ep,\alpha}^n h \ra \leq \|f\|\,\|h\|\,e^{-\ep^{\alpha} \lam^{\alpha n}}
\eean
Moreover, let $f,\,h \in G_{\ep}(L^{2}_{0}(\IT^{2d}))$ be smooth observables, where 
$G_{\ep}$ denotes Gaussian noise operator. Then the decay of correlations of unperturbed
operator $U_{F}$ i.e. of toral automorphism itself is still superexponential
\bean
C_{f,h}(n)= \la \bar{f}, U_{F} h \ra \leq \|G_{\ep}^{-1}f\|\,\|G_{\ep}^{-1}h\|\,e^{-\ep^2 \lam^{2n}}.
\eean
\end{prop}

%%%%%%%%%%%%%%%%%%%%%%%%%%%%%%%%%%%%%%%%%%%%%%%%%%%%%%%%%%%%%%%%%%%%%%%%%%%%

\subsection{Asymptotic arithmetic minimization problem}
\label{sA}

%%%%%%%%%%%%%%%%%%%%%%%%%%%%%%%%%%%%%%%%%%%%%%%%%%%%%%%%%%%%%%%%%%%%%%%%%%%%

In this section we prepare the ground for the proof of theorems stated in previous 
section. To this end we need to derive a concrete version of the general formula 
(\ref{Tnorm3p}) obtained in Section \ref{Opt} for any toral automorphism and
arbitrary noise kernel.
Using the fact that here we consider only $\alpha$-stable kernels the formula
(\ref{Tnorm3p}) can be rewritten as follows
\bea
\label{Tnorm3}
\|T_{\ep,\alpha}^{n}\| &=& \max_{0 \not=\bk \in \IZ^{d}} \prod_{l=1}^{n}\hat{g}_{\ep}(A^l\bk)=
e^{-\ep^{\alpha} \min_{0 \not=\bk \in \IZ^{d}} \sum_{l=1}^{n}|A^{l}\bk|^{\alpha}}. 
\eea
Hence, for toral automorphisms and $\alpha$-stable kernels, the calculation of the 
dissipation time reduces to the following nonlinear, asymptotic (large $n$) 
arithmetic minimization problem 
\bea 
\label{artmin0}
\min_{0 \not=\bk \in \IZ^{d}} \sum_{l=1}^{n}|A^{l}\bk|^{\alpha},
\eea
where $A\in SL(d,\IZ)$.
When $A$ is not ergodic the asymptotics of (\ref{artmin0}) is 
clearly of the order $O(n)$. Thus we will be only concerned with the ergodic case. 
For $d=2$ the problem (\ref{artmin0}) can be solved easily as follows. 
Consider first the case that $A$ is symmetric and $\alpha=2$.
From $\det(A)=1$ we see that eigenvalues are $\lam, \lam^{-1}$ with
$|\lam|>1$.
We have
\bean
&&\min_{0 \not =\bk \in \IZ^{d}} \sum_{l=1}^{2n+1}|A^{l}\bk|^{2}=
\min_{0 \not=\bk \in \IZ^{d}} \sum_{l=-n}^{n}|A^{l}\bk|^{2}\\
&=&\min_{0 \not=\bk=\bk_{1}+\bk_{2} \in \IZ^{d}}
\left( |\bk|^{2} + \sum_{l=1}^{n}|\lam|^{2l}|\bk_{1}|^{2}+|\lam|^{-2l}|\bk_{2}|^{2}
+\sum_{l=1}^{n}|\lam|^{-2l}|\bk_{1}|^{2}+|\lam|^{2l}|\bk_{2}|^{2}\right)\\
&=&
\min_{0 \not=\bk \in \IZ^{d}}\sum_{l=-n}^{n}|\lam|^{2l}|\bk|^{2}
=\sum_{l=-n}^{n}|\lam|^{2l}.
\eean
Hence there exist constants $C_{1}$ and $C_{2}$ such that 
\bean
C_{1}e^{h(A)n} \leq \min_{0 \not=\bk \in \IZ^{d}} \sum_{l=1}^{n}|A^{l}\bk|^{2}
\leq C_{2}e^{h(A)n}.
\eean
where $h(A)$ denotes the KS-entropy of $A$.
The estimates for the general case of non-symmetric $A$ and $\alpha \neq 2$
are similar. 

In higher dimensions, the solution to (\ref{artmin0}) is much more involved
because of the presence of different eigenvalues with absolute values
bigger than one. We have the following general estimate 
\begin{thm}
\label{thmart}
Let $A\in SL(d,\IZ)$ be ergodic.
There exist constants $C_{1}$ and $C_{2}$ such that for any $0<\del<1$ and sufficiently 
large $n$
\bea
\label{ulestimateA}
C_{1} e^{(1-\del)\alpha\tilde{h}(A) n} \leq
\min_{0 \not=\bk \in \IZ^{d}} \sum_{l=1}^{n}|A^{l}\bk|^{\alpha} \leq
C_{2}n e^{\alpha \hat{h}(A)n}
\eea
where as before $\hat{h}(A)$ denotes minimal dimensionally averaged entropy of $A$
and $\tilde{h}(A)$ denotes a constant satisfying $0<\tilde{h}(A)\leq \hat{h}(A)$,
with equality achieved for all diagonalizable matrices $A$.
\end{thm}
The question whether the equality $\tilde{h}(A)=\hat{h}(A)$ holds for all ergodic 
matrices remains open.

The proof of the theorem relies on nontrivial use of three number-theoretical results stated below.

\textbf{I. Minkowski's Theorem on linear forms}

{\em Let $L_{1},...,L_{d}$ be linearly independent linear forms on $\IR^{d}$ which are real
or occur in conjugate complex pairs. Suppose $a_{1},a_{2},...,a_{d}$ are real positive 
numbers satisfying $a_{1}a_{2}...a_{d}=1$ and $a_{i}=a_{j}$, whenever $L_{i}=\bar{L}_{j}$. 
Then there exists a nonzero integer vector
$\bk \in \IZ^{d}$ such that for every $j=1,...,d$,
\bea
\label{minko}
|L_{j}\bk| \leq Da_{j},
\eea
where $D=|\det[L_{1},...,L_{d}]|^{1/d}$.
}

Minkowski's Theorem on linear forms will be used to obtain a sharp upper bound
on the asymptotic solution of the arithmetic minimization problem.
The proof of the above theorem and its generalization to arbitrary lattices can be
found in \cite{Newman} (Chap. VI). 

\textbf{II. Schmidt's Subspace Theorem}

{\em Let $L_{1},...,L_{d}$ be  linearly independent linear forms on $\IR^{d}$ with real
or complex algebraic coefficients. Given $\del>0$, there are finitely many proper
rational subspaces of $\IR^{d}$ such that every nonzero integer vector $\bk$ with 
\bea
\label{Schmi}
\prod_{j=1}^{d}|L_{j}\bk|<|\bk|^{-\del}
\eea   
lies in one of these subspaces.
}

Schmidt's Subspace Theorem will be used in conjunction with
Van der Waerden's Theorem on arithmetic progressions (see below)
to obtain a sharp lower bound for the asymptotic solution
of the arithmetic minimization problem.
The proof of
Schmidt's Subspace Theorem can be found in \cite{Schmidt} (Theorem 1F, p. 153).

\begin{defin}
\label{exept}
For a given set of linear forms and for fixed $\del>0$, 
the smallest collection of proper rational subspaces of $\IR^{d}$ which
contain all nonzero integer vectors satisfying (\ref{Schmi}), is called the exceptional 
set and denoted by $E_{\del}$. 
\end{defin}

A main difficulty to be resolved in using Schmidt's Subspace Theorem
is to show that the minimizer of either the original problem (\ref{artmin0})
or an equivalent problem does not lie in the respective exceptional set which
is in general unknown.
We will pursue the latter route by using
Van der Waerden's Theorem on arithmetic progressions to
show that one can always construct an equivalent minimization problem
whose minimizer is guaranteed to lie outside the corresponding exceptional
set.
To this end
we note that Schmidt's Subspace Theorem is true when the standard lattice $\IZ^{d}$ 
is replaced by any other rational lattice, that is any 
lattice of the form $\Lambda=Q(\IZ^{d})$ where $Q\in GL(d,\IQ)$. 
Schmidt's subspace theorem  can be generalized  to this situation
by considering the set of new forms 
$\tilde{L}_{j}=L_{j}Q$. The fact that $Q \in GL(d,\IQ)$ 
implies immediately that $\tilde{L}_{j}$ are still linearly independent forms on 
$\IR^{d}$ with real or complex algebraic coefficients. 

\textbf{III. Van der Waerden's Theorem on arithmetic progressions}

{\em Let $k$ and $d$ be two arbitrary natural numbers. Then there exists a
natural number $n_*(k,d)$ such that, if an arbitrary segment of length $n\geq n_*$
of the sequence of natural numbers is divided in any manner into $k$
(finite) subsequences, then an arithmetic progression
of length $d$ appears in at least one of these subsequences. 
}

The original proof was published in \cite{vdW};
Lukomskaya's simplification can be found in \cite{Lukomskaya}.

Before presenting the proof of our main results
we state a number of technical facts
concerning the structure of toral automorphisms.

%%%%%%%%%%%%%%%%%%%%%%%%%%%%%%%%%%%%%%%%%%%%%%%%%%%%%%%%%%%%%%%%%%%%%%%%%%%%%%%%%%%%%%%%%%%%%%

\subsection{Algebraic structure of toral automorphisms}

%%%%%%%%%%%%%%%%%%%%%%%%%%%%%%%%%%%%%%%%%%%%%%%%%%%%%%%%%%%%%%%%%%%%%%%%%%%%%%%%%%%%%%%%%%%%%%

\label{sAA}
In this section we denote by $GL(d,\IQ)$ the group of nonsingular $d\times d $ matrices with
rational entries
or the group of linear operators on Euclidean space $\IR^{d}$, which are represented in standard 
basis by such matrices. We generally use the same symbol to denote both operator and its matrix.  

In the sequel a vector $x \in \IR^{d}$ will be called an integer (or integral) vector if all 
its components are integers, and similarly a rational, an algebraic vector if all its 
components are rational or respectively algebraic numbers.
The term {\em rational subspace of $\IR^{d}$} will then refer to a linear subspace of $\IR^{d}$
spanned by rational vectors (cf. \cite{Schmidt} p. 113).  

\begin{defin}
$A \in GL(d,\IQ)$ is called irreducible (over $\IQ$) if its characteristic polynomial 
is irreducible in $\IQ[x]$.
\end{defin}

\begin{lem}
\label{irred}
The following statements about a matrix $A \in GL(d,\IQ)$ are equivalent.

a) $A$ is irreducible.

b) $A$ does not possess any proper rational $A$-invariant subspaces of $\IR^{d}$.

c) No rational proper subspace of $\IR^{d}$ is contained in any proper 
   $A$-invariant subspace of $\IR^{d}$.

d) For any nonzero $\bq \in \IQ^{d}$ and any arithmetic progression of integer 
   numbers $n_{1},...,n_{d}$, the set 
   $\{A^{n_{1}}\bq,A^{n_{2}}\bq,...,A^{n_{d}}\bq\}$ forms a basis of $\IR^{d}$.

e) $A^\dagger$ is irreducible. 

f) No nonzero $\bq \in \IQ^{d}$ is orthogonal to any proper $A$-invariant 
   subspace of $\IR^{d}$.

g) No proper $A$-invariant subspace of $\IR^{d}$ is contained in any proper
   rational subspace of $\IR^{d}$.

\end{lem}

\begin{defin}
We say that operator $A \in GL(d,\IQ)$ is completely decomposable over $\IQ$ if there 
exists a rational basis of $\IR^{d}$ in which $A$ admits the following block diagonal form  
\bea
\label{blockdiag}
\begin{bmatrix}
A_{1} & 0     & ...& 0    \\
0     & A_{2} & ...& 0    \\
...   & ...   & ...& ...  \\
0     & 0     & ...& A_{r}   
\end{bmatrix}, 
\eea
where for each $j=1,...,r \leq d$, $A_{j} \in GL(d_{j},\IQ)$ is irreducible and 
$\sum_{j=1}^r d_{j}=d$.
\end{defin}

In general, any matrix $A \in GL(d,\IQ)$ admits a rational block diagonal 
representation $[A_{j}]_{j=1,...,r}$. The smallest rational blocks to 
which $A$ can be decomposed are called elementary divisor blocks.
The characteristic polynomial corresponding to any elementary divisor block is of the form 
$p^{m}$, where $p$ is an irreducible (over $\IQ$) polynomial
 (see, e.g., \cite{Dummit}).
Although elementary divisor blocks cannot be decomposed over $\IQ$ into smaller invariant blocks, 
some elementary divisor blocks may not be irreducible. This happens
iff $m>1$ iff
$A$ is not completely decomposable over $\IQ$. One has the following
elementary fact (see Section \ref{Ch3proofs} for a proof).  
\begin{prop}
\label{red}
 $A \in GL(d,\IQ)$ is completely decomposable over $\IQ$ iff $A$ is diagonalizable.
\end{prop}

However, even if $A\in GL(d,\IQ)$ is not completely decomposable, each elementary divisor block 
of $A$ can be uniquely represented (in a rational basis) in the following block upper triangular form
\bea
\label{but}
\begin{bmatrix}
B & C  \\
0 & D    
\end{bmatrix}, 
\eea
where $B$ is the unique rational irreducible sub-block associated with $A$-invariant
rational subspace of that elementary divisor and $C$, $D$ denote some rational matrices.

\begin{prop}
\label{distinct}
All the eigenvalues of an irreducible matrix $A\in GL(d,\IQ)$ are
distinct (complex) algebraic numbers. In particular all irreducible matrices are diagonalizable.
\end{prop}
The proofs of the above  propositions can be found in Appendix B.

\commentout{
\begin{defin}
We say that matrix $A\in GL(d,\IQ)$ is indecomposable if it does not possesses any proper 
elementary divisor blocks.
\end{defin}

It is easy to construct a nonergodic toral automorphism which has positive entropy. 
One can simply consider a toral automorphism composed of two invariant blocks:
one nonergodic and the other one ergodic. Obviously the resulting automorphisms is not 
indecomposable. The question whether there exists indecomposable nonergodic toral 
automorphism of positive entropy has a negative answer, as the following
proposition shows. 

\begin{prop}
\label{underg}
Let $F$ be an indecomposable toral automorphism. Then $F$ has positive entropy iff $F$ is ergodic.
\end{prop}

For the proofs of the above propositions we refer the reader to Appendix B.

\begin{cor}
 Let $F$ be an indecomposable and nonergodic toral automorphism. Then all eigenvalues
of $F$ are roots of unity. In particular its entropy is zero.
\end{cor}
}

Finally we note that since the leading coefficient and constant term of a characteristic 
polynomial of any toral automorphism are equal to 1, the only possible rational eigenvalues 
of such map are $\pm 1$ or $\pm i$. The latter fact implies that ergodic toral automorphisms 
do not possesses rational eigenvalues. Thus we have the following

\begin{cor}
\label{23dim}
 Let $F$ be an ergodic, two or three dimensional toral automorphism. Then $F$ is 
 irreducible (and hence diagonalizable). 
\end{cor}

%%%%%%%%%%%%%%%%%%%%%%%%%%%%%%%%%%%%%%%%%%%%%%%%%%%%%%%%%%%%%%%%%%%%%%%%%%%%%%%%%%%%%%%%%%%

\subsection{The Proof of Theorems \ref{thm2}, \ref{thm3} and \ref{thmart}}
\label{sAP}

%%%%%%%%%%%%%%%%%%%%%%%%%%%%%%%%%%%%%%%%%%%%%%%%%%%%%%%%%%%%%%%%%%%%%%%%%%%%%%%%%%%%%%%%%%%

This section is entirely devoted to the proofs of the main theorems of this chapter. 
We start with the proof of Theorem \ref{thmart}, which constitutes the main
ingredient in the proofs of Theorems \ref{thm2} and \ref{thm3}.
 
Let $[A_{j}]_{j=1,...,r}$ be a rational block-diagonal decomposition 
of $A$ into elementary divisor blocks. Since $A\in SL(d,\IZ)$, there exist 
a transition matrix $Q\in SL(d,\IQ)$ such that for every $l\in\IZ$, 
\bean
A^{l}=Q^{-1}([A_{j}])^{l}Q
\eean
and moreover each elementary divisor block $[A]_{j}$ is represented in its 
block upper triangular form (\ref{but}).   

The matrix $Q$ defines a new lattice $\Lambda=Q(\IZ^{d})$ and acts bijectively between 
this lattice and the standard lattice $\IZ^{d}$. Hence   
\bean 
\min_{0 \not=\bk \in \IZ^{d}} \sum_{l=1}^{n}|A^{l}\bk|^{\alpha}=
\min_{0 \not=\bk \in \IZ^{d}} \sum_{l=1}^{n}|Q^{-1}([A_{j}])^{l}Q\bk|^{\alpha}=
\min_{0 \not=\bq \in \Lambda} \sum_{l=1}^{n}|Q^{-1}([A_{j}])^{l}\bq|^{\alpha}.
\eean      
Moreover
\bean
\|Q\|^{-\alpha}|([A_{j}])^{l}\bq|^{\alpha}\leq
|Q^{-1}([A_{j}])^{l}\bq|^{\alpha}\leq
\|Q^{-1}\|^{\alpha}|([A_{j}])^{l}\bq|^{\alpha}, \quad\forall l, j, \alpha.
\eean
Now we decompose $\Lambda$ into the direct sum of lower dimensional sublattices $\Lambda_{j}$
corresponding to invariant blocks $[A_{j}]$. So that
\bea
\label{minimal}
\min_{0 \not=\bq \in \Lambda} \sum_{l=1}^{n}|([A_{j}])^{l}\bq|^{\alpha}=
\min_{j\in \{1,...,r\}}\min_{0 \not=\bq \in \Lambda_{j}} \sum_{l=1}^{n}|(A_{j})^{l}\bq|^{\alpha}.
\eea
Thus, without loss of generality, we may  specialize to the case that $A$ is 
already indecomposable over $\IQ$ i.e.  $A$ 
does not possesses any proper elementary divisor blocks.
To simplify the notation we will work with the standard lattice 
$\Lambda=\IZ^{d}$. According to  the
remarks following the statements of Minkowski's and Schmidt's 
Theorems  the proof can be easily adapted for
 any rational lattice $\Lambda=Q(\IZ^{d})$. 

Since the technique of the proof differs depending on diagonalizability of $A$
we consider two cases:

%%%%%%%%%%%%%%%%%%%%%%%%%%%%%%%%%%%%%%%%%%%%%%%%%%%%%%%%%%%%%%%%%%%%%%%%%%%%%%%%%%%%%%%%%%%%%%%%%%%

\subsection*{Diagonalizable case}

%%%%%%%%%%%%%%%%%%%%%%%%%%%%%%%%%%%%%%%%%%%%%%%%%%%%%%%%%%%%%%%%%%%%%%%%%%%%%%%%%%%%%%%%%%%%%%

Here we concentrate on the case when $A$ is diagonalizable and hence due 
to its in-decomposability irreducible (cf. Proposition \ref{red}).

We denote by $\lam_{j}$ ($j=1,...,d$) the eigenvalues of $A$.
Following Proposition \ref{distinct} we note that $\lam_{j}$ are distinct (possibly complex) algebraic 
numbers and hence there exists a basis (of $\IC^{d}$) $\{\bv_{j}\}_{j=1,...,d}$ composed of 
normalized algebraic eigenvectors corresponding to eigenvalues $\lam_{j}$.

We denote by $[P_{j}]_{j=1}^d$ the projections on $[\bv_{j}]$, and by $[L_{j}]$ the 
corresponding linear forms. It is easy to check that $[L_{j}]$ are given, in the Riesz identification, by 
the eigenvectors $[\bu_j]$ of the matrix $A^\dagger$ which are co-orthogonal to
$[\bv_j]$, i.e., $\la \bu_i,\bv_j\ra=0$ for $i\not=j$.
$[\bu_j]$ and $[\bv_j]$ are real or
occur in complex conjugate pairs. We have
\bean
 \bx=\sum_{j=1}^{d}P_{j}\bx=\sum_{j=1}^{d}(L_{j}\bx)\bv_{j}
=\sum_{j=1}^d\la \bx,\bu_j\ra \bv_j, \quad\forall\bx\in \IR^d.
\eean  

The equivalence between any two norms in a finite dimensional vector space, implies the existence 
of absolute constants $C_{1},C_{2}$ such that
\bean
 C_{1} \sum_{j=1}^{d}|P_{j}\bx|^{2} \leq |\bx|^{2} \leq C_{2} \sum_{j=1}^{d}|P_{j}\bx|^{2}.
\eean
Using the above inequalities, the monotonicity of a map $\bx \mapsto \bx^{\alpha}$ and 
an obvious inequality $(a+b)^{\alpha} \leq a^{\alpha}+ b^{\alpha}$, which holds for all 
positive $a,b$ and $\alpha \in (0,1]$ one obtains the following estimates
\bean
\sum_{l=1}^{n}|A^{l}\bk|^{\alpha} &\leq&
\sum_{l=1}^{n}\left(C_{2}\sum_{j=1}^{d}|P_{j}A^{l}\bk|^{2}\right)^{\alpha}=
C_{2}^{\alpha}\sum_{l=1}^{n}\left(\sum_{j=1}^{d}|\lam_{j}|^{2l}|P_{j}\bk|^{2}\right)^{\alpha}\\
&\leq& C_{2}^{\alpha}\sum_{l=1}^{n}\sum_{j=1}^{d}|\lam_{j}|^{\alpha l}|P_{j}\bk|^{\alpha}=
C_{2}^{\alpha}\sum_{j=1}^{d}\left(\sum_{l=1}^{n}|\lam_{j}|^{\alpha l}\right)|P_{j}\bk|^{\alpha}
\eean
and on the other hand
\bean
\sum_{l=1}^{n}|A^{l}\bk|^{\alpha} &\geq&
\left(\sum_{l=1}^{n}|A^{l}\bk|^{2}\right)^{\alpha} \geq
\left(\sum_{l=1}^{n} C_{1}\sum_{j=1}^{d}|P_{j}A^{l}\bk|^{2}\right)^{\alpha}\\
&=&C_{1}^{\alpha}\left(\sum_{l=1}^{n}\sum_{j=1}^{d}|\lam_{j}|^{2l}|P_{j}\bk|^{2}\right)^{\alpha}=
C_{1}^{\alpha}\left(\sum_{j=1}^{d}\left(\sum_{l=1}^{n}|\lam_{j}|^{2l}\right)|P_{j}\bk|^{2}\right)^{\alpha}.
\eean
Now we introduce some notation 
\bea
\label{lamhat}
\hat{\lam}_{j}&:=&\max\{1,|\lam_{j}|\},\\
\label{hatlamgeo}
\hat{\lam}_{geo}&:=&\left(\prod_{j=1}^{d}\hat{\lam}_{j} \right)^{1/d}. 
\eea

One can easily observe that there exists a constant $C$ such that
\bean
 C\hat{\lam}_{j}^{\alpha n}\leq \sum_{l=1}^{n}|\lam_{j}|^{\alpha l}
 \leq n\hat{\lam}_{j}^{\alpha n}.
\eean
In the sequel we do not distinguish between particular values of constants appearing 
in computations. The symbols $C_{1},C_{2},..$ are used to denote any generic
constants independent of $n$.

The normalization condition $|\bv_j|=1$ implies the following relation
\bea
\label{normalization}
|P_{j}\bx|=|L_{j}\bx|. 
\eea
Combining 
the above estimates one gets the following general bounds
\bea
\label{ulestimate}
C_{1}\left(\sum_{j=1}^{d} \hat{\lam}_{j}^{2 n}|L_{j}\bk|^{2}\right)^{\alpha} \leq
\sum_{l=1}^{n}|A^{l}\bk|^{\alpha} \leq
C_{2}n\sum_{j=1}^{d} \hat{\lam}_{j}^{\alpha n}|L_{j}\bk|^{\alpha}.
\eea
Therefore in order to estimate (\ref{artmin0}) it suffices, essentially, to estimate
\bea
\label{artmin2}
\min_{0 \not=\bk \in \IZ^{d}} \sum_{j=1}^{d}\hat{\lam}_{j}^{\alpha n}|L_{j}\bk|^{\alpha}.
\eea
We denote by ${\bz}_{n}$ the sequence of minimizers i.e. nonzero integral vectors
solving (\ref{artmin2}).

%%%%%%%%%%%%%%%%%%%%%%%%%%%%%%%%%%%%%%%%%%%%%%%%%%%%%%%%%%%%%%%%%%%%%%%%%%%%%%%%%%%%%%%%%%%%%%
\subsubsection*{Upper bound.}
%%%%%%%%%%%%%%%%%%%%%%%%%%%%%%%%%%%%%%%%%%%%%%%%%%%%%%%%%%%%%%%%%%%%%%%%%%%%%%%%%%%%%%%%%%%%%%

For the upper bound we assign to the set of linear forms $L_{j}$ the set $\mathcal{A}$ 
composed of all real vectors $\ba=(a_{1},...,a_{d})$ satisfying the conditions
$a_{j}>0$, for $j=1,...,d$ and $a_{i}=a_{j}$ whenever $L_{i}=\bar{L}_{j}$ and 
\bea
\label{constr}
\prod_{j=1}^{d}a_{j}=1.
\eea
From Minkowski's theorem on linear forms, we know that
for any $\ba\in\mathcal{A}$, there exists nonzero integral vector $\bk_{\ba}$ satisfying 
$|L_{j}\bk_{\ba}| \leq Da_{j}$, $j=1,...,d$, where $D=|\det[L_{1},...,L_{d}]|^{1/d}$.
 
Thus
\bea
\label{up1}
\sum_{j=1}^{d}\hat{\lam}_{j}^{\alpha n}|L_{j}\bk_{\ba}|^{\alpha} \leq 
D\sum_{j=1}^{d}\hat{\lam}_{j}^{\alpha n}a_{j}^{\alpha}.
\eea
The minimizing property of $\bz_{n}$ implies that for 
any $\ba \in \mathcal{A}$,
\bea
\label{up3}
\sum_{j=1}^{d}\hat{\lam}_{j}^{\alpha n}|L_{j}{\bz}_{n}|^{\alpha} \leq 
\sum_{j=1}^{d}\hat{\lam}_{j}^{\alpha n}|L_{j}\bk_{\ba}|^{\alpha}.
\eea
Thus combining (\ref{up1}) and (\ref{up3}), and applying the Lagrange multipliers 
minimization with the constraint (\ref{constr}) (and using the fact
that $\hat{\lam}_{i}=\hat{\lam}_{j}$ whenever $L_{i}=\bar{L}_{j}$), we get
\bea
\label{ubound1}
\sum_{j=1}^{d}\hat{\lam}_{j}^{\alpha n}|L_{j}{\bz}_{n}|^{\alpha} \leq 
D\min_{\ba\in\mathcal{A}}\sum_{j=1}^{d}\hat{\lam}_{j}^{\alpha n}a_{j}^{\alpha}=
dD\left(\prod_{j=1}^{d}\hat{\lam}_{j}^{\alpha n}\right)^{1/d}=
dD\hat{\lam}_{geo}^{2 \alpha n}.
\eea 
Thus the
following upper bound holds
\bea
\label{ubound}
 \min_{0 \not=\bk \in \IZ^{d}} \sum_{l=1}^{n}|A^{l}\bk|^{\alpha}\leq
 C_{2} n\hat{\lam}_{geo}^{2 \alpha n}.
\eea

%%%%%%%%%%%%%%%%%%%%%%%%%%%%%%%%%%%%%%%%%%%%%%%%%%%%%%%%%%%%%%%%%%%%%%%%%%%%%%%%%%%%%%%%%%%%%%
\subsubsection*{Lower bound.}
%%%%%%%%%%%%%%%%%%%%%%%%%%%%%%%%%%%%%%%%%%%%%%%%%%%%%%%%%%%%%%%%%%%%%%%%%%%%%%%%%%%%%%%%%%%%%%

Let $m$ denote an arbitrary natural number. Using the fact that $A$ acts
bijectively on $\IZ^{d}$ we can restate the minimization problem (\ref{artmin2}) 
in the following form 
\bea
\label{artmin3}
\min_{0 \not=\bk \in \IZ^{d}}  \sum_{j=1}^{d}\hat{\lam}_{j}^{\alpha n}|L_{j}\bk|^{\alpha} &=&
\min_{0 \not=\bk \in \IZ^{d}}  
\sum_{j=1}^{d}\hat{\lam}_{j}^{\alpha n}|L_{j}A^{-m}A^{m}\bk|^{\alpha} \\
&=&\min_{0 \not=\bk \in \IZ^{d}}  
\sum_{j=1}^{d}\hat{\lam}_{j}^{\alpha n}|\lam_{j}|^{-\alpha m}|L_{j}A^{m}\bk|^{\alpha}
\eea
That is
\bea
\label{eqref3}
\sum_{j=1}^{d}\hat{\lam}_{j}^{\alpha n}|L_{j}{\bz}_{n}|^{\alpha}=
\sum_{j=1}^{d}\hat{\lam}_{j}^{\alpha n}|\lam_{j}|^{-\alpha m}|L_{j}A^{m}\bz_{n}|^{\alpha}.
\eea
We choose arbitrary $\del>0$ and consider the exceptional set $E_{\del}$ 
(see Definition \ref{exept}) associated with the system of linear forms $[L_{j}]$. 
Since $[L_{j}]$ correspond to the eigen-pairs $[\bar{\lam}_j, \bu_j]$  of $A^\dagger$
they are linearly
independent linear forms with (real or complex) algebraic coefficients. Thus the subspace theorem 
asserts that $E_{\del}$ is a finite collection of proper rational subspaces of $\IR^{d}$.   
We denote by $k_{\del}$ the number of subspaces forming $E_{\del}$. 

Now we want to show that for all sufficiently large $n$ there exist an integer
$m\leq n$ such that $A^{m}\bz_{n}$ 
does not lie in any element of $E_{\del}$. 
To this end we assume to the contrary that all $A^{m}\bz_{n}$ lie in the subspaces forming $E_{\del}$ and we 
divide the sequence of natural numbers $1,...,n$ into $k_{\del}$ classes in such 
a way that two numbers $m_{1}$ and $m_{2}$ are in the same class if $A^{m_{1}}\bz_{n}$ and 
$A^{m_{2}}\bz_{n}$ lie in the same element of $E_{\del}$.
Now let $n_*(k_{\del},d)$ be the number given in the van der Waerden theorem and let
$n\geq n_*$. 
Then there exists an arithmetic progression $m_{1},...,m_{d}$ in one of these subsequences.
By Lemma \ref{irred} d) the set of vectors
$\{A^{m_{1}}\bz_{n},A^{m_{2}}\bz_{n},...,A^{m_{d}}\bz_{n}\}$ 
forms a basis of the whole space $\IR^{d}$,
which contradicts the fact that they lie in one fixed rational proper subspace.
Hence for any $\del>0$ and $n\geq n_*$ there exists $m_*\leq n$ such that 
$A^{m_*}{\bz}_{n}$ does not lie in any element of  $E_{\del}$.

Now, introducing the notation
\bea
\label{minrel1}
\hat{\bz}_{n}=A^{m_*}{\bz}_{n}
\eea
one concludes from (\ref{eqref3}) that for any $\del >0$ and all $n \geq n_*$
the following equality and estimate hold
\bea
\label{minrel2}
\sum_{j=1}^{d}\hat{\lam}_{j}^{\alpha n}|L_{j}{\bz}_{n}|^{\alpha}&=&
\sum_{j=1}^{d}\hat{\lam}_{j}^{\alpha n}|\lam_j|^{-\alpha m_{*}}|L_{j}\hat{\bz}_{n}|^{\alpha}\\
\label{constr1}
\prod_{j=1}^{d}|L_{j}\hat{\bz}_{n}| &\geq& \frac{1}{|\hat{\bz}_{n}|^{\del}}.
\eea
Inequality (\ref{constr1}) may be rewritten as 
\bea
\label{f}
  \prod_{j=1}^{d} |L_{j}\hat{\bz}_{n}|=\frac{1}{f(|\hat{\bz}_{n}|)^{\del}}
\eea
with some $f:\IR^+\to\IR^+$ such that $f(r)\leq r, \forall r>0$.

Using (\ref{minrel1}) and (\ref{ubound1}) we obtain the existence of 
a constant $\lam>1$ such that
\bea
\nonumber
f(|\hat{\bz}_{n}|) &\leq& 
|\hat{\bz}_{n}|=
|A^{m_*}{\bz}_{n}| \leq 
\hat{\lam}_{max}^{m_*}|{\bz}_{n}| \\
\label{minbound}
&\leq& 
\hat{\lam}_{max}^{n}\sum_{j=1}^{d}\hat{\lam}_{j}^{n}|L_{j}{\bz}_{n}| \leq
dD(\hat{\lam}_{max}\hat{\lam}_{geo})^{n} \leq
\lam^{n}.
\eea
Note that $\prod_{j}\lam_j=1$. So, by (\ref{f}) the quantities 
$B_{j,n}=\left(|\lam_{j}|^{-m_*}f(|\hat{\bz}_{n}|)^{\del/d}|L_{j}\hat{\bz}_{n}|\right)^{\alpha}$, $j=1,...,d$ 
satisfy the constraint
\bea
\label{newc}
\prod_{j=1}^{d} B_{j,n}=1,\quad\forall n>n_*.
\eea
Thus applying (\ref{minbound}) and the Lagrange multipliers minimization with the
constraint  (\ref{newc}) one gets
\bean
\sum_{j=1}^{d}\hat{\lam}_{j}^{\alpha n}|\lam_{j}|^{-\alpha m_{*}}|L_{j}\hat{\bz}_{n}|^{\alpha}
= f(|\hat{\bz}_{n}|)^{-\alpha\del/d}\sum_{j=1}^{d}\hat{\lam}_{j}^{\alpha n}B_{j,n}\geq 
  \lam^{-2 \alpha n\del/d}\hat{\lam}_{geo}^{\alpha n}=:  
  \hat{\lam}_{geo}^{\alpha n(1-\hat{\del})}.
\eean
This and equality (\ref{minrel2}) yields the following lower bound for (\ref{artmin2})
\bea
\label{dlbound}
\min_{0 \not=\bk \in \IZ^{d}} \sum_{j=1}^{d}\hat{\lam}_{j}^{\alpha n}|L_{j}\bk|^{\alpha} \geq 
\hat{\lam}_{geo}^{2 \alpha n(1-\hat{\del})}.
\eea

%%%%%%%%%%%%%%%%%%%%%%%%%%%%%%%%%%%%%%%%%%%%%%%%%%%%%%%%%%%%%%%%%%%%%%%%%%%%%%%%%%%%%%%%%%%

\subsection*{Non-diagonalizable case}

%%%%%%%%%%%%%%%%%%%%%%%%%%%%%%%%%%%%%%%%%%%%%%%%%%%%%%%%%%%%%%%%%%%%%%%%%%%%%%%%%%%%%%%%%%%%%%

We move on to the general case where $A$ is not irreducible (but,
as assumed at the beginning of the proof, indecomposable over $\IQ$).
We denote by $B$ the invariant irreducible sub-block of $A$ given by its block upper triangular
decomposition (\ref{but}) and by $S$ the rational invariant subspace associated with this block.
We note that $B$ as an irreducible matrix is diagonalizable.

%%%%%%%%%%%%%%%%%%%%%%%%%%%%%%%%%%%%%%%%%%%%%%%%%%%%%%%%%%%%%%%%%%%%%%%%%%%%%%%%%%%%%%%%%%%%%%
\subsubsection*{Upper bound.}
%%%%%%%%%%%%%%%%%%%%%%%%%%%%%%%%%%%%%%%%%%%%%%%%%%%%%%%%%%%%%%%%%%%%%%%%%%%%%%%%%%%%%%%%%%%%%%

Note that
\bea
 \min_{0 \not=\bk \in \IZ^{d}} \sum_{l=1}^{n}|A^{l}\bk|^{\alpha} \leq 
 \min_{0 \not=\bk \in S \cap \IZ^{d}} \sum_{l=1}^{n}|B^{l}\bk|^{\alpha}.
\eea
The corresponding upper bound (\ref{ubound}) for $B$ is then also an upper 
bound for the whole matrix $A$. We note that geometric average of $\hat{\lam}_j$ over $S$ is
equal to the geometric average of all $\hat{\lam}_j$ associated with matrix $A$ 
(i.e. over the whole space $\IR^{d}$).

%%%%%%%%%%%%%%%%%%%%%%%%%%%%%%%%%%%%%%%%%%%%%%%%%%%%%%%%%%%%%%%%%%%%%%%%%%%%%%%%%%%%%%%%%%%%%%
\subsubsection*{Lower bound.}
%%%%%%%%%%%%%%%%%%%%%%%%%%%%%%%%%%%%%%%%%%%%%%%%%%%%%%%%%%%%%%%%%%%%%%%%%%%%%%%%%%%%%%%%%%%%%%

According to our assumption $A$ is indecomposable and thus the characteristic polynomial
of $A$ is of the form $p^m$ for some irreducible $p$. All 
Jordan blocks of $A$ have the same size $m$ and different Jordan blocks correspond to distinct eigenvalues.
Denote by $b$ the number of the Jordan blocks in $A$ and 
by $\lam_{j}$, where $j=1,..,l$ all these distinct eigenvalues . Since each $\lam_{j}$ has algebraic multiplicity 
$m$, we get $d=mb.$
Let $\{\bv_{j,h}\}_{j=1,...,b;h=0,...,m-1}$ be a basis (of $\IC^{d}$) 
in which $A$ admits the Jordan canonical form. As usually $L_{j,h}$ will denote the corresponding linear forms.
Each  $\bv_{j,h}$ can be regarded as a generalized eigenvector of $A$ associated with 
an eigenvalue $\lam_{j}$. We assume that these generalized eigenvectors are ordered according to
their degree i.e. $\bv_{j,h}$ satisfies the equation $(A-\lam_{j}I)^{1+h}\bv_{j,h}=0$.   
Reordering the eigenvalues, if necessary, we can also assume that $\lam_{1}$ has the largest modulus among
all eigenvalues of $A$ and hence $\hat{\lam}_{1}=|\lam_{1}|$.
Let ${\bz}_{n}$ be the sequence of minimizers solving (\ref{artmin0}).
We first note that for each $n$ there exists 
$0 \leq h \leq m-1$ such that $L_{1,h}{\bz}_{n} \not =0$. 
Indeed, otherwise for all $h=0,...,m-1$, $L_{1,h}{\bz}_{n} =0$ and
consequently for any $n$ and $h$ $L_{1,h}A^{n}{\bz}_{n}=0$. The latter implies that the set of consecutive
iterations $\{{\bz}_{n},A^{1}{\bz}_{n},A^{2}{\bz}_{n},...\}$ spans a proper rational $A$-invariant 
subspace of $\IR^{d}$ which does not have any intersection with the subspace spanned by the generalized eigenvectors 
of $A$ associated with eigenvalue $\lam_{1}$. This clearly contradicts the irreducibility of $p$. 
Now, for given $n$ we denote by $h(n)$ the biggest index $h$ for which the condition $L_{1,h}{\bz}_{n} \not =0$
holds.

We have the following estimate
\bea
\label{est1}
 &&\hat{\lam}_{1}^{\alpha n}|L_{1,h(n)}{\bz}_{n}|^{\alpha} \leq
 \left(\sum_{j=1}^{b}\sum_{h=0}^{m-1}
 \left|\sum_{i=0}^{m-1-h}\lam_{j}^{n-i}\binom{n}{i}L_{j,h+i}{\bz}_{n}\right|^{2}\right)^{\alpha}\\
 &\leq&C_{1}|A^{n}{\bz}_{n}|^{\alpha} \leq
 C_{1}\sum_{l=1}^{n}|A^{l}{\bz}_{n}|^{\alpha} \leq
 C_{2}n\hat{\lam}_{geo}^{\alpha n},
\eea
where the last inequality follows from previously established upper bound.

From the Diophantine approximation and the assumption that $|L_{1,h(n)}{\bz}_{n}|\not=0$, 
there exists $\beta>0$ such that (see \cite{Schmidt} p. 164) 
\bea
\label{est2}
|L_{1,h(n)}{\bz}_{n}| \geq \frac{1}{|{\bz}_{n}|^{\beta}}.
\eea 
Thus combining (\ref{est1}) with (\ref{est2}) one gets
\bean
 \hat{\lam}_{1}^{\alpha n}|{\bz}_{n}|^{-\alpha\beta} \leq
 \hat{\lam}_{1}^{\alpha n}|L_{1,h(n)}{\bz}_{n}|^{\alpha} \leq 
 C_{2}n\hat{\lam}_{geo}^{\alpha n}.
\eean

After rearrangements one obtains the following 
lower bound estimate for (\ref{artmin0}) 
\bea
\label{nlbound}
\frac{C}{n}\hat{\lam}^{\alpha n} \leq 
C|{\bz}_{n}|^{\alpha} \leq 
\sum_{l=1}^{n}|A^{l}{\bz}_{n}|^{\alpha}=
\min_{0 \not=\bk \in \IZ^{d}} \sum_{l=1}^{n}|A^{l}\bk|^{\alpha},
\eea
where
\bean
 \hat{\lam}=\left(\frac{\hat{\lam}_{1}}{\hat{\lam}_{geo}}\right)^{1/\beta}.
\eean
We note that ergodicity of $A$ implies $\hat{\lam}_{1}>\hat{\lam}_{geo}>1$ 
(see (\ref{hatlamgeo}), (\ref{entropy}) and Proposition \ref{zeroentropy}) which ensures 
non-triviality of this lower bound.

Now in order to finish the proof is suffices to combine the estimates (\ref{ubound}),
(\ref{dlbound}) and (\ref{nlbound}), and note that 
\bean
\hat{\lam}_{geo}^{\alpha n}=e^{\alpha\frac{h(A)}{d}n}=e^{\alpha\hat{h}(A)n}
\eean
which yields (\ref{ulestimateA}). $\qquad \blacksquare$

%%%%%%%%%%%%%%%%%%%%%%%%%%%%%%%%%%%%%%%%%%%%%%%%%%%%%%%%%%%%%%%%%%%%%%%%%%%%%%%%%%%%%%%%%%%%

\subsection*{Proofs of Theorems \ref{thm2} and \ref{thm3}}

%%%%%%%%%%%%%%%%%%%%%%%%%%%%%%%%%%%%%%%%%%%%%%%%%%%%%%%%%%%%%%%%%%%%%%%%%%%%%%%%%%%%%%%%%%

We start with part i) of Theorem \ref{thm2} which follows as
a simple consequence of Theorem \ref{weakmix}. Indeed, it suffices to construct an  
eigenfunction of $U_{F}$ which belongs to 
$L^{2}_{0}(\IT^{d}) \cap H^{\alpha}(\IT^{d})$.
Directly from Proposition \ref{ergodicTA} one concludes that $F$, and
hence also $A$, possesses a root of unity in its spectrum. This means
that $A^{m}\bk_{0}=\bk_{0}$, for some $m$ and certain nonzero vector
$\bk_{0}$, which can be chosen to be an integer.
Now we define
\bean
f=\fb_{\bk_{0}}+\fb_{A\bk_{0}}+...+\fb_{A^{m-1}\bk_{0}}
\eean
Obviously $f\in L^{2}_{0}(\IT^{d}) \cap H^{\alpha}(\IT^{d})$, 
for any $\alpha$. To complete the proof it suffices to notice that
\bean
U_{F}f=\fb_{A\bk_{0}}+\fb_{A^{2}\bk_{0}}+...+\fb_{A^{m}\bk_{0}}
=\fb_{\bk_{0}}+\fb_{A\bk_{0}}+...+\fb_{A^{m-1}\bk_{0}}=f.\qquad \blacksquare
\eean  

Now we apply Theorem \ref{thmart} to prove part ii) and iii) of Theorem \ref{thm2} and
Theorem \ref{thm3}.

In order to determine the dissipation time of $T_{\ep,\alpha}$ one has to 
determine the asymptotics of $\|T_{\ep,\alpha}^{n}\|$ when $n$ goes to 
infinity. 
According to formula (\ref{Tnorm3}) this problem reduces 
to problem (\ref{artmin0}) which has been solved by Theorem \ref{thmart}).
This in view of this theorem there exist constants $C_{1}$ and $C_{2}$ such 
that for any $\del,\del'>0$ and sufficiently large $n$
\bean
\label{ulestimate1}
C_{1} e^{(1-\del)\alpha\tilde{h}(A) n} \leq
\min_{0 \not=\bk \in \IZ^{d}} \sum_{l=1}^{n}|A^{l}\bk|^{\alpha} \leq
C_{2}n e^{\alpha \hat{h}(A)n}\leq
C_{2}e^{(1+\del')\alpha \hat{h}(A)n}
\eean
Using formula (\ref{Tnorm3})
\bean
e^{-\ep^{\alpha}  C_{2}e^{(1+\del')\alpha \hat{h}(A)n}}  
\leq \|T_{\ep,\alpha}^{n}\| \leq 
e^{-\ep^{\alpha}  C_{1}e^{(1-\del)\alpha\tilde{h}(A)n}}.
\eean
Now when $n=\tau_{c}$, we have
\bean
C_{1} e^{(1-\del)\tilde{h}(A)\tau_{c}}
\leq \frac{1}{\ep} \leq
C_{2} e^{(1+\del') \hat{h}(A)\tau_{c}}  
\eean
and 
\bean
\frac{1}{(1+\del')\hat{h}(A)}\bigg(\ln(1/\ep) -\ln C_{2}\bigg)   
\leq \tau_{c} \leq
\frac{1}{(1-\del) \tilde{h}(A)}\bigg(\ln(1/\ep)-\ln C_{1}\bigg),
\eean
which proves part ii) of Theorem \ref{thm2} i.e. the logarithmic growth of 
dissipation time as a function of $\ep^{-1}$. 

Moreover, using the definition of dissipation rate constant
\bean
  R_{c}=\lim_{\ep \rightarrow 0} \frac{\tau_{c}}{\ln(1/\ep)}
\eean  
we obtain
\bean
\frac{1}{(1+\del')\hat{h}(A)} \leq 
R_{c} \leq 
\frac{1}{(1-\del) \tilde{h}(A)}.
\eean
Finally letting $\del \rightarrow 0$ and $\del' \rightarrow 0$  
we arrive at the following results:
\begin{itemize}
\item The general case - Theorem \ref{thm2} iii)
\bean
\frac{d}{\hat{h}(F)}\leq R_{c} 
\leq \frac{1}{\tilde{h}(F) }
\eean
\item The diagonalizable case - Theorem \ref{thm3}
\bean
 R_{c}=\frac{1}{ \hat{h}(F)}.
\eean
\end{itemize}
This completes the proof.
$\qquad \blacksquare$

%%%%%%%%%%%%%%%%%%%%%%%%%%%%%%%%%%%%%%%%%%%%%%%%%%%%%%%%%%%%%%%%%%%%%%%%%%%%%%%%%%%%%%%%%%%%
%%%%%%%%%%%%%%%%%%%%%%%%%%%%%%%%%%%%%%%%%%%%%%%%%%%%%%%%%%%%%%%%%%%%%%%%%%%%%%%%%%%%%%%%%%%%
%%%%%%%%%%%%%%%%%%%%%%%%%%%%%%%%%%%%%%%%%%%%%%%%%%%%%%%%%%%%%%%%%%%%%%%%%%%%%%%%%%%%%%%%%%

\section{Generalizations and Applications}
\label{GandA}

%%%%%%%%%%%%%%%%%%%%%%%%%%%%%%%%%%%%%%%%%%%%%%%%%%%%%%%%%%%%%%%%%%%%%%%%%%%%%%%%%%%%%%%%%%%%
%%%%%%%%%%%%%%%%%%%%%%%%%%%%%%%%%%%%%%%%%%%%%%%%%%%%%%%%%%%%%%%%%%%%%%%%%%%%%%%%%%%%%%%%%%
%%%%%%%%%%%%%%%%%%%%%%%%%%%%%%%%%%%%%%%%%%%%%%%%%%%%%%%%%%%%%%%%%%%%%%%%%%%%%%%%%%%%%%%%%%%%

In this section we generalize or apply our main results i.e. Theorems \ref{thm2} and \ref{thm3}
in the following situations
\begin{itemize}
\item[1.] Coarse grained dynamics (Section \ref{coarse}).
\item[2.] Affine toral maps (Section \ref{aAT}).
\item[3.] Degenerate noises (Section \ref{degnoise}).
\item[4.] Kinematic dynamo problem (Section \ref{dynamo}).
\end{itemize}

%%%%%%%%%%%%%%%%%%%%%%%%%%%%%%%%%%%%%%%%%%%%%%%%%%%%%%%%%%%%%%%%%%%%%%%%%%%%%%%%%%%%%%%%%%%%%%%%%%%
%%%%%%%%%%%%%%%%%%%%%%%%%%%%%%%%%%%%%%%%%%%%%%%%%%%%%%%%%%%%%%%%%%%%%%%%%%%%%%%%%%%%%%%%%%%%%%%%%%%

\subsection{Dissipation time of coarse-grained dynamics}
\label{coarse}

%%%%%%%%%%%%%%%%%%%%%%%%%%%%%%%%%%%%%%%%%%%%%%%%%%%%%%%%%%%%%%%%%%%%%%%%%%%%%%%%%%%%%%%%%%%%%%%%
%%%%%%%%%%%%%%%%%%%%%%%%%%%%%%%%%%%%%%%%%%%%%%%%%%%%%%%%%%%%%%%%%%%%%%%%%%%%%%%%%%%%%%%%%%%%%%%%

The uncertainties in the initial preparation and the final measurement of
the noiseless system give rise to non-cumulative random perturbations
to the system. The dynamics of such systems can be modeled by the coarse-grained
family of noisy operators $\tilde{T}^{(n)}_{\ep,\alpha}$ introduced in Section \ref{Noevop}. 
In this section we compute corresponding dissipation time $\tilde{\tau}_{c}$.
We show that, remarkably, for ergodic toral automorphisms
both dissipation times are asymptotically the same, despite considerable difference in
the structure of the corresponding noisy dynamical systems. 

To prove this result one can represent the action of $U_{F}$ or more generally $U_{F}^{n}$
in the Fourier series along the lines introduced in Section ref{Opt}
\ben
 U_{F}^{n}\be_{\mf{\bk}}=\sum_{0 \not = \bk' \in \IZ^{d}}u_{\bk,\bk'}^{(n)}\be_{\bk'},
\een
where $u_{\bk,\bk'}^{(1)}$ coincides with $u_{\bk,\bk'}$ defined previously
(cf. (\ref{uk})) and
\[
u^{(n)}_{\bk,\bk'}=\sum_{0\neq \bk_1,...,\bk_{n-1}\in \IZ^d} u_{\bk,\bk_1}u_{\bk_1,\bk_2}...u_{\bk_{n-1},\bk'}
\]
which satisfies
\bea
\label{ul3}
\sum_{0 \not = \bk' \in \IZ^{d}}|u^{(n)}_{\bk,\bk'}|^{2}=1,\quad \forall\, n,\,\,\bk.
\eea
Then 
\bean
\tilde{T}_{\ep,\alpha}^{(n)}\be_{\bk_{0}}&=&
G_{\ep,\alpha}U_F^{n}G_{\ep,\alpha}\be_{\bk_{0}}=
G_{\ep,\alpha}U_F^{n}e^{-\ep|\bk_{0}|^{2}} \be_{\bk_{0}}=
e^{-\ep|\bk_{0}|^{2}}G_{\ep,\alpha}\sum_{0 \not = \bk_{n} \in \IZ^{d}}
u^{(n)}_{\bk_{0},\bk_{n}}\be_{\bk_{n}}\\
&=&e^{-\ep(|\bk_{0}|^{2}+|\bk_{n}|^{2})}
\sum_{0 \not = \bk_{n} \in \IZ^{d}}u^{(n)}_{\bk_{0},\bk_{n}}\be_{\bk_{n}}.
\eean

Now we define
\ben
S_n(\bk_{n})=\{\bk_{0} \in \IZ^{d}\backslash \{0\} : u^{(n)}_{\bk_{0},\bk_{n}}\not = 0\}.
\een
Similar computations to these performed in Section \ref{Opt}
give the following  general upper bound for $\|\tilde{T}_{\ep,\alpha}^{(n)}\|$
\bea
\label{uppernorm1}
\|\tilde{T}_{\ep,\alpha}^{(n)}f\|^{2} \leq
\sum_{0 \not = \bk_{n} \in \IZ^{d}}
\sum_{\bk_{0} \in S_n(\bk_{n})}
|\hat{f}(\bk_{0})|^{2}
\sum_{\bk_{0} \in S_n(\bk_{n})}
|u_{\bk_{0},\bk_{n}}^{(n)}|^{2}.
\eea
For a toral automorphism one easily sees that
\bea
u_{\bk_{0},\bk_{n}}^{(n)}=
e^{-\ep(|\bk_{0}|^{\alpha}+|A^{n}\bk_{0}|^{\alpha})}\del_{\bk_{n},A^{n}\bk_{0}}
\eea
and hence
\ben
\|\tilde{T}_{\ep,\alpha}^{(n)}\| = 
e^{-\ep \min_{0 \not=\bk \in \IZ^{d}}(|\bk_{0}|^{\alpha}+|A^{l}\bk_{0}|^{\alpha})}. 
\een
The arithmetic minimization problem (\ref{artmin0}) corresponding to
the dissipation time of $\hat{T}_{\ep,\alpha}$ now becomes
\bea 
\label{artminA1}
\min_{0 \not=\bk \in \IZ^{d}}\left(|\bk|^{\alpha}+|A^{n}\bk|^{\alpha}\right).
\eea
The key observation is that, by the same arguments as before,
similar estimates to these given in (\ref{ulestimate}) hold
\bea
\label{ulestimate2}
C_{1}\left(\sum_{j=1}^{d} \hat{\lam}_{j}^{2 n}|L_{j}\bk|^{2}\right)^{\alpha} \leq
|\bk|^{\alpha}+|
A^{n}\bk|^{\alpha} \leq
C_{2}\sum_{j=1}^{d} \hat{\lam}_{j}^{\alpha n}|L_{j}\bk|^{\alpha}.
\eea
The remaining computations are the same verbatim so 
the dissipation time of $T_{\ep,\alpha}$ and the family $\tilde{T}_{\ep,\alpha}^{(n)}$ are 
equal asymptotically.

%%%%%%%%%%%%%%%%%%%%%%%%%%%%%%%%%%%%%%%%%%%%%%%%%%%%%%%%%%%%%%%%%%%%%%%%%%%%%%%%%%%%%%%%%%%%%%%%
%%%%%%%%%%%%%%%%%%%%%%%%%%%%%%%%%%%%%%%%%%%%%%%%%%%%%%%%%%%%%%%%%%%%%%%%%%%%%%%%%%%%%%%%%%%%%%%%

\subsection{Affine toral maps}
\label{aAT}

%%%%%%%%%%%%%%%%%%%%%%%%%%%%%%%%%%%%%%%%%%%%%%%%%%%%%%%%%%%%%%%%%%%%%%%%%%%%%%%%%%%%%%%%%%%%%%%%
%%%%%%%%%%%%%%%%%%%%%%%%%%%%%%%%%%%%%%%%%%%%%%%%%%%%%%%%%%%%%%%%%%%%%%%%%%%%%%%%%%%%%%%%%%%%%%%%

In this section we present a slight generalization of the main results
concerning the asymptotics of the dissipation time of toral automorphisms. 
Namely, we consider here general affine transformations of the torus.
The term {\em affine transformations} will be used here to refer to  
homeomorphisms of the torus with zero periodic 
but not necessary zero constant part (cf. Section \ref{prelim}) i.e. 
transformations of the form $\tilde{F}=F+\bc$, where $F$ is a toral 
automorphism and $\bc$ is a constant shift vector.
 
We begin with a short discussion of the ergodicity of affine
transforms.
 
The relation between ergodicity of a given affine
transform $\tilde{F}$ and associated with it toral 
automorphism $F$ is summarized in
the following proposition (for the proof we refer to appendix B)
\begin{prop}
\label{affinerg}
Let $F$ be any toral automorphism. Then

i) If $F$ is ergodic then $\tilde{F}$ is also ergodic.

ii) If $F$ is not ergodic then $\tilde{F}$ is ergodic iff $1$ is the only
root of unity in the spectrum of $F$ and $\bc\cdot\bk\not\in\IZ^{d}$ 
for any integer eigenvector $\bk$ of $F^{\dagger}$.
\end{prop}
{\bf Proof.} i) Assume $F$ is ergodic and for some $\bc$, $\tilde{F}=F+\bc$ is not ergodic.
Then there exists non-constant $f\in L_{0}^{2}(\IT^{d})$ satisfying 
$f=f\circ \tilde{F}$ or in the Fourier representation 
\bea
\label{inv}
\sum_{\bk \in \IZ^{d}}\hat{f}(\bk)\be_{\bk}=\sum_{\bk \in \IZ^{d}}
 e^{2 \pi i A^{-1}\bk\cdot \bc}\hat{f}(A^{-1}\bk)\be_{\bk}
\eea
where $A=F^{\dagger}$. 
Comparing the absolute values of the coefficients we get 
\bea
\label{inv1}
|\hat{f}(\bk)|=|\hat{f}(A^{-n}\bk)|
\eea
for any integer $n$ and any $\bk$.
However, ergodicity of $F$ implies that $A^{-n}\bk\not=\bk$ for all $\bk\not=0$, which 
contradicts our assumption that $f\in L_{0}^{2}(\IT^{d})$.

ii) We will use the following fact, which can be proved by
simple application of rational canonical decomposition. 
For any $A\in SL(d,\IZ)$ the following conditions are equivalent

a) $A$ possesses in its spectrum a root of unity not equal to one.

b) There exists nonzero $\bk \in \IZ^{d}$ and a positive integer $n$ such
that $\bk+A\bk+...+A^{n-1}\bk=0$.
 
Now assume that $1$ is the only root of unity in spectrum of $F$ (and hence
of $A$) and $\bc\cdot\bk\not\in\IZ^{d}$ for any integer eigenvector $\bk$ of 
$A$, and that both $F$ and $\tilde{F}$ are not ergodic. 
The latter assumption implies the 
existence of a non-constant $f\in L_{0}^{2}(\IT^{d})$ satisfying 
equations (\ref{inv}) and (\ref{inv1}). 
Relation (\ref{inv1}) clearly implies that if
$\hat{f}(\bk)\not= 0$ then $A^{n}\bk=\bk$ for some $n$. Moreover, 
since $1$ is the only root of unity in spectrum of $A$,
we have, in view of b) that $A\bk=\bk$. Thus the only possible non-constant
invariant functions of $\tilde{F}$ are single Fourier modes 
$\be_{\bk}$ corresponding to integer eigenvectors of $A$. 
But if such a Fourier mode is invariant under $\tilde{F}$ then
directly from (\ref{inv}) one concludes that $e^{2 \pi i \bk\cdot \bc}=1$
or equivalently $\bk\cdot \bc \in\IZ^{d}$, for some integer eigenvector of $A$. 
To prove the converse we assume that $F$ is not ergodic and consider two
cases:

Case 1. $A$ possesses in its spectrum a root of unity not equal to one.
In this case according to condition b) there exists nonzero 
$\bk \in \IZ^{d}$ and a positive integer $n$ such
that $\bk+A\bk+...+A^{n-1}\bk=0$, which implies in particular that 
$A^{n}\bk=\bk$ and $A\bk\not=\bk$. Now we define the function
\bean
f=\be_{\bk}+e^{2\pi i \bk \cdot \bc}\be_{A\bk}+...+
e^{2\pi i \left(\sum_{l=0}^{n-2}A^{l}\bk \right)\cdot \bc}\be_{A^{n-1}\bk}
\eean
which clearly satisfies the condition $f=f\circ \tilde{F}$.
This proves that $\tilde{F}$ is not ergodic.

Case 2. There exists integer eigenvector of $A$ such that $\bk\cdot \bc \in\IZ^{d}$. 
Then clearly for such $\bk$, $f=e_{\bk}$ is $\tilde{F}$-invariant and hence again 
$\tilde{F}$ is not ergodic. $\qquad \blacksquare$

We recall that $\mf{c}=(c_{1},..,c_{d})$ generates ergodic shift
on the torus iff $1,c_{1},..,c_{d}$ are linearly independent over rationals.
Thus as a direct consequence of the above proposition we get
\begin{cor}
\label{corerg}
If $F$ is not ergodic and $1$ is the only
root of unity in the spectrum of $F$ then $\tilde{F}$ is ergodic for all
vectors $\bc$ generating ergodic shifts on the torus.
\end{cor}

Now we are in a position to state and prove the generalization of 
Theorem \ref{thm2} from Section \ref{MainT} to the case of affine transforms
(the corresponding generalizations of Theorem \ref{thm3} and 
Corollary \ref{cat} are straightforward)  

\begin{thm}
\label{thm2'}

Let $\tilde{F}$ be any affine transformation on the torus $\IT^{d}$, 
$F$ associated with $\tilde{F}$ toral automorphism and 
$T_{\ep,\alpha}=G_{\ep,\alpha}U_{\tilde{F}}$. Then
 
i) $T_{\ep,\alpha}$ has simple dissipation time iff $F$ is not ergodic.

ii) $T_{\ep,\alpha}$ has logarithmic dissipation time iff $F$ is ergodic. 

iii) If $T_{\ep,\alpha}$ has logarithmic dissipation time then the dissipation rate constant 
     satisfies the following constraint
\bean
\frac{1}{\hat{h}(F)} \leq R_{c} \leq \frac{1}{\tilde{h}(F)},
\eean
where $\tilde{h}(F)\leq\hat{h}(F)$ is certain positive constant. 
\end{thm}

\begin{rem}
\label{remerg}
The dissipation time of an affine transformation $\tilde{F}$ is determined 
by ergodic properties of its linear part $F$ and hence not by
ergodic properties of $\tilde{F}$ itself. In particular all ergodic affine
transformations associated with nonergodic toral automorphisms
(cf. Proposition \ref{affinerg}) have simple dissipation time.     
\end{rem}

\textbf{Proof of Theorem \ref{thm2'}}
Specializing the general calculations of dissipation time presented 
in Section \ref{Opt} to the case of
affine transformations $\tilde{F}=F+\bc$, with nonzero $\bc$, 
one easily finds the following counterparts of formulas (\ref{linear}) 
and (\ref{linear1})
\bean
u_{\bk,\bk'}&=&e^{2\pi i \bk \cdot \bc}\del_{A\bk,\bk'}, \\
\ml{U}_{n}(\bk_{0},\bk_{n})
&=&e^{2\pi i \left(\sum_{l=0}^{n-1}A^{l}\bk\right) \cdot \bc}
e^{-\ep^{\alpha}\sum_{l=1}^{n}|A^{l}\bk|^{\alpha}}\del_{A^{n}\bk_{0},\bk_{n}}.
\eean 
Now, in order to determine the dissipation time of 
$T_{\ep,\alpha}=G_{\ep,\alpha}U_{\tilde{F}}$ one has to determine the 
asymptotics of $\|T_{\ep,\alpha}^{n}\|$ as $n$ goes to infinity.  
According to the above formulas and formulas (\ref{uppernorm}) and (\ref{Tnorm3})
from Sections \ref{Opt} and \ref{sA} the value of $\|T_{\ep,\alpha}^{n}\|$ does 
not depend on $\bc$, which reduces the proof to the case we already considered i.e. $\bc=0$.
$\qquad \blacksquare$

%%%%%%%%%%%%%%%%%%%%%%%%%%%%%%%%%%%%%%%%%%%%%%%%%%%%%%%%%%%%%%%%%%%%%%%%%%%%%%%%%%%%%%%%%%%%
%%%%%%%%%%%%%%%%%%%%%%%%%%%%%%%%%%%%%%%%%%%%%%%%%%%%%%%%%%%%%%%%%%%%%%%%%%%%%%%%%%%%%%%%%%%%

\subsection{Degenerate noise}
\label{degnoise}

%%%%%%%%%%%%%%%%%%%%%%%%%%%%%%%%%%%%%%%%%%%%%%%%%%%%%%%%%%%%%%%%%%%%%%%%%%%%%%%%%%%%%%%%%%%%%%%%
%%%%%%%%%%%%%%%%%%%%%%%%%%%%%%%%%%%%%%%%%%%%%%%%%%%%%%%%%%%%%%%%%%%%%%%%%%%%%%%%%%%%%%%%%%%%%%%%

In this section we compute the dissipation time for non-strictly contracting generalizations 
of $\alpha$-stable transition operators. 
Instead of considering standard $\alpha$-stable kernels of the form 
(\ref{Akernel}) one can allow for some degree of degeneracy of noise in chosen directions by 
introducing the following family of noise kernels
\bea
\label{noker}
g_{\ep,\alpha,B}(\bx):=\sum_{\bk \in \IZ^{d}}e^{-|\ep B\bk|^{\alpha}}\be_{\bk}(\bx),
\eea
Where $B$ denotes any $d\times d$ matrix with $\det B=0$. 

We denote by $G_{\ep,\alpha,B}$ the noise operator associated with $g_{\ep,\alpha,B}$.  
The degeneracy of $B$ immediately implies that $\|G_{\ep,\alpha,B}\|=1$ and hence the
general considerations of sections 1 and 2 do not apply here.
The answer to the question whether or not the dissipation time 
is finite depends on the choice of matrix $B$. 

For simplicity we concentrate on the case when $B$ is diagonalizable.

We call the eigenvector of $B$ nondegenerate if it corresponds
to nonzero eigenvalue. 
\begin{thm}
Let $F$ be any toral automorphism and 
$T_{\ep,\alpha,B}=G_{\ep,\alpha,B}U_{F}$. 
Assume that $B$ is diagonalizable. Then
 
i) If all nondegenerate eigenvectors of $B^{*}$ lie in one proper invariant subspace of $F$ then 
dissipation does not take place i.e. $\tau_{c}=\infty$.

ii) Otherwise the following statements hold.

a) $T_{\ep,\alpha}$ has simple dissipation time iff $F$ is not ergodic.

b) $T_{\ep,\alpha}$ has logarithmic dissipation time iff $F$ is ergodic. 

c) If $T_{\ep,\alpha,B}$ has logarithmic dissipation time then the dissipation rate constant satisfies the following bounds
\bean
\frac{1}{\hat{h}(F)} \leq R_{c} \leq \frac{1}{\tilde{h}(F)},
\eean
with some constant $\tilde{h}(F)\leq\hat{h}(F)$. The equality is achieved for all diagonalizable automorphisms $F$. 
\end{thm}
\textbf{Proof.}

We continue to use the convention $A=F^{\dagger}$.
The general formula derived previously for $\|T_{\ep,\alpha}^{n}\|$ (see (\ref{Tnorm3})), will now
take the form 
\bea
\label{Tnorm3B}
\|T_{\ep,\alpha,B}^{n}\| &=& \sup_{0 \not=\bk \in \IZ^{d}} e^{-\ep^{\alpha} \sum_{l=1}^{n}|BA^{l}\bk|^{\alpha}} = 
\exp{-\ep^{\alpha} \underset{0 \not=\bk \in \IZ^{d}}{\inf} \sum_{l=1}^{n}|BA^{l}\bk|^{\alpha}}. 
\eea
Thus we need to estimate
\bean
%\label{Binf}
\inf_{0 \not=\bk \in \IZ^{d}} \sum_{l=1}^{n}|BA^{l}\bk|^{\alpha}.
\eean
To this end we denote by $\mu_{j}$ ($j=1,...,d$) 
the eigenvalues of $B$
and we construct a basis (of $\IC^{d}$) $\{\bv_{j}\}_{j=1,...,d}$ composed of 
normalized eigenvectors corresponding to eigenvalues $\mu_{j}$.
We denote by ${P_{j}}_{j=1,...,d}$ the set of eigen-projections on ${\bv_{j}}$, and by ${L_{j}}$ the 
set of corresponding linear forms, given by the eigenvectors $\bu_j$ of $B^{\dagger}$, which are
of course co-orthogonal to ${\bv_j}$, i.e. $\la \bu_i,\bv_j\ra=0$ for $i\not=j$.
We have
\bean
 \bx=\sum_{j=1}^{d}P_{j}\bx=\sum_{j=1}^{d}(L_{j}\bx)\bv_{j}
=\sum_{j=1}^d\la \bx,\bu_j\ra \bv_j, \quad\forall\bx\in \IR^d.
\eean  
In subsequent computations the symbols $C_{1}$, $C_{2}$ denote some
absolute constants values of which are subject to change during calculations. 

We consider two cases.

i) All nondegenerate eigenvectors of $B^{\dagger}$ lie in one proper subspace of $F$.
We have the following estimates
\bean
|BA^{l}\bk|^{2}&\geq& 
C_{1} \sum_{j=1}^{d}|P_{j}BA^{l}\bk|^{2} =
C_{1} \sum_{j=1}^{d}|\mu_{j}|^{2}|P_{j}A^{l}\bk|^{2}\\ &=&
C_{1} \sum_{j=1}^{d}|\mu_{j}|^{2}|\la A^{l}\bk,\bu_j\ra|^{2} =
C_{1} \sum_{j=1}^{d}|\mu_{j}|^{2}|\la \bk,F^{l}\bu_j\ra|^{2}
\eean
and 
\bean
|BA^{l}\bk|^{2}&\leq& 
C_{2} \sum_{j=1}^{d}|P_{j}BA^{l}\bk|^{2}=
C_{2} \sum_{j=1}^{d}|\mu_{j}|^{2}|P_{j}A^{l}\bk|^{2} \\ &=&
C_{2} \sum_{j=1}^{d}|\mu_{j}|^{2}|\la A^{l}\bk,\bu_j\ra|^{2}=
C_{2} \sum_{j=1}^{d}|\mu_{j}|^{2}|\la \bk,F^{l}\bu_j\ra|^{2}
\eean
Since at least one of $\mu_{j}$ is zero and all nondegenerate vectors $\bu_j$ lie in a proper invariant 
subspace of $F$, one easily sees that for each fixed $n$ 
\bean
\inf_{0 \not=\bk \in \IZ^{d}}\sum_{l=1}^{n}|BA^{l}\bk|^{\alpha}=
\inf_{0 \not=\bk \in \IZ^{d}}\sum_{l=1}^{n}\sum_{j=1}^{d}|\mu_{j}|^{\alpha}|\la \bk,F^{l}\bu_j\ra|^{\alpha}=0.
\eean
ii) In this case we have the following upper bound
\bea
\label{ubinf}
\inf_{0 \not=\bk \in \IZ^{d}} \sum_{l=1}^{n}|BA^{l}\bk|^{\alpha}
\leq \|B\|^{\alpha}\inf_{0 \not=\bk \in \IZ^{d}} \sum_{l=1}^{n}|A^{l}\bk|^{\alpha}
= \|B\|^{\alpha}\min_{0 \not=\bk \in \IZ^{d}} \sum_{l=1}^{n}|A^{l}\bk|^{\alpha}. 
\eea
In order to provide an appropriate lower bound
we note that the set of vectors $\{F^{h}\bu_{j}\}$, where $1\leq h \leq d$ and $j$ runs through the
indices of all nondegenerate eigenvectors of $B$, spans the whole space 
(otherwise all nondegenerate $\bu_{j}$ would lie in one proper invariant subspace of $F$). 
We denote by $\{F^{h_{i}}\bu_{j_{i}}\}$ ($1\leq i \leq d$) a basis extracted from the above set.
We can define now a new norm $|\cdot|_{\bu}$ on $\IR^{d}$ by
\bean
|\bx|^{2}_{\bu}=\sum_{i=1}^{d}|\la \bx,F^{h_{i}}\bu_{j_{i}} \ra|^{2}
\eean
and compute
\bean
\sum_{l=1}^{dn}|BA^{l}\bk|^{\alpha}
&=&\sum_{l=0}^{n-1}\sum_{h=1}^{d}|BA^{dl+h}\bk|^{\alpha}
\geq \sum_{l=0}^{n-1}\sum_{h=1}^{d}C_{1}\sum_{j=1}^{d}|P_{j}BA^{dl+h}\bk|^{\alpha}\\
&=&C_{1}\sum_{l=0}^{n-1}\sum_{h=1}^{d}\sum_{j=1}^{d}|\mu_{j}|^{\alpha}|P_{j}A^{dl+h}\bk|^{\alpha}
\geq C_{1} \sum_{l=0}^{n-1}\sum_{i=1}^{d}|\la A^{dl+h_{i}}\bk,\bu_{j_{i}} \ra|^{\alpha}\\
&=&C_{1}\sum_{l=0}^{n-1}\sum_{i=1}^{d}|\la A^{dl}\bk,F^{h_{i}}\bu_{j_{i}} \ra|^{\alpha}
=C_{1}\sum_{l=0}^{n-1}|A^{dl}\bk|^{\alpha}_{\bu}
\eean
Using the equivalence between norms $|\cdot|$ and $|\cdot|_{\bu}$ and combining
(\ref{ubinf}) with the above estimate we get
\bean
C_{1}\min_{0 \not=\bk \in \IZ^{d}} \sum_{l=0}^{n-1}|A^{dl}\bk|^{\alpha}\leq
\inf_{0 \not=\bk \in \IZ^{d}}\sum_{l=1}^{dn}|BA^{l}\bk|^{\alpha}\leq
\|B\|^{\alpha}\min_{0 \not=\bk \in \IZ^{d}} \sum_{l=1}^{dn}|A^{l}\bk|^{\alpha}. 
\eean
This together with the obvious fact that $\hat{h}(A^{d})=d\hat{h}(A)$ and the general estimate
(\ref{ulestimateA}) reduces the proof back to the nondegenerate case considered in the previous section. 
$\qquad \blacksquare$

%%%%%%%%%%%%%%%%%%%%%%%%%%%%%%%%%%%%%%%%%%%%%%%%%%%%%%%%%%%%%%%%%%%%%%%%%%%%%%%%%%%%%%%%%%%%%%%%
%%%%%%%%%%%%%%%%%%%%%%%%%%%%%%%%%%%%%%%%%%%%%%%%%%%%%%%%%%%%%%%%%%%%%%%%%%%%%%%%%%%%%%%%%%%%%%%%

\subsection{Time scales in kinematic dynamo}
\label{dynamo}

%%%%%%%%%%%%%%%%%%%%%%%%%%%%%%%%%%%%%%%%%%%%%%%%%%%%%%%%%%%%%%%%%%%%%%%%%%%%%%%%%%%%%%%%%%%%%%%%
%%%%%%%%%%%%%%%%%%%%%%%%%%%%%%%%%%%%%%%%%%%%%%%%%%%%%%%%%%%%%%%%%%%%%%%%%%%%%%%%%%%%%%%%%%%%%%%%

In this section we briefly discuss the connection between the dissipation time 
and some characteristic time scales associated with kinematic dynamo,
which concerns the generation of
electromagnetic fields by  
mechanical motion. 
For a general setup and discussion we refer the reader to \cite{CG} and
\cite{KY} and 
references therein.
Here we restrict ourselves only to necessary definitions.

Let $\bB\in L^{2}_{0}(\IT^{d},\IR^{d})$ denote periodic, zero mean and divergence 
free magnetic field and let $F$ be the time-1 map associated with the fluid velocity. 
We define the push-forward map 
\bean
F_{*}\bB(\bx)=dF(F^{-1}(\bx))\bB(F^{-1}(\bx)).
\eean
The noisy push-forward map $P_{\ep,\alpha}$
on $L_{0}^2(\IT^{d},\IR^{d})$ 
is then given by
\bea
\label{Tdyn}
P_{\ep,\alpha}\bB:=G_{\ep,\alpha}F_{*}\bB,
\eea   
where the convolution (the action of $G_{\ep,\alpha}$) is applied component-wise.

It is said that the kinematic dynamo action (positive dynamo effect) occurs if
the dynamo growth rate is positive i.e.
\bean
R_{dyn}=\lim_{n\rightarrow \infty}\frac{1}{n}\ln\|P_{\ep,\alpha}^{n}\|>0.
\eean
Moreover if
\bean
\lim_{\ep\rightarrow 0}\lim_{n\rightarrow \infty}\frac{1}{n}\ln\|P_{\ep,\alpha}^{n}\|>0,
\eean
then the dynamo action is said to be fast; otherwise it is slow. 
The anti-dynamo action takes place if
\bean
\lim_{n\rightarrow \infty}\frac{1}{n}\ln\|P_{\ep,\alpha}^{n}\|<0.
\eean
Now we introduce the {\em threshold time} scale as 
\bean
n_{th}=\max \{n : \|P_{\ep,\alpha}^{n}\|>e\,\, \hbox{such that}\,\,
\|P_{\ep,\alpha}^{n-1}\| \,\,\hbox{or}\,\,
\|P_{\ep,\alpha}^{n+1}\| \leq e\}.
\eean 
The threshold time $n_{th}(\ep)$ is of order $O(1)$ as $\ep\rightarrow 0$ for all
fast kinematic dynamo systems. In the case of anti-dynamo action, $n_{th}(\ep)$ captures
the longest time scale on which the generation of the magnetic field still takes place.
Finally $n_{th}(\ep)$ is not defined for systems with do not exhibit any growth of 
magnetic field throughout the evolution. 
In the case of anti-dynamo
we consider the time scale on which the generation of the
magnetic field achieves its maximal value
\bean
n_{p}=\min\{n: \|P_{\ep,\alpha}^{n}\|=\sup_{m}\|P_{\ep,\alpha}^{m}\|\}.
\eean
which is called the {\em peak time} of the anti-dynamo action.

Our next theorem establishes the relation between $n_{p}$, $n_{th}$ and $\tau_{c}$ for toral automorphisms.
Thus $dF=F$ and
\bean
P_{\ep,\alpha}\bB=g_{\ep,\alpha}*F(\bB\circ F^{-1}).
\eean
\begin{thm}
Let $F$ be any toral automorphism.
Then

i) If $F$ is nonergodic and has positive entropy then for all 
$0<\ep<R_{c}\ln\rho_{F}$ the fast dynamo action takes place
with dynamo growth rate satisfying 
\bean
R_{dyn}= \ln\rho_{F}-\ep R_{c}^{-1}\overset{\ep\rightarrow 0}\longrightarrow \ln\rho_{F}>0,
\eean
where $\rho_{F}$ denotes the spectral radius of $F$.
The threshold time $n_{th}$ is of order $O(1)$ and 
if $F$ is diagonalizable then $n_{th}\approx [R_{dyn}^{-1}]$. 

ii) If $F$ is nonergodic and has zero entropy then anti-dynamo action occurs and 
for nondiagonalizable $F$, 
\bean
n_{p}\sim \frac{n_{th}}{\ln(n_{th})}
\sim \tau_{c}
\approx R_{c}\frac{1}{\ep}.
\eean
Moreover there exists a constant $0< \gamma \leq d$ such that 
$\|P_{\ep,\alpha}^{n_{p}}\|\sim (1/\ep)^{\gamma}$.
If $F$ is diagonalizable then $\|P_{\ep,\alpha}^{n}\|$ is strictly decreasing (in $n$) and, hence,
$n_{p}=0$ and $n_{th}$ is not defined.
   
iii) If $F$ is ergodic then anti-dynamo action occurs and
$n_{p}\approx \tau_{c}.$
In particular if $F$ is diagonalizable then
\bean
n_{p} &\approx& n_{th}-R_{c}\ln(n_{th})
\approx R_{c}\ln(1/\ep)\\
&=& \frac{1}{\alpha\hat{h}(F)}\ln(1/\ep)
\eean
and 
\bea
\label{anti}
\|P_{\ep,\alpha}^{n_{p}}\|\sim (1/\ep)^{\frac{\ln\rho_{F}}{\alpha \hat{h}(F)}}.
\eea
\end{thm}

We see that even in the case of anti-dynamo action the magnetic field
can still grow to relatively large magnitude when the noise
is small (power-law in $1/\ep$).

\textbf{Proof.}
Representing the initial magnetic field $\bB=(\bb_{1},...,\bb_{d})$ in Fourier basis
\bean
\bB=\sum_{0 \not = \bk \in \IZ^{d}}\hat{\bB}(\bk)e_{\bk}
\eean
one obtains
\bean
P_{\ep,\alpha}\bB=\sum_{0 \not = \bk \in \IZ^{d}}F\hat{\bB}(\bk)
e^{-\ep^{\alpha} |A\bk|^{\alpha}}e_{A\bk},
\eean
where we set $A=(F^{-1})^{\dagger}$.
After $n$ iterations
\bean
P_{\ep,\alpha}^{n}\bB=\sum_{0 \not = \bk \in \IZ^{d}}F^{n}\hat{\bB}(\bk)
e^{-\ep^{\alpha} \sum_{l=1}^{n}|A^{l}\bk|^{\alpha}}
e_{A^{n}\bk}.
\eean
Thus
\bean
\|P_{\ep,\alpha}^{n}\bB\|^{2} &\leq& 
\sum_{0 \not = \bk \in \IZ^{d}}|F^{n}\hat{\bB}(\bk)|^{2}
e^{-2\ep^{\alpha} \sum_{l=1}^{n}|A^{l}\bk|^{\alpha}}\\
&\leq& \max_{0 \not=\bk \in \IZ^{d}}e^{-2\ep^{\alpha} \sum_{l=1}^{n}|A^{l}\bk|^{\alpha}} 
\sum_{0 \not = \bk \in \IZ^{d}}|F^{n}\hat{\bB}(\bk)|^{2}\\
&=&e^{-2\ep^{\alpha} \min_{0 \not=\bk \in \IZ^{d}}\sum_{l=1}^{n}|A^{l}\bk|^{\alpha}} 
|F^{n}\bB|^{2} \\ &=& e^{-2\ep^{\alpha} \sum_{l=1}^{n}|A^{l}\bk_{n}|^{\alpha}} |F^{n}\bB|^{2} \\
&\leq& e^{-2\ep^{\alpha} \sum_{l=1}^{n}|A^{l}\bk_{n}|^{\alpha}} \|F^{n}\|^{2}|\bB|^{2},
\eean
where $\bk_{n}$ denotes a solution of the minimization problem 
\bean
\min_{0 \not=\bk \in \IZ^{d}}\sum_{l=1}^{n}|A^{l}\bk|^{\alpha}.
\eean
The above calculation provides the following upper bound
\bean
\|P_{\ep,\alpha}^{n}\| &\leq& e^{-\ep^{\alpha} \sum_{l=1}^{n}|A^{l}\bk_{n}|^{\alpha}} \|F^{n}\|.
\eean
On the other hand let $\bv_{n}$ denote a unit vector satisfying $\|F^{n}\|=|F^{n}\bv|$. 
One immediately sees that the above upper bound for $\|P_{\ep,\alpha}^{n}\|$ is
achieved for magnetic field of the form $\bB=\bv_{n}e_{\bk_{n}}$.
Thus
\bea
\|P_{\ep,\alpha}^{n}\|= e^{-\ep^{\alpha} \sum_{l=1}^{n}|A^{l}\bk_{n}|^{\alpha}} \|F^{n}\|.
\eea
Now we consider the cases mentioned in the statement of the theorem

i) Nonergodic, nonzero entropy case. 

For any nonergodic map we have
\bean
\sum_{l=1}^{n}|A^{l}\bk_{n}|^{\alpha} \approx R_{c}^{-1} n.
\eean
This implies the following asymptotics
\bea
\label{na}
\|P_{\ep,\alpha}^{n}\|\approx e^{-\ep^{\alpha} R_{c}^{-1} n} \|F^{n}\|\approx 
e^{(-\ep^{\alpha} R_{c}^{-1} + \ln\rho_{F})n + c_{1}\ln n +c_{2}},
\eea
where $c_{1},c_{2}\geq 0$ are constants (both equal $0$ iff $F$ is diagonalizable). 
Thus for $\ep<R_{c}{\ln\rho_{F}}$ we have
\bean
R_{dyn}= \ln\rho_{F}-\ep^{\alpha} R_{c}^{-1} \overset{\ep\rightarrow 0}\longrightarrow \ln\rho_{F}>0.
\eean
The threshold time is clearly of order $O(1)$ and in
diagonalizable case can be written as 
\bean
n_{th}\approx \frac{1}{\ln\rho_{F}-\ep^{\alpha} R_{c}^{-1}} 
\overset{\ep \rightarrow 0}\longrightarrow \frac{1}{\ln\rho_{F}}.
\eean

ii) Nonergodic, zero entropy case. 

In this case $\ln\rho_{F}=0$. Thus if $F$ is nondiagonalizable then 
(\ref{na}) reads
\bean
\|P_{\ep,\alpha}^{n}\|\approx e^{-\ep^{\alpha} R_{c}^{-1} n} \|F^{n}\|\approx 
e^{-\ep^{\alpha} R_{c}^{-1}n + c_{1}\ln n +c_{2}},
\eean
with $0<c_{1}\leq d$.
This immediately yields
\bean
n_{p}\approx  R_{c}\frac{c_{1}}{\ep},\qquad \frac{n_{th}}{\ln(n_{th})} \sim \frac{1}{\ep}.
\eean
And moreover $\|P_{\ep,\alpha}^{n_{p}}\|\sim (1/\ep)^{c_{1}}$.

If $F$ is diagonalizable then $\|F^{n}\|=1$ and in
this case $\|P_{\ep,\alpha}^{n}\|\approx e^{-\ep^{\alpha} R_{c}^{-1} n}$ which implies 
$n_{p}=0$. 

iii) If $F$ is diagonalizable, then from (\ref{ulestimateA}) we know that for 
any $0<\del<1$ and sufficiently large $n$
\bea
\label{est}
\lam_{-\del}^{n}=e^{(1-\del)\alpha\hat{h}(A) n} \leq
\min_{0 \not=\bk \in \IZ^{d}} \sum_{l=1}^{n}|A^{l}\bk|^{\alpha} \leq
e^{(1+\del)\alpha \hat{h}(A)n}=\lam_{+\del}^{n}.
\eea
Thus for large $n$ we have
\bean
\max_{n} e^{-\ep^{\alpha} \lam_{+\del}^{n}}\rho_{F}^{n} \leq \max_{n}\|P_{\ep,\alpha}^{n}\|
\leq \max_{n} e^{-\ep^{\alpha} \lam_{-\del}^{n}}\rho_{F}^{n}.
\eean
We obtain the following constraints for  $n_{p}$
\bean
\frac{1}{\ln \lam_{+\del}}\ln\left(\frac{\ln \rho_{F}}{\ln \lam_{+\del}}\right)
+\frac{1}{\ln \lam_{-\del}}\ln\left(\frac{1}{\ep}\right)
\leq n_{p}\leq \frac{1}{\ln \lam_{-\del}}\ln\left(\frac{\ln \rho_{F}}{\ln \lam_{-\del}}\right)
+\frac{1}{\ln \lam_{-\del}}\ln\left(\frac{1}{\ep}\right).
\eean
This gives
\bean
\frac{1}{\ln \lam_{+\del}} \leq \lim_{\ep \rightarrow 0} \frac{n_{p}}{\ln(1/\ep)}
\leq\frac{1}{\ln \lam_{-\del}}.
\eean
Now since $\lam_{\pm \del} \rightarrow e^{\alpha\hat{h}(F)}$ for $\del\rightarrow 0$ 
the above estimation yields the following asymptotics
\bean
n_{p} \approx \frac{1}{\hat{h}(F)}\ln(1/\ep)\approx \tau_{c}
,\qquad n_{th}-R_{c}\ln(n_{th})\approx \tau_{c}.
\eean
Similar asymptotic estimates (except for the constant) hold
for nondiagonalizable $F$.
$\qquad \blacksquare$

%%%%%%%%%%%%%%%%%%%%%%%%%%%%%%%%%%%%%%%%%%%%%%%%%%%%%%%%%%%%%%%%%%%%%%%%%%%%%%%%%%%%%%%%%%%%
%%%%%%%%%%%%%%%%%%%%%%%%%%%%%%%%%%%%%%%%%%%%%%%%%%%%%%%%%%%%%%%%%%%%%%%%%%%%%%%%%%%%%%%%%%%%
%%%%%%%%%%%%%%%%%%%%%%%%%%%%%%%%%%%%%%%%%%%%%%%%%%%%%%%%%%%%%%%%%%%%%%%%%%%%%%%%%%%%%%%%%%%%

\section{Dissipation time of general $C^{3}$ Anosov maps}
\label{Anosov}

%%%%%%%%%%%%%%%%%%%%%%%%%%%%%%%%%%%%%%%%%%%%%%%%%%%%%%%%%%%%%%%%%%%%%%%%%%%%%%%%%%%%%%%%%%%%
%%%%%%%%%%%%%%%%%%%%%%%%%%%%%%%%%%%%%%%%%%%%%%%%%%%%%%%%%%%%%%%%%%%%%%%%%%%%%%%%%%%%%%%%%%%%
%%%%%%%%%%%%%%%%%%%%%%%%%%%%%%%%%%%%%%%%%%%%%%%%%%%%%%%%%%%%%%%%%%%%%%%%%%%%%%%%%%%%%%%%%%%%

We recall that a diffeomorphism $F:\IT^d\mapsto \IT^d$ is called
Anosov if it is uniformly hyperbolic:
there exist constants $A>0$ and $0<\lam_{s}<1<\lam_{u}$ such that
at each $\bx\in\IT^d$ the tangent space $T_{\bx}\IT^d$
admits the direct sum decomposition $T_{\bx}\IT^d= E^{s}_{\bx} \oplus E^{u}_{\bx}$ 
into stable and unstable subspaces such that for every $n\in \IN$,
\bean
(D_{\bx}F)(E^{s}_{\bx})=E^{s}_{F\bx}, \qquad \|(D_{\bx}F^n)_{|E^{s}_{\bx}}\|\leq A \lam_{s}^n;\\
(D_{\bx}F)(E^{u}_{\bx})=E^{u}_{F\bx}, \qquad \|(D_{\bx}F^{-n})_{|E^{u}_{\bx}}\|\leq A \lam_{u}^{-n}.
\eean
These inequalities have obvious consequences on the expansion rates of $F$ and $F^{-1}$, for 
instance they imply $\|DF\|_\infty^n\geq\|DF^n\|_\infty\geq A^{-1}\lam_u^n$. As a consequence,
the quantities of interest in Theorem~\ref{nln}, {\em i)} satisfy
\bean
\|DF\|_\infty\wedge\|DF^{-1}\|_\infty&\geq& \lam_u\wedge\lam_s^{-1},\\
\mu_F\wedge\mu_{F^{-1}}&\geq& \lam_u\wedge\lam_s^{-1}.
\eean 
All these expansion rates are $>1$, so both noisy and coarse-grained dissipation times admit
logarithmic lower bounds as in Eq.~\eqref{nlb}.

To obtain upper bounds, we use the mixing properties of this dynamics.
Exponential mixing has been proved for Anosov diffeomorphisms of regularity $C^{1+\eta}$ ($0<\eta<1$)
by Bowen \cite{Bow}, using
symbolic dynamics; the exponential decay is then valid for H\"older 
observables in $C^{\eta'}$ for some $0<\eta'<\eta$.

Because we are interested in the noisy dynamics as well, we also
refer to a more recent work \cite{BKL} concerning $C^3$ Anosov maps on $\IT^d$, which bypasses
symbolic dynamics. The authors
construct an {\em ad hoc} invariant Banach space $\cB$ of generalized functions on the phase space, 
such that the Perron-Frobenius operator is quasicompact on this space. The difficulty compared to the
case of expanding maps, is that the space $\cB$ explicitly depends
on the (un)stable foliations generated by the map $F$ on $\IT^d$. Vaguely speaking, the elements of $\cB$ are
``smooth'' along this unstable direction $E^{u}_{\bx}$, but can be singular (``dual of smooth'') 
along the stable foliation $E^{s}_{\bx}$. 
The space $\cB$ is
the completion of $C^1(\IT^d)$ with respect to a weaker norm $\|\cdot\|_{\cB}$ adapted to these
foliations. 
The definition of that norm is given in terms of a parameter $0<\beta<1$, the choice of which
depends on the {\sl regularity} of the unstable foliation. 
In general, the latter is only 
H\"older continuous on the torus, with some exponent $0<\tau<2$ (note that the regularity of the foliations
has little to do with the regularity of the map itself: a real-analytic map may have
foliations being only H\"older). Then, one must take $\beta<\tau\wedge 1$, and the authors prove
that the essential spectrum of $P_F$ on $\cB$ has a radius smaller than 
$r_1=\max(\lambda_u^{-1},\lambda_s^{\beta})$. This upper bound
is sharper if $\beta$ can be taken close to $1$, that is, if the foliation is at least $C^1$. 
This is the case for smooth area-preserving Anosov maps on $\IT^2$, for which the foliations have regularity
$C^{2-\delta}$ for any $\delta>0$ \cite{HK}. 
The operator $P_F$ may have isolated eigenvalues $1>|\lambda_i|>r_1$, corresponding to eigenstates in $\cB$ 
which are genuine distributions, outside $L^2_0$. As opposed to the radius $r_1$, there is (to our
knowledge) no simple
general upper bound for largest resonance $|\lambda_1|$ 
in terms of $(\lambda_u,\ \lambda_s)$.

The space $C^1(\IT^d)$ can be embedded continuously in both $\cB$ and its dual $\cB^*$, 
so that one can take $s=s_*=1$ in Eq.~\eqref{expon}. Therefore, for any
$1>\sigma>\max(|\lam_1|,r_1)$, there is some constant $C>0$ such that for any $f$, $g\in C_0^1(\IT^d)$,
\bea
\label{corr-anosov}
\forall n>0,\qquad |C_{f,g}(n)|\leq C\,\|g\|_{C^1}\,\|f\|_{C^1}\,\sigma^n.
\eea
In the proof of Theorem~\ref{ccln} (Step 3), for the case $s=s_*=1$ we only need to assume that
the noise generating function $g$ is $C^1$ with fast-decaying first derivatives. The fast decay
implies that the first moment of $g$ is finite (that is, one can take $\alpha\geq 1$).

The results of \cite{BKL} also concern the noisy dynamics. 
If the unstable foliation has regularity $C^{1+\eta}$ with
$\eta>0$ (for instance for any $C^3$  Anosov 
diffeomorphism on the $2$-dimensional torus), and assuming the noise generating
function $g\in C^2(\IR^d)$ with {\em compact support}, the authors prove
the strong spectral stability of the Perron-Frobenius operator $P_F$ on the space $\cB$ defined
with the parameter $\beta<\eta$. 
Therefore, the estimate \eqref{corr-anosov} also applies to
the noisy correlation function $C^{\ep}_{f,g}(n)$ as long as $\ep$ is small enough. 

Using these estimates and applying Theorem \ref{nln} (Section \ref{LB}) and
Corollary \ref{correl-upper} (Section \ref{UB}) we obtain the following 
result regarding the dissipation time of $C^3$ Anosov maps on the torus:

\begin{thm}
\label{TAnosov}
Let $F$ be a volume preserving $C^{3}$ Anosov diffeomorphisms on $\IT^d$ 
and let $g$ be a $C^1$ noise generating function with fast decay at infinity.

I) Then there exist $\mu\geq\lam_u\wedge\lam_s^{-1}$, $0<\sigma<1$ and
 $\tilde{C}>0$ such that
the dissipation time of the coarse-grained dynamics satisfies
\bean
  \frac{1}{\ln \mu }\ln(\ep^{-1}) -\tilde{C} \leq
  \tilde{\tau}_c \leq \frac{d+2}{|\ln \sigma|} \ln(\ep^{-1}) + \tilde{C},
\eean

II) If in addition $F$ has $C^{1+\eta}$-regular foliations  
and $g\in C^2(\IR^{d})$ is compactly 
supported, then the  dissipation time of the noisy evolution satisfies for some $C>0$ and
small enough $\ep$:
\bean
  \frac{1}{\ln \|DF\|_\infty}\ln(\ep^{-1}) - C 
\leq  {\tau}_c \leq \frac{d+2}{|\ln \sigma|}\ln(\ep^{-1}) + C
\eean
\end{thm}

%%%%%%%%%%%%%%%%%%%%%%%%%%%%%%%%%%%%%%%%%%%%%%%%%%%%%%%%%%%%%%%%%%%%%%%%%%%%
%%%%%%%%%%%%%%%%%%%%%%%%%%%%%%%%%%%%%%%%%%%%%%%%%%%%%%%%%%%%%%%%%%%%%%%%%%%%
%%%%%%%%%%%%%%%%%%%%%%%%%%%%%%%%%%%%%%%%%%%%%%%%%%%%%%%%%%%%%%%%%%%%%%%%%%%%

\section{Examples and general comments on other chaotic systems}
\label{examples}

%%%%%%%%%%%%%%%%%%%%%%%%%%%%%%%%%%%%%%%%%%%%%%%%%%%%%%%%%%%%%%%%%%%%%%%%%%%%%%%%%%%%%%%%%%%%
%%%%%%%%%%%%%%%%%%%%%%%%%%%%%%%%%%%%%%%%%%%%%%%%%%%%%%%%%%%%%%%%%%%%%%%%%%%%%%%%%%%%%%%%%%%%
%%%%%%%%%%%%%%%%%%%%%%%%%%%%%%%%%%%%%%%%%%%%%%%%%%%%%%%%%%%%%%%%%%%%%%%%%%%%%%%%%%%%%%%%%%%%

%%%%%%%%%%%%%%%%%%%%%%%%%%%%%%%%%%%%%%%%%%%%%%%%%%%%%%%%%%%%%%%%%%%%%%%%%%%%%%%%%%%%%%%%%%%%
%%%%%%%%%%%%%%%%%%%%%%%%%%%%%%%%%%%%%%%%%%%%%%%%%%%%%%%%%%%%%%%%%%%%%%%%%%%%%%%%%%%%%%%%%%%%

\subsection{Noiseless correlations}

%%%%%%%%%%%%%%%%%%%%%%%%%%%%%%%%%%%%%%%%%%%%%%%%%%%%%%%%%%%%%%%%%%%%%%%%%%%%%%%%%%%%%%%%%%%%
%%%%%%%%%%%%%%%%%%%%%%%%%%%%%%%%%%%%%%%%%%%%%%%%%%%%%%%%%%%%%%%%%%%%%%%%%%%%%%%%%%%%%%%%%%%%

\label{noiseless}
Let us first consider the correlation functions $C_{f,g}(n)$ for the noiseless dynamics. A common
route to prove that a map $F$ is mixing consists in studying the Perron-Frobenius 
operator $P_F$ on some cleverly selected space $\cB$ of densities
(this may be a Banach or Fr\'echet space).
The objective is to prove that the spectrum of $P_F$ on $\cB$ is of \emph{quasicompact} type. More
precisely, one shows that $P_F$ admits $1$ as simple eigenvalue (with the constant eigenfunction), 
and that the radius of its essential
spectrum is smaller than some $0<r_1<1$. The spectrum outside the disk of radius $r_1$ is
made of a finite number of
isolated, finitely-degenerate eigenvalues $\{\lambda_i\}$ of moduli
$r_1<|\lambda_i|<1$ (usually called Ruelle-Pollicott resonances). 
We order these eigenvalues according to their decreasing moduli, so that the
largest modulus is $|\lambda_1|$. This spectral structure of $P_F$ on $\cB$ implies that for any 
$1>\sigma>\max(|\lambda_1|,r_1)$ 
(if the largest resonance $\lambda_1$ is semisimple, one may take $\sigma=|\lambda_1|$), 
there is some constant $C=C(\sigma)$ such that
for any $f\in\cB_0$, $g\in\cB^*_0$:
\bea
\label{expon1}
|C_{f,g}(n)|&=&|\la g,P_F^n f\ra_{\cB^*,\cB}|\non\\
&\leq& \|g\|_{\cB^*}\|P_F^n\|_{\cB_0}\|f\|_{\cB}\non\\
&\leq& C\,\|g\|_{\cB^*}\,\|f\|_{\cB}\;\sigma^n.
\eea

The choice of the space $\cB$ is crucial
here: in general the spectrum of $P_F$ on $L^2_0$ intersects the unit circle, 
so there is no chance to prove 
exponential decay within the $L^2$ framework.
$\cB$ can be a Banach space of bounded variation, H\"older or $C^{s}$ functions which embeds 
continuously in $L^2$ (this is the choice made for $F$ uniformly expanding, see Section~\ref{expanding}); it may also
be a space of generalized functions lying outside $L^2$ (case of Anosov maps, see Section~\ref{Anosov}). 

In general, there exist H\"older exponents $0\leq s_*\leq s$ such that $C^s$  (resp. $C^{s_*}$) 
embeds continuously in $\cB$ (resp. its dual $\cB^*$). As a result, the upper bound \eqref{expon1} induces
that for any $f\in C_0^s$, $g\in C_0^{s_*}$ and any $n>0$, 
\bea
\label{expon}
|C_{f,g}(n)|\leq C\,\|g\|_{C^{s_*}}\,\|f\|_{C^{s}}\;\sigma^n.
\eea
This is the form of upper bound we used in Theorem~\ref{ccln}. When the radius $\sigma$ is given
by a resonance $\lambda_1$, it can often not be decreased when one takes observables of higher regularity. 
In this case, it is preferable to
take in the above estimate the weakest norms possible, that is take $s$ and $s_*$ as small as possible 
(to obtain better upper bounds in Corollary~\ref{correl-upper}).

This strategy of proof has been applied to several types of maps \cite{B,Bal-decay}, including
the (noninvertible) expanding maps and the Anosov or Axiom-A diffeomorphisms on a
compact manifold. We will give some details and examples of these two types of maps in the next
subsections.  Exponential decay of correlations has also been proved (using various methods) 
for piecewise expanding maps on the
interval, some nonuniformly hyperbolic/expanding maps, like ``good'' logistic maps on the interval, or
``good'' H\'enon maps;  some expanding or hyperbolic maps with singularities. In those cases, the decay
rate may have no (obvious) spectral interpretation, as opposed to the formalism described above
\cite{Bal-decay}.

Different types of decay, like the stretched exponential $C_{f,g}(n)\leq Kr^{n^\xi}$ with $0<r<1$ and 
$0<\xi<1$ have been proved for some Poincar\'e maps of hyperbolic flows and some random nonuniformly
hyperbolic systems; yet, it seems that the bound may not be optimal in some of these cases, but rather
due to the method of the proof. On the opposite, the polynomial decay $C_{f,g}(n)\lesssim n^{-\beta}$ was
shown to be optimal for some ``intermittent'' systems, like a
one-dimensional map expanding everywhere except at a fixed ``neutral'' point (such maps are sometimes
called ``almost expanding'' or ``almost hyperbolic'').

\textbf{Remark:}
In general, an expanding or Anosov map does not preserve the Lebesgue measure, so the first
step before dealing with correlations is to precise the invariant measure with respect to which
one wants to study the ergodic properties. In the ``nice'' cases, one can prove the existence 
and uniqueness of a ``physical measure'', which is ergodic for the map $F$, and then study the correlation
functions with respect to this measure. The formalism of Perron-Frobenius operators applies also to this
general case, the physical measure being related to the simple eigenstate of $P_F$ (on the space $\cB$)
associated with the eigenvalue $1$. As stated in the Introduction, in this work we only consider maps 
for which the physical measure is the Lebesgue measure.

%%%%%%%%%%%%%%%%%%%%%%%%%%%%%%%%%%%%%%%%%%%%%%%%%%%%%%%%%%%%%%%%%%%%%%%
%%%%%%%%%%%%%%%%%%%%%%%%%%%%%%%%%%%%%%%%%%%%%%%%%%%%%%%%%%%%%%%%%%%%%%%

\subsection{Noisy correlations}

%%%%%%%%%%%%%%%%%%%%%%%%%%%%%%%%%%%%%%%%%%%%%%%%%%%%%%%%%%%%%%%%%%%%%%%%%%%%%%%%%%%%%%%%%%%%
%%%%%%%%%%%%%%%%%%%%%%%%%%%%%%%%%%%%%%%%%%%%%%%%%%%%%%%%%%%%%%%%%%%%%%%%%%%%%%%%%%%%%%%%%%%%

There exist fewer results on the decay of correlations for stochastic perturbations of deterministic maps, 
like our noisy evolution $T_\ep$. In general, one wants to prove some sort of 
\emph{strong stochastic stability}, that is stability of the invariant measure, and of 
the rate of decay of the correlations in the limit when the noise parameter $\ep$ vanishes. 

In the case of exponential mixing, this strong stochastic stability implies that for
small enough $\ep$, the upper bound \eqref{expon} is ``stable'' when switching on the noise:
for small enough $\ep>0$ there exists a radius $\sigma_\ep\epto \sigma$ such that for any 
$f\in C^{s}_0$, $g\in C^{s_*}_0$,
\bea
\label{noisy-expon}
\forall n>0,\qquad |C^\ep_{f,g}(n)|\leq C\,\|g\|_{C^{s_*}}\,\|f\|_{C^{s}}\,\sigma_\ep^n.
\eea

This stability has been proved for smooth uniformly expanding
maps \cite{BalYou}, some nonuniformly expanding or piecewise expanding maps
(see the review in \cite{Bal-decay} or the book \cite{B}).
It has been shown also for uniformly hyperbolic (Anosov) maps on the $2$-dimensional torus \cite{BKL}. 
The proof uses perturbation theory: one shows that the isolated
eigenvalues of $G_\ep\circ P_F$ on $\cB$ of moduli $|\lambda_{i,\ep}|>r_1$ (and the associated
eigenstates) vary continuously w.r.to $\ep$ when $\ep\to 0$. 
Therefore, one can choose a rate $\sigma_\ep>\max(|\lambda_{1,\ep}|,r_1)$ which is a continuous function
of $\ep$.

\medskip

In the next two sections, we describe in some detail the exponential decay of correlations
for smooth uniformly expanding maps and Anosov diffeomorphisms on the torus.

%%%%%%%%%%%%%%%%%%%%%%%%%%%%%%%%%%%%%%%%%%%%%%%%%%%%%%%%%%%%%%%%%%%%%%%
%%%%%%%%%%%%%%%%%%%%%%%%%%%%%%%%%%%%%%%%%%%%%%%%%%%%%%%%%%%%%%%%%%%%%%%

\subsection{Smooth uniformly expanding maps}
\label{expanding}

%%%%%%%%%%%%%%%%%%%%%%%%%%%%%%%%%%%%%%%%%%%%%%%%%%%%%%%%%%%%%%%%%%%%%%%%%%%%%%%%%%%%%%%%%%%%
%%%%%%%%%%%%%%%%%%%%%%%%%%%%%%%%%%%%%%%%%%%%%%%%%%%%%%%%%%%%%%%%%%%%%%%%%%%%%%%%%%%%%%%%%%%%

Let $F$ be a $C^{s+1}$ map on $\IT^d$ (with $s\geq 0$). 
Assume that there exists $\lam>1$ such that for any $\bx\in\IT^d$ and any $\bv$ in the
tangent space $T_{\bx}\IT^d$, $\|DF(\bx)\bv\|\geq\lam \|\bv\|$ (we assume that $\lam$ is the
largest such constant). Such a map is called
uniformly expanding. In general, it admits a unique absolutely continuous invariant
probability measure; here we will restrict ourselves to maps for which this measure is the Lebesgue measure.

Ruelle \cite{Ru89} proved that the Perron-Frobenius operator $P_F$ of such a map is quasicompact on the space
$C^s(\IT^d)$, and that its essential spectrum is contained inside the disk of radius $r_1=\lam^{-s}$. In general, 
one has little information on the possible discrete spectrum outside this disk (upper bounds on the
decay rate have been obtained in the case of an expanding map of 
regularity $C^{1+\eta}$ \cite{B}).  
Strong stochastic stability for such maps was proved in \cite{BalYou}, with a 
more general definition of the noise than the one we gave.

For all these cases, one can take $s_*=0$, since the 
continuous functions are continuously embedded in any space $(C^s)^*$. 

\subsection*{Case of a linear expanding map}
We describe the simplest example possible for such a map, namely the angle-doubling map on $\IT^1$ defined as
$F(x)=2x\bmod 1$. This map is real analytic, with uniform expansion rate $\lam=2$. Due to its linearity, the
dynamics of this map (as well as its noisy version) 
is simple to express in the basis of Fourier modes $\be_k(x)=e^{2i\pi kx}$:
\bean
\forall k\in\IZ,\quad U_{F}\be_{k}&=&\be_{2k}\\
\Longrightarrow T_\ep\be_{k}&=&\hat g(\ep k)\be_{2k}\\
\Longrightarrow T^n_\ep\be_{k}&=&\big[\prod_{j=1}^{n}\hat g(\ep 2^j k)\big]\be_{2^n k}
\eean
The computation is even simpler for the coarse-grained propagator:
$$
\tilde T^{(n)}_\ep \be_{k}=\hat g(k)\hat g(2^n k)\be_{2^n k}.
$$
To fix the ideas, we take for the noise generating function $\hat g(\xi)=e^{-|\xi|^\alpha}$ for some $0<\alpha\leq 2$.
One easily checks that for any $n\geq 1$, 
\bean
\|T_\ep^n\|&=&\|T_\ep^n \be_1\|=\exp\bigg\{-\ep^\alpha \frac{2^{n\alpha}-1}{1-2^{-\alpha}}\bigg\},\\
\|\tilde{T}^{(n)}_\ep\|&=&\exp\{-\ep^\alpha (2^{n\alpha}+1)\}.
\eean
For any $\ep>0$, these decays are super-exponential: the spectrum of $T_\ep$ on $L^2_0$
is reduced to $\{0\}$ for any $\ep>0$ (the spectrum of $U_F$ is the full unit disk).
From there, we get explicit expressions for both dissipation times:
\bea
\label{exact-asympt}
{\tau}_{c} =\frac{1}{\ln 2}\ln(\ep^{-1})+\cO(1),
\qquad \tilde{\tau}_{c}=\frac{1}{\ln 2}\ln(\ep^{-1})+\cO(1).
\eea
For this linear map, $2=\|DF\|_\infty=\mu_F$, so this estimate is in agreement
with the lower bounds \eqref{nlb}, the latter being sharp if $\alpha\in [1,2]$.
On the other hand, $\ln 2$ is also equal with the Kolmogorov-Sinai (K-S) entropy $h(F)$ of $F$.
Therefore, for this linear map the dissipation rate constant exists, and is equal to $\frac{1}{h(F)}$.

To compare these exact asymptotics with the upper bounds of Corollary~\ref{correl-upper}, we estimate the
correlation functions $C_{f,g}(n)$ on the spaces $C^{s}(\IT^1)$. We give below a short proof in the
case $s>\half$. We will use the following Fourier estimates \cite{Zyg}:
$$
\exists C>0,\quad,\forall f\in C_0^{s}(\IT^1),\quad \forall k\neq 0,\qquad 
|\hat f(k)|\leq C\frac{\|f\|_{C^{s}}}{|k|^{s}}.
$$
Therefore, writing the correlation function as a Fourier series, we get:
\bea
\|P_F^n f\|^2&=&\sum_{0\not =k\in \IZ} |\hat f(2^n k)|^2
\leq\sum_{0\not = k\in \IZ}\lp C\frac{\|f\|_{C^{s}}}{|2^n k|^{s}}\rp^2\non\\
\Longrightarrow \|P_F^n f\|&\leq&C'\,\frac{\|f\|_{C^{s}}}{(2^{s})^n}.\label{angle-d-spectrum}
\eea
This estimate yields a decay of the correlation function 
as in Eq.~\eqref{expon}, with a rate $\sigma=2^{-s}$ and $s_*=0$. One can check that this rate is
sharp for functions in $C^{s}$: indeed, any $z\in\IC$, $|z|<2^{-s}$ is an eigenvalue
of $P_F$ on that space.
Applying the Corollary~\ref{correl-upper},{\em I ii)}, we
get that for any $s\geq 0$, there exists a constant $\tilde{c}$ such that
\bea
\label{angle-d-upper}
\tilde{\tau}_{c}\leq \frac{1+s}{s\ln 2}\ln(\ep^{-1})+\tilde{c}
\eea
for sufficiently small $\ep$. The exact dissipation rate
constant $1/\ln 2$ is recovered only for large $s$.

A straightforward computation shows that the estimate
\eqref{angle-d-spectrum} also holds if one replaces $P_F$ by $P_F\circ G_\ep$; hence the noisy
correlation function dynamics satisfies the same uniform upper bound as the noiseless one, with the
decay rate
$\sigma_\ep=2^{-s}$. As a result, the upper bound on ${\tau}_{c}$ given by 
Corollary~\ref{correl-upper},{\em II ii)} is the same as in  Eq.~\eqref{angle-d-upper}.

%%%%%%%%%%%%%%%%%%%%%%%%%%%%%%%%%%%%%%%%%%%%%%%%%%%%%%%%%%%%%%%%%%%%%%%%%%%%%%%%%%%%%%%%%%%%%%%%%%%%%%%%%%
%%%%%%%%%%%%%%%%%%%%%%%%%%%%%%%%%%%%%%%%%%%%%%%%%%%%%%%%%%%%%%%%%%%%%%%%%%%%%%%%%%%%%%%%%%%%%%%%%%%%%%%%%%%
%%%%%%%%%%%%%%%%%%%%%%%%%%%%%%%%%%%%%%%%%%%%%%%%%%%%%%%%%%%%%%%%%%%%%%%%%%%%%%%%%%%%%%%%%%%%%%%%%%%%%%%%%%%

\section{Technical proofs}
\label{Ch3proofs}

%%%%%%%%%%%%%%%%%%%%%%%%%%%%%%%%%%%%%%%%%%%%%%%%%%%%%%%%%%%%%%%%%%%%%%%%%%%%%%%%%%%%%%%%%%%%%%%%%%%%%%%%%%
%%%%%%%%%%%%%%%%%%%%%%%%%%%%%%%%%%%%%%%%%%%%%%%%%%%%%%%%%%%%%%%%%%%%%%%%%%%%%%%%%%%%%%%%%%%%%%%%%%%%%%%%%%%
%%%%%%%%%%%%%%%%%%%%%%%%%%%%%%%%%%%%%%%%%%%%%%%%%%%%%%%%%%%%%%%%%%%%%%%%%%%%%%%%%%%%%%%%%%%%%%%%%%%%%%%%%%%

\subsection*{Proof of Proposition \ref{irred}.}

For the purposes of the proof we use the following abbreviation
\begin{itemize}
\item $PRS(\IR^{d})$ - proper rational subspace of $\IR^{d}$.

\item $PIS(A,\IR^{d})$ - proper $A$-invariant subspace of $\IR^{d}$.

\item $PRIS(A,\IR^{d})$ - proper rational $A$-invariant subspace of $\IR^{d}$.
\end{itemize}
\begin{description}
\item[a) $\Rightarrow $ b)]. Suppose there exists $PRIS(A,\IR^{d})$ $S_{1}$. Let $A_{1}$ be a matrix 
representing invariant rational block associated with $S_{1}$. Then $A_{1}$ is rational 
matrix and its characteristic polynomial $P_{1}$ belongs to $\IQ[x]$. 
Let $P$ denote the characteristic polynomial of $A$. Then $P=P_{1}P_{2}$ and since both 
$P,P_{1}\in \IQ[x]$ then also $P_{2}\in \IQ[x]$, which means $P$ and hence $A$ is not irreducible. 

\item[b) $\Rightarrow $ c)]. Assume there exists $PRS(\IR^{d})$ $S$ contained in $PIS(A,\IR^{d})$ $V$. 
Take any rational vector $\bq\in S$ and let $d_{0}=dimV$ then the set 
$\{\bq,A\bq,..,A^{d_{0}-1}\bq\}$ spans $PRIS(A,\IR^{d})$.

\item[c) $\Rightarrow$ d)]. Assume that for given $\bq$ and an arithmetic sequence $n_{1},...,n_{d}$, the set 
$S=\{A^{n_{1}}\bq,A^{n_{2}}\bq,...,A^{n_{d}}\bq\}$ does not form a basis. Since for some fixed
integer $r$, $n_{l}=n_{1}+(l-1)r$, we have $A^{n_{l}}\bq=(A^{r})^{l-1}A^{n_{1}}\bq=(A^{r})^{l-1}\hat{\bq}$, 
where $\hat{\bq}=A^{n_{1}}\bq$. Now consider the biggest subset 
$S_{0}=\{\hat{\bq},A^{r}\hat{\bq},(A^{r})^{2}\hat{\bq},...,(A^{r})^{d_{0}-1}\hat{\bq}\}$ such 
that $d_{0}<d$ and $S_{0}$ is linearly independent. Obviously $S_{0}$ spans a $PRIS(A^{r})$ which
is also a $PRIS(A)$.

\item[d) $\Rightarrow $ a)]. 
Suppose that characteristic polynomial $P$ of $A$ is not irreducible in $\IQ[x]$. 
Then $P=P_{1}P_{2}$, with $P_{1},P_{2}\in \IQ[x]$. From the Cayley-Hamilton theorem we get that
$0=P(A)=P_{1}(A)P_{2}(A)$. Hence for any nonzero rational vector $\bq$, either 1) $P_{2}(A)\bq=0$ 
or 2) $\hat{\bq}:=P_{2}(A)\bq \not = 0$ and $P_{1}(A)\hat{\bq}=0$. Since $\max\{deg(P_{1},P_{2})\}<d$,
there exists a nonzero rational vector $\tilde{\bq}$ (namely $\bq$ or $\hat{\bq}$) such that 
the set of iterations $\{\tilde{\bq},A\tilde{\bq},A^{2}\tilde{\bq},...,A^{d-1}\tilde{\bq}\}$ does 
not form a basis of $\IR^{d}$.

\item[e) $\Rightarrow $ f)]. Assume there exist nonzero $\bq \in \IQ^{d}$ orthogonal to certain 
$PIS(A,\IR^{d})$ $V$. Then for any $n$ and any $f\in V$, $\la (A^{\dagger})^{n}\bq, f \ra= \la \bq, A^{n}f \ra  = 0$ 
and hence $S=\{\bq,A^{\dagger}\bq,(A^{\dagger})^{2}\bq,...,(A^{\dagger})^{d-1}\bq\}$, cannot form a basis, which in view 
of equivalence a) $\Leftrightarrow$ d) implies reducibility of $A^{\dagger}$. 

\item[f) $\Rightarrow$ g)]. Suppose there exists $PIS(A,\IR^{d})$ $V$ contained in certain 
$PRS(\IR^{d})$ $S$. Since $S$ is rational, $S^{\perp}$ is also rational. Consider any rational 
vector $\bq \in S^{\perp}$, then $\la \bq,f \ra = 0$ for any $f \in V$.

\item[g) $\Rightarrow$ b)]. If there exists $PRIS(\IR^{d})$, then this subspace is $A$-invariant and 
contained in $PRS(\IR^{d})$ i.e
in itself. 
\end{description}

Now since b) is equivalent to a) it is enough to establish the equivalence
between a) and e) to complete the proof. But the latter equivalence is obvious
in view of the fact that $A$ and $A^{\dagger}$ have the same characteristic
polynomial.
$\qquad \blacksquare$

\subsection*{Proof of Proposition \ref{zeroentropy}.}

Suppose $A$ is a toral automorphism of zero entropy. The latter property is 
equivalent
to the fact that all the eigenvalues of $A$ are of modulus $1$. Let $P_{A}$ be a
 characteristic
 polynomial of $A$. Consider any irreducible over $\IZ$ factor $P$ of polynomial
 $P_{A}$ and construct
a toral automorphism $B$ such that its characteristic polynomial is equal to $P$
 .
Obviously all the eigenvalues of $B$ are also the eigenvalues of $A$, and each 
 eigenvalue of
$A$ can be found among eigenvalues of some matrix $B$ of this type.
 Irreducibility of $P$ implies irreducibility and hence diagonalizability of $B$.

 Thus for any nonzero vector $\bk \in \IZ^{d}$ and any positive integer $n$ the
 following estimate holds
 $|B^{n}\bf{k}| \leq |\bf{k}|$,
which implies the existence (for each $\bf{k}$) of some integer $r$ such that $
 B^{r}\bf{k}=\bf{k}$.

The latter shows that all the eigenvalues of $B$ (and hence also of $A$) are 
roots
 of unity.
$\qquad \blacksquare$

\subsection*{Proof of Proposition \ref{red}.}

We first show that irreducible polynomials $P\in\IQ[x]$ do not have repeated roots.
Indeed suppose $\lam$ is a root of $P$ of multiplicity greater that 1, then $\lam$ is
also a root of a derivative polynomial $P'\in\IQ[x]$. Since the minimal polynomial of $\lam$
must divide both $P$ and $P'$ and $deg(P')<deg(P)$ one immediately concludes that $P$ is not irreducible.
Now, suppose $A \in GL(d,\IQ)$ is completely decomposable over $\IQ$ and let (\ref{blockdiag}) be its 
block diagonal decomposition into irreducible blocks.
Each $P_{A_{j}}$, as a characteristic polynomial of $A_{j}$, is irreducible over $\IQ$ and hence does
not possesses repeated roots, which implies diagonalizability of each $A_{j}$ and hence of $A$.
On the other hand if $A$ is diagonalizable then its minimal polynomial does not possesses repeated 
roots, which implies that all characteristic polynomials associated with elementary divisors are 
(first powers of) irreducible polynomials. 
This implies irreducibility of each block in representation (\ref{blockdiag}). 
$\qquad \blacksquare$

\subsection*{Proof of Proposition \ref{distinct}.}

Let $P_{A}$ be the characteristic polynomial of an irreducible matrix $A \in GL(d,\IQ)$.
Since $P_{A}$ is an irreducible element of $\IQ[x]$ it does not possesses repeated roots
(see the proof of Proposition \ref{red}).
$\qquad \blacksquare$

\subsection*{Proof of Proposition \ref{supexp}}
Combining formula \ref{Tnorm3} and Theorem \ref{thmart} in case of
$\alpha$-stable noises we get that for any $\del>0$
\bea
\label{norm}
\| T^{n}_{\ep,\alpha}\|\leq e^{-\ep^{\alpha}e^{(1-\del)\alpha\hat{h}(F)n}}.
\eea
Using this estimate one immediately gets
\bean
C^{\ep}_{f,h}(n)=\la \bar{f}, T^{n}_{\ep,\alpha}h \ra \leq \|f\|\|h\|\| T^{n}_{\ep,\alpha}\| 
\leq \|f\|\|h\|e^{-\ep^{\alpha} \lam^{\alpha n}}.
\eean
Now let $f=G_{\ep}f_{0}$ and $h=G_{\ep}g_{0}$. Since the estimate (\ref{norm}) holds also in 
coarse-grained version, we have for $\alpha=2$
\bean
C_{f,h}(n)&=&\la \bar{f}, U_{F}^{n}h \ra = \la G_{\ep}\bar{f_{0}}, U_{F}^{n}G_{\ep}h_{0} \ra
=\la \bar{f_{0}}, \tilde{T}^{(n)}_{\ep}h_{0} \ra \\
&\leq& \|f_{0}\|\|h_{0}\|\| \tilde{T}^{(n)}_{\ep}\| \leq \|f_{0}\|\|h_{0}\|e^{-\ep^2 \lam^{2n}}.
\qquad \blacksquare
\eean

\commentout{
\bigskip
\textbf{Proof of Proposition \ref{underg}.}

We already know that ergodic automorphisms have positive entropy.
Assume $F$ is indecomposable and let $P^{m}$ be its characteristic
polynomial, where $P$ is irreducible over $Q$.
Let $B$ be any toral automorphism with characteristic polynomial $P$.
$B$ is irreducible and has positive entropy. Suppose $B$ is not 
ergodic. Take integral vector $k$ which satisfies
$B^{r}\bk=\bk$ for certain $r$. Then for any $n$ $|B^{n}\bk|<C$ which means
$\bk$ has no components in expanding eigenspace of $B$.
Therefore the set of consecutive iterates ${\bk,A\bk,A^{2}\bk,...}$ spans 
proper rational $B$-invariant subspace, which contradicts irreducibility
of $B$. Now ergodicity of $B$ clearly implies ergodicity of $F$. 
$\qquad \blacksquare$
}

%%%%%%%%%%%%%%%%%%%%%%%%%%%%%%%%%%%%%%%%%%%%%%%%%%%%%%%%%%%%%%%%%%%%%%%%%%%%%%%%%%%%%%%%%%%%%%%%%%%%%%%%%%
%%%%%%%%%%%%%%%%%%%%%%%%%%%%%%%%%%%%%%%%%%%%%%%%%%%%%%%%%%%%%%%%%%%%%%%%%%%%%%%%%%%%%%%%%%%%%%%%%%%%%%%%%%%
%%%%%%%%%%%%%%%%%%%%%%%%%%%%%%%%%%%%%%%%%%%%%%%%%%%%%%%%%%%%%%%%%%%%%%%%%%%%%%%%%%%%%%%%%%%%%%%%%%%%%%%%%%%
%%%%%%%%%%%%%%%%%%%%%%%%%%%%%%%%%%%%%%%%%%%%%%%%%%%%%%%%%%%%%%%%%%%%%%%%%%%%%%%%%%%%%%%%%%%%%%%%%%%%%%%%%%
%
%
%                  *****************      P  A  R  T    II         ******************
%
%
%%%%%%%%%%%%%%%%%%%%%%%%%%%%%%%%%%%%%%%%%%%%%%%%%%%%%%%%%%%%%%%%%%%%%%%%%%%%%%%%%%%%%%%%%%%%%%%%%%%%%%%%%%
%%%%%%%%%%%%%%%%%%%%%%%%%%%%%%%%%%%%%%%%%%%%%%%%%%%%%%%%%%%%%%%%%%%%%%%%%%%%%%%%%%%%%%%%%%%%%%%%%%%%%%%%%%%
%%%%%%%%%%%%%%%%%%%%%%%%%%%%%%%%%%%%%%%%%%%%%%%%%%%%%%%%%%%%%%%%%%%%%%%%%%%%%%%%%%%%%%%%%%%%%%%%%%%%%%%%%%%
%%%%%%%%%%%%%%%%%%%%%%%%%%%%%%%%%%%%%%%%%%%%%%%%%%%%%%%%%%%%%%%%%%%%%%%%%%%%%%%%%%%%%%%%%%%%%%%%%%%%%%%%%%

\newpage
\thispagestyle{empty}

$ $

\newpage
\markright{ \rm \normalsize Part II. \hspace{0.25cm} Quantum Mechanics}
$ $
\vskip 2.7in
\begin{center}
\begin{huge}
\textbf{Part II}
\vskip 0.5in
\textbf{Quantum Mechanics}
\end{huge}
\end{center}
\pagestyle{fancy}
\lhead[\rightmark]{\rightmark}
\rhead[\thepage]{\thepage}
\cfoot{}
\thispagestyle{fancy}

%%%%%%%%%%%%%%%%%%%%%%%%%%%%%%%%%%%%%%%%%%%%%%%%%%%%%%%%%%%%%%%%%%%%%%%%%%%%%%%%%%%%%%%%%%%%%%%%%%%%%%%%%%%
%%%%%%%%%%%%%%%%%%%%%%%%%%%%%%%%%%%%%%%%%%%%%%%%%%%%%%%%%%%%%%%%%%%%%%%%%%%%%%%%%%%%%%%%%%%%%%%%%%%%%%%%%%

\newpage

%take care of chapter title page 
\markright{ \rm \normalsize Chapter 4. \hspace{0.25cm} Interludium}
\chapter{Interludium}
\label{Inter}
\thispagestyle{fancy}

%this time the markers must be reset after \chapter [I'm not sure why but this wokrs]
\markboth{ \rm \normalsize Chapter \thechapter . \hspace{0.25cm} Interludium}
{ \rm \normalsize Chapter \thechapter . \hspace{0.25cm} Interludium}
\lhead[\thepage]{\rightmark}
\rhead[\leftmark]{\thepage}
\cfoot{} 

\parindent=1cm

This chapter - different in its form from the rest of the work - is meant as an 
introduction and historical overview of the development of the idea of quantum 
mechanics on the torus.
We put an emphasis on the semiclassical analysis and we select only most
important contributions to the subject directly pertaining to the problems
considered in this work.
Hence, although a lot of other important topics are going to be omitted, this introduction 
should be helpful for someone who wants to enter the field
or at least understand some of its results without prior familiarity with it and without 
an intention to study all subsidiary contributions and developments which have been introduced 
during the now almost 25-years long history of the subject. 

In the late seventies and early eighties both mathematicians and physicists 
realized the importance of and started to discuss seriously the problem of chaoticity in quantum 
dynamical systems \cite{Ch,Za}. 

By that time the analysis of chaotic behavior of classical discrete-time dynamical systems on compact 
phase spaces (represented usually by a torus) constituted one of the most developed and best 
understood branches of the general theory of dynamical systems \cite{Arnold, HK,CoFoSi}. In fact, even today,
most of the well-known and deeply understood examples of fully chaotic systems are described in terms
of toral maps (ergodic toral automorphisms, Baker and sawtooth maps, angle doubling maps, general Bernoulli 
shifts etc.). 
 
Despite some controversy concerning physical interpretation (periodicity in momentum), from a
mathematical point of view it became clear that chaotic toral maps should provide a convenient testing 
ground for a newly born field of study, now usually referred to as 'quantum chaos'.
The quantization on the torus became at that point a necessity. As in the general case of
quantization of any dynamical system the procedure is highly non-unique and from the very beginning 
different approaches had been taken and developed.
Here we consider the two most popular ones, described respectively in finite and infinite dimensional
settings. We start with the finite dimensional case. 

In 1980 M.V. Berry and J.H. Hannay \cite{HB} made a first breakthrough in the area by introducing 
the notion of the finite dimensional quantum Hilbert space associated with a classical 
symplectic 2-torus. 
The dimension of the space in their model depends on
the value of the Planck constant which itself is restricted to reciprocals of integers. The
later condition results from the assumption, made by the authors, that quantum mechanical wave 
functions of any system on the torus should be periodic in both position and momentum representations. 
This assumption restricts the set of admissible pure states of the systems to periodic
Dirac delta combs with equally spaced spikes. The distance between two neighboring delta functions  
in the comb is equal to the Planck constant (the smallest possible quantum resolution of the
phase space) and since the whole comb is 1-periodic ('lives' on the torus) the constant can 
assume only inverse integer values $\hbar=1/N$.
Each delta comb can be identified with a set of its $N$ complex amplitudes, i.e., the
strengths associated to its 'delta spikes'. The corresponding quantum Hilbert space
of all pure states is then $N$-dimensional and can be identified with $\IC^{N}$.
  
This way the {\em kinematic} step of the quantization (i.e., description of the quantum phase
space) had been completed. The next step was to define the quantum {\em dynamics} on that space. 
Hannay and Berry chose to work with the 
well-known from the classical setting \cite{Arnold} discrete-time cat map dynamics.
The assumption of the strict periodicity of their quantum phase space restricted in a 
considerable way the class of quantizable maps. The authors came to the conclusion that
in order to be quantizable a cat map had to satisfy a so-called 'checkerboard condition',
i.e., its matrix had to assume one of the following forms
$\left[
\begin{smallmatrix}
e & o \\
o & e 
\end{smallmatrix}
\right]$
or
$\left[
\begin{smallmatrix}
o & e \\
e & o
\end{smallmatrix}
\right]$,
where '$e$' and '$o$' denote respectively even and odd integer entries. 

Under this assumption the authors constructed the quantum propagator ($N\times N$ unitary matrix) associated
with a given map. One of the key points of choosing the cat map dynamics for quantization lay in the fact 
that the Hamiltonian of the classical map is in this case quadratic (see Section \ref{CatPlane} in Appendix
\ref{A}) and hence the semiclassical 
approximation to the Green function of the corresponding Schr\"odinger equation on $\IR^{2}$ is 
exact \cite{Men}. This allowed the authors to define the quantum propagator on
the torus as a 'projection' of the standard propagator on the plane. The projection, which in this case
amounts to the discretization (applying the propagator to a delta comb) and periodization 
(wrapping the result around the torus) 
was performed formally by averaging of the kernel of the planar quantum propagator over all integer winding 
numbers associated with a classical path on the tours. This way the authors constructed 
a discrete and finite kernel ($N\times N$ matrix) representing the unitary quantum propagator of a toral map. 
The entries of the matrix were expressed in terms of oscillatory Gauss sums and
the method introduced considerable number-theoretical complications as well as  some difficult to resolve
normalization issues. It thus became clear that
except for some trivial cases the explicit formula for the matrix of the propagator was unattainable.
Nevertheless, the authors were able to prove periodicity of such a quantum propagator, i.e.,
the existence of an integer $P(N)$ such that $U^{P(N)}=e^{i\phi(N)} 1\!\!1$. This result yielded the following
important information regarding the spectrum of $U$: each eigenvalue $e^{i\alpha_{j}}$ of $U$ is constrained to
take one of the possible $P(N)$ values specified in terms of their eigenangles
\bea
\label{angles}
\alpha_{j}=\frac{2\pi j + \phi(N)}{P(N)}.
\eea
Not all of the values have to be assumed however, and on the other hand, some eigenvalues may be 
(and in fact usually are) highly degenerate.
Thus in order to determine the spectrum one needs to find the period P(N) and the multiplicities $d_{j}$ 
associated with each $\alpha_{j}$. 
Hannay and Berry conjectured (and supported the conjecture with numerical evidence) that for 
regular maps $P(N)$ would be either bounded (elliptic case) or grow linearly with $N$ (parabolic case) and 
that it would grow 'in average' as $N$ but erratically in the chaotic (i.e., hyperbolic) case.

This line of research had been continued by Tabor \cite{Tabor} and further developed in 1991 by J.P. Keating \cite{K1,K2}. 
Using probabilistic number-theoretical approach supported by numerical computations Keating argued
\cite{K1} that for chaotic cat maps $P(N)$ grows in fact slightly sublinearly (see below)
and indeed highly erratically - the fluctuations themselves being of the order of $N$.
More precisely, Keating studied the asymptotic behavior of 
the average order of $P(N)$ denoted by $\la P(N)\ra$ and defined as 
\bean
\la P(N) \ra = \frac{1}{N}\sum_{n=1}^{N}P(n).
\eean
The reason to study $\la P(N)\ra$ instead of $P(N)$ is that as opposed to $P(N)$ the cumulant 
function $C(N):=\sum_{n=1}^{N}P(n)$ behaves very regularly and grows in $N$ 'almost' like $N^2$.
The normalized cumulant, i.e., the averaged order $\la P(N) \ra$ is then particularly suitable
for asymptotic analysis. The result obtained by Keating states that for arbitrary small
$\del>0$ and arbitrary big $\rho>0$,
\bea
\label{K}
 N^{1-\del} \lesssim   \la P(N) \ra \lesssim (\ln N)^{-\rho} N.
\eea
From this Keating concluded that in average $P(N)/N$ tends in fact to zero, although at a very slow rate.

As to the precise behavior of $P(N)$ itself (for large $N$) not much can be said because of its 
chaotic dependence on $N$. The size of the fluctuations of $P(N)$ had been estimated analytically
by Kurlberg and Rudnick in \cite{KR1,KR2}. The result states that there exist constants $c$ and $C$ such 
that for every $N$
\bea
\label{KR}
c\ln(N)<P(N)<C N \ln(\ln(N)).
\eea
It is worth pointing out that the bounds are sharp. In particular it has been proved recently in \cite{FNDB}
that for every hyperbolic cat map there exists a sequence $\{N_k\}$ such that $P(N_k)\sim \ln (N_k)$.
As will be described later in more details the existence of such sequences plays a crucial role
in the semiclassical analysis of ergodic properties of quantum cat maps.

Once some knowledge about the properties of the quantum period function had been accumulated the next step 
was to study the degeneracies $d_{j}$ of each eigenvalue to obtain the information about the spectrum of 
$U$. 

It is easy to show (see \cite{K2}) that $d_{j}$ can be represented in terms of the trace of the 
powers of the propagator $U$:
\bea
\label{e-d}
d_{j}=\frac{1}{P(N)}\sum_{k=1}^{P(N)}\Tr(U^{k})e^{-k\alpha_{j}}.
\eea
In \cite{K2} Keating estimated the value of $d_{j}$ for cat maps by deriving an exact version 
of the Gutzwiller's formula \cite{Gu}
\bea
\label{Ke-Gu}
\Tr(U^k)=\frac{1}{\sqrt{R_{k}}}\sum_{m,n\in \cP} \exp \lp \frac{i\pi N}{R_{k}} 
(c^k m^2 - b^k n^2 +2 (d^k-1)mn )\rp,
\eea
where 
$F=\left[\begin{smallmatrix}
a & b \\
c & d
\end{smallmatrix}\right]$,
$R_{k}=\det(F^k-1)=2-\Tr F^k$ and $\cP$ denotes the fundamental parallelogram formed by the action of the 
planar linear map $F^k-I$ on the unit square. This parallelogram encompasses all the representatives
of the periodic orbits of $F$ of period $k$ (all nonequivalent fixed points of $F^{k}$).  

Thus just like in the case of the general Gutzwiller's formula, (\ref{Ke-Gu}) relates the trace of $U^{k}$
and hence in view of (\ref{e-d}) and (\ref{angles}) the distribution of the eigenvalues (the spectrum) 
of the quantum propagator $U$ to its average over periodic orbits of the classical map 
(i.e., over the fixed points of $F^k$). It has however two important features which distinguish it from 
the general Gutzwiller's formula. 
First of all, the summation does not constitute merely a semiclassical approximation but
gives the exact value of the trace for all (allowed on the torus) values of Planck constant $\hbar=1/N$.
Secondly, the formula is explicit. Thus in this particular case the most important feature of Gutzwiller's
formula, i.e., the fact that it translates 'hard-to-compute' iterations of the quantum propagator into
'easy-to-compute' powers of the classical map, can be used to compute exactly the spectrum of $U$.
For a fixed and small value of $N$ this can be done numerically.
For large $N$ (which is important for semiclassical analysis) the number of periodic orbits proliferates 
exponentially and the numerical approach becomes impractical. But the formula remains useful
for asymptotic analysis especially since the behavior of the periodic orbits of classical cat maps
is now well understood (for further details in this direction we refer to \cite{PV} and \cite{K1}). 

Having described the first stages of the development of the finite dimension quantization on the torus 
we now compare it with another approach - the infinite dimensional one, developed almost simultaneously
but by different authors and for different reasons.

Just like the spectral approach to quantum chaoticity provided the main motivation for the introduction
of the finite dimensional quantization on the torus, the infinite dimensional approach was initiated during 
the attempts to extend the notion of the entropy and K-property from classical ergodic theory to the quantum 
setting or more generally to the case of dynamical systems defined on noncommutative algebras. 

Below we briefly discuss the role which the study of quantum entropy played in the development of the 
infinite dimensional noncommutative mechanics on the torus. We want to stress that the discussion is not meant 
to summarize the whole (much longer) history of the notion of quantum entropy but to mention the developments relevant to
quantization on the torus. In particular we will not have space here to discuss (but instead take for granted) 
von Neumann's definition of the entropy of the density matrix \cite{vN1,vN2}, its extension to normal states on von 
Neumann algebras (obtained after the work by Araki \cite{Ar}), Lieb's results on strong subadditivity and WYD 
conjecture \cite{Li1,Li2} etc. For a comprehensive overview of these notions and results we refer to \cite{OP}.  

In 1975 A. Connes and E. St{\o}rmer \cite{CS} introduced the notion of the entropy for automorphisms
on certain noncommutative von Neumann algebras, namely on a hyperfinite type $I\!I_{1}$ factor (see Definition \ref{type}).  
These algebras occur in the context of quantization on the torus. 
As was suggested by the authors, the main difficulty in generalizing the notion of the classical KS-entropy 
to non abelian setting seemed to lay in the fact that two finite dimensional subalgebras of the nonabelian 
von Neumann algebra need not generate a finite dimensional algebra (for a simple example see \cite{NT89}), and 
hence there was no immediate analogue of the classical operation of the refinement $P \vee Q$ of two finite partitions 
$P,Q$ of a phase space - a key notion used in the classical setting. We will see later in the discussion that 
an appropriate analogue in fact can be constructed by means of operational partition of unity but it
leads to a different notion of quantum entropy [so-called ALF-entropy].
The solution the authors provided at that time was the following: instead of replacing the notion of the refinement of two
partitions by its quantum analogue, the notion of its entropy 
$h_{KS}(P \vee Q)$ was replaced by a nonabelian counterpart $H(P,Q)$.
The way the replacement can be made is, as one can expect, non-unique and different choices 
led to nonequivalent notions of quantum entropy. An interesting feature is that different but
classically equivalent approaches can still yield nonequivalent quantum counterparts.  
Indeed, Connes and St{\o}rmer constructed a replacement for $h_{KS}(P \vee Q)$ based on its original
definition, while independently but at almost the same time Emch \cite{Em} introduced another replacement, 
based on a (classically equivalent) approach via conditional entropy. The resulting quantum notions did not
coincide. In the original work by Connes and St{\o}rmer the noncommutativity of the 
algebra was 'overshadowed' to some extent by the requirement that the invariant state of the dynamics be tracial. 
In this case the state itself does not see the noncommutativity of the underlying algebraic structure.
The generalization to nontracial states and to arbitrary $C^*$- and $W^*$ algebras was achieved by Connes, 
Narnhofer and Thirring in 1987 \cite{CNT} and resulted in the now well-known CNT entropy (in 1992 
Sauvageot and Thouvenot introduced an alternative definition \cite{ST} and showed that for hyperfinite 
algebras all three, i.e., CS, CNT and 'ST' entropies coincide). 

The main idea of the CNT construction is to introduce an abelian model of the noncommutative algebra and to transfer
via a completely positive unital (CPU) map the invariant state and the dynamics from the nonabelian to the abelian setting. 
After appropriate corrections (taking into account the 'entropy defect' introduced by the CPU map) the entropy 
is computed in a classical manner.   

After the definitions had been established the authors of the newly introduced notions started to look 
for examples of dynamical systems on which the notions could be tested and compared. As in 
the previous (finite dimensional) case the well known examples of classical chaotic 
toral maps once again came to the attention. 
At that time, however, it seemed obvious that an infinite dimensional model is absolutely
necessary [that it is not exactly the case will become clear later in the discussion]. The 
main argument was that, just like in the classical case, in principle, the system can have 
nonzero entropy only if the generator (Koopman operator) of the dynamics possesses a continuous spectrum 
(cf. \cite{CoFoSi}). The latter requirement obviously calls for an infinite dimensional algebra of 
observables [this was the main reason why CS and CNT entropies were introduced in infinite dimensional settings].  
 
Thus in 1991 Benatti, Narnhofer and Sewell \cite{BNS} introduced a non-commutative version of the Arnold 
cat map. Unlike in the case of the finite dimensional quantization, the authors didn't pay attention 
to the underlying physical space of pure states of the system and instead began the construction 
by introducing the appropriate noncomutative algebra of (quantum) observables. The most natural choice
was the discrete (countably generated) Weyl-Heisenberg algebra spanned by the elements  
$W_{\bk}$ (Weyl translations) indexed over the integral lattice $\IZ^{2d}$ and satisfying the standard
canonical commutation relations (CCR)
\bea
\label{firstMCCR}
W_{\bk}W_{\bm}&=& e^{\pi i h \bk\wedge \bm}W_{\bk+\bm}, \\ 
W_{\bk}W_{\bm}&=& e^{2\pi i h \bk\wedge \bm}W_{\bm}W_{\bk}.
\eea
[The detailed technical description including the explanation of the seemingly nonstandard placement for
the Planck constant will be given in the next Chapter].

After introducing the cat map dynamics and an appropriate tracial state corresponding to the 
Lebesgue measure of a classical system the authors constructed the GNS representation, and hence
in particular the Hilbert space ($(l^{2}(\IZ^2))$) of states for the system. It turned out that the 
unitary map implementing the quantum automorphism on this space coincided with the Koopman operator 
of the classical map. Since the result turned out to be independent of the value of the Planck
constant it became clear that the model would not be suitable for semiclassical considerations.

Nevertheless, the model still seemed adequate for the computations of CS-CNT entropies. 
Indeed, the above mentioned algebra is hyperfinite and either reduces to a tensor product of 
$L^{\infty}(\IT^{2})$ and a finite matrix algebra (if $h$ is rational) or is irreducible and of type $I\!I_{1}$ 
in irrational case (cf. Definition \ref{type}). 
In either case both CS and CNT entropies are well defined and coincide. 
In the rational case the CS entropy of a quantum cat map was shown 
\cite{KL} to coincide with a classical one. But the result did not contribute any new information as far as the quantum
setting is concerned. Indeed, the cat map dynamics factorizes into a purely classical part acting on $L^{\infty}(\IT^{2})$ 
with its classical entropy and a finite dimensional quantum remainder with trivially vanishing quantum entropy. 

In hope to obtain some nontrivial information one needed to look at 'irrational-$h$' case. 
It soon became clear, however, that in this case exceptionally strong properties of clustering and 
asymptotic abelianness \cite{NT94,Na,Ne} were necessary to secure the possibility of non-vanishing CNT entropy.
Without entering into the details of the definitions, let us only recall that the notion of strong 
asymptotic abelianness was introduced in \cite{BN} and that it was basically shown in \cite{AN} that its lack
implies zero CNT entropy. It was then found \cite{NT95} that indeed, except for a possibly measure zero set of 
the values of the Planck constants, the CNT-entropy of the cat map is in fact zero.  
On top of that, in the course of the investigations, it turned out \cite{NST} that using
the above mentioned property one can show that CNT-entropy fails to be additive w.r.t. the
tensor products. Thus although CS-CNT entropies had been and are successfully applied to different
dynamical systems another notion of noncommutative entropy was still needed. 

In 1994 R. Alicki and M. Fannes \cite{AF94} introduced a new approach to quantum dynamical entropy.
Finite partitions used in the definition of the classical KS entropy had been replaced by the authors by
(also finite) operational partitions of unity in corresponding quantum algebra. Such partitions could be
evolved with a dynamics and composed among themselves yielding at each step finer but always finite 
new operational partitions. Using the standard notion of the von Neumann entropy of a quantum state 
and applying the procedure similar to the one known from the classical setting the authors constructed
a new quantum dynamical entropy, now usually referred to as ALF entropy (in recognition of the connection of
 their idea to an earlier work by G. Lindblad \cite{L}). 

The definition was clearly compatible with a classical one in abelian case. Moreover it was  
shown \cite{AAFT96} that, e.g., in the case of toral automorphisms in arbitrary dimensions the computations 
of the classical KS entropy using the operational approach could be significantly simplified in comparison
to the traditional approach \cite{Yuz}. The entropy has also been successfully computed for quantized cat maps 
in the infinite dimensional setting \cite{AAFT95} with the result identical to the classical one for all values 
of Planck constant (the result is consistent with the fact that the generator of the dynamics is independent 
of its value). The latter example indicated that the notion differs significantly from CS-CNT entropies.
In fact there even exist examples of systems for which $H_{ALF}=\infty$ while $H_{CNT}=0$ (see \cite{AN}).  
The two entropies differ also in the cases where the latter one is nonzero (e.g., for the shifts on quantum
spin chains \cite{AF94,AN}).
Nevertheless, similarly as all previously mentioned quantum dynamical entropies, ALF entropy also vanishes 
for any fixed finite-dimensional quantum dynamical system - the phenomena known as {\em saturation}. 
The word {\em saturation} is a key notion here and it is worth to pause for a while and explain the source 
of the phenomenon in more detail.

In the classical setting KS entropy is defined as an infinite-time limit of the appropriately rescaled 
entropy production which occurs during the evolution of the system. Since the classical phase space can be 
refined into arbitrary small partitions, the rescaled entropy production converges to a well-defined limit
and that limit is understood as the entropy of the system (in some, but not most interesting, cases the
limit may be infinite).

In the finite dimensional quantum setting, with the phase-space resolution constrained by the value
of the Planck constant, it is clear that after an initial stage of positive entropy production
the growth of the entropy of the evolved partition comes to an end as further iterations
do not introduce any new information. When this stage is achieved (i.e., the saturation takes place), 
the rescaling factor dominates the asymptotics and the infinite-time limit is zero.

The above analysis suggests the following solution. In order to get nontrivial entropy results in the finite
dimensional setting, instead of taking an infinite-time limit with a fixed Planck constant, one should
consider simultaneous semiclassical and long time limits computing the entropy production for each
value of positive but eventually vanishing Planck constant only up to certain 'saturation' 
time. If the limits are taken in this particular way and order, the infinite time limit should be nonzero
and should provide truly valuable information about chaotic behavior of the underlying system.
In particular, the longest time scale on which the above procedure recovers classical entropy can be
thought of as a kind of ``breaking'' time, after which classical and quantum evolutions no longer agree.
Some recent preprints \cite{ALPZ,BCCFV,BCZ} confirm the
usefulness of the idea. In particular in \cite{BCCFV} the authors compute semiclassical limit of both CNT and
ALF entropy productions (on logarithmic in $\hbar$ time scales) for cat maps and prove that both notions lead 
to the same classical result. In particular the results remain in full agreement with our results (see Theorem \ref{QDTthm})
providing the base for the interpretation of the quantum dissipation rate constant as quantum dynamical entropy 
of the system (at least in 2-dimensional setting). It is also worthwhile to note at this point that the agreement holds 
exactly in the situation when the dissipation time does not exceeds the ``breaking'' time of the system.    

The problem of ``breaking'' time is in some sense universal and emerges in all aspects of semiclassical 
analysis of quantum chaos. In order to understand it better we need to come back for a while to the 
description of the finite dimensional quantization procedure.

After initial works and promising results by Hannay, Berry, Tabor and Keating a need arose to clarify 
technical issues of finite dimensional quantization and in particular to generalize the scheme to allow 
quantization of at least all toral automorphisms (not only cat maps satisfying the 'checkerboard' condition) 
but eventually also Anosov maps and in general all canonical maps (symplectic diffeomorphisms) on the torus.

The first major breakthrough in this direction came in 1993 with a paper by Mirko Degli Esposti \cite{E}
in which the author suggested a representation-theoretical approach to the finite dimensional quantization
(for more modern treatment see \cite{EG,EGI}).
Just like in the infinite dimensional case the key ingredient of the quantization was the Weyl representation
of the discrete Heisenberg group, with the distinction that the representation was considered here
over finite dimensional Hilbert space $L^2(\IT,\mu)$, with purely atomic measure $\mu$.
The resulting $^*$-algebras are indexed by rational values of $h$. Unlike the standard 
Weyl representation over the usual infinite dimensional quantum Hilbert space $L^{2}(\IR)$,
where by the Stone-von Neumann theorem there is essentially unique irreducible infinite dimensional 
representation, for any fixed rational $h$ there are infinitely many irreducible Weyl 
representations  over finite dimensional space (in particular $L^2(\IT,\mu)$). The representations
are indexed by a parameter $\btht\in \IT^{2}$ and
all share the same dimension $N$ (if $h=p/q$ then $N$ is the smallest integer such that $Nh$ is an integer \cite{GH}).
Using this ``extra space'' provided by the non-uniqueness of the infinite dimensional representations  
the author was able to remove the old checkerboard quantization condition and hence proved that for every cat map
(i.e., 2-dim hyperbolic automorphism of the torus) there exists a finite dimensional Weyl representation on 
which the corresponding quantum Koopman operator acts as an (inner) $^*$-automorphism (see Section \ref{catmap}). 
It was soon realized that the abstract Hilbert space $L^2(\IT,\mu)$ can be considered as a 
generalization of the original Hannay and Berry space in a sense that strict periodicity of a wave function is replaced 
by a weaker condition of $\tht$-quasiperiodicity defined as follows 
\bean
\psi(q+m)=e^{2\pi i \tht_{p}m}\psi(q),
\eean
where $m$ is an integer and $\tht_{p}\in \IT$. An analogous condition is required for the momentum representation
(with a corresponding Bloch angle $\tht_{q}\in \IT$).
For a given cat map 
$F=\left[ \begin{smallmatrix}
a & b \\
c & d
\end{smallmatrix} \right]$
the appropriate quantization condition on the joint Bloch angle $\btht=(\tht_{q},\tht_{p})$ has been established 
(see e.g. {\cite{E,KMR}})
\bean
\frac{N}{2}
\begin{pmatrix}
a \cdot b \\
c \cdot d
\end{pmatrix}
+F\btht=\btht \mod 1,
\eean
and after the work of Keating, Mezzadri and Robbins \cite{KMR} it is usually referred to as 
'quantum boundary condition'. In the same paper the authors generalized the quantization scheme
to arbitrary canonical maps on the torus (some additional technical issues of the multiplicativity
of the unitary propagators implementing the dynamics had been addressed later 
by Mezzadri, Kurlberg and Rudnick \cite{M,KR1}, but since they do not pertain directly
to our study, the results will not be described here).
 
In the meantime and afterward, different authors introduced finite dimensional quantization procedures
for almost all other (e.g., discontinuous) well-known classical systems on the torus (see, e.g., \cite{BV, Sa, SV}
for quantization of the Baker map and \cite{DBEG} for quantization of the Sawtooth map).

Thus, after almost twenty years of development, the finite dimensional quantization on the torus had finally been 
established as a general and rigorous procedure and successfully applied to a large variety of toral maps. 

In 1996 A. Bouzouina and S. De Bi\`{e}vre \cite{BozDB} introduced an elegant and highly efficient mathematical 
formalism and notational setting in which the procedure could be described in a natural and transparent way.
The authors started with a standard Weyl quantization on the usual Hilbert space $L^{2}(\IR)$,
extended unitary Weyl translation operators (generators of the algebra) to the 
space of all tempered distributions $\cS'(\IR)$ and then restricted their action to the
space of quasiperiodic wave functions. 
The construction seems more natural then previous ones since instead of starting from an abstract representation
theory, the appropriate algebra of quantum observables is derived directly from the traditional one. Thanks to
this, any quantum observable corresponds in an natural way to its classical counterpart through its
decomposition in the quantum Fourier basis formed by Weyl translation operators.

The method also provided a transparent link between the finite and the infinite dimensional settings (although
the authors did not address this issue in their work directly). Indeed, if we denote by $\cA$ the infinite dimensional
algebra of the model introduced by Benatti, Narnhofer and Sewell \cite{BNS} and restrict the values of the Planck 
constant to rationals then in the formalism introduced by Bouzouina and De Bi\`{e}vre we simply have
(cf. Section \ref{ModelII})
\bean
\cA=\int^{\oplus}_{\IT^{2}}\cA_{N}(\btht), 
\eean
where $\cA_{N}(\btht)$ denotes finite dimensional algebras of observables over the spaces of $\btht$-quasiperiodic
wave functions (and in Degli Esposti's formalism $\cA_{N}(\btht)$ run through all the irreducible representations
of the discrete Heisenberg group as $\btht$ varies in $\IT^{2}$).
Moreover as will be shown in Sections \ref{QTK} and \ref{QTD} both finite and infinite dimensional quantizations
can be constructed within the framework similar to the one introduced by Bouzouina and De Bi\`{e}vre, provided one 
defines in an appropriate way the Hilbert spaces of pure states to which the action of the Weyl operators 
is to be restricted.

After clarifying quantization issues we can now return to the discussion of the breaking time. 
We did not define this notion precisely yet
and in fact there is still no agreement in the literature on how it should be defined and even named.
In an attempt to formalize this notion let us recall that according to Bohr's correspondence principle, 
the quantum evolution of any observable should approximate its classical evolution better and better as 
the Planck constant tends to zero. In mathematical literature the principle is expressed in terms of 
the so-called Egorov property (\cite{Eg,Ho,Ro}), which states that for any smooth classical observable $f$ and 
the classical evolution operator $U$, and for their corresponding quantizations $Op(f)$, $\cU$ 
the following estimate holds
\bea
\label{Egorov1}
||\cU^t Op(f) - Op(U^t(f))||\leq C_{t,f}\hbar.
\eea
In the above estimate the constant in general depends in an essential way on both $t$ and $f$. The
key point is that it is $\hbar$-independent and hence in the classical limit 
the quantum and the classical evolutions ``commute''. We want to stress that even in case of very reasonable 
quantization procedures the Egorov property cannot be taken for granted and has to be 
formally proved (see remarks after Theorem 4.2 in \cite{BozDB} for an appropriate counterexample, cf. also \cite{MR};
for the most up to date results on the dependence of $C_{t,f}$ on $t$ and $f$, see \cite{BR}).

Using Egorov property we can introduce one possible definition of the breaking time
by asking for the largest $t=t(\hbar)$ for which RHS of (\ref{Egorov1}) remains bounded. The corresponding
time scale $\tau_{E}$ is usually referred to as the {\em Ehrenfest time} which 
 had been introduced many years ago \cite{Ch,Za} when physicists conjectured that for regular systems $\tau_{E}$ 
should diverge as a function of $\hbar$ 
at a power-law rate, while for chaotic ones, at a logarithmic rate. It turns out that the conjecture, although
intuitively appealing, is extremely hard to verify on rigorous grounds, even 
for one of the simplest possible models of quantized fully chaotic systems; i.e., toral cat maps.
Most of the results confirm the agreement between the quantum and the classical evolutions up to the logarithmic
times. There is however still no general result stating that after the logarithmic time 
is passed, the quantum system diverges in its behavior from its classical counterpart. 

Below we briefly review the results obtained in this matter up to the present time. 
The first interesting observation is that the problem of estimating the Ehrenfest time is closely 
related to the problem of determining the semiclassical behavior of the eigenstates of the quantum 
system, on the one hand, and with the problem of finding the quantum period function on the other.  
The first problem (semiclassical behavior of eigenstates) is usually approached from the
following two, slightly 
different perspectives. In the first approach one hopes to prove so called Quantum Unique 
Ergodicity (QUE), which states that {\em all} eigenstates of a chaotic quantum propagator
equidistribute (i.e., converge weakly to the Lebesgue measure) in the classical limit.
In the second approach one relaxes a little bit the above requirement and aims at proving
the so-called Schnirelman property (\cite{Sch,CdV}), which states that if a quantum system has ergodic classical
limit, then {\em most} of its eigenstates (with given energy level) equidistribute in the
classical phase-space when $\hbar \rightarrow 0$.
QUE results are very difficult to prove. The first example of QUE 
for a quantum dynamical system with ergodic classical limit has been presented by Marklof and Rudnick
in their paper \cite{MR}, published in 2000. The example is however very specific. The authors consider
irrational skew translations of the two-torus. The key point which simplifies the matter considerably
here is that the classical limit is not only ergodic but also uniquely ergodic, meaning that there is no
other than Lebesgue measure on which quantum eigenstates can possibly concentrate on in the classical
limit. Thus the problem of proving QUE in this particular case was essentially reduced to the problem
of proving the Egorov property. Moreover irrational translations, although ergodic, are not weakly mixing and
hence do not represent a typical example of a chaotic system. In fact QUE has not been proved yet 
for any quantized fully chaotic system \cite[p.4]{BonDB2}, and to the contrary in many 
cases, including as we will see below all cat maps, it simply does not hold. 

As far as cat maps are concerned the ``closest'' results to QUE had been proved 
either under the very strong additional restriction that the classical limit ($\hbar=1/N \rightarrow 0$) 
is taken only over special sequences of reciprocals of primes $N$ (and under the assumption that
generalized Riemann Hypothesis holds) \cite{EGI}, or under weaker conditions 
on $\hbar$ (density one sequence in $\IN$) but with quite restrictive number-theoretical 
conditions on admissible cat maps \cite{KR2}, cf. also \cite{KR1}. 
In both cases restrictions aim at exactly the same point - to ensure the absence or sufficiently slow growth of
the degeneracies of the corresponding eigenvalues, which corresponds to long (of order $N$)
periods of the quantum propagator. Rapidly growing degeneracies of the
eigenspaces of the propagator caused by short (e.g., of order $\ln N$) quantum periods existing for some particular 
values of the Planck constants allowed for construction of the sequences of eigenstates which concentrate in the semiclassical 
limit on measures with a nontrivial pure-point component. The existence of such sequences - the phenomenon referred to as 
{\em strong scarring} - has recently been proved by Faure, Nonnenmacher and De Bi\`{e}vre in \cite{FNDB} 
for an {\em arbitrary} hyperbolic toral automorphism in 2-dim (for some results in higher dimension see \cite{BonDB2}).
Thus QUE does not hold in these cases. The scarred eigenstates cannot concentrate however totally on pure-point measures.  
In \cite{FN} Faure and Nonnenmacher show that the Lebesgue measure component of the support of any such state must account
for at least $1/2$ of the total weight (and the bound is sharp).
Moreover, the sequences of scarred eigenstates are very exceptional (in the sense that to construct them one needs to choose
these exceptional values of $\hbar$ for which the quantum propagator has minimal, i.e., logarithmic in $\hbar$ period). 
The conclusion is that although QUE in general fails, some version of the Schnirelman property usually holds for general chaotic 
quantum systems. For cat maps the property had been proved already in \cite{BozDB}. It has been also confirmed in a 
wide variety of other contexts (e.g. for ergodic geodesic flows on compact Riemannian manifolds \cite{Sch, Z0, CdV}, 
for ergodic billiards \cite{GL,ZZ}, and recently for maps with mixed dynamics \cite{MOK}). 

As far as the Ehrenfest time is concerned the discovery of the scarred states turned out to be directly connected with the 
construction of an example in which the breakdown of the classical-quantum correspondence happens sharply on a 
logarithmic scale. 
The appropriate example was constructed within the framework of cat maps by Bonechi and De Bi\`{e}vre in \cite{BonDB1}. 
The main idea was to consider the action of the quantum propagator on coherent states supported on the region in the phase 
space of the diameter not exceeding $\sqrt{\hbar}$. For times tending to infinity but no faster than 
$\tau_{E}=(2\gamma)^{-1}\ln(\hbar^{-1})$  ($\gamma$ is the Lyapunov exponent of the cat map) 
the whole support of the state must shrink and in the classical limit
the state's evolution concentrates on the classical orbit of its center - 
the behavior shared by both classical and quantum evolutions
(the Wigner function of the coherent state converges weakly to the delta function).
On time scales slightly longer than $\tau_{E}$ strong mixing properties of the classical map prevail and cause 
the support of the coherent state to stretch and the resulting semiclassical limit corresponds to a constant
function over the whole phase-space (the Wigner function of the coherent state converges to $1$).
For classical dynamics this scenario will continue regardless the length of the time scale. To the contrary,
the quantum system is periodic and the initial concentration phenomena must repeat itself once the period of the 
map is completed. At that point quantum and classical evolutions depart from each other. It is then enough to find a sequence
of Planck constants for which the quantum period function is of order $\ln(\hbar)$. This is exactly the minimal
possible period and as was mentioned above the appropriate sequence can be constructed for any cat map yielding
the desired logarithmic asymptotics of the breaking time. 

A few words of caution are necessary here. First of all, the above described phenomena need not reflect 
general properties of quantum systems including cat maps. Indeed, according to earlier
described results by Keating, Kurlberg and Rudnick (see \ref{K} and \ref{KR}) for a great majority
of the values of $\hbar$, quantum period is much longer than $\ln(\hbar)$ and hence for 'most' of
classical limits the phenomenon may not be visible on this time scale at all. As the time progresses more and more
trajectories will start to complete their periods and the breaking of the correspondence, understood as a
statistical phenomenon, may happen on longer time scales.
   
It is also important to distinguish here between two different semiclassical approaches to the
question of determining when the breakdown between classical and quantum mechanics occurs. 
The above-described results belong to the category of direct 'quantum-classical' limits.

Another approach is to study the validity of semiclassical approximations to quantum propagators.
Here the situation is different and the correspondence on times scales much longer than
logarithmic have been observed in several cases. For numerical results in these directions
see \cite{TH1,TH2,OTH}. These results cannot shed any light on the question of the
breaking time for cat maps, however, since in this particular case the semiclassical approximation to
the quantum propagator is exact and hence no breaking time can be observed through this analysis.

A general conclusion is that depending on the observed property and the level of the intensity of the 
phenomenon on which the correspondence is tested one can expect that a whole spectrum 
of different breaking times may exist ranging from logarithmic to power-law scales in $\hbar$. 
In Section \ref{SAQDT}, Corollary \ref{QC} (point IV) we give an example of the situation when the quantum and
classical noisy toral automorphisms behave in the same way from their dissipative properties 
point of view, while the evolutions have already exceeded the Ehrenfest time. 
It would be of particular interest now to see whether 'CNT-entropic' breaking time coincides with 'ALF-entropic' 
breaking time and how these two are related to Ehrenfest and dissipation time scales. 
The results which we are going to derive in the next chapter are closely related to this question.
The complete solution is however still elusive.

%%%%%%%%%%%%%%%%%%%%%%%%%%%%%%%%%%%%%%%%%%%%%%%%%%%%%%%%%%%%%%%%%%%%%%%%%%%%%%%%%%%%%%%%%%%%%%%%%%%%%%%%%%%%
%%%%%%%%%%%%%%%%%%%%%%%%%%%%%%%%%%%%%%%%%%%%%%%%%%%%%%%%%%%%%%%%%%%%%%%%%%%%%%%%%%%%%%%%%%%%%%%%%%%%%%%%%%%%
%%%%%%%%%%%%%%%%%%%%%%%%%%%%%%%%%%%%%%%%%%%%%%%%%%%%%%%%%%%%%%%%%%%%%%%%%%%%%%%%%%%%%%%%%%%%%%%%%%%%%%%%%%%%

\newpage
\markright{  \rm \normalsize Chapter 5. \hspace{0.25cm} Quantum dissipation time}

\chapter{Quantum dissipation time}
\label{QuDiTi}
\thispagestyle{fancy}

\lhead[\thepage]{\rightmark}
\rhead[\leftmark]{\thepage}
\cfoot{}

%%%%%%%%%%%%%%%%%%%%%%%%%%%%%%%%%%%%%%%%%%%%%%%%%%%%%%%%%%%%%%%%%%%%%%%%%%%%%%%%%%%%%%%%%%%%%%%%%%%%%%%%%%%%
%%%%%%%%%%%%%%%%%%%%%%%%%%%%%%%%%%%%%%%%%%%%%%%%%%%%%%%%%%%%%%%%%%%%%%%%%%%%%%%%%%%%%%%%%%%%%%%%%%%%%%%%%%%%
%%%%%%%%%%%%%%%%%%%%%%%%%%%%%%%%%%%%%%%%%%%%%%%%%%%%%%%%%%%%%%%%%%%%%%%%%%%%%%%%%%%%%%%%%%%%%%%%%%%%%%%%%%%%

\parindent=0cm

The main goal of this chapter is to introduce the notion of quantum dissipation time and 
to study its semiclassical and noisy asymptotics for quantized toral maps. The first 
step is to construct appropriate quantum model of the phase space (Section \ref{QTK})
and of the dynamics (Section \ref{QTD}) so that it can be considered as a quantization 
of the classical systems studied in Part I. 
In the second step (Section \ref{SAQDT}) we will then focus on definitions and semiclassical 
analysis of the corresponding quantum notions.

%%%%%%%%%%%%%%%%%%%%%%%%%%%%%%%%%%%%%%%%%%%%%%%%%%%%%%%%%%%%%%%%%%%%%%%%%%%%%%%%%%%%%%%%%%%%%%%%%%%%%%%%%%%%
\section{Quantization on the torus. The kinematics}
\label{QTK}
%%%%%%%%%%%%%%%%%%%%%%%%%%%%%%%%%%%%%%%%%%%%%%%%%%%%%%%%%%%%%%%%%%%%%%%%%%%%%%%%%%%%%%%%%%%%%%%%%%%%%%%%%%%%

In this section we present in a systematic and rigorous way kinematic step of the quantization 
of classical systems on the torus. That is, we construct appropriate Hilbert spaces of pure
states and algebras of observables. In the following section we will concentrate on 
the dynamical step i.e. on quantization of canonical toral maps. 
The presentation is based on the approach by Bouzouina and De Bi\'evre \cite{BozDB} and
generalizes it in two directions:

1. We quantize the systems with arbitrary phase-space dimension (for equivalent approaches to quantization
in multidimensional setting see e.g. \cite{RSO,GH}).

2. The quantization scheme is extended in such a way that both finite and infinite dimensional 
   settings are included as particular cases.

Throughout this chapter we will use the terms Model I and Model II to refer respectively to finite 
and infinite dimensional quantization schemes. 

%%%%%%%%%%%%%%%%%%%%%%%%%%%%%%%%%%%%%%%%%%%%%%%%%%%%%%%%%%%%%%%%%%%%%%%%%%%%%%%%%%%%%%%%%%%%%%%%%%%%%%%%%%%%

\subsection{Weyl quantization on $\IR^d$}
\label{WQ}

%%%%%%%%%%%%%%%%%%%%%%%%%%%%%%%%%%%%%%%%%%%%%%%%%%%%%%%%%%%%%%%%%%%%%%%%%%%%%%%%%%%%%%%%%%%%%%%%%%%%%%%%%%%%

We briefly recall the standard Weyl quantization of 
systems with $d$ degrees of freedom and with $\IR^{2d}$ as a classical 
phase space (for systematic presentation see \cite{AF}).

As before, $h$ will denote the Planck constant. Whenever convenient
we will also use the notation $\hbar=\frac{h}{2\pi}$. 

Consider usual quantized position and
momentum operators 
\bean
\bQ=(Q_{1},...,Q_{d}), \qquad \bP=(P_{1},...,P_{d})
\eean
on the Hilbert space $L^{2}(\IR^d)$ of square integrable wave functions 
\bean
Q_{j}\psi(\bx)=\bx_{j} \psi(\bx), \qquad
P_{j}\psi(\bx)=-i\hbar \frac{\pa \psi}{\pa \bx_{j}}(\bx).
\eean
The domains of these operators contain the Schwartz
space $\ml{S}(\IR^d)$ and hence are dense in $L^{2}(\IR^d)$.
Although all are essentially selfadjoint, they do not possess 
a common domain of selfadjointness.
In Weyl quantization one considers exponentiated versions 
of the above operators, which helps to 
avoid the domain problems and, more importantly, introduces in a
natural way the notion of quantum translation operators
\bean
U_{\bq}=e^{-\frac{i}{\hbar}\bq \cdot \bP}, \qquad 
V_{\bp}=e^{\frac{i}{\hbar}\bp\cdot \bQ}.
\eean
The explicit action of $U_{\bq}$ and $V_{\bp}$ on $L^{2}(\IR^d)$ is given by
\bean
(U_{\bq}\psi)(\bx)=\psi(\bx-\bq), \qquad 
(V_{\bp}\psi)(\bx)=e^{\frac{i}{\hbar}\bp\cdot \bx}\psi(\bx).
\eean 
The commutation relations for $U_{\bq}$ and $V_{\bp}$ follow easily 
\bean
U_{\bq}V_{\bp}\psi(\bx)=e^{\frac{i}{\hbar}\bp\cdot(\bx-\bq)}\psi(\bx-\bq), \qquad
V_{\bp}U_{\bq}\psi(\bx)=e^{\frac{i}{\hbar}\bp\cdot\bx}\psi(\bx-\bq),
\eean
and hence
\bea
\label{HC}
U_{\bq}V_{\bp}=e^{-\frac{i}{\hbar} \bq\cdot \bp}V_{\bp}U_{\bq}.
\eea
Using BCH formula and the fact that $[\bp\cdot\bQ,\bq\cdot\bP]= i\hbar\bq\cdot\bp \pmb{1}$ we also get
\bean
U_{\bq}V_{\bp}=e^{-\frac{i}{2\hbar}\bq \cdot \bp}e^{\frac{i}{\hbar}(\bp\cdot \bQ-\bq\cdot \bP)}.
\eean

Let $\bv=(\bq,\bp)\in \IR^{2d}$. The Weyl translation operators $T_{\bv}$ are defined as the following 
symmetrized products of $U_{\bq}$ and $V_{\bp}$
\bean
T_{\bv}=e^{\frac{i}{2\hbar} \bq\cdot \bp}U_{\bq}V_{\bp}=
e^{-\frac{i}{2\hbar} \bq\cdot \bp}V_{\bp}U_{\bq}=
e^{\frac{i}{\hbar}(\bp\cdot \bQ-\bq\cdot \bP)}.
\eean
Using the notation convention $\bX=(\bQ,\bP)$ and $\bv\wedge\bX= \bp\cdot \bQ - \bq\cdot \bP$ the 
Weyl operators can be written in a compact form
\bea
\label{WeylTr}
T_{\bv}=e^{\frac{i}{\hbar} \bv \wedge \bX}.
\eea
$T_{\bv}$ can naturally be extended to the space of
tempered distributions $\cS'(\IR^d)$.

The explicit action of $T_{\bv}$ on a (generalized) wave function $\psi\in \cS'(\IR^d)$ is given by 
\bea
\label{WeylAc}
(T_{\bv}\psi)(\bx)=e^{\frac{i}{2\hbar}\bq\cdot\bp} e^{\frac{i}{\hbar}\bp \cdot (\bx - \bq)} \psi (\bx- \bq)
=e^{\frac{i}{\hbar}\bp \cdot (\bx - \bq/2)} \psi (\bx- \bq)
\eea
and yields the following property 
\bea
\label{CCR1}
T_{\bv}T_{\bv'}= e^{\frac{i}{2\hbar} \bv\wedge \bv'}T_{\bv+\bv'},
\eea
which in turn implies Weyl-type Canonical Commutation Relations 
\bea
\label{CCR}
T_{\bv}T_{\bv'}=e^{\frac{i}{\hbar} \bv\wedge \bv'}T_{\bv'}T_{\bv}.
\eea

%%%%%%%%%%%%%%%%%%%%%%%%%%%%%%%%%%%%%%%%%%%%%%%%%%%%%%%%%%%%%%%%%%%%%%%%%%%%%%%%%%%%%%%%%%%%%%%%%%%%%

\subsection{The space of pure states on $\IT^{d}$.}
\label{HS}

%%%%%%%%%%%%%%%%%%%%%%%%%%%%%%%%%%%%%%%%%%%%%%%%%%%%%%%%%%%%%%%%%%%%%%%%%%%%%%%%%%%%%%%%%%%%%%%%%%%%%

In this section we determine a quantum analog of the notion of periodic phase-space
of a classical system by introducing the space of generalized quasi-periodic wave
functions (distributions).

The idea of quantization on the torus will be reflected here in the requirement that the
wave function be quasiperiodic in its position or momentum representation.

A distribution $\psi\in \cS'(\IR^d)$ is called quasiperiodic in position 
representation if there exists a constant $\btht_{p}\in\IT^{d}$ such that for any
$\bm_{1}\in \IZ^d$
\bean
\psi(\bq+\bm_{1})&=&e^{2\pi i \btht_{p}\cdot \bm_{1}}\psi(\bq).
\eean
The set of all such distributions will be denoted by $\cS^{(q)}_{h}(\btht_{p})$.

Similarly a distribution $\psi\in \ml{S}'(\IR^d)$ is called quasiperiodic in momentum 
representation if there exists a constant $\btht_{q}\in\IT^{d}$ such that for any 
$\bm_{2}\in \IZ^d$
\bean
(\ml{F}_{h}\psi)(\bp+\bm_{2})&=&e^{-2\pi i \btht_{q}\cdot \bm_{2}}(\ml{F}_{h}\psi)(\bp),
\eean
where $\ml{F}_{h}$ denotes the quantum Fourier transform 
\bea
(\ml{F}_{h}\psi)(\bp)=
\frac{1}{h^{d/2}}\int_{\IR^{d}}\psi(\bq)e^{-2\pi i\frac{\bq \cdot \bp}{h}}d\bq.
\eea
The corresponding set will be denoted by $\cS^{(p)}_{h}(\btht_{q})$. 

In our approach the space of all admissible pure states denoted by $\cH_{h}(\btht)$ 
of the quantum system on the torus will always be understood as a linear 
space generated by (not necessary all) elements of $\cS^{(q)}_{h}(\btht_{p}) \cup \cS^{(p)}_{h}(\btht_{q})$ 
and equipped, according to usual requirements of quantum mechanics, with some 
Hilbert structure. The choice of the space will depend on the model one intends to work with.

We note that if $\psi\in \cS^{(q)}_{h}(\btht_{p}) $ then its momentum representation 
$\ml{F}_{h}\psi$ has a discrete uniformly $h$-spaced and $\btht_{p}$-shifted support 
and hence is represented by a Dirac delta comb (or more generally brush) of the form 
\bean
\ml{F}_{h}\psi(\bp)=h^{d/2}\sum_{\bs\in \IZ^{d}}c_{\bs}\del_{h(\bs+\btht_{p})}(\bq),
\eean
where $c=\{c_{\bs}\}_{\bs\in \IZ^d}$ denotes a sequence of complex numbers.
Thus any distribution $\psi \in \cS^{(q)}_{h}(\btht_{p})$ is uniquely
determined by a triple $(c,h,\btht_{p})$. We will sometimes write $\psi=(c,h,\btht_{p})$. 
Similar remark applies to any element of $\cS^{(p)}_{h}(\btht_{q})$.  

One easily notices that for any $\psi\in \cS^{(q)}_{h}(\btht_{p})$ and any 
$\bm_{1}\in \IZ^{d}$
\bea
\label{Tmpsiq}
T_{(\bm_{1},0)}\psi = e^{-2\pi i \btht_{p}\cdot \bm_{1}}\psi
\eea
Similarly for any $\psi\in \cS^{(p)}_{h}(\btht_{q})$ and any 
$\bm_{2} \in \IZ^{d}$
\bea
\label{Tmpsip}
T_{(0,\bm_{2})}\psi = e^{2\pi i \btht_{q}\cdot \bm_{2}} \psi. 
\eea

That is, the spaces  $\cS^{(q)}_{h}(\btht_{p})$ and $\cS^{(p)}_{h}(\btht_{q})$ consists of eigenstates 
of integral translations.

%%%%%%%%%%%%%%%%%%%%%%%%%%%%%%%%%%%%%%%%%%%%%%%%%%%%%%%%%%%%%%%%%%%%%%%%%%%%%%%%%%%%%%%%%%%%%%%%%%%

\subsection{The algebra of observables on $\IT^{d}$.}
\label{q-algebra}

%%%%%%%%%%%%%%%%%%%%%%%%%%%%%%%%%%%%%%%%%%%%%%%%%%%%%%%%%%%%%%%%%%%%%%%%%%%%%%%%%%%%%%%%%%%%%%%%%%%

In this and subsequent sections we will frequently use the following terminology.

\begin{defin}
An algebra $\cA$ is called
 
i) $^*$-algebra if it is equipped with an involution $^*:\cA \mapsto \cA$, $A^{**}=A$, $A\in \cA$.

ii) $B^*$-algebra if it has a Banach space structure and $\|AB\|\leq \|A\|\|B\|$. 

iii) $C^*$-algebra if it is a $B^*$-algebra that satisfies $\|A^*A\|=\|A\|^2$, for all $A\in \cA$.

iv) $H^*$-algebra if it is a $B^*$-algebra with an inner product satisfying 
\bean
\la A,BC^* \ra=\la AC,B \ra =\la C,A^*B \ra
\eean
(e.g. $\la A,B \ra=\tau(A^{*}B)$ if $\tau$ is a faithful tracial state). 

v) $W^*$-algebra (von Neumann algebra) if $\cA''=\cA$, where $\cA''$ is a bicommutant.
\end{defin}

The Weyl quantization on the torus will consist in the restriction of the action
of Weyl translations from the whole $\cS'(\IR^d)$ to $\cH_{h}(\btht)$.

This restriction is well defined only if translations $T_{\bv}$ preserve 
$\cS^{(q)}_{h}(\btht_{p})$ and $\cS^{(p)}_{h}(\btht_{q})$ spaces. 
In view of (\ref{Tmpsiq}) and (\ref{Tmpsip}) the condition is equivalent to the 
commutativity of $T_{\bv}$ with integer translations.
 The latter condition can be stated as follows
\begin{prop} $ $

$T_{\bv}:\cS^{(q)}_{h}(\btht_{p}) \mapsto \cS^{(q)}_{h}(\btht_{p})$ 
iff $\bv\in \IR^{d}\times h\IZ^{d}$.

$T_{\bv}:\cS^{(p)}_{h}(\btht_{q}) \mapsto \cS^{(p)}_{h}(\btht_{q})$ 
iff $\bv\in h\IZ^{d}\times \IR^{d} $.
\end{prop}
\begin{rem}
\label{presboth}
If $\cH_{h}(\btht)$ has nontrivial intersection with 
 $\cS^{(q)}_{h}(\btht_{p})$ and $\cS^{(p)}_{h}(\btht_{q})$
 then $T_{\bv}:\cH_{h}(\btht) \mapsto \cH_{h}(\btht)$ 
iff $\bv\in h\IZ^{2d}$.
\end{rem}

Motivated by the above considerations we define microscopic quantum phase-space 
translations acting on the space $\cH_{h}(\btht)$ 
\bea
\label{micro}
W_{\bk}:=T_{h\bk}=e^{2\pi i \bk\wedge \bX}.
\eea
The operators $W_{\bk}$ are indexed by the elements of the integral lattice $\IZ^{2d}$ and 
can be thought of as quantum counterparts of classical Fourier modes.
The corresponding commutation relations are now given by
\bea
\label{MCCR}
W_{\bk}W_{\bm}&=& e^{\pi i h \bk\wedge \bm}W_{\bk+\bm}, \\ 
W_{\bk}W_{\bm}&=& e^{2\pi i h \bk\wedge \bm}W_{\bm}W_{\bk}.
\eea

Note that the placement of the Planck constant has changed after this rescaling 
(cf. (\ref{CCR1}) and (\ref{CCR})).

The formal algebra of observables of our quantum system is generated by the set of 
operators $\{W_{\bk}\}_{\bk\in\IZ^{2d}}$ 
acting on $\cH_{h}(\btht)$ and will be denoted by $\cA_{h}(\btht)$. 
We note that $\cA_{h}(\btht)$ is a $^*$-algebra with the involution defined by 
$W_{\bk}^{*}=W_{-\bk}$. 
Equipped with the standard operator norm $\cA_{h}(\btht)$ would become a $C^{*}$-algebra
and its weak closure - a $W^{*}$-algebra. 

The reference state defined on the generators of $\cA_{h}(\btht)$ by 
\bean
\tau(W_{\bk})=\del_{\bk,0} 
\eean
corresponds to the standard Lebesgue measure in the classical system and can 
be uniquely extended to a tracial state on the whole algebra
$\cA_{h}(\btht)$.

In our approach, instead of studying $\cA_{h}(\btht)$ as a $C^{*}$ or $W^{*}$-algebras, 
 we prefer to take advantage of its natural $H^{*}$-algebraic structure, 
associated with the state $\tau$. If $\tau$ is faithful the inner product
on  $\cA_{h}(\btht)$ is introduced as follows
\bean
\la A, B \ra = \tau(A^{*}B).
\eean
In particular
\bean
\la W_{\bk}, W_{\bm} \ra =\del_{\bk,\bm}
\eean 
and hence automatically $\{W_{\bk}\}_{\bk\in \IZ^{2d}}$ becomes an orthonormal basis for $\cA_{h}(\btht)$.

This completes the discussion of the kinematic step of our general quantization scheme.
Below we derive as particular cases two models described in Chapter \ref{Inter}. We
call them here Model I and Model II.

%%%%%%%%%%%%%%%%%%%%%%%%%%%%%%%%%%%%%%%%%%%%%%%%%%%%%%%%%%%%%%%%%%%%%%%%%%%%%%%%%%%%%%%%%%%%%%%%%%%%%%%%%

\subsection{Model I}
\label{ModelI}

%%%%%%%%%%%%%%%%%%%%%%%%%%%%%%%%%%%%%%%%%%%%%%%%%%%%%%%%%%%%%%%%%%%%%%%%%%%%%%%%%%%%%%%%%%%%%%%%%%%%%%%%%

In this section we derive the original Hannay-Berry model including all its later developments
and generalizations described in Chapter \ref{Inter}. To this end we set
\bean
\cH^{I}_{h}(\btht):=\cS^{(q)}_{h}(\btht_{p}) \cap \cS^{(p)}_{h}(\btht_{q}).
\eean
Thus a distribution $\psi\in \ml{S}'(\IR^d)$ belongs to $\cH^{I}_{h}(\btht)$ 
iff for any $\bm\in \IZ^{2d}$,
\bean
\psi(\bq+\bm_{1})&=&e^{2\pi i \btht_{p}\cdot \bm_{1}}\psi(\bq),\\
(\ml{F}_{h}\psi)(\bp+\bm_{2})&=&e^{-2\pi i \btht_{q}\cdot \bm_{2}}(\ml{F}_{h}\psi)(\bp).
\eean

The parameters $\btht_{p},\btht_{q}$ are sometimes called Floquet or Bloch angels.
It is well known (see e.g. \cite{BozDB}) that the set of all distributions
satisfying the above mutual quasiperiodicity conditions contains nonzero elements iff
\bea
\label{BS}
h=\frac{1}{N},
\eea
where $N\in \IZ_{+}$. In literature (\ref{BS}) is sometimes referred to as 
the Bohr-Sommerfield condition. Thus whenever we discuss the Model I
we assume that (\ref{BS}) holds.

We pause for a while to introduce some notation.
Let
\bean
J_{-}:=
\begin{bmatrix}
0 & -I \\
I & 0
\end{bmatrix}, \qquad
J_{+}:=
\begin{bmatrix}
0 & I \\
I & 0
\end{bmatrix},
\eean
where $I$ denotes the identity matrix on $\IR^{d}$.

For any pair of vectors $\bv, \bv' \in \IR^{2d}$ we set
\bean
\bv \wedge \bv' := \bv J_{-} \bv', \qquad
\bv \vee \bv' := \bv J_{+} \bv'. 
\eean

Of course $\bv \wedge \bv'$ is a standard symplectic
product on $\IR^{2d}$.

Using the above notation quasiperiodicity conditions can be naturally encoded in terms
of the action of translation operators

\begin{prop}
\label{Tkpsi}
Let $\psi\in  \ml{S}'(\IR^d)$, then $\psi\in \cH^{I}_{h}(\btht)$ iff for all $\bm \in \IZ^{2d}$,
\bea
\label{q-P}
T_{\bm}\psi=e^{2\pi i \lp \frac{N}{4} \bm\vee \bm + \bm\wedge \btht \rp }\psi.
\eea 
\end{prop}
\textbf{Proof.}
Obviously if (\ref{q-P}) holds then $\psi$ is $\btht$-quasiperiodic
 (consider $\bm=(\bm_{1},0)$ and $\bm=(0,\bm_{2})$). 
On the other hand if
$\psi\in \cH^{I}_{h}(\btht)$ then
\bean
T_{\bm}\psi&=&T_{(\bm_{1},0)+(0,\bm_{2})}\psi
=e^{-\pi i N (\bm_{1},0)\wedge (0,\bm_{2})} 
T_{(\bm_{1},0)}T_{(0,\bm_{2})}\psi\\
&=&e^{2\pi i (N/4) \bm \vee \bm }  e^{-2\pi i \btht_{p}\cdot \bm_{1}} 
e^{2\pi i \btht_{q}\cdot \bm_{2}}\psi=e^{2\pi i \lp N/4 \bm\vee \bm + \bm\wedge \btht \rp}\psi. 
\qquad \blacksquare
\eean

The general form of an element of $\cH^{I}_{h}(\btht)$ can easily be determined. One finds (for detailed
derivation see Section \ref{CanQuant} of Appendix \ref{A}) 
that $\psi \in \cH^{I}_{h}(\btht)$ is necessary a quasiperiodic Dirac delta comb (brush) of the form
\bea
\label{QPDC}
\psi(\bq)=\frac{1}{N^{d/2}}\sum_{\bs\in \IZ^{d}/N}c_{\bs}\del_{\bs+\btht_{q}/N}(\bq),
\eea
where $c_{\bs}$ is a quasiperiodic sequence of arbitrary numbers 
supported on $\IZ^{d}/N$ lattice and satisfying $c_{\bs+\bn}=e^{2\pi i \btht_{p}\cdot \bn}c_{\bs}$. 
Thus, although $\cS^{(q)}_{h}(\btht_{p})$ and $\cS^{(q)}_{h}(\btht_{p})$ are not Hilbert spaces,
$\cH^{I}_{h}(\btht)$ can naturally be identified with the Hilbert space $\IC^{N^d}$,
by introducing the following $L^2$-norm
\bean
\|\psi\|_{2}^{2}:=\frac{1}{N^{d}}\sum_{\bs\in \IQ^{d}_{N}}|c_{\bs}|^{2}.
\eean

We note that, the crucial difference between the present model and the
one considered in the next section lies in the
fact that full quasiperiodicity of the state space
$\cH^{I}_{h}(\btht)$ implies also quasiperiodicity of quantum Fourier modes.
Indeed we have
\begin{prop}
For any $\bm\in\IZ^{2d}$,
\bea
\label{Wq-P}
W_{\bk +N\bm}=e^{2\pi i \al(\bk,\bm,\btht)}W_{\bk},
\eea
where
\bean
\al(\bk,\bm,\btht)=\frac{1}{2}\bk\wedge\bm + \frac{N}{4} \bm\vee\bm + \bm\wedge\btht.
\eean
\end{prop}
\textbf{Proof.}
Using Proposition \ref{Tkpsi} we have for any $\psi\in \cH^{I}_{h}(\btht)$, 
\bean
W_{\bk +N\bm}\psi &=& T_{\bk/N +\bm}\psi=e^{\pi i \bk \wedge\bm}T_{\bk/N}T_{\bm}\psi\\
&=& 
  e^{\pi i \bk \wedge\bm}e^{2\pi i (N/4 \bm\vee \bm + \bm\wedge \btht)T_{\bk/N}}\psi
=e^{2\pi i \al(\bk,\bm,\btht)}W_{\bk}\psi. \qquad \blacksquare
\eean

The algebra of observables of Model I will be denoted by $\cA^{I}_{N}(\btht)$.
Due to the quasiperiodicity of the
set of its generators $\cA^{I}_{N}(\btht)$
is finite dimensional and as a linear space can be identified with 
$\cL(\cH^{I}_{h}(\btht)) \cong \cM_{N^{d}\times N^{d}} \cong \IC^{N^{2d}}$. 

We note that $\cA^{I}_{N}(\btht)$ is still a $^*$-algebra with the involution defined by $W_{\bk}^{*}=W_{-\bk}$. 
The fact that operation $*$ is consistent with the quasiperiodic structure of the set of generators follows 
from the property $\al(-\bk,-\bm,\btht)=-\al(\bk,\bm,\btht) \mod 1$.

The tracial state on 
$\cA^{I}_{N}(\btht)$ can be defined now explicitly  
\bean
\tau(A)=\frac{1}{N^d} \Tr A. 
\eean

And as above $\{W_{\bk}, \bk\in \IZ_{N}^{2d}\}$ ($\IZ_{N}^{2d}:=\IZ^{2d}\mod N$) becomes 
an orthonormal, but this time finite basis for an $H^{*}$-structure on $\cA^{I}_{N}(\btht)$.

The $H^{*}$-norm on $\cA^{I}_{N}(\btht)$ will be denoted by $\|\cdot\|_{HS}$. 
One needs to keep in mind that $\|\cdot\|_{HS}$ does not coincide with
the standard operator norm, hence $\cA^{I}_{N}(\btht)$ is not (considered here as) a $C^{*}$-algebra.

Now we choose the fundamental domain of periodicity $\IZ_{N}^{2d}$ for our quantum 
Fourier lattice. The choice centered around the origin 
seems to be the most natural one (cf. \cite{N}). That is, we assume that if 
$\bk=(k_{1},...,k_{2d})\in \IZ_{N}^{2d}$ then for every $j\in \{1,...,2d\}$
\bean
k_{j}\in
\begin{cases}
\{-N/2+1,...,N/2\}, & \qquad \text{for $N$ even} \\ 
\{-(N-1)/2+1,...,(N-1)/2\}, & \qquad \text{for $N$ odd}.
\end{cases}
\eean
The set $\{W_{\bk}, \bk\in \IZ_{N}^{2d}\}$ forms
an orthonormal basis for $\cA_{N}(\btht)$.

To any classical observable $f\in C^{\infty}(\IT^{2d})$, 
or more generally $f\in L^{2}(\IT^{2d})$, with $\sum_{\bk}|\hat{f}(\bk)|<\infty$ 
there corresponds
an element of $\cA^{I}_{N}(\btht)$ i.e. its Weyl quantization, denoted by $Op_{N}(f)$, and
defined in terms of its Fourier expansion
\bean
Op_{N}(f)=\sum_{\bk \in \IZ^{2d}}\hat{f}(\bk)W_{\bk}=\sum_{\bk \in \IZ_{N}^{2d}} 
\lp \sum_{\bm \in \IZ^{2d}}e^{2\pi i \al(\bk,\bm,\btht)}\hat{f}(\bk+N\bm) \rp W_{\bk}.
\eean
The map $Op_{N}$ is not invertible. One can 
define however the isometry $W^{P}:\cA^{I}_{N}(\btht) \mapsto  L^{2}(\IT^{2d})$ which
associates with each observable $A\in\cA^{I}_{N}(\btht)$ its polynomial Weyl symbol
\bean
W^{P}(A)=\sum_{\bk \in \IZ_{N}^{2d}}a_{\bk}w_{\bk},
\eean
where $w_{k}:=e^{2\pi i \bk\wedge\bx}$ denote classical Fourier modes and 
$a_{k}=\la W_{k}, A \ra$. The operators $Op_{N}$ and $W^{P}$ are inverse
of each other when the domain of $Op_{N}$ and the codomain of $W^{P}$ are 
restricted to $\cI_{N}=W^{P} \circ Op_{N}(L^{2}(\IT^{2d}))$.

%%%%%%%%%%%%%%%%%%%%%%%%%%%%%%%%%%%%%%%%%%%%%%%%%%%%%%%%%%%%%%%%%%%%%%%%%%%%%%%%%%%%%%%%%%%%%%%%%%%%%%%%%

\subsection{Model II}
\label{ModelII}
%%%%%%%%%%%%%%%%%%%%%%%%%%%%%%%%%%%%%%%%%%%%%%%%%%%%%%%%%%%%%%%%%%%%%%%%%%%%%%%%%%%%%%%%%%%%%%%%%%%%%%%%%

In their original paper \cite{BNS} the authors did not specify any particular
Hilbert space of pure states for their model and instead performed the
quantization starting on the algebraic level. 

In order to emphasize a close link between the two models within the
framework considered here, we first construct the physical Hilbert space 
for this model and then show that the natural restriction of 
Weyl quantization on $\cS'(\IR^{d})$ to this space yields the algebra of observables 
considered in \cite{BNS}.

First we define the following two Hilbert spaces
\bean
\cH^{(q)}_{h}(\btht_{p}):=\{(c,h,\btht_{p}) \in \cS^{(q)}_{h}(\btht_{p}) : c\in l^{2}(\IZ^d)\}, \\
\cH^{(p)}_{h}(\btht_{q}):=\{(c,h,\btht_{q}) \in \cS^{(p)}_{h}(\btht_{q}) : c\in l^{2}(\IZ^d)\}.  
\eean 
The values of the parameters $\btht_{p}$ and $\btht_{q}$ do not play any significant role 
in this model and hence we will be working with the following spaces
\bean
\cH^{(q)}_{h}:=\cH^{(q)}_{h}(0), \qquad
\cH^{(p)}_{h}:=\cH^{(p)}_{h}(0).
\eean
As as we proved in the previous section, if 
$(c,h,0)\in \cS^{(q)}_{h}(0) \cap \cS^{(p)}_{h}(0)$
then either $c\equiv 0$ or $c \not \in l^{2}(\IZ^d)\}$, since any such $c$ must be periodic. 
Thus $\cH^{(q)}_{h} \cap \cH^{(q)}_{h}=\{0\}$.

We then define
\bean
\cH^{II}_{h}:=\cH^{(q)}_{h} \oplus \cH^{(p)}_{h}.
\eean
In opposition to $\cH^{I}_{h}$, the space $\cH^{II}_{h}$ is infinite dimensional.

Now similarly as in Model I the quantization consists in the
restriction of the standard Weyl quantization on $\cS'(\IR^{d})$ to the
space $\cH^{II}_{h}$.  
The construction of the space $\cH^{II}_{h}$ insure that the set of 
all admissible quantum translations is discrete and coincides with the family
of operators ${W_{\bk}}_{\bk\in\IZ^{2d}}$ considered 
in the previous section (see Remark \ref{presboth}). 

As it was mentioned above (cf. (\ref{MCCR})) $W_{\bk}$ satisfy the relations
\bean
W_{\bk}W_{\bm}&=& e^{\pi i h \bk\wedge \bm}W_{\bk+\bm}.
\eean
The Planck constant $h$ plays the role of deformation parameter $\theta$ 
considered in \cite{BNS} (unrelated, of course, to our $\btht$).
Putting $h:=2\theta$ one recovers exactly the relations
assumed in \cite{BNS}. The algebra of observables of model II $\cA^{II}_{h}$ can now be defined as a 
$^*$-algebra generated by elements $W_{\bk}$.
Taking the weak closure (the bicommutant) in $\cB(\cH^{II}_{h})$
yields a $W^{*}$-algebra isomorphic to the algebra of observables introduced in \cite{BNS}.

For all $h\geq 0$, $\cA^{II}_{h}$ is always infinite dimensional and hyperfinite i.e. is
generated by an ascending sequence of finite dimensional algebras.
However the internal structure of $\cA^{II}_{h}$ depends in an essential way on
$h$ is rational or not. Before we discuss the classification of $\cA^{II}_{h}$
we need to introduce some definitions \cite{BrRo}.

\begin{defin}
\label{type} $ $

i) A $W^{*}$-algebra is called a factor if its center is trivial (the algebra is
irreducible).

ii) A factor is of type $I$ if it is isomorphic to $\cB(\cH)$ for some Hilbert
space $\cH$. 

iii) If $dim(\cH)=N$ then the factor is called to be of type $I_{N}$.

iv) Two projections $P_{1},P_{2}$ in a $W^{*}$-algebra are said to be equivalent if there
exists an element $A$ in the algebra such that $P_{1}=A^*A$ and $P_{2}=AA^*$.  

v) Projection is said to be finite if it is not equivalent to any proper
subprojection of itself.

vi) A factor is called type $I\!I_{1}$ if the identity operator $1 \!\! 1$ is finite.
\end{defin}

If $h$ is rational then $\cA^{II}_{h}$ factorizes into its nontrivial
commutative center and a type $I_{N^{d}}$ factor (matrix algebra) 
\bean
\cA^{II}_{h}=L^{2}(T^{2d})\otimes \cM_{N^{d}\times N^{d}}
\eean
where $N$ is the smallest integer such that $h N\in \IZ_{+}$ 
(see \cite{KL} and \cite{GH} for transparent proofs).
Moreover it can be shown (cf. \cite{BozDB}) that in this case
The algebras of both models are related by the following decomposition
formula
\bean
\cA^{II}_{h}=\int^{\oplus}_{\IT^{2}}\cA^{I}_{N}(\btht), 
\eean

If $h$ is irrational $\cA^{II}_{h}$ is a factor of type $I\!I{1}$ (cf. \cite{S}).

In both cases $\cA_{h}$ admits a faithful tracial state, which in case
of rational $h$ factorizes into standard Lebesgue measure on $\IT^{2d}$
and the normalized trace on $\cM_{N^d \times N^d}$. 

Moreover it can be shown that 

We end this section with a few remarks regarding the choice of the Hilbert space
for the Model II. 

\begin{rem}
The choice of the physical space for Model II is not unique. One can consider e.g. the 
space $\tilde{\cH}^{II}_{h}:=\cH^{(q)}_{h} \otimes \cH^{(p)}_{h}$
and the algebra generated by elements $\cW_{\bk}=W_{\bk} \otimes W_{\bk}$.
\end{rem}

\begin{rem}
In contrast to the case of Model I, for all irrational $h$, the algebra of Model II
is of type $I\!I{1}$ is never isomorphic to any full (i.e. type I) algebra $\cB(\cH_{h})$ 
regardless the choice of the Hilbert space $\cH_{h}$.
\end{rem}

%%%%%%%%%%%%%%%%%%%%%%%%%%%%%%%%%%%%%%%%%%%%%%%%%%%%%%%%%%%%%%%%%%%%%%%%%%%%%%%%%%%%%%%%%%%%%%%%%%%%%%%%%

\section{Quantization on the torus. The dynamics}
\label{QTD}

%%%%%%%%%%%%%%%%%%%%%%%%%%%%%%%%%%%%%%%%%%%%%%%%%%%%%%%%%%%%%%%%%%%%%%%%%%%%%%%%%%%%%%%%%%%%%%%%%%%%%%%%%

In this section we perform the dynamical step in the process of quantization on the torus.
We consider arbitrary canonical toral maps i.e. area and orientation 
preserving homeomorphisms on $\IT^{2d}$. 

Let $\Phi$ denote such a map.
As was already mentioned in Section \ref{prelim}, $\Phi$ can be
decomposed into the product of three maps $\Phi=F\circ t_{\bv} \circ \Phi_{1}$, where
$F\in SL(2d,\IZ)$ is a symplectomorphism, $t_{\bv}$ denotes a classical translation by vector $\bv$ and 
$\Phi_{1}(\bx)=\bx+p(\bx)$, where $p$ is an arbitrary zero-mean, continuous and periodic function.

We assume that $\Phi_{1}$ represents a
time-1 flow map associated with a periodic Hamiltonian. In $2$-dimensional
case this assumption is equivalent (cf. \cite{KMR}) to the above requirement
that $p$ be of zero mean (w.r.t. the Lebesgue measure on $\IT^{2d}$).
This can always be achieved by replacing $P$ with $p'=p-\la p\ra$ and
adjusting accordingly translational component of $\Phi$.

To quantize $\Phi$ one first quantizes 
$F$, $t_{\bv}$ and $\Phi_{1}$ separately. The quantization
of $\Phi$ is then defined as a composition of corresponding quantum  $^*$-automorphisms
\bean
\cU_{\Phi}=\cU_{F} \cU_{t_{\bv}} \cU_{1}.
\eean

The quantization procedure will be prescribed in such a way that the
correspondence principle will hold. In the case of Model I this will be expressed
in terms of so called Egorov property, which states that for every 
$f\in C^{\infty}(\IT^{2d})$ there exists $C>0$ such that
\bea
\label{Egorov}
\|\cU(Op_{N}(f))-Op_{N}(U f)\|\leq \frac{C}{N},
\eea
where $U f=f \circ \Phi$ is a classical Koopman operator of $\Phi$.

%%%%%%%%%%%%%%%%%%%%%%%%%%%%%%%%%%%%%%%%%%%%%%%%%%%%%%%%%%%%%%%%%%%%%%%%%%%%%%%%%%%%%%%%%%%%%%%%%%%%%%%%%

\subsection{Quantization of symplectomorphisms}
\label{catmap}

%%%%%%%%%%%%%%%%%%%%%%%%%%%%%%%%%%%%%%%%%%%%%%%%%%%%%%%%%%%%%%%%%%%%%%%%%%%%%%%%%%%%%%%%%%%%%%%%%%%%%%%%%

Here we describe the quantization of toral symplectomorphism i.e. symplectic automorphism on $\IT^{2d}$.
In classical setting the action of a toral automorphism on the algebra of classical observables was defined
by means of the Koopman operator given by $U_{F}f=f\circ F$, where  $f$ was an arbitrary observable, 
usually assumed to be an element of $L^{\infty}(\IT^{2d})$ or $C^{\infty}(\IT^{2d})$. In 
our approach however it was more convenient to consider the action of $U_{F}$ on a slightly
bigger space $L^{2}(\IT^{2d})$. The natural Hilbert structure of the later
allowed us to determine the dynamics by specifying it on the basis of classical Fourier modes 
$w_{\bk}(\bx)=e^{2\pi i \bk\wedge \bx}$
\bean
(U_{F}w_{\bk})(\bx)=e^{2\pi i \bk \wedge F\bx}= e^{2\pi i \bk J_{-} F\bx}
= e^{2\pi i J_{-} F^{\dagger} J^{\dagger}_{-} \bk \wedge \bx},
\eean 
where $F^{\dagger}$ denotes the transposed map.
Setting $F'=J_{-} F^{\dagger} J^{\dagger}_{-}$ we get
\bean
U_{F}w_{\bk}=w_{F'\bk}.
\eean

The most natural way to define the quantum counterpart of this dynamics is
to consider the formal superoperator version $\cU_{F}$ of the classical Koopman 
operator
\bea
\label{QCatMap}
\cU_{F}W_{\bk}=W_{F'\bk}.
\eea
Such dynamics will be well defined i.e. will define a $^*$-automorphism of 
$\cA_{h}(\btht)$ only if the action of $\cU_{F}$ 
is consistent with its algebraic structure (Weyl commutation relations).
We note that
\bean
\cU_{F}(W_{\bk}W_{\bm})
&=&e^{\pi i h\bk\wedge\bm}\cU_{F}(W_{\bk+\bm})=
e^{\pi i h\bk\wedge\bm}W_{F'\bk+F'\bm}
\\&=&e^{\pi i h\bk\wedge\bm}e^{-\pi i h F'\bk\wedge F'\bm}
W_{F'\bk}W_{F'\bm}\\
&=& e^{2\pi i \frac{h}{2}(\bk\wedge\bm-F'\bk\wedge F'\bm)}
\cU_{F}(W_{\bk})\cU_{F}(W_{\bm}).
\eean
Thus in order for $\cU_{F}$ to be a $^{*}$-automorphism the following condition has to be satisfied
\bea
\label{quc}
F'\bk\wedge F'\bm = \bk\wedge\bm \mod \frac{2}{h}.
\eea
Since the condition (\ref{quc}) has to be valid for arbitrary small $h$ 
we have
\begin{prop}
If a map $F\in SL_{\pm}(2d,\IZ)$ is quantizable then it is symplectic.
\end{prop}

Symplecticity is then necessary for quantization regardless the
model i.e. the space $\cH_{h}(\btht)$ one choses to work with. 
We note that for symplectic maps $F'=F^{-1}$.

Depending on the choice of $\cH_{h}(\btht)$ the condition may be
also sufficient. Indeed, this is exactly the case in Model II. 
In view of the lack of additional conditions on the 
generators of the algebra $\cA^{II}_{h}$ there are no quantization 
restrictions other than symplecticity of the map.

In some cases, however, the structure of $\cH_{h}(\btht)$ impose additional
relations on the generators of the algebra $\cA_{h}(\btht)$ and then additional
conditions are needed.
This is the case in Model I, where the assumption of qasiperiodicity in both 
position and momentum resulted in qasiperiodicity of algebra. 

Thus in this case we have to ensure the compatibility of the action of 
$\cU_{F}$ with the quasiperiodic structure of $\cA^{I}_{N}(\btht)$. To this end 
we note that for a given quantum Fourier mode $W_{\bk+N\bm}$,
one can compute the value of $\cU_{F}W_{\bk+N\bm}$ in the following two,
in general different, ways: 

- on one hand using linearity of $F'$ we have
\bean
\cU_{F}W_{\bk+N\bm}=W_{F'\bk+N F'\bm}=e^{2\pi i \al(F'\bk,F'\bm,\btht)}W_{F'\bk}
\eean

- on the other hand, by linearity of $\cU_{F}$ 
\bean
\cU_{F}W_{\bk+N\bm}=e^{2\pi i \al(\bk,\bm,\btht)}\cU_{F}W_{\bk}=e^{2\pi i \al(\bk,\bm,\btht)}W_{F'\bk}.
\eean
That is, for every $\bk$ and $\bm$ in $\IZ^{2d}$ we must have
\bea
\label{alqu}
\al(F'\bk,F'\bm,\btht)=\al(\bk,\bm,\btht) \mod 1.
\eea
The map $F$ will be called quantizable in Model I if for every $N\in \IZ_{+}$ there exists 
$\btht\in \IT^{2d}$
(possibly depending on N) such that (\ref{alqu}) holds for all $\bk$ and $\bm$ in $\IZ^{2d}$.

Below we summarize the quantization condition in Model I for general toral automorphisms. 
\begin{prop}
\label{HBq}
A toral automorphism represented by $F\in SL_{\pm}(2d,\IZ)$ is quantizable iff it is symplectic. 
For any given $h=N^{-1}$, the corresponding $\btht$ has to satisfy 
the following condition:  
\bea
\label{qbc}
\frac{N}{2}
\begin{pmatrix}
A \cdot B \\
C \cdot D
\end{pmatrix}
+F\btht=\btht \mod 1,
\eea
where $A,B,C,D$ denote block-matrix elements of F, that is
\bean
F=\begin{bmatrix}
A & B \\
C & D
\end{bmatrix}
\eean
and $A\cdot B$ denotes a contraction of
two matrices into a (column) vector, defined as follows
\bean
(A\cdot B)_{i}=\sum_{j}A_{ij}B_{ij}.
\eean
\end{prop}

The existence of solutions of equations of the type (\ref{qbc}) is easy to establish.
We note that if $N$ is even then one can simply choose $\btht=0$. The same solution
can be chosen whenever the vector 
\bean
\begin{pmatrix}
A \cdot B \\
C \cdot D
\end{pmatrix}
\eean
has all even components (and this condition reduces to Hannay and Berry's 'checkerboard'
condition stated in \cite{HB} in $d=1$ case).
Otherwise one considers two cases. 
If $F-I$ is invertible then for any $\bk\in\IZ^{2d}$
\bean
\btht=(F-I)^{-1}\lp
\frac{N}{2}
\begin{pmatrix}
A \cdot B \\
C \cdot D
\end{pmatrix}
+\bk\rp.
\eean
There are exactly $|\det(F-I)|$ distinct solutions. In particular the solution is unique if
$F-I\in SL_{\pm}(2d,\IZ)$ (in $d=1$ case $\det(F-I)=2-\Tr F$, hence uniqueness holds iff
 $\Tr F=1$ or $3$).
If $F-I$ is singular one can decompose the matrix $F$ into an identity and
nonsingular block and construct an appropriate $\btht$ by applying the above considerations 
to each block separately.

We want remark that in view of the defining condition (\ref{QCatMap}), the
Egorov property \eqref{Egorov} is automatically satisfied (with no error term). 

We end this section with a few comments about the above quantization conditions.
The conditions were imposed to ensure that $\cU_{F}$ is a $^{*}$-automorphism of the
algebra $\cA^{I}_{N}(\btht)$. 

In Section \ref{Ch5proofs} we will prove the following simple
\begin{prop}
\label{inner}
Any $^*$-automorphism on finite matrix algebra is inner.
\end{prop}
Thus in view of this proposition
our conditions ensure the existence of a physical 
quantum propagator $\hat{U}_{F}$ implementing the dynamics 
on the underlying physical space 
\bean
\cU_{F}A=ad(\hat{U}_{F})A= \hat{U}^{*}_{F}A\hat{U}_{F}.
\eean
It can be shown that $\hat{U}_{F}$ (which is only determined up to the
phase factor) coincides with the propagator introduced by Hannay and Berry in 
their original paper \cite{HB}. For further details regarding this construction we refer the reader 
to \cite{M}.

We also note that there exists a geometric interpretation of these quantization conditions. 
Indeed, consider the evolution of the Wigner transform of a given wave function under the cat map dynamics. 
It is Well known that the Wigner function evolves according to a classical map (see Appendix \ref{B}) and
forms a $(2N)^d\times(2N)^d$ periodic delta brush supported
on the half-integer lattice (if N is length of the side of the fundamental domain of its periodicity).
Symplecticity insures that the evolved delta brush represents once again a Wigner function.

In case of odd $N$ or when the
wave function is $\btht$-quasiperiodic ($\btht\not = 0$) which means
that the supporting lattice of Wigner function is 
shifted by $\btht$ on the coordinate plane, one wants to ensure that
this supporting lattice remains on the same place throughout the
evolution. The latter property is equivalent to condition (\ref{qbc}), and can
be thought of as the conservation of the initial 'quantum boundary conditions' 
(see \cite{KMR}).     

%%%%%%%%%%%%%%%%%%%%%%%%%%%%%%%%%%%%%%%%%%%%%%%%%%%%%%%%%%%%%%%%%%%%%%%%%%%%%%%%%%%%%%%%%%%%%%%%

\subsection{Quantization of translations}
\label{Qtran}

%%%%%%%%%%%%%%%%%%%%%%%%%%%%%%%%%%%%%%%%%%%%%%%%%%%%%%%%%%%%%%%%%%%%%%%%%%%%%%%%%%%%%%%%%%%

As explained in Section \ref{WQ}, a translation $t_{\bv}$ 
is quantized on $L^2(\IR^d)$ through a Weyl operator $T_{\bv}$. We have noticed
that such quantum translations act inside the algebra $\cA^{I}_{N}(\btht)$ only if
$\bv\in h\IZ^{2d}$. If this condition is not satisfied then quantization depends
on the model. In case of Model I there are several possibilities to
quantize such translations \cite{BozDB}. We will choose the prescription given in \cite{MR}:
we take the vector $\bv^{(N)}\in N^{-1}\IZ^{2d}$ closest to $\bv$ (in Eudlidean distance),
which can be obtained by taking, for each $j=1,\ldots,2d$,  the
component $v^{(N)}_j=\frac{[Nv_j]}{N}$,
where $[x]$ denotes the closest integer to $x$. 
One then quantizes $t_{\bv}$ on $\cH_{N}(\btht)$ through the restriction of $T_{\bv^{(N)}}$ on
that space (this is the same operator as $W_{[N\bv]}(N,\btht)$). 
The corresponding 
$^*$-automorphism on $\cA_{N}(\btht)$ is given by 
\bean
\cU_{t_{\bv}}=ad(T_{\bv^{(N)}}).
\eean
It was proved in \cite{MR} that the Egorov property \eqref{Egorov} holds for this quantization.
In case of Model II the situation is simpler. Even though $T_{\bv}$ itself does not act as an
inner *-automorphism on $\cA^{II}_{\hbar}$ it introduces an external *-automorphism on this
algebra
\bea
\label{exter}
\cU_{t_{\bv}}W_{\bk}:= T^{*}_{\bv}W_{\bk}T_{\bv}=e^{-2\pi i \bv \wedge \bk}W_{\bk}
\eea
and this automorphism can be taken as a quantization of $t_{\bv}$. Note that
definition given by \eqref{exter} cannot be applied in Model I since the action of $\cU_{t_{\bv}}$ 
is not consistent with quasiperiodic structure of $\cA^{I}_{N}(\btht)$.

%%%%%%%%%%%%%%%%%%%%%%%%%%%%%%%%%%%%%%%%%%%%%%%%%%%%%%%%%%%%%%%%%%%%%%%%%%%%%%%%%%%%%%%%%%%

\subsection{Quantization of time-1 flow maps of periodic Hamiltonians} 
\label{time1}

%%%%%%%%%%%%%%%%%%%%%%%%%%%%%%%%%%%%%%%%%%%%%%%%%%%%%%%%%%%%%%%%%%%%%%%%%%%%%%%%%%%%%%%%%%%

We present here the quantization in case of Model I.
Let $\Phi_{1}$ denote the time-1 flow map associated with a periodic Hamiltonian $H(\bz,t)$,
meaning that $\Phi_t:\IT^{2d}\to\IT^{2d}$ satisfies the Hamilton equations:
\bean
\frac{\pa \Phi_{t}(\bz)}{\pa t}=\nabla^{\perp} H( \Phi_{t}(\bz),t),\qquad
\Phi_{0}=I.
\eean
To quantize $\Phi_{1}$, one applies the Weyl quantization to the Hamiltonian $H(t)$, 
obtaining a time-dependent Hermitian operator $Op_{N,\btht}(H(t))$. From there, one constructs
the time-1 quantum propagator on $\cH_{N}(\btht)$
associated with the Schr\"odinger
equation for $Op_{N,\btht}(H(t))$: 
$$
U_{N,\btht}(\Phi_1):=\cT\, e^{-2\pi i N \int_{0}^{1} Op_{N,\btht}(H(t))\, dt}
$$
($\cT$ represents the time ordering).
As above, the corresponding $^*$-automorphism on  
$\cA^{I}_{N}(\btht)$ is defined by
\bean
\cU_1A=ad(U_{N,\btht}(\Phi_1))\,A= U^{*}_{N,\btht}(\Phi_1) A U_{N,\btht}(\Phi_1).
\eean
For such a propagator, the Egorov property is holds, yet with a nonzero 
error (see \cite{Ro}, Theorem IV.10 for the case of a time-independent Hamiltonian on
$\IR^{2d}$).

%%%%%%%%%%%%%%%%%%%%%%%%%%%%%%%%%%%%%%%%%%%%%%%%%%%%%%%%%%%%%%%%%%%%%%%%%%%%%%%%%%%%%%%%%%%
%%%%%%%%%%%%%%%%%%%%%%%%%%%%%%%%%%%%%%%%%%%%%%%%%%%%%%%%%%%%%%%%%%%%%%%%%%%%%%%%%%%%%%%%%%%

\section{Quantum noise}
\label{Qnoise}
%%%%%%%%%%%%%%%%%%%%%%%%%%%%%%%%%%%%%%%%%%%%%%%%%%%%%%%%%%%%%%%%%%%%%%%%%%%%%%%%%%%%%%%%%%%
%%%%%%%%%%%%%%%%%%%%%%%%%%%%%%%%%%%%%%%%%%%%%%%%%%%%%%%%%%%%%%%%%%%%%%%%%%%%%%%%%%%%%%%%%%%

In this section we construct quantum noise operator for our system.
We first recall some standard facts regarding quantum noise in general (cf. \cite{AF},\cite{CN}).
 
The influence of a noise on a quantum system is described through the interaction 
between the system and its environment. If the system is in a state given by a density 
matrix $\rho$ and the state of the environment is $\rho_{env}$ then the state of the open  
system (universe) is given by their tensor product $\rho\otimes \rho_{env}$. As a principle
the quantum evolution of the whole universe is assumed to be unitary. Thus there 
exists a unitary operator $U$ such that the evolved jointed density is given by 
$U\rho\otimes \rho_{env}U^{*}$. 
The noisy quantum evolution $\Gamma$ of the small system is recovered by 
tracing out the environment
\bea
\label{uni}
\Gamma(\rho)=\Tr_{env}(U\rho\otimes \rho_{env}U^{*}).
\eea 
In general, $\Gamma$ is not unitary. It is however always (I) trace preserving and
(II) completely positive. Complete positivity implies positivity which together with trace preserving 
property ensures that $\Gamma$ maps densities into densities. 

Usually the environment is not specified explicitly and the noisy evolution
of the system is described entirely in terms of some abstract operator $\Gamma$ 
acting only on the small system. Although positivity and trace preserving property would 
suffice to insure that such operator preserves densities, it would not insure its
representability in the form (\ref{uni}).
The sufficient condition for such representation to exists is complete positivity (which provides robustness of 
positivity of $\Gamma$ w.r.t. tensor products). 
Any quantum noise operator is thus characterized by and should satisfy properties (I) and (II).

By Kraus theorem condition (I) is equivalent to the fact that $\Gamma$ admits the following operation-sum
representation
\bea
\label{I}
\Gamma(\rho)=\sum_{k}G_{k}\rho G_{k}^{*},
\eea      
where $G_{k}$, called operation elements, are arbitrary bounded operators.
The above representation is not unique and in some cases it is
useful to use continuous parameter (integral) representation
\bea
\label{Iint}
\Gamma(\rho)=\int_{k}dk G_{k}\rho G_{k}^{*},
\eea      

The corresponding dynamics on the observables of the system is given by
\bean
A \mapsto \sum_{k}G_{k}^{*}AG_{k} \qquad A \mapsto \int_{k}dk G^{*}_{k}A G_{k}.
\eean  

Condition (II) is then equivalent to the requirement that the family $G_{k}$ constitutes 
operational partition of unity
\bea
\label{II}
\sum_{k}G_{k}^{*}G_{k}=Id.
\eea

We now proceed with the construction of the quantum analog of the  
classical noise operator introduced in Section \ref{Cnoise}. 
We start with Model I and follow the quantization method used in \cite{N}. 
For given noise generating density $g$ we will use the following notation
\bean
g_{\ep}(\bx):=\frac{1}{\ep^{2d}}g\left(\frac{\bx}{\ep}\right),\qquad
\tilde{g}_{\ep}(\bx):=\sum_{\bn\in \IZ^{2d}} g_{\ep}(\bx+\bn).
\eean
Also the following notation regarding Fourier transform will be utilized
\bean
%\label{FT1}
\hat{g}(\bxi)&:=& \int_{\IR^{2d}}f(\bx)e^{-2\pi i \bx \cdot \bxi} d\bx, \\
\hat{\tilde{g}}(\bk)&:=& \int_{\IT^{2d}}\tilde{g}(\bx)e^{-2\pi i \bx \cdot \bk} d\bx, \\
\tilde{\hat{g}}(\bx)&:=& \sum_{\bn\in \IZ^{2d}} \hat{g}(\bx+\bn).
\eean
It is easy to check that $\hat{\tilde{g}}_{\ep}(\bk)=\hat{g}_{\ep}(\bk)=\hat{g}(\ep\bk)$ 
for $\bk\in \IZ^{2d}$ and hence
\bean
\tilde{g}_{\ep}(\bx)=\sum_{\bk\in \IZ^{2d}}\hat{g}(\ep\bk)\be_{\bk}(\bx).
\eean

Let us note that the classical noise operator
$G_{\ep}$ on $L^{2}(\IT^{2d})$ can be represented as follows
\bean
G_{\ep}f 
=\int_{\IT^{2d}}\tilde{g}_{\ep}(\bv)(t_{\bv}f)(\bx)d\bv,
\eean
where, as before, $t_{\bv}$ denotes the Frobenius-Perron operator of the
classical phase space translation.
A natural way of quantizing the noise operator is to formally replace
$t_{\bv}$ with quantum phase space translations $T_{\bv}$.
However due to the discrete character of quantum translations one
has to discretize the classical operator before quantization
\bean
G_{\ep} \mapsto \frac{1}{N^{2d}} \sum_{\bk\in \IZ_{N}^{2d}} 
\tilde{g}_{\ep} \lp\frac{\bk}{N}\rp  P_{\frac{\bk}{N}}
\eean
and then quantize it by introducing the following superoperator
\bean
\cG_{\ep,N}:=
\frac{1}{Z}\sum_{\bk\in \IZ_{N}^{2d}}\tilde{g}_{\ep}  \lp\frac{\bk}{N}\rp ad(T_{\frac{\bk}{N}})=
\frac{1}{Z}\sum_{\bk\in \IZ_{N}^{2d}}\tilde{g}_{\ep} \lp\frac{\bk}{N}\rp ad(W_{\bk}),
\eean
where, as always, $ad(W_{\bk})A=W^{*}_{\bk} A W_{\bk}$ for any $A\in  \cA^{I}_{N}(\btht)$.

The role of the prefactor $\frac{1}{Z}$ is to insure that $\cG_{\ep,N}$ is trace 
preserving. One can easily check (see Appendix \ref{pqnoise}) that 
$Z=N^{2d}\tilde{g}_{\ep N}(0)$ and that moreover
\begin{prop}
\label{qnoise}
$\cG_{\ep,N}$ is a completely positive trace preserving map and admits the following 
spectral representation on $\cA^{I}_{N}(\btht)$ 
\bea
\label{spG}
\cG_{\ep,N}A=\sum_{\bk\in \IZ_{N}^{2d}} \gamma_{\ep N}(\bk^{\perp}) a_{\bk} W_{\bk},
\eea
where
\bea
\gamma_{\ep N}(\bxi) := \frac{\tilde{\hat{g}}_{\ep N} (N^{-1}\bxi)}{\tilde{\hat{g}}_{\ep N}(0)},
\qquad
A &:=&\sum_{\bk\in \IZ_{N}^{2d}}a_{\bk}W_{\bk}.
\eea
\end{prop}
Defining the subalgebra of observables orthogonal to identity 
\bean
\cA^{0}_{N}(\btht)=\{A\in \cA^{I}_{N}(\btht): a_{0}=0\}
\eean 
and introducing the superoperator norm w.r.t. $H^{*}$-norm on $\cA^{0}_{N}(\btht)$ 
\bean
\|\cG_{\ep,N}\|:=\sup_{\|A\|_{HS}=1} \|\cG_{\ep,N}A\|_{HS}
\eean
one immediately gets
\bean
\|\cG_{\ep,N}\|=\max_{0\not=\bk\in \IZ_{N}^{2d}}\gamma_{\ep N}(\bk),
\eean
which in particular means that the quantum noise operator acts as a strict contraction on 
$\cA^{0}_{N}(\btht)$.

We end this section with a brief comment regarding the quantization of the noise operator in
the case of Model II. In this case due to the lack of quasiperiodicity conditions and
the uniform in $\bv$ representation of the action of quantum translations on $\cA^{II}_{h}$ 
(see \eqref{exter}) the quantum noise operator acts in an isomorphic fashion to the classical one:
i.e. we can introduce a continuous version of Kraus noise operator
\bean
\cG_{\ep,h}:=\int_{\IR^{2d}}g_{\ep}(\bv)ad(T_{\bv})d\bv,
\eean
which establishes uniform w.r.t. $\hbar\in \IR_{+}$ isometry between classical and quantum noise operators.

%%%%%%%%%%%%%%%%%%%%%%%%%%%%%%%%%%%%%%%%%%%%%%%%%%%%%%%%%%%%%%%%%%%%%%%%%%%%%%%%%%%%%%%%%%%%%%%%%%%%%%%%%%
%%%%%%%%%%%%%%%%%%%%%%%%%%%%%%%%%%%%%%%%%%%%%%%%%%%%%%%%%%%%%%%%%%%%%%%%%%%%%%%%%%%%%%%%%%%%%%%%%%%%%%%%%%

\section{Semiclassical analysis of quantum dissipation time}
\label{SAQDT}

%%%%%%%%%%%%%%%%%%%%%%%%%%%%%%%%%%%%%%%%%%%%%%%%%%%%%%%%%%%%%%%%%%%%%%%%%%%%%%%%%%%%%%%%%%%%%%%%%%%%%%%%%%
%%%%%%%%%%%%%%%%%%%%%%%%%%%%%%%%%%%%%%%%%%%%%%%%%%%%%%%%%%%%%%%%%%%%%%%%%%%%%%%%%%%%%%%%%%%%%%%%%%%%%%%%%%

For any canonical map of the torus $\Phi$, the full noisy quantum dynamics $\cT_{\ep,N}$ 
is defined (in case of Model I) on $\cA^{0}_{N}(\btht)$ by the composition 
$\cT_{\ep,N}:=\cG_{\ep}\cU_{\Phi}$.
We will also consider coarse-grained family of quantum operators defined as follows
\bea
\label{qcoarse}
\tilde{\cT}_{\ep,N}^{(n)}&:=&\cG_{\ep}\cU_{\Phi}^{n}\cG_{\ep}.
\eea
We can now introduce the notion of quantum dissipation time.
\bea
\label{ndiss}
\tau_{q}(\ep,N):=\min\{n \in \IZ_{+}:\|\cT_{\ep,N}^{n}\| < e^{-1}\},\\
\tilde{\tau}_{c}(\ep,N):=\min\{n \in \IZ_{+}:\|\tilde{\cT}_{\ep,N}^{(n)}\| < e^{-1}\}
\eea
Similarly as in classical case the dissipation time provides an intermediate
scale between initial stage of the evolution (where conservative dynamics
dominates due to the assumption of negligible contribution from
the noise term) and final stage when the noise has already driven the system
to its final equilibrium (maximally mixed) state. On the dissipation time scale
the contributions from these competing terms are, roughly speaking, balanced.
In the following sections we analyse the behavior of the dissipation time for
fixed quantum system and for its semiclassical limit. 
To avoid any confusion we will reserve the
symbols $T_{\ep}$, $\tilde{T}_{\ep}$, $\tau_{c}(\ep)$, $\tilde{\tau}_{c}(\ep)$ 
for classical quantities. 
We note that in case on Model II, due to the above established isometry between quantum and classical 
propagators (Section \ref{catmap}) and noise operators (Section \ref{Qnoise}), quantum dissipation
times for this model coincide with their classical counterparts (and do not depend on the value
of the Planck constant). Thus in this case no further analysis is necessary. On the opposite, as we 
will see in the following section, in case of Model I, quantum dynamics differs considerably from
its classical version and in order to recover similarities it will be necessary to consider
appropriate semiclassical limit. 

The main goal of the present section is the semiclassical analysis of the dissipation time of 
noisy evolution of quantum toral maps. Since semiclassical analysis
is meaningful only in the case of Model I, form now on all the considerations are restricted to
this case. In order to better understand the need of semiclassical 
analysis in the case of Model I we start with some simple considerations regarding the behavior of the
dissipation time for finite quantum systems with fixed $N$.

%%%%%%%%%%%%%%%%%%%%%%%%%%%%%%%%%%%%%%%%%%%%%%%%%%%%%%%%%%%%%%%%%%%%%%%%%%%%%%%%%%%%%%%%%%%%%%%%%%%%%%%%%%
%%%%%%%%%%%%%%%%%%%%%%%%%%%%%%%%%%%%%%%%%%%%%%%%%%%%%%%%%%%%%%%%%%%%%%%%%%%%%%%%%%%%%%%%%%%%%%%%%%%%%%%%%%

\subsection{Dissipation time in the 'quantum limit'}
\label{DTQL}

%%%%%%%%%%%%%%%%%%%%%%%%%%%%%%%%%%%%%%%%%%%%%%%%%%%%%%%%%%%%%%%%%%%%%%%%%%%%%%%%%%%%%%%%%%%%%%%%%%%%%%%%%%
%%%%%%%%%%%%%%%%%%%%%%%%%%%%%%%%%%%%%%%%%%%%%%%%%%%%%%%%%%%%%%%%%%%%%%%%%%%%%%%%%%%%%%%%%%%%%%%%%%%%%%%%%%

We denote by $\cU_{\Phi}$ the quantum Koopman operator associated with a canonical map $\Phi$
on the torus $\IT^{2d}$.  
Since $\dim \cA^{I}_{N}(\btht)<\infty$ there exists a unitary matrix (quantum propagator) $U_{N}$ 
implementing $\cU_{\Phi}$ on $\cH_{N}(\btht)$ (cf. Proposition \ref{inner}).
That is
\bean
\cU_{\Phi}A=ad(U_{N})=U_{N}^{*}AU_{N}, \qquad A\in \cA^{I}_{N}(\btht).
\eean
Since $U_{N}$ is unitary, there exists a basis of its eigenfunctions $\psi_{k}^{(N)}\in \cH_{N}(\btht)$. 
For each such function, $\cU_{\Phi}|\psi_{k}^{(N)}\ra\la\psi_{k}^{(N)}|=|\psi_{k}^{(N)}\ra\la\psi_{k}^{(N)}|$.
Moreover $\cU_{\Phi}\mathbf{1}=\mathbf{1}$, and hence we have
\begin{prop}
\label{qKoopman}
The degeneracy of unity in the spectrum of any quantum Koopman operator on $\cA^{I}_{N}(\btht)$ 
is at least of order $N^{d}$. 
For any fixed value of $N$ the corresponding quantum system is nonergodic.
\end{prop}

In Section \ref{NWM}, we showed that the classical dissipation time 
behaves in a power-law fashion in $\ep$ if the Koopman operator has a nontrivial eigenfunction
(if it possesses a modicum of regularity). In finite dimensional quantum setting all observables are
'smooth', since they are represented by a finite Fourier series. 
Thus one should expect that the existence of nontrivial pure point spectrum of the quantum
propagator should lead to slow dissipation. The following Proposition formalizes this
intuition.

\begin{prop}\label{Nfixed}
Assume that the noise generating density $g$ decays sufficiently fast at infinity: 
$\exists \gamma>2d$  s.t. $|g(x)|= \cO(|x|^{-\gamma})$ as $|x|\to\infty$. 

Then, for any $N>0$ and any $\btht$, the quantum noise operator on $\cA_N(\btht)$ satisfies
\bea
\label{estimate1}
\|(1-\cG_{\ep,N})\|\leq C\,(\ep N)^\gamma.
\eea
This bound is useful in the limit $\ep N\to 0$.

As a result, the quantum dissipation time 
associated with any quantized map $\cU_{\Phi}$ is bounded from below as $\tau_q(\ep,N)\geq 
C(\ep N)^{-\gamma}$, where $C>0$ is independent of the classical map $\Phi$. Besides, in this
regime the coarse-grained quantum dynamics does not undergo dissipation, meaning that
$\tilde{\tau}_{q}(\ep,N)= \infty$.
\end{prop}
\textbf{Proof.} We use the RHS of the explicit expressions for the eigenvalues 
$\gamma_{\ep, N}(\bk)$ of $\cG_{\ep,N}$. From the decay assumptions on $g$, we see that
in the limit $\ep N\to 0$, 
$$
\sum_{\bn\in\IZ^{2d}-0} g\bigg(\frac{\bn}{\ep N}\bigg)\leq 
(\ep N)^\gamma\sum_{\bn\in\IZ^{2d}-0} \frac{1}{|\bn|^\gamma}.
$$
The sum on the RHS converges because $\gamma>2d$. Therefore, we get
$\gamma_{\ep, N}(\bk)=1+\cO\big((\ep N)^\gamma\big)$ uniformly w.r.to $\bk\in \IZ_{N}^{2d}$.
Since $\cG_{\ep,N}$ is Hermitian, this yields the estimate \eqref{estimate1}.

The lower bound for the quantum dissipation time then follows from the same
considerations as in the proof of Theorem \ref{weakmix}. $\qquad \blacksquare$

In view of the above Proposition, the only nontrivial information regarding chaoticity of quantum systems and
asymptotics of the dissipation time can be retrieved in appropriate semiclassical limit.
We thus turn now to semiclassical analysis of quantum maps. 

Following the notation introduced in Section \ref{q-algebra}
we denote by $\Pi_{\cI_{N}}$ an orthogonal Galerkin-type projection of $L^2_{0}(\IT^{2d})$
onto its subspace $\cI_{N}$. It is easy to see that the map 
\bean
\sigma_{N}:\cB(\cA^{0}_{N}(\btht)) \ni \cT \mapsto W^{P}\cT Op_{N} \Pi_{\cI_{N}} \in \cB(L^2_{0}(\IT^{2d}))
\eean
defines an isometric embedding of a finite dimensional quantum algebra 
of superoperators $\cB(\cA^{0}_{N}(\btht))$
into infinite dimensional classical one $\cB(L^2_{0}(\IT^{2d}))$. 

It has been shown in \cite{N} (see Lemma 1 and its proof there)
that for any fixed $\ep>0$, the operators $\sigma_{N}(\cT_{\ep,N})$, which are isometric to $\cT_{\ep,N}$,
converge uniformly (i.e. in the norm of $\cB(L^2_{0}(\IT^{2d}))$) as $N\rightarrow \infty$ to the
classical operator $T_{\ep}$. 
This implies in particular that for any fixed $\ep>0$, and $n\in \IN$
the sequence $\sigma_{N}(\cT^n_{\ep,N})$ converges to $T^{n}_{\ep}$ in $\cB(L^2_{0}(\IT^{2d}))$ as 
$N\rightarrow \infty$. 

The above result, valid for arbitrary quantizable canonical toral map and any compact
noise operator, implies that in appropriate semiclassical regime $N\rightarrow \infty$ one recovers
classical behavior of the dissipation time. More precisely we have

\begin{prop}\label{AQC}
For any quantizable, canonical map $F$ on the torus and any noise generating function $g$,
quantum dissipation time coincides asymptotically in sufficiently fast classical limit
$N=N(\ep)\rightarrow \infty$ as $\ep\to 0$ with its classical counterpart.
\end{prop}

\textbf{Proof.}
We have $\|\cT^{n}_{\ep,N}\|=\|\sigma_{N}(\cT^n_{\ep,N})\|\leq
\|\sigma_{N}(\cT^n_{\ep,N})-T^n_{\ep}\|+\|T^n_{\ep}\|$.
Now for given $\ep$ we set $n=\tau_{c}(\ep)$ and choose $N=N(\ep)$
such that $\|\sigma_{N}(\cT^n_{\ep,N})-T^n_{\ep}\|<\ep$.
Thus $\|\cT^{n}_{\ep,N}\|\leq \ep + e^{-1}$.
On the other hand 
\bean
\|T^n_{\ep}\|\leq \|\sigma_{N}(\cT^n_{\ep,N})-T^n_{\ep}\|+ \|\sigma_{N}(\cT^n_{\ep,N})\|
=\|\sigma_{N}(\cT^n_{\ep,N})-T^n_{\ep}\|+\|\cT^{n}_{\ep,N}\|
\eean
Thus for $n=\tau_{c}(\ep)-1$ and we have
$e^{-1}\leq \ep + \|\cT^{n}_{\ep,N}\|$, which together with the previous inequality
establishes the desired result. $\qquad \blacksquare$

The above statement, despite its generality gives, no information about the actual 
behavior of $\tau_{q}$ unless the behavior of the classical one is known. 

The behavior of classical dissipation time of general nonlinear maps
has been analyzed in Part I of this work. In 
particular we can use here our result regarding the logarithmic asymptotics of the dissipation time 
 of Anosov systems (see Theorem \ref{TAnosov}).  Joining this result with the above Proposition we get

\begin{cor}
\label{QTAnosov}
Let $F$ be a volume preserving $C^{3}$ Anosov diffeomorphisms on $\IT^d$ 
and let $g$ be a $C^1$ noise generating function with fast decay at infinity.
Then there exist $A_{1},A_{2}>0$ and $\tilde{C}>0$ such that

I) Quantum dissipation time of the coarse-grained dynamics satisfies in 
sufficiently fast semiclassical and small noise limit $\ep(N) N \rightarrow \infty$
the following estimate
\bean
  A_{1} \ln(\ep^{-1}(N)) -\tilde{C} \leq
  \tilde{\tau}_*(\ep,N) \leq A_{2} \ln(\ep^{-1}(N)) + \tilde{C},
\eean

II) If in addition $F$ has $C^{1+\eta}$-regular foliations  
and $g\in C^2(\IR^{d})$ is compactly 
supported, then in the above specified semiclassical limit the dissipation time of the noisy 
evolution satisfies for some $C>0$
\bean
  A_{1}\ln(\ep^{-1}(N)) - C 
\leq  {\tau}_*(\ep,N) \leq A_{2}\ln(\ep^{-1}) + C
\eean
\end{cor}

One has to note here however that the crucial from the semiclassical point of view question 
about the 
regime i.e. the relation between $N$ and $\ep$ for which the above asymptotics holds remains 
in this general setting open. In the next section we answer this question in full details in 
the case of linear Anosov systems. 

%%%%%%%%%%%%%%%%%%%%%%%%%%%%%%%%%%%%%%%%%%%%%%%%%%%%%%%%%%%%%%%%%%%%%%%%%%%%%%%%%%%%%%%%%%%%%%%%%%%%%%%%%%
%%%%%%%%%%%%%%%%%%%%%%%%%%%%%%%%%%%%%%%%%%%%%%%%%%%%%%%%%%%%%%%%%%%%%%%%%%%%%%%%%%%%%%%%%%%%%%%%%%%%%%%%%%

\subsection{Classical limit for quantum toral symplectomorphisms}
\label{CLTS}

%%%%%%%%%%%%%%%%%%%%%%%%%%%%%%%%%%%%%%%%%%%%%%%%%%%%%%%%%%%%%%%%%%%%%%%%%%%%%%%%%%%%%%%%%%%%%%%%%%%%%%%%%%
%%%%%%%%%%%%%%%%%%%%%%%%%%%%%%%%%%%%%%%%%%%%%%%%%%%%%%%%%%%%%%%%%%%%%%%%%%%%%%%%%%%%%%%%%%%%%%%%%%%%%%%%%%

In this section we analyze linear maps projected on the torus (generalized cat maps). 
In this case all computations can be carried out explicitly. 

To focus attention and avoid unnecessary notational and computational complications
we restrict the considerations of this subsection to Gaussian noises (the
assumption follows the made in classical case, where only $\alpha$-stable laws
were considered). 
Thus we set $\hat{g}(\bk)=e^{-|\bk|^{2}}$.
 
It is easy to see that, as in classical case, for any $A\in \cA^{0}_{N}(\btht)$ 
and any symplectic toral automorphism
\bean
\cT_{\ep,N}^{n}A=\sum_{0\not=\bk\in \IZ^{2d}_{N}} a_{k}\prod_{l=1}^{n}
\gamma_{\ep N}((F^{-l}\bk)^{\perp}) W_{F^{-n}\bk}
\eean
This yields
\bea
\label{nd}
\|\cT_{\ep,N}^{n}\|=\max_{0\not=\bk\in \IZ^{2d}_{N}}
\prod_{l=1}^{n}\gamma_{\ep N}(F^{-l}\bk)=
\max_{0\not=\bk\in \IZ^{2d}_{N}}
\prod_{l=1}^{n}\gamma_{\ep N}(F^{l}\bk),
\eea
In the above computations we have used the fact 
that the dissipation time does not depend on the direction of time.
Similarly in coarse grained case we get
\bea
\label{cnd}
\|\tilde{\cT}_{\ep,N}^{(n)}\|=\max_{0\not=\bk\in \IZ^{2d}_{N}}
\gamma_{\ep N}(\bk)\gamma_{\ep N}(F^{n}\bk).
\eea
We are now in a position to state the main theorem of the present section.

\begin{thm}
\label{QDTthm}
For arbitrary symplectic, ergodic and diagonalizable $F\in SL(2d,\IZ)$ and Gaussian noise, 
one has the following estimates

I)
\bean
\tau_{q}(\ep,N) \geq \tau_{c}(\ep),\qquad
\tau_{q}(\ep,N) \geq \frac{2}{d} \frac{1}{(N\ep)^{2}},\qquad
\tilde{\tau}_{q}(\ep,N) \geq \tilde{\tau}_{c}(\ep),
\eean
uniformly in $N$.

II) There exists $M>0$ (cf. \ref{M}) such that
\bean
\tau_{q}(\ep,N) \approx \tau_{c}(\ep) \approx \frac{1}{\hat{h}(F)}\ln(\ep^{-1})
, \qquad \ep \rightarrow 0\, , \ep N > M,
\eean
Moreover for any $\beta$ satisfying
\bea
\label{beta}
\beta>\frac{\ln\|F\|}{\hat{h}(F)}+1
\eea
one has 
\bean
\tilde{\tau}_{q}(\ep,N) &\approx& \tilde{\tau}_{c}(\ep)\approx \frac{1}{\hat{h}(F)}\ln(\ep^{-1})
, \qquad \ep \rightarrow 0\, ,\ep^{\beta}N>1,
\eean
where $\hat{h}(F)$ is a constant equal to minimal dimensionally averaged K-S entropy 
of $F$ (cf. \cite{FW}).

III) In the limit $\ep N \rightarrow 0$, the 
quantum coarse-grained dynamics does not undergo dissipation i.e. 
$\tau_{q}(\ep,N)= \infty$. The noisy dynamics undergoes slow (power-law) dissipation. 
\end{thm}

As a direct corollary of the above theorem we get the following relation between 
spatial (small $\ep$ and small $\hbar$) and time (dissipation and Ehrenfest)
scales for classical-quantum correspondence of noisy quantum dynamics in case of
toral automorphisms. 

\begin{cor}[Dissipation vs. Ehrenfest times]
\label{QC} $ $

Under the assumptions of Theorem \ref{QDTthm} the following relations hold 

I)
\bean
 \ep\hbar^{-1} \rightarrow \infty  \Rightarrow \tau_{q}(\ep,N) \lesssim \tau_{E},
\eean

II) There exists $M>0$ (see \ref{M}) such that
\bean   
\ep\hbar^{-1} \sim C>M \Rightarrow \tau_{q}(\ep,N) \sim \tau_{E},
\eean 

III) There exists $M_{0}\geq d^{-1/2}$ such that
\bean
\ep\hbar^{-1} \sim C< M_{0},\quad  \tilde{\tau}_{*}(\ep,N)= \infty,
\eean 

IV) If $N \ep (\ln(\ep^{-1}))^{1/2} \ll 1$ and the classical dynamics
has logarithmic dissipation time (i.e. $F$ is ergodic) then 
\bean
\tau_{q}(\ep,N) \gg \tau_{c}(\ep) \gg \tau_{E}.
\eean
\end{cor}
\textbf{Proof of Corollary.}
Statements I) and II) follow immediately from statement II) of the above
theorem (and its proof). Statement III) is a direct consequence of the estimate 
\bean
\|\tilde{\cT}_{\ep,N}^{(n)}\|\geq e^{-d(\ep N)^2} \geq e^{-1},
\eean
which holds true (for all $n$) whenever $\ep N<M_{0}:=d^{-1/2}$.

The last statement follows from the following estimates.
First we note that
\bean
\tau_{q}(\ep,N)\geq \frac{2}{d} \frac{1}{(N\ep)^{2}} \gg \ln(\ep^{-1}) \sim \tau_{c}(\ep)
\eean
But we also have
\bean
\tau_{c}(\ep) \sim \ln(\ep^{-1}) \geq \ln(\ep^{-1}) - \frac{1}{2}\ln\ln(\ep^{-1})
\gg \ln N \sim \tau_{E}.
\eean
This completes the proof of the corollary   
$\qquad \blacksquare$

To prove the theorem we will need the following estimate (for the proof see Appendix \ref{est}).
\begin{lem} 
\label{GES} $ $

For any $\bxi\in \IR^{2d}$ denote by $\tilde{\bxi}$ the unique vector
satisfying $\tilde{\bxi}_{j}\in (-1/2,1/2]$ for all $j=1,...,2d$ and
$\tilde{\bxi}=\bxi \mod 1$. 
There exists a constant $C>0$ such that for all $\sigma>0$ and all $\bxi\in \IR^{2d}$, 
\bea
\label{pGauss}
\hat{g}_{\sigma} (\bxi) \leq \hat{g}_{\sigma} (\tilde{\bxi}) \leq 
\gamma_{\sigma}(N\bxi) \leq 
\frac{\hat{g}_{\sigma}(\tilde{\bxi})}{\tilde{\hat{g}}_{\sigma}(0)}
+Ce^{-\frac{1}{4}\sigma^2}
\leq
\hat{g}_{\sigma}(\tilde{\bxi})+Ce^{-\frac{1}{4}\sigma^2}
\eea
\end{lem}

\textbf{Proof of Theorem.}

We start with the proofs of statements I) and III).
The proof that quantum dissipation time is never shorter than classical one
 follows from (\ref{nd}) (\ref{cnd}), Lemma \ref{GES} and the
fact that $\hat{g}_{\ep}(\bxi)\leq 1$. Indeed, we have
\bean
\|\cT_{\ep,N}^{n}\|=\max_{0\not=\bk\in \IZ^{2d}_{N}}
\prod_{l=1}^{n}\gamma_{\ep N}(F^{l}\bk)
\geq \sup_{0\not=\bk\in \IZ^{2d}}
\prod_{l=1}^{n}\hat{g}_{\ep}(F^{l}\bk)=
\|T_{\ep}^{n}\|.
\eean
Similar estimation holds in coarse-grained case.
To show that the remaining assertion in I) and that III) hold
we first consider coarse-grained version.
Denoting $\bxi_{0}:=N^{-1}\bk$ and 
$\bxi_{n}:=N^{-1}F^{n}\bk$, we have
for any $\bk\in \IZ^{2d}_{N}$,
\bean
\|\tilde{\cT}_{\ep,N}^{(n)}\| &\geq&   
\gamma_{\ep N}(\bk)\gamma_{\ep N}(F^{n}\bk)
\geq \hat{g}_{\ep N} (\tilde{\bxi_{0}}) \hat{g}_{\ep N} (\tilde{\bxi_{n}})\\
&=& e^{-(\ep N)^2(|\tilde{\bxi_{0}}|^2+|\tilde{\bxi_{n}}|^2)}
\geq e^{-d(\ep N)^2}. 
\eean
And obviously $e^{-d(\ep N)^2}\rightarrow 1$ as $\ep N\rightarrow 0$.

In noisy case, continuing the notation $\bxi_{l}:=N^{-1}F^{l}\bk$, we have
\bean
\|\cT_{\ep,N}^{n}\| \geq   
\prod_{l=1}^{n}\gamma_{\ep N}(F^{l}\bk)
\geq
\prod_{l=1}^{n}\hat{g}_{\ep N} (\tilde{\bxi_{l}})
= e^{-(\ep N)^2\sum_{l=1}^{n}|\tilde{\bxi_{l}}|^2}
\geq
e^{-(\ep N)^2 \frac{dn}{2}}.
\eean
Thus 
\bean
\tau_{q}(\ep,N) \geq \frac{2}{d}\frac{1}{(\ep N)^{2}}.
\eean

Now we pass to the proof of the statement II). 

The lower bounds for both noisy and coarse-grained versions 
follow from the general estimate established in point I) and results obtained in classical
setting. 

We turn now to upper bound computations.
First we consider coarse-grained version.
 
In view of (\ref{cnd}) we have to estimate from above the following product
\bea
\label{prod}
\|\tilde{\cT}_{\ep,N}^{(n)}\|=\max_{0\not=\bk\in \IZ^{2d}_{N}}
\gamma_{\ep N}(\bk)\gamma_{\ep N}(F^{n}\bk).
\eea

Given $\beta$ satisfying (\ref{beta}) we fix $\del_{\beta}>0$. 
According to the assumption that $\ep^{\beta} N >1$, for all
$0<\del<\del_{\beta}$, all sufficiently small $\ep>0$ and all sufficiently big $N$ there
exists $n\in \IN$ such that
\bea
\label{Et}
\frac{1}{(1-\del)\hat{h}(F)}\ln(\ep^{-1})< n <\frac{1}{(1-\del)\hat{h}(F)\ln\|F\|}\ln (N/2).
\eea

Now, given $\bk_{0}\in \IZ^{2d}_{N}$ we consider two cases

a) $\bk_{0}$ generates classical orbit i.e. $\bk_{0},F^{n}\bk_{0}\in \IZ^{2d}_{N}$. 
Then in view of Theorem \ref{thmart}, for any $\del'<\del$, for sufficiently small $\ep$ and 
any $n$ satisfying (\ref{Et}) we have
\bea
\label{classmin}
 |\bk_{0}|^{2}+|F^{n}\bk_{0}|^{2} \geq 
 \min_{0 \not = \bk \in \IZ^{2d}}(|\bk|^{2}+|F^{n}\bk|^{2})>e^{2(1-\del')\hat{h}(F)n}
>e^{2(1-\del)\hat{h}(F)n}
\eea
Thus for any such $\bk_{0}$, there exists $l\in\{0,n\}$ such that
\bean
|F^{l}\bk_{0}|>e^{(1-\del)\hat{h}(F)n}.
\eean
Using (\ref{pGauss}) we arrive at the following upper bound
\bean
\gamma_{\ep N}(\bk)\gamma_{\ep N}(F^{n}\bk_{0})<
\gamma_{\ep N}(F^{l}\bk_{0})
\leq e^{-\ep^2 e^{2(1-\del)\hat{h}(F)n}}+Ce^{-\frac{1}{4}(\ep N)^2} < e^{-1},
\eean
where the last inequality holds for sufficiently big $\ep N$.

b) $\bk_{0}$ generates non-classical orbit i.e. $F^{n}\bk_{0} \not \in \IZ^{2d}_{N}$. 
In this case, $|\bk_{0}|>e^{(1-\del)\hat{h}(F)n}$.
Indeed otherwise, in view of RHS of (\ref{Et}) we would have
\bean
|F^{n}\bk_{0}|\leq \|F^{n}\|e^{(1-\del)\hat{h}(F)n} \leq \frac{N}{2},
\eean 
which would imply $F^{n}\bk \in \IZ^{2d}_{N}$.

Thus similarly as in previous case we have for sufficiently big $\ep N$,  
\bean
\gamma_{\ep N}(\bk_{0})\gamma_{\ep N}(F^{n}\bk_{0})<
\gamma_{\ep N}(\bk_{0})
< e^{-\ep^2 e^{2(1-\del)\hat{h}(F)n}}+Ce^{-\frac{1}{4}(\ep N)^2} < e^{-1}.
\eean
The above two cases exhaust all possible values of $\bk_{0}$. We then
conclude that
\bean
\|\tilde{\cT}_{\ep,N}^{(n)}\|<e^{-1},
\eean
which in view of the definition of dissipation time completes the proof 
in coarse-grained case.

Now we consider fully noisy case i.e. we need to estimate from above the following product
\bea
\|\cT_{\ep,N}^{(n)}\|=\max_{0\not=\bk\in \IZ^{2d}_{N}}
\prod_{l=1}^{n}\gamma_{\ep N}(F^{l}\bk).
\eea
Let $C$ denote the constant of the RHS of (\ref{pGauss}) and define
\bea
\label{M}
M:=\max \left\{4e^{\hat{h}(F)}\|F\|, \sqrt{4\ln\lp \frac{C}{e^{-1}-e^{-e^{\hat{h}(F)}}}\rp} \right\}.
\eea
We fix $0<\del<1/2$. Using the fact that $\ep N > M$ we obtain for every $\ep>0$ the existence 
of $n\in \IN$ such that
\bea
\label{Et1}
\frac{1}{(1-\del)\hat{h}(F)}\ln(\ep^{-1})< n<n+1 
<\frac{1}{(1-\del)\hat{h}(F)}\ln \lp \frac{N}{2\|F\|}\rp.
\eea

Now, once again, given $\bk_{0}\in \IZ^{2d}_{N}$ we consider two cases.

a) $\bk_{0}$ generates classical orbit i.e. $F^{l}\bk_{0}\in \IZ^{2d}_{N}$ for $l=1,...,n$. 
Then in view of Theorem \ref{thmart}, for any  $0<\del'<\del$, all sufficiently 
small $\ep$, and corresponding $n$ satisfying (\ref{Et1}) we have
\bea
\label{cm}
\sum_{l=1}^{n}|F^{l}\bk_{0}|^{2} \geq
\min_{0 \not = \bk \in \IZ^{2d}} \sum_{l=1}^{n}|F^{l}\bk|^{2}>
e^{2(1-\del')\hat{h}(F)(n+1)}>e^{2(1-\del)\hat{h}(F)(n+1)}.
\eea
Thus for any such $\bk_{0}$, there exists $l_{0}\in\{1,...,n\}$ such that
\bean
|F^{l_{0}}\bk_{0}|>e^{(1-\del)\hat{h}(F)(n+1)}.
\eean

b) $\bk_{0}$ generates non-classical orbit i.e. there exists $1 \leq l\leq n$ 
such that $F^{l}\bk_{0} \not \in \IZ^{2d}_{N}$. Let $l_{0}$ be the largest exponent
such that $F^{l_{0}}\bk_{0} \in \IZ^{2d}_{N}$ but $F^{l_{0}+1}\bk_{0} \not \in \IZ^{2d}_{N}$.
Then
\bea
\label{ncm}
|F^{l_{0}}\bk_{0}|>e^{(1-\del)\hat{h}(F)(n+1)}.
\eea
Indeed, otherwise in view of the RHS of (\ref{Et1}) one would have
\bean
|F^{l_{0}+1}\bk_{0}|\leq \|F\| |F^{l_{0}}\bk_{0}| \leq \|F\|e^{(1-\del)\hat{h}(F)(n+1)}
<\frac{N}{2},
\eean
which would imply $F^{l_{0}+1}\bk_{0} \in \IZ^{2d}_{N}$.

Thus in both cases, using (\ref{pGauss}) and (\ref{M}), we get
\bean
\prod_{l=1}^{n}\gamma_{\ep N}(F^{l}\bk) &<& \gamma_{\ep N}(F^{l_{0}}\bk)
\leq e^{-\ep^2 e^{2(1-\del)\hat{h}(F)(n+1)}}+ Ce^{-\frac{1}{4}(\ep N)^2} \\
&<& e^{-e^{\hat{h}(F)}}+Ce^{-\frac{M^2}{4}}<e^{-1}.
\eean
This completes the proof of statement II) and the whole theorem.
$\qquad \blacksquare$

%%%%%%%%%%%%%%%%%%%%%%%%%%%%%%%%%%%%%%%%%%%%%%%%%%%%%%%%%%%%%%%%%%%%%%%%%%%%%%%%%%%%%%%%%
%%%%%%%%%%%%%%%%%%%%%%%%%%%%%%%%%%%%%%%%%%%%%%%%%%%%%%%%%%%%%%%%%%%%%%%%%%%%%%%%%%%%%%%%%
%%%%%%%%%%%%%%%%%%%%%%%%%%%%%%%%%%%%%%%%%%%%%%%%%%%%%%%%%%%%%%%%%%%%%%%%%%%%%%%%%%%%%%%%%

\section{Technical proofs}
\label{Ch5proofs}

\subsection*{Proof of Proposition \ref{HBq}}
In the proof we follow the approach presented in \cite{RSO}.

Since we already know that symplecticity is necessary for quantization it is enough to
prove that for given $h=N^{-1}$ and $\btht$ the condition (\ref{qbc}) is satisfied 
iff $\cU_{F}$ is a $^{*}$-automorphism of the algebra $\cA^{I}_{N}(\btht)$ which in
view of (\ref{alqu}) and assumed symplecticity is equivalent to
\bea
\label{qbcr}
\frac{N}{4} F'\bm\vee F'\bm + F'\bm\wedge\btht=\frac{N}{4} \bm\vee\bm + \bm\wedge\btht \mod 1.
\eea
If $N$ is even then for all $\bm$, $\frac{N}{4} \bm\vee\bm \in \IZ$ and hence the above
condition is trivially satisfied for $\btht=0$.
If $N$ is odd we first note that due to the following identities
\bean
F'\bm\vee F'\bm&=&(J_{-}F^{\dagger}J_{-}^{\dagger})\bm J_{+}J_{-}F^{\dagger}J_{-}^{\dagger}\bm\\
&=&\bm J_{-}FJ_{-}J_{+}J_{-}F^{\dagger}J_{-}\bm=\bm J_{-}FJ_{+}F^{\dagger}J_{-}\bm \\
 F'\bm\wedge\btht&=& \bm\wedge F\btht,
\eean
condition (\ref{qbcr}) can be rewritten as 
\bea
\label{qbcr1}
\frac{N}{4} \bm\vee\bm + \frac{N}{4}\bm J_{-}FJ_{+}F^{\dagger}J_{-}\bm + \bm\wedge F\btht=
 \bm\wedge\btht \mod 1
\eea
To see that (\ref{qbcr}) implies (\ref{qbc}) we take $\bm=\be_{j}$ ($j \in \{1,...,2d\}$),
where $\be_{j}$ denote the standard basis vectors,
use the fact that $\be_{j}\vee \be_{j}=0$ and get
\bea
\label{qbcr2}
\frac{N}{4}\be_{j} J_{-}FJ_{+}F^{\dagger}J_{-}\be_{j} + \be_{j}\wedge F\btht
=\be_{j}\wedge\btht \mod 1.
\eea
Now we note that
\bean
\frac{N}{4} \be_{j} J_{-}FJ_{+}F^{\dagger}J_{-}\be_{j}&=&
 \frac{N}{4}\be_{j}
\begin{bmatrix}
-CD^{\dagger}-DC^{\dagger} & CB^{\dagger}+DA^{\dagger} \\
-AD^{\dagger}-BC^{\dagger} & -AB^{\dagger}+BA^{\dagger} 
\end{bmatrix}\be_{j}\\
&=& -\frac{N}{2}
\begin{pmatrix}
 C\cdot D \\
 A\cdot B
\end{pmatrix}
\cdot \be_{j}
\eean
and
\bean
-\frac{N}{2}
\begin{pmatrix}
 C\cdot D \\
 A\cdot B
\end{pmatrix}
\cdot \be_{j}=
\frac{N}{2} \be_{j}\wedge
\begin{pmatrix}
 A\cdot B \\
 C\cdot D
\end{pmatrix} \mod 1.
\eean
Substituting the above expression in (\ref{qbcr2}) immediately yields (\ref{qbc}).

Now we sketch the proof of the opposite implication.
First we rearrange the LHS of (\ref{qbcr1})
\bean
 &&\frac{N}{4}\sum_{i,j=1}^{2d}m_{i}m_{j} \be_{i}J_{+}\be_{j} 
+\frac{N}{4}\sum_{i,j=1}^{2d}m_{i}m_{j} \be_{i} J_{-}FJ_{+}F^{\dagger}J_{-}\be_{j} 
+ \sum_{i=1}^{2d} m_{i}\be_{i}\wedge F\btht\\
&=&
\frac{N}{4}\sum_{i\not = j=1}^{2d}m_{i}m_{j} \be_{i} (J_{-}FJ_{+}F^{\dagger}J_{-}+J_{+})\be_{j} \\
&+&\frac{N}{4}\sum_{i=1}^{2d}m^{2}_{i} \be_{i} J_{-}FJ_{+}F^{\dagger}J_{-}\be_{i}
+ \sum_{i=1}^{2d} m_{i}\be_{i}\wedge F\btht
\eean
Now we note that due to the symplecticity of $F$
\bean
\frac{N}{4}\sum_{i\not = j=1}^{2d}m_{i}m_{j} \be_{i} J_{-}FJ_{+}F^{\dagger}J_{-}\be_{j}=
\frac{N}{4}\sum_{i\not = j=1}^{2d}m_{i}m_{j} \be_{i} J_{+}\be_{j} \mod 1
\eean
Thus it is enough to use the following obvious identity
\bean
\sum_{i=1}^{2d}m^{2}_{i} \be_{i} J_{-}FJ_{+}F^{\dagger}J_{-}\be_{i}=
\sum_{i=1}^{2d}m_{i} \be_{i} J_{-}FJ_{+}F^{\dagger}J_{-}\be_{i} \mod 2,
\eean
to conclude in view of the first part of the proof the equivalence between (\ref{qbcr})
and (\ref{qbc}). $\qquad \blacksquare$.

\subsection*{Proof of Proposition \ref{inner}.} 

Denote by $\{\be_{j}\}$ ($j=1,...,N$) the 
standard basis of $\IR^N$ and let $E_{jk}:=|\be_{j}\ra\la\be_{k}|$,
$F_{jk}:=\Gamma(E_{jk})$. $\Gamma$ is inner if there exists $U\in \cM_{N}$ such that
$
F_{jk}=U^{*}|\be_{j}\ra\la\be_{k}|U=|U^{*}\be_{j}\ra\la U^{*}\be_{k}|
=|\fb_{j}\ra\la\fb_{k}|
$,
where $\fb_{j}=U^{*}\be_{j}$.
It is then enough to show that there exists an orthonormal basis 
$\{\fb_{j}\}$ such that 
$
F_{jk}=|\fb_{j}\ra\la\fb_{k}|
$. 
First we note that
$F_{jj}=\Gamma(E_{jj})=\Gamma(E_{jj}^*)=F_{jj}^*$,
$F_{jj}=\Gamma(E_{jj})=\Gamma(E_{jj}^2)=F_{jj}^2$ and 
$F_{jj}F_{kk}=\Gamma(E_{jj}E_{kk})=\del_{jk}F_{jj}$,
which implies that $F_{jj}$ form a set of N mutually
orthogonal rank-1 projections. Thus there exists an 
orthonormal basis $\{\fb_{j}\}$ such that 
$F_{jj}=|\fb_{j}\ra\la\fb_{j}|$.
Moreover since $F_{jk}=F_{jj}F_{jk}F_{kk}$ one also
finds that $F_{jk}=|\fb_{j}\ra\la\fb_{k}|$.
$\qquad \blacksquare$.

\subsection*{Proof of Proposition \ref{qnoise}}
\label{pqnoise}
Since by definition $\cG_{\ep,N}$ is already given in Kraus form, it is completely
positive \cite{CN,AF}. Trace preservation follows from the normalization. 
The value of the normalization constant can be found as follows
\bean
Z=\sum_{\bk\in \IZ^{2d}_{N}} \tilde{g}_{\ep}(N^{-1}\bk)=
\sum_{\bk\in \IZ^{2d}} g_{\ep}(N^{-1}\bk)
=N^{2d}\sum_{\bk\in\IZ^{2d}}g_{\ep N}(\bk)= N^{2d}\tilde{g}_{\ep N}(0).
\eean
Now using periodicity of $ad(W_{\bk}$ we get
\bean
\cG_{\ep,N}&=& 
\frac{1}{Z}\sum_{\bk\in \IZ^{2d}_{N}}\tilde{g}_{\ep} \lp\frac{\bk}{N}\rp ad(W_{\bk})=
\frac{1}{N^{2d}\tilde{g}_{\ep N}(0)}\sum_{\bk\in \IZ^{2d}}g_{\ep} \lp\frac{\bk}{N}\rp ad(W_{\bk})\\
&=&\frac{1}{\tilde{g}_{\ep N}(0)}\sum_{\bk\in\IZ^{2d}}g_{\ep N}(\bk)ad(W_{\bk}).
\eean
$\cG_{\ep,N}$ can thus be diagonalized in the basis $\{W_{\bk}\}$
\bean
\cG_{\ep,N}W_{\bk_{0}}
=\frac{1}{\tilde{g}_{\ep N}(0)}\sum_{\bk\in\IZ^{2d}}g_{\ep N}(\bk)
e^{2\pi i N^{-1} \bk_{0}\wedge \bk}W_{\bk_{0}}
=\frac{\tilde{\hat{g}}_{\ep N}(N^{-1}\bk_{0}^{\perp})}
{\tilde{\hat{g}}_{\ep N}(0)}W_{\bk_{0}},
\eean
where in the last inequality we used Poisson summation formula and
symmetricity of $g_{\ep N}(\bk)$.

Hence for any $\bk \in \IZ^{2d}$, $
\cG_{\ep,N}W_{\bk}= \gamma_{\ep N}(\bk^{\perp})W_{\bk}$,
which yields spectral representation of $\cG_{\ep,N}$.
$\qquad \blacksquare$

\subsection*{Proof of Lemma \ref{GES}}
%\label{est}
Let $\bk_{1/2}=(1/2,0,...,0)\in \IR^{2d}$. There exists a constant $C$ depending only on $d$ 
such that
\bean
\tilde{\hat{g}}_{\sigma}(\tilde{\bxi})
&\leq& \hat{g}_{\sigma}(\tilde{\bxi})+ 
C\tilde{\hat{g}}_{\sigma}(\bk_{1/2})
=\hat{g}_{\sigma}(\tilde{\bxi})+ C\sum_{\bn\in\IZ^{2d}}e^{-\sigma^2|\bk_{1/2}+\bn|^{2}}\\
&=&
\hat{g}_{\sigma}(\tilde{\bxi})+ Ce^{-\frac{\sigma^2}{4}}
\sum_{n\in\IZ}e^{-\sigma^2 n(n+1)}
\sum_{\bm\in\IZ^{2d-1}}e^{-\sigma^2|\bm|^2}.
\eean
Now
\bean
\sum_{n\in\IZ}e^{-\sigma^2 n(n+1)}=2\sum_{n\geq 0}e^{-\sigma^2 n(n+1)}\leq
2\sum_{n\geq 0}e^{-\sigma^2 n^2}=\sum_{n\in\IZ}e^{-\sigma^2 n^2}+1
\leq 2\sum_{n\in\IZ}e^{-\sigma^2 n^2}.
\eean
Thus
\bean
\tilde{\hat{g}}_{\sigma}(\tilde{\bxi})&\leq& 
\hat{g}_{\sigma}(\tilde{\bxi})+ 2Ce^{-\frac{\sigma^2}{4}}
\sum_{\bn\in\IZ^{2d}}e^{-\sigma^2|\bn|^2}
= \hat{g}(\sigma\tilde{\bxi})+ 2C\tilde{\hat{g}}_{\sigma}(0)e^{-\frac{\sigma^2}{4}}
\eean
and finally
\bean
\gamma_{\sigma}(N\bxi)=
\frac{\tilde{\hat{g}}_{\sigma}(\tilde{\bxi})}{\tilde{\hat{g}}_{\sigma}(0)}
\leq
\frac{\hat{g}_{\sigma}(\tilde{\bxi})}{\tilde{\hat{g}}_{\sigma}(0)}
+2Ce^{-\frac{1}{4}\sigma^2}
\leq
\hat{g}_{\sigma}(\tilde{\bxi})+2Ce^{-\frac{1}{4}\sigma^2}.
\eean

To prove the other estimate we consider a splitting of the lattice $\IZ^{2d}$ into 
three pairwise disjoint sets 
$\IZ^{2d}_{-}, \{0\},\IZ^{2d}_{+}$ such that $\bn\in\IZ^{2d}_{+}$ iff $-\bn\in\IZ^{2d}_{-}$.
With this notation we have
\bean
\tilde{\hat{g}}_{\sigma}(\bxi)&=&\tilde{\hat{g}}_{\sigma}(\tilde{\bxi})=
\sum_{\bn \in \IZ^{2d}}e^{-\sigma^2|\tilde{\bxi}+\bn|^2}
= e^{-\sigma^2|\tilde{\bxi}|^2}\sum_{\bn \in \IZ^{2d}}
e^{-\sigma^2(|\tilde{\bxi}+\bn|^2-|\tilde{\bxi}|^2)}\\
&=& e^{-\sigma^2|\tilde{\bxi}|^2}\sum_{\bn \in \IZ^{2d}}
e^{-\sigma^2|\bn|^2}e^{-2\sigma^2\tilde{\bxi}\cdot\bn}\\
&=& e^{-\sigma^2|\tilde{\bxi}|^2}\lp 1+ \sum_{\bn \in \IZ_{+}^{2d}}
e^{-\sigma^2|\bn|^2} \lp e^{-2\sigma^2\tilde{\bxi}\cdot\bn}+e^{2\sigma^2\tilde{\bxi}\cdot\bn}\rp\rp.
\eean
Now using the fact that $e^{-2\sigma^2\tilde{\bxi}\cdot\bn}+e^{2\sigma^2\tilde{\bxi}\cdot\bn}\geq 2$
we get
\bean
\tilde{\hat{g}}_{\sigma}(\bxi)\geq 
e^{-\sigma^2|\tilde{\bxi}|^2}\sum_{\bn \in \IZ^{2d}}e^{-\sigma^2|\bn|^2}
=\hat{g}_{\sigma}(\tilde{\bxi})\tilde{\hat{g}}_{\sigma}(0).
\eean
Thus in particular
$\gamma_{\ep N}(N\bxi) \geq \hat{g}_{\ep N}(\tilde{\bxi})$. $\qquad \blacksquare$

%%%%%%%%%%%%%%%%%%%%%%%%%%%%%%%%%%%%%%%%%%%%%%%%%%%%%%%%%%%%%%%%%%%%%%%%%%%%%%%%%%%%%%%%%%%%%%%%%%%%%%%%%%%
%%%%%%%%%%%%%%%%%%%%%%%%%%%%%%%%%%%%%%%%%%%%%%%%%%%%%%%%%%%%%%%%%%%%%%%%%%%%%%%%%%%%%%%%%%%%%%%%%%%%%%%%%%%
%   E N D    O F    T H E     M A I N     B O D Y     O F     T H E      D I S E R T A T I O N            % 
%%%%%%%%%%%%%%%%%%%%%%%%%%%%%%%%%%%%%%%%%%%%%%%%%%%%%%%%%%%%%%%%%%%%%%%%%%%%%%%%%%%%%%%%%%%%%%%%%%%%%%%%%%%
%%%%%%%%%%%%%%%%%%%%%%%%%%%%%%%%%%%%%%%%%%%%%%%%%%%%%%%%%%%%%%%%%%%%%%%%%%%%%%%%%%%%%%%%%%%%%%%%%%%%%%%%%%%

%%%%%%%%%%%%%%%%%%%%%%%%%%%%%%%%%%%%%%%%%%%%%%%%%%%%%%%%%%%%%%%%%%%%%%%%%%%%%%%%%%%%%%%%%%%%%%%%%%%%%%%%%%%

\appendix

%%%%%%%%%%%%%%%%%%%%%%%%%%%%%%%%%%%%%%%%%%%%%%%%%%%%%%%%%%%%%%%%%%%%%%%%%%%%%%%%%%%%%%%%%%%%%%%%%%%%%%%%%%%

%%%%%%%%%%%%%%%%%%%%%%%%%%%%%%%%%%%%%%%%%%%%%%%%%%%%%%%%%%%%%%%%%%%%%%%%%%%%%%%%%%%%%%%%%%%%%%%%%%%%%%%%%%%
%%%%%%%%%%%%%%%%%%%%%%%%%%%%%%%%%%%%%%%%%%%%%%%%%%%%%%%%%%%%%%%%%%%%%%%%%%%%%%%%%%%%%%%%%%%%%%%%%%%%%%%%%%%
%%%%%%%%%%%%%%%%%%%%%%%%%%%%%%%%%%%%%%%%%%%%%%%%%%%%%%%%%%%%%%%%%%%%%%%%%%%%%%%%%%%%%%%%%%%%%%%%%%%%%%%%%%%

\renewcommand{\chaptermark}[1]{\markboth{ \rm \normalsize Appendix \thechapter . \hspace{0.25cm} #1 }{}}

\newpage
\markright{  \rm \normalsize Appendix A. \hspace{0.25cm}
The dynamics of cat maps}
\chapter{The dynamics of cat maps}
\label{A}
\thispagestyle{fancy} 

\lhead[\thepage]{\rightmark}
\rhead[\leftmark]{\thepage}
\cfoot{}

%%%%%%%%%%%%%%%%%%%%%%%%%%%%%%%%%%%%%%%%%%%%%%%%%%%%%%%%%%%%%%%%%%%%%%%%%%%%%%%%%%%%%%%%%%%%%%%%%%%%%%%%%%%
%%%%%%%%%%%%%%%%%%%%%%%%%%%%%%%%%%%%%%%%%%%%%%%%%%%%%%%%%%%%%%%%%%%%%%%%%%%%%%%%%%%%%%%%%%%%%%%%%%%%%%%%%%%
%%%%%%%%%%%%%%%%%%%%%%%%%%%%%%%%%%%%%%%%%%%%%%%%%%%%%%%%%%%%%%%%%%%%%%%%%%%%%%%%%%%%%%%%%%%%%%%%%%%%%%%%%%%

For completeness we briefly recall in this appendix the most important
facts regarding classical cat map dynamics on the plane and its canonical
quantization. For simplicity we present the material in $2$-dimensional setting.

%%%%%%%%%%%%%%%%%%%%%%%%%%%%%%%%%%%%%%%%%%%%%%%%%%%%%%%%%%%%%%%%%%%%%%%%%%%%%%%%%%%%%%%%%%%%%%%%%%%%%%%%%%%
%%%%%%%%%%%%%%%%%%%%%%%%%%%%%%%%%%%%%%%%%%%%%%%%%%%%%%%%%%%%%%%%%%%%%%%%%%%%%%%%%%%%%%%%%%%%%%%%%%%%%%%%%%%

\section{Classical dynamics of cat maps}
\label{CatPlane}

%%%%%%%%%%%%%%%%%%%%%%%%%%%%%%%%%%%%%%%%%%%%%%%%%%%%%%%%%%%%%%%%%%%%%%%%%%%%%%%%%%%%%%%%%%%%%%%%%%%%%%%%%%%
%%%%%%%%%%%%%%%%%%%%%%%%%%%%%%%%%%%%%%%%%%%%%%%%%%%%%%%%%%%%%%%%%%%%%%%%%%%%%%%%%%%%%%%%%%%%%%%%%%%%%%%%%%%

In this section we describe both discrete and continuous-time classical 
cat map dynamics on the standard phase-plane $\IR^{2}$. In continuous
setting we derive the Hamiltonian, the Lagrangian, the Euler-Lagrange
equations and the action of the cat map dynamics. This will constitute
the basis for canonical quantization presented in the next section.

%%%%%%%%%%%%%%%%%%%%%%%%%%%%%%%%%%%%%%%%%%%%%%%%%%%%%%%%%%%%%%%%%%%%%%%%%%%%%%%%%%%%%%%%%%%%%%%%%%%%%%%%%%%
\subsection*{Discrete dynamics}
%\label{Disdyn}
%%%%%%%%%%%%%%%%%%%%%%%%%%%%%%%%%%%%%%%%%%%%%%%%%%%%%%%%%%%%%%%%%%%%%%%%%%%%%%%%%%%%%%%%%%%%%%%%%%%%%%%%%%%

The classical mechanics of a cat map with $1$ degree
of freedom is defined on $2$ dimensional Euclidean phase space $\IR^{2}$. 
Any point in the phase space is denoted by $(q,p)$ and represents respectively position and
momentum coordinates. The initial state of the system is denoted by $(q_{0},p_{0})$.
The discrete time dynamics is generated by any ergodic toral automorphism $A\in SL(2,\IZ)$

\bean
\begin{bmatrix}
q_{1} \\
p_{1} \\
\end{bmatrix}
=A
\begin{bmatrix}
q_{0}\\
p_{0}\\
\end{bmatrix}
=\begin{bmatrix}
a & b\\
c & d\\
\end{bmatrix}
\begin{bmatrix}
q_{0}\\
p_{0}\\
\end{bmatrix}.
\eean

In the case of $1$ degree of freedom ($d=1$), considered here, the ergodicity of $A$ is equivalent to its 
hyperbolicity and is determined by the condition $|\Tr A|>2$. For simplicity we will even assume that $A$ 
is positive definite i.e. $\Tr A>2$. 

%%%%%%%%%%%%%%%%%%%%%%%%%%%%%%%%%%%%%%%%%%%%%%%%%%%%%%%%%%%%%%%%%%%%%%%%%%%%%%%%%%%%%%%%%%%%%%%%%%%%%%%%%%%
\subsection*{Continuous dynamics} 
%\label{Condym}
%%%%%%%%%%%%%%%%%%%%%%%%%%%%%%%%%%%%%%%%%%%%%%%%%%%%%%%%%%%%%%%%%%%%%%%%%%%%%%%%%%%%%%%%%%%%%%%%%%%%%%%%%%%

Continuous version of the dynamics is defined by suspension over the discrete time sequence
\bean
\begin{bmatrix}
q(t) \\
p(t) \\
\end{bmatrix}
=A^{t}
\begin{bmatrix}
q_{0}\\
p_{0}\\
\end{bmatrix}.
\eean
In order to provide an explicit formula for this dynamics we 
need to compute $A^{t}=e^{t\ln A}$. We will use the following lemma
\begin{lem}
Let $A\in \cM_{n}$ and $m_{A}(z)=\sum_{i=1}^{d}(z-\lam_{i})^{m_{i}}$ be a polynomial
which satisfies $m_{A}(A)=0$ ($m_{A}$ can be e.g. the minimal or the characteristic polynomial of $A$). 
There exist matrices $M_{i,j}\in \cM_{n}$ such that for any sufficiently differentiable function $f$
\bean
f(A)=\sum_{i=1}^{d}\sum_{j=o}^{m_{d}-1}f^{(j)}(\lam_{i})M_{i,j}.
\eean
\end{lem}

We will apply the lemma twice. First we apply it to the cat map $A$ and the function $\ln$. 
We denote by $\lam$ the largest eigenvalue of $A$ and take $m_{A}:=(x-\lam)(x-\lam^{-1})$.
Thus for any continuous function f we have $f(A)=f(\lam)M_{+}+f(\lam^{-1})M_{-}$.
To find $M_{\pm}$ we take $f_{\pm}=x-\lam^{\pm 1}$ and get
\bean
A-\lam^{\pm 1} 1\!\! 1 = (\lam^{\mp 1}-\lam^{\pm 1})M_{\mp 1} \Rightarrow 
M_{\pm}=\frac{1}{\lam^{\pm 1}-\lam^{\mp 1}}(A-\lam^{\mp 1} 1\!\! 1)
\eean
hence
\bean
f(A)=\frac{f(\lam)-f(\lam^{-1})}{\lam-\lam^{-1}}A+\frac{f(\lam^{-1})\lam-f(\lam)\lam^{-1}}{\lam-\lam^{-1}} 1\!\! 1.
\eean
In particular
\bea
\label{lnA}
\ln(A)=\frac{\ln\lam}{\lam-\lam^{-1}}(2A-\Tr A 1\!\! 1)=
\frac{\ln\lam}{\lam-\lam^{-1}}
\begin{bmatrix}
a-d & 2b\\
2c & d-a
\end{bmatrix}.
\eea
The prefactor can be written in many different ways, e.g.,
\bean
\frac{\ln\lam}{\lam-\lam^{-1}}=\frac{h(A)}{\lam-\lam^{-1}}=\frac{h(A)}{\sqrt{(\Tr A)^{2}-4}}
=\frac{\sinh^{-1}(\half\sqrt{(\Tr A)^{2}-4})}{\sqrt{(\Tr A)^{2}-4}},
\eean
where $h(A)$ denotes the KS entropy of $A$.

Now we apply the lemma to the matrix $\ln A$,
denoting by $\mu_{1}=\ln \lam =h(A)$ and $\mu_{2}=-\ln \lam =-h(A)$ its eigenvalues. We have 
$m_{\ln A}=(x-h(A))(x+h(A))$ and for any $f$ and 
$f(\ln A)=f(h(A))M_{+}+f(-h(A))M_{-}$.
To find $M_{\pm}$ in this case we take first $f_{\pm}=x \pm h(A)$
\bean
\ln A \pm h(A)=\pm 2 h(A)M_{\pm} \Rightarrow M_{\pm}=\frac{\pm 1}{2h(A)}\ln A + \half  1\!\! 1,
\eean
hence
\bean
f(\ln A)
=\frac{1}{h(A)}\frac{f(h(A))-f(-h(A))}{2}\ln A
+\frac{f(h(A))+f(-h(A))}{2} 1\!\! 1.
\eean

In particular for any even function $f$, $f(\ln A) = f(h(A))1\!\! 1$ 
(e.g. $(\ln A)^{2}=h^{2}(A)1\!\! 1$) and for any odd function $f$, 
$f(\ln A) = \frac{f(h(A))}{h(A)}\ln A$. Taking $f(x)=e^{x}$
we finally get
\bea
\label{At}
A^{t}=\frac{1}{h(A)}\sinh(h(A)t)\ln A+\cosh(h(A)t)1\!\! 1.
\eea
From the above formula we see that $A^{t}$ can
be considered as a solution of a Cauchy problem
for the following matrix-valued ODE
\bean
&&\ddot{X}-h^{2}(A)X=0,\\
&&X(0)=1\!\! 1,\quad \dot{X}(0)=\ln A.
\eean  
The fact that $A^{t}$ indeed satisfies this
equation can also be verified directly using
the identity $(\ln A)^{2}=h^{2}(A)1\!\! 1$ and the obvious fact
that $\ddot{A^{t}}=(\ln A)^{2}A^{t}$.
This equation will appear later as the Euler-Lagrange
equation for this dynamics.

Combining (\ref{At}) with (\ref{lnA}) we arrive at
\bean
A^{t}=\frac{\sinh(h(A)t)}{\sqrt{(\Tr A)^{2}-4}}
\begin{bmatrix}
a-d & 2b\\
2c & d-a
\end{bmatrix}
+\cosh(h(A)t)1\!\! 1.
\eean
Now we write down the explicit
equations of continuous version of cat map dynamics
\bean
q(t)&=&\left(\frac{\sinh(h(A)t)}{\sqrt{(\Tr A)^{2}-4}}(a-d) +\cosh(h(A)t) \right)q(0)+
2b\frac{\sinh(h(A)t)}{\sqrt{(\Tr A)^{2}-4}}p(0)\\
p(t)&=&2c\frac{\sinh(h(A)t)}{\sqrt{(\Tr A)^{2}-4}}q(0)+
\left(\frac{\sinh(h(A)t)}{\sqrt{(\Tr A)^{2}-4}}(d-a) +\cosh(h(A)t) \right)p(0).
\eean

%%%%%%%%%%%%%%%%%%%%%%%%%%%%%%%%%%%%%%%%%%%%%%%%%%%%%%%%%%%%%%%%%%%%%%%%%%%%%%%%%%%%%%%%%%%%%%%%%%%%%%%%%%%
\subsection*{Hamiltonian and Lagrangian}
%%%%%%%%%%%%%%%%%%%%%%%%%%%%%%%%%%%%%%%%%%%%%%%%%%%%%%%%%%%%%%%%%%%%%%%%%%%%%%%%%%%%%%%%%%%%%%%%%%%%%%%%%%%

To find the Hamiltonian of the system we compute 
\bean
\begin{bmatrix}
\dot{q}(t) \\
\dot{p}(t) \\
\end{bmatrix}
=\ln A A^{t}
\begin{bmatrix}
q_{0}\\
p_{0}\\
\end{bmatrix}=
\ln A
\begin{bmatrix}
q(t)\\
p(t)\\
\end{bmatrix}.
\eean
In horizontal vector notation this can be rewritten as
$(\dot{q},\dot{p})=\ln A (q,p)$.

The Hamiltonian $H(p,q)$ satisfies $(\dot{q},\dot{p})=\nabla^{\perp}H(q,p)=(H_{p}(q,p),-H_{q}(q,p))$.
Thus $\nabla^{\perp}H=\ln A$ and
\bean
H=\frac{h(A)}{\sqrt{(\Tr A)^{2}-4}}
\begin{bmatrix}
-c & \half(a-d)\\
\half(a-d) & b
\end{bmatrix},
\eean
which gives
\bea
\label{CatHam}
H(q,p)=\frac{h(A)}{\sqrt{(\Tr A)^{2}-4}}(bp^{2}-cq^{2}+(a-d)pq).
\eea
We note that the Hamiltonian is time independent.

The explicit form of Hamilton equations reads
\bean
\dot{q}(t)&=&\frac{\pa H}{\pa p}=\frac{h(A)}{\sqrt{(\Tr A)^{2}-4}}((a-d)q(t)+ 2bp(t))\\
\dot{p}(t)&=&-\frac{\pa H}{\pa q}=\frac{h(A)}{\sqrt{(\Tr A)^{2}-4}}(2cq(t)+(d-a)p(t)).
\eean
We can now find the Lagrangian using the Legendre transform
\bea
\label{Lag}
L(q,\dot{q})=p\dot{q}-H(q,p)
\eea
We note that since $H$ is time independent, $L$ will not depend explicitly on time either and
we drop the time variable.
The use of the first Hamilton equation, yields
\bean
p=\frac{1}{2b}\left( \frac{\sqrt{(\Tr A)^{2}-4}}{h(A)}\dot{q}-(a-d)q\right).
\eean
Now inserting this formula for $p$ into (\ref{Lag}) and performing necessary
transformations and rearrangements one arrives at
\bean
L(q,\dot{q})=\frac{1}{4b}\left(\sqrt{(\Tr A)^{2}-4}h(A)q^{2}-2(a-d)\dot{q}q
+\sqrt{(\Tr A)^{2}-4}(h(A))^{-1}\dot{q}^{2} \right).
\eean

%%%%%%%%%%%%%%%%%%%%%%%%%%%%%%%%%%%%%%%%%%%%%%%%%%%%%%%%%%%%%%%%%%%%%%%%%%%%%%%%%%%%%%%%%%%%%%%%%%%%%%%%%%%
\subsection*{Euler-Lagrange equation}
%\label{ELEq}
%%%%%%%%%%%%%%%%%%%%%%%%%%%%%%%%%%%%%%%%%%%%%%%%%%%%%%%%%%%%%%%%%%%%%%%%%%%%%%%%%%%%%%%%%%%%%%%%%%%%%%%%%%%

Now we find the Euler-Lagrange equation for cat map dynamics. 
The general form of the Euler-Lagrange equations can be written as
\bean
\frac{d}{dt}\frac{\pa L}{\pa \dot{q}}=\frac{\pa L}{\pa q}.
\eean
Using the explicit formula for $L$ one gets
\bean
\frac{d}{dt}\frac{\pa L}{\pa \dot{q}}&=&\frac{1}{4b}\left(\sqrt{(\Tr A)^{2}-4}(h(A))^{-1}2\ddot{q}
-2(a-d)q\right),\\
\frac{\pa L}{\pa \dot{q}}&=&\frac{1}{4b}\left(\sqrt{(\Tr A)^{2}-4}(h(A))2q
-2(a-d)\dot{q}\right).
\eean
which immediately yields the following Euler-Lagrange equation (Cauchy problem) for our system
\bean
&&\ddot{q}-h^{2}(A)q=0,\\
&&q(0)=q_{0}, \quad \dot{q}(0)=v_{0},
\eean
where $v_{0}$ denotes the initial velocity.

%%%%%%%%%%%%%%%%%%%%%%%%%%%%%%%%%%%%%%%%%%%%%%%%%%%%%%%%%%%%%%%%%%%%%%%%%%%%%%%%%%%%%%%%%%%%%%%%%%%%%%%%%%%
\subsection*{The action}
%\label{Action}
%%%%%%%%%%%%%%%%%%%%%%%%%%%%%%%%%%%%%%%%%%%%%%%%%%%%%%%%%%%%%%%%%%%%%%%%%%%%%%%%%%%%%%%%%%%%%%%%%%%%%%%%%%%

Now we want to compute the action along the classical path of our dynamics.
In terms of Lagrangian the action is defined as
\bea
\label{action}
S(q_{1},q_{0})=S(q(t_{1}),q(t_{0}))=\int_{t_{0}}^{t_{1}}L(q(t),\dot{q}(t))dt.
\eea
Here we derive only equations for action and compute it for particular
choice of time interval, i.e., $t_{0}=0$ and $t_{1}=1$, that is, we find
the action associated with the discrete dynamics.
To derive action equations we use once again the Legendre transform \eqref{Lag}
in conjunction with Euler-Lagrange equations to get
\bean
\frac{\pa L}{ \pa \dot{q}}=p, \quad \frac{\pa L}{ \pa q}=\dot{p}.
\eean
Now from (\ref{action}) we get
\bean
\frac{\pa S}{\pa q_{1}}\dot{q}(t_{1})=\frac{d}{dt_{1}}S(q(t_{1}),q(t_{0}))=L(q(t_{1}),\dot{q}(t_{1})),\\
\frac{\pa S}{\pa q_{0}}\dot{q}(t_{0})=\frac{d}{dt_{0}}S(q(t_{1}),q(t_{0}))=-L(q(t_{0}),\dot{q}(t_{0})).
\eean
Dropping the explicit dependence on time
and taking derivative w.r.t. $\dot{q}$ we obtain
\bean
\frac{\pa S}{\pa q_{1}}=\frac{\pa L(q_{1},\dot{q})}{ \pa \dot{q}}=p_{1},\quad
\frac{\pa S}{\pa q_{0}}=-\frac{\pa L(q_{0},\dot{q})}{ \pa \dot{q}}=-p_{0}.
\eean
Hence the action equations are
\bean
p_{1}=\frac{\pa S}{\pa q_{1}},\quad
p_{0}=-\frac{\pa S}{\pa q_{0}}.
\eean
Using these equations one easily checks that for our discrete dynamics
\bea
\label{Cataction}
S(q_{1},q_{0})=\frac{1}{2b}(aq_{0}^{2}-2q_{0}q_{1}+dq_{1}^{2}).
\eea
Indeed
\bean
\frac{\pa S(q_{1},q_{0})}{\pa q_{1}}&=&\frac{1}{b}(dq_{1}-q_{0})=
\frac{1}{b}(d(aq_{0}+bp_{0})-q_{0})\\
&=&\frac{(da-1)}{b}q_{0}+dp_{0}=cq_{0}+dp_{0}=p_{1},\\
\frac{\pa S(q_{1},q_{0})}{\pa q_{0}}&=&\frac{1}{b}(aq_{0}-q_{1})=
\frac{1}{b}(aq_{0}-aq_{0}-bp_{0})=-p_{0}.
\eean

%%%%%%%%%%%%%%%%%%%%%%%%%%%%%%%%%%%%%%%%%%%%%%%%%%%%%%%%%%%%%%%%%%%%%%%%%%%%%%%%%%%%%%%%%%%%%%%%%%%%%%%%%%%
%%%%%%%%%%%%%%%%%%%%%%%%%%%%%%%%%%%%%%%%%%%%%%%%%%%%%%%%%%%%%%%%%%%%%%%%%%%%%%%%%%%%%%%%%%%%%%%%%%%%%%%%%%%

\section{Canonical quantization of cat maps}
\label{CanQuant}

%%%%%%%%%%%%%%%%%%%%%%%%%%%%%%%%%%%%%%%%%%%%%%%%%%%%%%%%%%%%%%%%%%%%%%%%%%%%%%%%%%%%%%%%%%%%%%%%%%%%%%%%%%%
%%%%%%%%%%%%%%%%%%%%%%%%%%%%%%%%%%%%%%%%%%%%%%%%%%%%%%%%%%%%%%%%%%%%%%%%%%%%%%%%%%%%%%%%%%%%%%%%%%%%%%%%%%%

%%%%%%%%%%%%%%%%%%%%%%%%%%%%%%%%%%%%%%%%%%%%%%%%%%%%%%%%%%%%%%%%%%%%%%%%%%%%%%%%%%%%%%%%%%%%%%%%%%%%%%%%%%%
\subsection*{Notation}
%\label{Not}
%%%%%%%%%%%%%%%%%%%%%%%%%%%%%%%%%%%%%%%%%%%%%%%%%%%%%%%%%%%%%%%%%%%%%%%%%%%%%%%%%%%%%%%%%%%%%%%%%%%%%%%%%%%

Throughout the rest of the appendix the following notation will be used
\bean
\IZ_{N}&=&\IZ\mod N=\{0,1,...,N-1\},\\
\IQ_{N}&=&(\IZ/N)\mod 1=\left\{0,\frac{1}{N},...,\frac{N-1}{N}\right\}.
\eean

Continuous quantum Fourier transform will be denoted by $\ml{F}_{h}$, i.e.
\bea
\label{ACQFT}
\ml{F}_{h}(\psi)(\bp)=
\frac{1}{h^{d/2}}\int_{\IR^{d}}\psi(\bq)e^{-2\pi i\frac{\bq \cdot \bp}{h}}d\bq.
\eea

With this normalization, Dirac delta function (on $\IR^{d}$) and periodic Dirac delta comb (on $\IT^{d}$)
can be written as 
\bean
&&\del_{\ba}(\bq)=\frac{1}{h^{d}}\int_{\IR^{d}}e^{\frac{2\pi i}{h} (\bq-\ba)\bp} d\bp
=\int_{\IR^{d}}e^{2\pi i (\bq-\ba)\bp} d\bp,\\
&&\tilde{\del}_{\ba}(\bq):=\sum_{\bn\in \IZ^{d}}\del_{\ba+\bn}(\bq)
=\sum_{\bn\in \IZ^{d}}e^{2\pi i\bn\cdot(\bq-\ba)}.
%&&\ml{F}_{h}(\del_{\ba})(\bp)=\frac{1}{h^{d/2}}e^{-2\pi i \frac{\ba \cdot \bp}{h}}\\
%&&\ml{F}_{h}(e^{2\pi i \ba \cdot \bq})(\bp)=h^{-d/2}\del_{\ba}\left(\frac{1}{h}\bp\right)=
%h^{d/2}\del_{h\ba}(\bp).
\eean
For any periodic function $\psi$ with period 1 and any vector $\bk \in h\IZ^{d}$ with $h=1/N$ and $N\in\IN$
we denote by $\hat{\psi}(\bk)$ its quantum Fourier coefficients defined as
\bea
\label{PQFT}
\hat{\psi}(\bk)=\frac{1}{h^{d/2}}\int_{\IT^{d}}\psi(\bq)e^{-2 \pi i \frac{\bk \cdot \bq}{h}}d\bq.
\eea
Then the inverse transform (i.e. the Fourier representation of $\psi$) is given by 
\bea
\label{PIQFT}
\psi(\bq)=h^{d/2} \sum_{\bk \in h\IZ^{d}}\hat{\psi}(\bk)e^{2 \pi i \frac{\bk \cdot \bq}{h}}.
\eea
The normalization in (\ref{PQFT}) implies that the following relation between 
continuous and periodic Fourier Transforms holds
\bea
\label{CPQFT}
\ml{F}_{h}(\psi)(\bp)= h^d \sum_{\bk \in h\IZ^{d}} \hat{\psi}(\bk)\del_{\bk}(\bp)
\eea
Parseval identity in this setting reads
\bean
%\label{perpar}
\la \phi, \psi \ra_{L^2(\IT^{d})} = \int_{\IT^{d}} \cc{\phi}(\bq) \psi(\bq) d\bq = h^d 
\sum_{\bk \in h\IZ^{d}} \cc{\hat{\phi}}(\bk)\hat{\psi}(\bk)= \la \hat{\phi}, \hat{\psi} \ra_{L^2(h\IZ^{d})}.
\eean
For a sequence $c=\{c_{\bs}\}_{\bs\in\IQ^{d}_{N}}$ we define its discrete Fourier transform 
$\hat{c}=\{\hat{c}_{\bk}\}_{\bk\in\IQ^{d}_{N}}$ as
\bea
\label{DQFT}
 \hat{c}_{\bk}=\frac{1}{N^{d/2}}\sum_{\bs \in\IQ^{d}_{N}}c_{\bs}e^{-2\pi i N\bs\cdot\bk}.
\eea
Then
\bea
\label{DIQFT}
 c_{\bs}=\frac{1}{N^{d/2}}\sum_{\bk \in\IQ^{d}_{N}}\hat{c}_{k}e^{2\pi i N\bk\cdot\bs}.
\eea
In discrete case the Parseval identity takes the standard form
\bean
\la c,d \ra =\sum_{\bs \in\IQ^{d}_{N}}\cc{c}_{\bs}d_{\bs}=\sum_{\bk \in \IQ^{d}_{N}}\cc{\hat{c}}_{\bk}\hat{d}_{\bk}
=\la \hat{c},\hat{d} \ra.
\eean
The relation between continuous and discrete Fourier transforms for periodic delta combs 
and $h=1/N$ is given by 
\bean
%\label{QFTDC}
\ml{F}_{h}\left(\sum_{\bs \in \IZ^{d}/N}c_{\bs}\del_{\bs}\right)=
\sum_{\bk \in \IZ^{d}/N}\hat{c}_{\bk}\del_{\bk}.
\eean
Finally we note that
\bea
\label{QFRDC}
\sum_{\bs \in \IZ^{d}/N}c_{\bs}\del_{\bs}=N^{d/2}\sum_{\bk \in \IZ^{d}/N}\hat{c}_{\bk}e^{2 \pi i N k\cdot q}
\eea

%%%%%%%%%%%%%%%%%%%%%%%%%%%%%%%%%%%%%%%%%%%%%%%%%%%%%%%%%%%%%%%%%%%%%%%%%%%%%%%%%%%%%%%%%%%%%%%%%%%%%%%%%%%
\subsection*{Quasi-periodic wave functions}
%\label{Qperiod}
%%%%%%%%%%%%%%%%%%%%%%%%%%%%%%%%%%%%%%%%%%%%%%%%%%%%%%%%%%%%%%%%%%%%%%%%%%%%%%%%%%%%%%%%%%%%%%%%%%%%%%%%%%%

Now we are in a position to determine the space of all quasiperiodic wave functions.
Let $\btht=(\btht_{q},\btht_{p})\in\IT^{2d}$. The wave function $\psi$ is 
$\btht$-quasiperiodic (cf. \cite{HB,BozDB}) if for all $\bm=(\bm_{1},\bm_{2})\in \IZ^{2d}$
\bean
\psi(\bq+\bm_{1})=e^{2\pi i \btht_{p}\cdot \bm_{1}}\psi(\bq), \qquad
(\ml{F}_{h}\psi)(\bp+\bm_{2})=e^{-2\pi i \btht_{q}\cdot \bm_{2}}(\ml{F}_{h}\psi)(\bp).
\eean
The second condition gives
\bean
\frac{1}{h^{d/2}}\int_{\IR^{d}}
e^{2\pi i (\btht_{q}-\bq/h)\cdot \bm_{2}}
\psi(\bq)e^{-2\pi i\frac{\bq \cdot \bp}{h}}d\bq=
(\ml{F}_{h}\psi)(\bp).
\eean
Using the bijective property of the Fourier transform on $\cS'(\IR^{d})$ we get
\bean
e^{2\pi i (\btht_{q}-\bq/h)\cdot \bm_{2}} \equiv 1,
\eean
which gives $\btht_{q}-\bq/h \in \IZ^{d}$. Hence the only
possible values of $\bq$ for which $\psi\not=0$ are determined
by condition $\bq\in h\IZ^{d}+h\btht_{q}$.
The wave $\psi$ is then necessarily a delta comb of the form
\bean
\psi(\bq)=\sum_{\bs\in h\IZ^{d}+h\btht_{q}}a_{\bs}\del_{\bs}(\bq).
\eean
Now
\bean
\psi(\bq+\bm_{1})=\sum_{\bs\in h\IZ^{d}+h\btht_{q}}a_{\bs}\del_{\bs}(\bq+\bm_{1})
=\sum_{\bs\in h\IZ^{d}-\bm_{1}+h\btht_{q}}a_{\bs+\bm_{1}}\del_{\bs}(\bq)
\eean
Thus quasiperiodicity of $\psi$ implies that $h\IZ^{d}-\bm_{1}=h\IZ^{d}$
and $a_{\bs+\bm_{1}}=e^{2\pi i \btht_{p}\cdot \bm_{1}}a_{\bs}$. The former condition
implies the existence of $N\in\IZ$ such that $h=1/N$ and hence
\bea
\label{AQPDC}
\psi(\bq)=\frac{1}{N^{d/2}}\sum_{\bs\in \IZ^{d}/N}c_{\bs}\del_{\bs+\btht_{q}/N}(\bq),
\eea
where $c_{\bs}=N^{d}a_{\bs+\btht_{q}/N}$ is a quasi-periodic sequence of arbitrary numbers 
supported on $\IZ^{d}/N$ lattice satisfying $c_{\bs+\bm_{1}}=e^{2\pi i \btht_{p}\cdot \bm_{1}}c_{\bs}$. 
Throughout the appendix the following normalization will be used
\bea
\label{Appnormal}
\|\psi\|_{2}^{2}=\frac{1}{N^{d}}\sum_{\bs\in \IQ^{d}_{N}}|c_{\bs}|^{2}=1.
\eea
This enables us to identify the space of all admissible quasiperiodic wave functions
with the Hilbert space $\IC^{N}$.

%%%%%%%%%%%%%%%%%%%%%%%%%%%%%%%%%%%%%%%%%%%%%%%%%%%%%%%%%%%%%%%%%%%%%%%%%%%%%%%%%%%%%%%%%%%%%%%%%%%%%%%%%%%
\subsection*{Quantum propagator}
%\label{Qprop}
%%%%%%%%%%%%%%%%%%%%%%%%%%%%%%%%%%%%%%%%%%%%%%%%%%%%%%%%%%%%%%%%%%%%%%%%%%%%%%%%%%%%%%%%%%%%%%%%%%%%%%%%%%%

In this section, following \cite{HB} and \cite{Men}, we recall
the explicit formula for the planar quantum cat map propagator.
We denote by $U(\bq_{1},\bq_{0})$ the kernel of
the propagator i.e. the Green function
for the Schr\"odinger equation for quantum cat map
dynamics on $\IR^{2}$. 
In order to find $U(\bq_{1},\bq_{0})$ we can either
quantize the Hamiltonian (derived in Section \ref{CatPlane}) and
solve the corresponding Schr\"odinger equation, or use the fact that the 
Hamiltonian is quadratic which implies that the semiclassical 
approximation to the quantum propagator is exact \cite{Men}
and hence the kernel can be directly expressed in terms of the 
classical action given by formula \eqref{Cataction} in section \ref{CatPlane}.  
Applying the second method we immediately get
\bean
U(\bq_{1},\bq_{0}) &=& 
\left(\frac{i}{2\pi \hbar}\frac{1}{b}\right)^{d/2}
\exp\left(\frac{i}{\hbar}\frac{1}{2b}
(a\bq_{0}^{2}-2\bq_{0}\bq_{1}+d\bq_{1}^{2})\right)\\
&=&\left(\frac{Ni}{b}\right)^{d/2}
\exp\left(\frac{N\pi i}{b}
(a\bq_{0}^{2}-2\bq_{0}\bq_{1}+d\bq_{1}^{2})\right).
\eean

The quantum evolution $\psi_{1} = U\psi_{0}$ of an initial wave function $\psi_{0}$
is then given by
\bean
 \psi_{1}(\bq_{1}) = \int_{\IR^{d}}U(\bq_{1},\bq_{0})\psi_{0}(\bq_{0})d\bq_{0}.
\eean

In particular, for an initial wave in form of a $\btht$-quasiperiodic delta comb
\bean
\psi_{0}(\bq)=\sum_{\bs\in\IZ^{d}/N}c_{\bs}\del_{\bs+\btht_{q}/N}(\bq),
\eean
the evolution is given by 
\bea
\label{Qp}
\psi_{1}(\bq_{1})&=&\int_{\IR^{d}}U(\bq_{1},\bq_{0})
\sum_{\bs\in\IZ^{d}/N}c_{\bs}\del_{\bs+\btht_{q}/N}(\bq_{0})d\bq_{0}\\
\nonumber
&=&\left(\frac{iN}{b}\right)^{d/2}
\sum_{\bs\in\IZ^{d}/N}c_{\bs}\exp\left(\frac{\pi i N}{b}
(a(\bs+\btht_{q}/N)^{2}-2(\bs+\btht_{q}/N)\bq_{1}+d\bq_{1}^{2})\right).
\eea

We will use the above formula in Appendix \ref{B} where we show that the 
evolution of the Wigner transform of $\psi_{0}$ is
given by the Frobenius-Perron operator associated with a classical
cat map.

%%%%%%%%%%%%%%%%%%%%%%%%%%%%%%%%%%%%%%%%%%%%%%%%%%%%%%%%%%%%%%%%%%%%%%%%%%%%%%%%%%%%%%%%%%%%%%%%%%%%%%%%%%%
%%%%%%%%%%%%%%%%%%%%%%%%%%%%%%%%%%%%%%%%%%%%%%%%%%%%%%%%%%%%%%%%%%%%%%%%%%%%%%%%%%%%%%%%%%%%%%%%%%%%%%%%%%%
%%%%%%%%%%%%%%%%%%%%%%%%%%%%%%%%%%%%%%%%%%%%%%%%%%%%%%%%%%%%%%%%%%%%%%%%%%%%%%%%%%%%%%%%%%%%%%%%%%%%%%%%%%%
%%%%%%%%%%%%%%%%%%%%%%%%%%%%%%%%%%%%%%%%%%%%%%%%%%%%%%%%%%%%%%%%%%%%%%%%%%%%%%%%%%%%%%%%%%%%%%%%%%%%%%%%%%%

\newpage
\markright{  \rm \normalsize Appendix B. \hspace{0.25cm} Wigner transform}
\chapter{Wigner transform}
\label{B}
\thispagestyle{fancy}

\markboth{  \rm \normalsize Appendix B. \hspace{0.25cm} Wigner transform}
{  \rm \normalsize Appendix B. \hspace{0.25cm} Wigner transform}
\lhead[\thepage]{\rightmark}
\rhead[\leftmark]{\thepage}
\cfoot{}

%%%%%%%%%%%%%%%%%%%%%%%%%%%%%%%%%%%%%%%%%%%%%%%%%%%%%%%%%%%%%%%%%%%%%%%%%%%%%%%%%%%%%%%%%%%%%%%%%%%%%%%%%%%
%%%%%%%%%%%%%%%%%%%%%%%%%%%%%%%%%%%%%%%%%%%%%%%%%%%%%%%%%%%%%%%%%%%%%%%%%%%%%%%%%%%%%%%%%%%%%%%%%%%%%%%%%%%
%%%%%%%%%%%%%%%%%%%%%%%%%%%%%%%%%%%%%%%%%%%%%%%%%%%%%%%%%%%%%%%%%%%%%%%%%%%%%%%%%%%%%%%%%%%%%%%%%%%%%%%%%%%

This appendix is devoted to the Wigner transform. By means of distribution
theory we derive an explicit form of its discrete version for wave 
functions assuming quasiperiodic delta comb form. We show that the
discrete Wigner function takes the form of a uniformly spaced Dirac
``delta brush'', which is periodic and supported on the $2N\times 2N$ 
rational grid on the unit torus. We derive several important properties
of the discrete Wigner transform and show that it provides
a clear geometric interpretation of the quantum cat maps evolution
(Proposition \ref{evol}). We conclude the appendix with a simple proof
of the equivalence between canonical and algebraic finite dimensional
quantizations of the cat map dynamics.

We start by recalling briefly standard definition of the Wigner transform.
In order to simplify the notation we will use the symbol 
$\bx=(\bq,\bp)$ to denote the phase space variables and the symbol
$\bk=(\bk_{1},\bk_{2})$ to denote the conjugate variables in the frequency space. 

Let $\psi$ denote a wave function. The associated Wigner transform 
$W_{\psi}(\bq,\bp)$ is given by
\bean
W_{\psi}(\bx)&=&W_{\psi}(\bq,\bp)=
%\label{Wigner1}
\frac{1}{h^{d}}\int_{\IR^{d}}
\psi\left(\bq+\frac{\bq'}{2}\right)\cc{\psi}\left(\bq-\frac{\bq'}{2}\right)
e^{-\frac{2 \pi i}{h}\bp\cdot\bq'}d\bq'\\
&=&\frac{1}{h^{d}}\int_{\IR^{d}}
\rho\left(\bq+\frac{\bq'}{2},\bq-\frac{\bq'}{2}\right)
e^{-2\pi i \frac{\bp\cdot\bq'}{h}}d\bq',
\eean
where $\rho(\bu,\bv)=|\psi\ra\la\psi|=\psi\otimes\cc{\psi}=\psi(\bu)\cc{\psi}(\bv)$
is the density kernel associated with the wave function $\psi$.

In its most general form the Wigner transform is defined for any tempered distribution 
$\psi\in \cS'(\IR^n)$ or density $\rho\in \cS'(\IR^{2n})$. Using this fact one can derive 
a useful integral version of the Wigner transform. To this end, let $\phi$ denote an 
arbitrary test function from $\cS(\IR^{d})$ and apply the 
change of variables
\bean
\bu=\bq+\frac{\bq'}{2},\qquad
\bv=\bq-\frac{\bq'}{2};
\eean
to get
\bean
\int_{\IR^{2d}}W_{\rho}(\bx)\phi(\bx)d\bx
&=&\frac{1}{h^d}\int_{\IR^{3d}}
\rho\left(\bq+\frac{\bq'}{2},\bq-\frac{\bq'}{2}\right)
e^{-\frac{2\pi i}{h} \bp\cdot\bq'}\phi(\bx) d\bq'd\bx\\
&=&\frac{1}{h^d}\int_{\IR^{3d}}
\rho(\bu,\bv)e^{-\frac{2\pi i}{h} \bp\cdot(\bu-\bv)}
\phi\left(\frac{\bu+\bv}{2},\bp\right)d\bu d\bv d\bp\\
&=&\frac{1}{h^d}\int_{\IR^{4d}}\rho(\bu,\bv)
\del_{\bq}\left(\frac{\bu+\bv}{2}\right) e^{-\frac{2\pi i}{h} \bp\cdot(\bu-\bv)}
\phi(\bx)d\bu d\bv d\bx.
\eean
From above computations we see that the kernel of the Wigner transform is given by the 
continuous family of integral Fano operators $S_{\bx}$
\bean
S_{\bx}(\bu,\bv):=\del_{\bq}\left(\frac{\bu+\bv}{2}\right)e^{\frac{2\pi i}{h}\bp\cdot(\bu-\bv)}.
\eean
Thus the Wigner transform $W_{\rho}(\bx)$ of $\rho$ admits the following kernel representation
\bean
W_{\rho}(\bx)&=&
\frac{1}{h^d}\Tr \rho S_{\bx}=\frac{1}{h^d}\int_{\IR^{2d}}\rho(\bu,\bv)S_{\bx}(\bv,\bu)d\bv d\bu\\
&=&\frac{1}{h^d}\int_{\IR^{2d}}\rho(\bu,\bv)\del_{\bq}\left(\frac{\bu+\bv}{2}\right)
e^{-\frac{2\pi i}{h}\bp\cdot(\bu-\bv)}d\bu d\bv
\eean

%%%%%%%%%%%%%%%%%%%%%%%%%%%%%%%%%%%%%%%%%%%%%%%%%%%%%%%%%%%%%%%%%%%%%%%%%%%%%%%%%%%%%%%%%%%%%%%%%%%%%%%%%%%
\subsection*{Discrete Wigner function}
%\label{DisWigner}
%%%%%%%%%%%%%%%%%%%%%%%%%%%%%%%%%%%%%%%%%%%%%%%%%%%%%%%%%%%%%%%%%%%%%%%%%%%%%%%%%%%%%%%%%%%%%%%%%%%%%%%%%%%

In this section we construct a discrete version of a Wigner transform. 

Using the kernel representation and applying it to a quasiperiodic distributional wave function $\psi$ of
the form \eqref{AQPDC} with $h=N^{-1}$ we get
\bea
\label{Wdist}
W_{\psi}(\bx)=\sum_{\bs_{1},\bs_{2}\in\IZ^{d}/N}c_{\bs_{1}}\bar{c}_{\bs_{2}}
\del_{\frac{\bs_{1}+\bs_{2}}{2}+\frac{\btht_{q}}{N}}(\bq)e^{-2\pi i N \bp\cdot(\bs_{1}-\bs_{2})}.
\eea

Our main goal is to show that \eqref{Wdist} assumes a periodic ``delta brush'' structure.
To this end we first change summation indexes in (\ref{Wdist}). 
\bean
\br=\frac{\bs_{1}+\bs_{2}}{2} \quad \bs=\frac{\bs_{1}-\bs_{2}}{2}
\eean
Note that the above transformation does not merely undo the change of variables performed 
in deriving the kernel representation of $W_{\psi}$ (the Jacobian this time is non-unital).
\bean
W_{\psi}(\bx)&=&\sum_{\br,\bs\in\IZ^{d}/2N}c_{\br+\bs}\bar{c}_{\br-\bs}
\del_{\br+\frac{\btht_{q}}{N}}(\bq)e^{-2\pi i 2N \bp\bs}\\
&=&\sum_{\br\in\IZ^{d}/2N}\del_{\br+\frac{\btht_{q}}{N}}(\bq)
\sum_{\bs\in\IZ^{d}/2N}
c_{\br+\bs}\bar{c}_{\br-\bs}e^{-2\pi i 2N \bp\bs}
\eean
(We use a convention that $c_{q}=0$ whenever $q\not \in \IZ/N$.)
Next we have
\bean
&&\sum_{\bs\in\IZ^{d}/2N}
c_{\br+\bs}\bar{c}_{\br-\bs}e^{-2\pi i 2N \bp\bs}
=\sum_{\bt\in\IQ^{d}_{2N}}c_{\br+\bt}\bar{c}_{\br-\bt}e^{-2\pi i 2N \bp\bt}
\sum_{\bs\in\IZ^{d}}e^{-2\pi i 2N(\bp-\frac{\btht_{p}}{N})\cdot\bs}\\
&=&\sum_{\bt\in\IQ^{d}_{2N}}c_{\br+\bt}\bar{c}_{\br-\bt}e^{-2\pi i 2N \bp\bt}
\sum_{\bs\in\IZ^{d}}\del_{\bs}(2N(\bp-\frac{\btht_{p}}{N}))\\
&=&\sum_{\bt\in\IQ^{d}_{2N}}c_{\br+\bt}\bar{c}_{\br-\bt}e^{-2\pi i 2N \bp\bt}
\frac{1}{(2N)^{d}}\sum_{\bs\in\IZ^{d}/2N}\del_{\bs+\frac{\btht_{p}}{N}}(\bp)
\eean
Thus
\bean
W_{\psi}(\bx)&=&
\frac{1}{(2N)^{d}}\sum_{\br,\bs\in\IZ^{d}/2N}
\sum_{\bt\in\IQ^{d}_{2N}}c_{\br+\bt}\bar{c}_{\br-\bt}
e^{-2\pi i 2N (\bs+\frac{\btht_{p}}{N})\cdot\bt}
\del_{\br+\frac{\btht_{q}}{N}}(\bq)\del_{\bs+\frac{\btht_{p}}{N}}(\bp)\\
&=&\frac{1}{(2N)^{d}}
\sum_{\br,\bs,\bt\in\IQ^{d}_{2N}}c_{\br+\bt}\bar{c}_{\br-\bt}
e^{-2\pi i 2N (\bs+\frac{\btht_{p}}{N})\cdot\bt}
\tilde{\del}_{\br+\frac{\btht_{q}}{N}}(\bq)\tilde{\del}_{\bs+\frac{\btht_{p}}{N}}(\bp).
\eean

The above formulas provide a clear geometric interpretation of the Wigner transform
of a quasiperiodic delta comb. A few remarks are in order here. First of all, the 
resulting Wigner function is strictly periodic (even though the original wave need
not be) and is supported on the grid with a mesh spacing of the size of $h/2$ (half of
the corresponding resolution for the wave). The support of the Wigner transform
forms a $2N\times 2N$ lattice centered at (shifted from the origin by) $\btht$.
The later property explains the interpretation of $\btht$ as Bloch or Floquet ``angles''
(cf. Section \ref{ModelI}) and the term ``quantum boundary conditions'' coined in 
\cite{KMR} in the context of the quantization condition \ref{qbc} (see Proposition 
\ref{HBq} in Section \ref{catmap}).

Now we want to find discrete Fourier coefficients of $W_{\psi}$. Taking the Fourier
transform in this case requires some care, since one has to adjust the value of
the Planck constant adequately. Taking into account the support of $W_{\psi}$,
its discrete Fourier transform agrees with a distributional one if the latter is
taken with the value of the Planck constant equal to $h/2$. Indeed,
for any $\bk\in\IZ^{2d}/2N$, we have 
\bea
\label{FW2}
\hat{W}_{\psi}(\bk)=\frac{1}{(2N)^{2d}}
\sum_{\br,\bs,\bt\in\IQ^{d}_{2N}}c_{\br+\bt}\bar{c}_{\br-\bt}
e^{-2\pi i 2N (\bs+\frac{\btht_{p}}{N})\bt}
e^{-2\pi i 2N\bk_{1}\cdot(\br+\frac{\btht_{q}}{N})}e^{-2\pi i 2N\bk_{2}\cdot(\bs+\frac{\btht_{p}}{N})}.
\eea
The above formula can be simplified as follows
\bean
\hat{W}_{\psi}(\bk)&=&\frac{1}{(2N)^{2d}}
\sum_{\br,\bt\in\IQ^{d}_{2N}}c_{\br+\bt}\bar{c}_{\br-\bt}
e^{-2\pi i 2N\bk_{1}\cdot(\br+\frac{\btht_{q}}{N})}
\sum_{\bs\in\IQ^{d}_{2N}}e^{-2\pi i 2N(\bs+\frac{\btht_{p}}{N})\cdot(\bt+\bk_{2})}\\
&=&\frac{1}{(2N)^{d}}\sum_{\br\in\IQ^{d}_{2N}}c_{\br-\bk_{2}}\bar{c}_{\br+\bk_{2}}
e^{-2\pi i 2N\bk_{1}\cdot(\br+\frac{\btht_{q}}{N})}.
\eean
For further simplification one needs to note that half of
the coefficients are zero, since $c_{\br}$ are supported on $\IZ/N$. Moreover due to 
the quasiperiodicity of $c_{\br}$ the product $c_{\br}\bar{c}_{\br}$ is periodic.
We can thus apply the change of indices $\br=\bt-\bk_{2}$ to obtain
\bean
\hat{W}_{\psi}(\bk)&=&\frac{1}{(2N)^{d}}\sum_{\br\in\IQ^{d}_{2N}-\bk_{2}}c_{\br}\bar{c}_{\br+2\bk_{2}}
e^{-2\pi i2N\bk_{1}(\br+\bk_{2}+\frac{\btht_{q}}{N})}\\
&=&\frac{1}{(2N)^{d}}\sum_{\br\in\IQ^{d}_{2N}}c_{\br}\bar{c}_{\br+2\bk_{2}}
e^{-2\pi i2N\bk_{1}(\br+\bk_{2}+\frac{\btht_{q}}{N})}\\
&=&\frac{1}{(2N)^{d}}e^{-2\pi i2N\bk_{1}(\bk_{2}+\frac{\btht_{q}}{N})}
\sum_{\br\in\IQ^{d}_{N}}c_{\br}\bar{c}_{\br+2\bk_{2}}
e^{-2\pi iN\bk_{1}\cdot\br}.
\eean 
Thus we conclude that the discrete Fourier coefficients of $W_{\psi}$ are given by 
the so-called discrete ambiguity function.
\bea
%\label{FW3}
\hat{W}_{\psi}(\bk)&=&\frac{1}{(2N)^{d}}\sum_{\br\in\IQ^{d}_{2N}}c_{\br-\bk_{2}}\bar{c}_{\br+\bk_{2}}
e^{-2\pi i 2N\bk_{1}\cdot(\br+\frac{\btht_{q}}{N})}\\
&=&\frac{1}{(2N)^{d}}\sum_{\br\in\IQ^{d}_{N}}c_{\br}\bar{c}_{\br+2\bk_{2}}
e^{-2\pi i2N\bk_{1}(\br+\bk_{2}+\frac{\btht_{q}}{N})}.
\eea

From the above formula we see that $\hat{W}(\bk)$ is 1-quasiperiodic
in both variables and is supported on $\IZ^{d}/2N$ lattice.

%%%%%%%%%%%%%%%%%%%%%%%%%%%%%%%%%%%%%%%%%%%%%%%%%%%%%%%%%%%%%%%%%%%%%%%%%%%%%%%%%%%%%%%%%%%%%%%%%%%%%%%%%%%
\subsection*{Properties of the Discrete Wigner function}
%\label{propWig}
%%%%%%%%%%%%%%%%%%%%%%%%%%%%%%%%%%%%%%%%%%%%%%%%%%%%%%%%%%%%%%%%%%%%%%%%%%%%%%%%%%%%%%%%%%%%%%%%%%%%%%%%%%%

It is well known that in the continuous setting the Wigner function can be viewed as a quantum
counterpart of the joint phase space density for the position and momentum variables.
For any $\psi \in L^{2}(\IR^d)$ satisfying normalization condition $\|\psi\|_{2}=1$ one has
\bean
\int_{\IR^d}W_{\psi}(\bx)d\bp=|\psi(\bq)|^{2}, \quad 
\int_{\IR^d}W_{\psi}(\bx)d\bq=|\cF_{h}\psi(\bp)|^{2}, \quad
\int_{\IR^{2d}}W_{\psi}(\bx)d\bx= 1,
\eean
The Wigner function is always real but need not be nonnegative 
(one can easily note that $W_{\psi}(\pmb{0})=-\lp\frac{2}{h} \rp^d \|\psi(\bq)\|^2_{2}$ is negative for any odd $\psi$).
However, the following useful property (Parseval identity) holds
\bean
0 \leq \la W_{\psi_{1}} , W_{\psi_{2}} \ra_{L^{2}(\IR^{2d})} = 
\frac{1}{h^d}|\la \psi_{1}, \psi_{2}\ra_{L^{2}(\IR^{d})}|^2 \leq \frac{1}{h^d}.
\eean
Not all of these properties can be generalized in any obvious way to the whole $\cS'(\IR)$ since the waves from
outside of $L^{2}(\IR)$ are no longer normalizable.  
Nevertheless, as we show in this section, all these properties are preserved for quasiperiodic waves. 
There are also some interesting characteristic properties of discrete Wigner function (not shared
by the standard continuous version). Below we mention two of them

I. $W(\bx)$ is completely determined by $(2N)^{2d}$ matrix of the
strengths of its delta functions.
\bean
w_{\br,\bs}=\frac{1}{(2N)^{d}}
\sum_{\bt\in\IQ^{d}_{2N}}c_{\br+\bt}\bar{c}_{\br-\bt}
e^{-2\pi i 2N (\bs+\frac{\btht_{p}}{N})\cdot\bt}.
\eean

II. In general all $(2N)^{2d}$ coefficients may be nonzero, but at the same time the coefficients 
attached to a vertices of a $\half$-box may differ at most in sign. 
\bean
w_{\br+\frac{\be_{i}}{2},\bs}&=&\frac{1}{(2N)^{d}}\sum_{\bt\in\IQ^{d}_{2N}}
c_{\br+\frac{e_{i}}{2}+\bt}\bar{c}_{\br+\frac{\be_{i}}{2}-\bt}
e^{-2\pi i 2N (\bs+\frac{\btht_{p}}{N})\cdot\bt}\\
&=&\frac{1}{(2N)^{d}}\sum_{\bt\in\IQ^{d}_{2N}}c_{\br+\bt}\bar{c}_{\br-\bt}
e^{-2\pi i 2N(\btht_{p_{i}}/2N)}e^{-2\pi i 2N (\bs+\frac{\btht_{p}}{N})\cdot(\bt-\frac{\be_{i}}{2})}\\
&=&e^{\pi i 2N \bs\cdot \be_{i}}\frac{1}{(2N)^{d}}
\sum_{\bt\in\IQ^{d}_{2N}}c_{\br+\bt}\bar{c}_{\br-\bt}
e^{-2\pi i 2N (\bs+\frac{\btht_{p}}{N})\cdot\bt}=(-1)^{2N\bs\cdot \be_{i}}w_{\br,\bs}
\eean
and
\bean
w_{\br,\bs+\frac{\be_{i}}{2}}&=&\frac{1}{(2N)^{d}}
\sum_{\bt\in\IQ^{d}_{2N}}c_{\br+\bt}\bar{c}_{\br-\bt}
e^{-2\pi i 2N (\bs+\frac{\btht_{p}}{N}+\frac{\be_{i}}{2})\bt}\\
&=&\frac{1}{(2N)^{d}}\sum_{\bt\in\IQ^{d}_{2N}}
e^{-\pi i 2N\bt\cdot\be_{i}}c_{\br+\bt}\bar{c}_{\br-\bt}
e^{-2\pi i 2N (\bs+\frac{\btht_{p}}{N})\bt}\\
&=&\frac{1}{(2N)^{d}}\sum_{\bt\in\IQ^{d}_{2N}}(-1)^{2N\bt\cdot\be_{i}}
c_{\br+\bt}\bar{c}_{\br-\bt}
e^{-2\pi i 2N (\bs+\frac{\btht_{p}}{N})\bt}.
\eean
Now it suffices to note that $c_{\br+\bt}$ are zero whenever $\br\cdot\be_{i}$ 
and $\bt\cdot\be_{i}$ are of different parity which gives
\bean
w_{\br,\bs+\half}=(-1)^{2N\br\cdot\be_{i}}\frac{1}{(2N)^{d}}
\sum_{\bt\in\IQ^{d}_{2N}}c_{\br+\bt}\bar{c}_{\br-\bt}
e^{-2\pi i 2N (\bs+\frac{\btht_{p}}{N})\bt}=(-1)^{2N\br\cdot\be_{i}}w_{\br,\bs}.
\eean

Next we derive the properties regarding marginal distributions 
%(the integrals are replaced by averages)

III. $\bq$-marginal projection.
\bean
\sum_{\bs\in\IQ^{d}_{2N}}w_{\br,\bs}&=&
\frac{1}{(2N)^{d}}\sum_{\bs\in\IQ^{d}_{2N}}
\sum_{\bt\in\IQ^{d}_{2N}}c_{\br+\bt}\bar{c}_{\br-\bt}
e^{-2\pi i 2N (\bs+\frac{\btht_{p}}{N})\bt}\\
&=&\frac{1}{(2N)^{d}}\sum_{\bt\in\IQ^{d}_{2N}}c_{\br+\bt}\bar{c}_{\br-\bt}
\sum_{\bs\in\IQ^{d}_{2N}}e^{-2\pi i 2N (\bs+\frac{\btht_{p}}{N})\bt}
=c_{\br}\bar{c}_{\br}=|c_{\br}|^{2}
\eean

IV. $\bp$-marginal projection results in $|\hat{c}_{\bs}|^{2}$ (we omit the
proof, which is in this case a bit more technically involved).

We end this section by deriving the discrete version of the Parseval 
identity for the Wigner function. That is, we want to prove the following identity

V.
\bean
\sum_{\bk\in\IQ^{2d}_{2N}} \hat{W}_{\psi_{1}}(\bk)\hat{W}_{\psi_{2}}(\bk)
=N^d \left| \frac{1}{N^{d}} \sum_{\br\in\IQ^{d}_{N}} \cc{c}_{\br} d_{\br} \right|^{2}.
\eean 
In the prove we use the Fourier transform of $W_{\psi}$ derived in the previous section. 
\bean
&&\sum_{\bk\in\IQ^{2d}_{2N}} \hat{W}_{\psi_{1}}(\bk) \hat{W}_{\psi_{2}}(\bk)\\
&=&\frac{1}{(2N)^{2d}}
\sum_{\bk\in\IQ^{2d}_{2N}}
\sum_{\br,\bs\in\IQ^{d}_{N}}c_{\br}\cc{c}_{\br+2\bk_{2}}
e^{-2\pi i2N\bk_{1}\cdot(\br+\bk_{2} + \frac{\btht_{q}}{N})}
\cc{d}_{\bs}d_{\bs+2\bk_{2}}
e^{2\pi i2N\bk_{1}\cdot(\bs+\bk_{2} + \frac{\btht_{q}}{N})}\\
&=&\frac{1}{(N)^{2d}}\sum_{\bk\in\IQ^{2d}_{N}}
\sum_{\br,\bs\in\IQ^{d}_{N}}c_{\br}\cc{c}_{\br+\bk_{2}}
e^{-2\pi iN\bk_{1}\cdot \br}
\cc{d}_{\bs}d_{\bs+\bk_{2}}
e^{2\pi iN\bk_{1}\cdot \bs}\\
&=&\frac{1}{N^{2d}}\sum_{\bk_{2},\br,\bs\in\IQ^{d}_{N}}
c_{\br}\bar{c}_{\br+\bk_{2}}\bar{d}_{\bs}d_{\bs+\bk_{2}}
\sum_{\bk_{1}\in\IQ^{d}_{N}}e^{-2\pi iN\bk_{1}\cdot(\br-\bs)}\\
&=&\frac{1}{N^{2d}}\sum_{\bk_{2},\br,\bs\in\IQ^{d}_{N}}
c_{\br}\bar{c}_{\br+\bk_{2}}\bar{d}_{\bs}d_{\bs+\bk_{2}}N^d\del(\br-\bs)
%&=&\frac{1}{N^{d}}\sum_{\br\in\IQ^{d}_{N}} {c}_{\br}\cc{d}_{\br}
%\sum_{\bk_{2}\in\IQ^{d}_{N}}\bar{c}_{\br+\bk_{2}}d_{\br+\bk_{2}}
=N^d \left| \frac{1}{N^{d}} \sum_{\br\in\IQ^{d}_{N}} 
\cc{c}_{\br} d_{\br} \right|^{2}.
\eean

%%%%%%%%%%%%%%%%%%%%%%%%%%%%%%%%%%%%%%%%%%%%%%%%%%%%%%%%%%%%%%%%%%%%%%%%%%%%%%%%%%%%%%%%%%%%%%%%%%%%%%%%%%%
\subsection*{Cat map evolution of the Wigner function}
%\label{EvWig}
%%%%%%%%%%%%%%%%%%%%%%%%%%%%%%%%%%%%%%%%%%%%%%%%%%%%%%%%%%%%%%%%%%%%%%%%%%%%%%%%%%%%%%%%%%%%%%%%%%%%%%%%%%%

In this section we show the the Wigner function associated with
a quasiperiodic delta comb evolves classically under the cat map
dynamics. Indeed, we have the following
\begin{prop}
\label{evol}
Let $U$ denote the quantum propagator associated with a cat map $F$. For any
initial $\btht$-quasiperiodic delta wave $\psi_{0}$, the Wigner transform
of the evolved wave $\psi_{1}=U\psi_{0}$ satisfies the property
$W_{\psi_{1}}=W_{\psi_{0}}\circ F^{-1}$.
\end{prop}
\textbf{Proof}.
The prove is obtained by a direct application of formula \eqref{Qp} (see Section \ref{CanQuant})
which gives
\bean
&&W_{\psi_{1}}(\bx)=
\int_{\IR^{d}}
\left(\frac{Ni}{b}\right)^{d/2}\sum_{\bs_{1}}c_{\bs_{1}}
e^{\frac{N\pi i}{b}
\left(a(\bs_{1}+\frac{\btht_{q}}{N})^{2}-2(\bs_{1}+\frac{\btht_{q}}{N})\left(\bq+\frac{\bq'}{2}\right)+
d\left(\bq+\frac{\bq'}{2}\right)^{2}\right)}\\
&&\left(-\frac{Ni}{b}\right)^{d/2}\sum_{\bs_{2}}\bar{c}_{\bs_{2}}
e^{-\frac{N\pi i}{b}
\left(a(\bs_{2}+\frac{\btht_{q}}{N})^{2}-2(\bs_{2}+\frac{\btht_{q}}{N})\left(\bq-\frac{\bq'}{2}\right)+
d\left(\bq-\frac{\bq'}{2}\right)^{2}\right)}
e^{-2 \pi i N\bp\bq'}d\bq'
\eean
From the above we can represent $W_{\psi_{1}}(\bx)$ by
\bean
&&\frac{N^d}{b}\sum_{\bs_{1},\bs_{2}}c_{\bs_{1}}\bar{c}_{\bs_{2}}
\int_{\IR^{d}}e^{\frac{2\pi i N}{b}
\left(a\frac{\bs_{1}^{2}-\bs_{2}^{2}}{2}+a(\bs_{1}-\bs_{2})\frac{\btht_{q}}{N}-(\bs_{1}-\bs_{2})\bq 
-\frac{\bs_{1}+\bs_{2}}{2}\bq'-\bq'\frac{\btht_{q}}{N}+d\bq\bq'-b\bp\bq'\right)}d\bq'\\
&=&\sum_{\bs_{1},\bs_{2}}c_{\bs_{1}}\bar{c}_{\bs_{2}}
\int_{\IR^{d}}
e^{2\pi i\bq'\left(d\bq-b\bp-\left(\frac{\bs_{1}+\bs_{2}}{2}+\frac{\btht_{q}}{N}\right)\right)}d\bq'
e^{\frac{2 \pi i N}{b}
\left(a\frac{\bs_{1}^{2}-\bs_{2}^{2}}{2}+a(\bs_{1}-\bs_{2})\frac{\btht_{q}}{N}-(\bs_{1}-\bs_{2})\bq\right)}.
\eean
Thus using the spectral resolution of the Dirac delta comb we get
\bean
&&W_{\psi_{1}}(\bx)
=\sum_{\bs_{1},\bs_{2}}c_{\bs_{1}}\bar{c}_{\bs_{2}}
\del_{\frac{\bs_{1}+\bs_{2}}{2}+\frac{\btht_{q}}{N}}(d\bq-b\bp)
e^{\frac{2\pi i N}{b}
\left(a\left(\frac{\bs_{1}+\bs_{2}}{2}+\frac{\btht_{q}}{N}\right)-\bq\right)(\bs_{1}-\bs_{2})}\\
&=&\sum_{\bs_{1},\bs_{2}}c_{\bs_{1}}\bar{c}_{\bs_{2}}
\del_{\frac{\bs_{1}+\bs_{2}}{2}+\frac{\btht_{q}}{N}}(d\bq-b\bp)
e^{2\pi iN\left(\frac{a}{b}\left(\frac{\bs_{1}+\bs_{2}}{2}+\frac{\btht_{q}}{N}\right)
-\frac{ad-bc}{b}\bq\right)(\bs_{1}-\bs_{2})}\\
&=&\sum_{\bs_{1},\bs_{2}}c_{\bs_{1}}\bar{c}_{\bs_{2}}
\del_{\frac{\bs_{1}+\bs_{2}}{2}+\frac{\btht_{q}}{N}}(d\bq-b\bp)
e^{2\pi iN\left(\frac{a}{b}\left(\frac{\bs_{1}+\bs_{2}}{2}+\frac{\btht_{q}}{N}\right)
-\frac{a}{b}\left((\frac{\bs_{1}+\bs_{2}}{2}+\frac{\btht_{q}}{N}\right)+b\bp+c\bq\right)
(\bs_{1}-\bs_{2})}\\
&=&\sum_{\bs_{1},\bs_{2}}c_{\bs_{1}}\bar{c}_{\bs_{2}}
\del_{\frac{\bs_{1}+\bs_{2}}{2}+\frac{\btht_{q}}{N}}(d\bq-b\bp)
e^{2\pi iN(-c\bq+a\bp)(\bs_{1}-\bs_{2})}=
W_{\psi_{0}}\circ F^{-1}(\bx)
\eean

%%%%%%%%%%%%%%%%%%%%%%%%%%%%%%%%%%%%%%%%%%%%%%%%%%%%%%%%%%%%%%%%%%%%%%%%%%%%%%%%%%%%%%%%%%%%%%%%%%%%%%%%%%%
\subsection*{Wigner function, Weyl quantization and ambiguity function}
%\label{WeylWig}
%%%%%%%%%%%%%%%%%%%%%%%%%%%%%%%%%%%%%%%%%%%%%%%%%%%%%%%%%%%%%%%%%%%%%%%%%%%%%%%%%%%%%%%%%%%%%%%%%%%%%%%%%%%

We end this appendix by recalling briefly the relation between the Wigner function,
the Weyl quantization and the ambiguity function. We have
\begin{prop}
\label{WWA}
For any classical observable $f\in L^{2}(\IR^{2d})$ and any wave function $\psi\in L^{2}(\IR^{2d})$,
the expectation of the Weyl quantization $Op(f)$ of $f$ satisfies
\bean
\la \psi, Op(f)\psi \ra = \int_{R^{2d}}f(\bx)W_{\psi}(\bx)d\bx.
\eean
In particular, the expectation of any Weyl translation operator $T_{\bk}$ (cf. \eqref{WeylTr}, Section \ref{WQ}) 
is given by the ambiguity function $A_{\psi}(\bk)$ (the inverse Fourier transform of the Wigner function) 
associated with a wave $\psi$
\bean
\la \psi, T_{\bk} \psi \ra = A_{\psi}(\bk)=\int_{R^{2d}}W_{\psi}(\bx)
e^{\frac{2\pi i}{h} \bk \wedge \bx}d\bx.
\eean
\end{prop}

\textbf{Proof.}
Weyl translations generate the whole algebra of observables. It is thus enough
to prove the second statement. Using the formula \eqref{WeylAc} for the explicit action
of Weyl translations on wave functions derived in section \ref{WQ} we get
\bean
\la \psi, T_{\bk} \psi \ra &=& \int_{R^{d}}\cc{\psi}(\bq)T_{k}\psi(\bq)d\bq=
\int_{R^{d}}\cc{\psi}(\bq)e^{\frac{2 \pi i}{h}\bk_{2}\cdot(\bq-\bk_{1}/2)}\psi(\bq-\bk_{1})d\bq \\
&=&
\int_{R^{d}}\psi(\bq-\bk_{1}/2)\cc{\psi}(\bq+\bk_{1}/2))e^{\frac{2\pi i}{h}\bk_{2}\cdot \bq}d\bq=A_{\psi}(\bk)
\eean
On the other hand
\bean
\int_{R^{2d}}W_{\psi}(\bx)e^{2\pi i \frac{\bk \wedge \bx}{h}}d\bx &=&
\frac{1}{h^d}\int_{R^{3d}}\psi(\bq+\bq'/2)\cc{\psi}(\bq-\bq'/2))
e^{-\frac{2\pi i}{h}(\bp\cdot \bq'-\bk_{2}\cdot \bq + \bk_{1}\cdot \bp)}d\bq'd\bx\\
&=&
\int_{R^{d}}\psi(\bq-\bk_{1}/2)\cc{\psi}(\bq+\bk_{1}/2))e^{\frac{i}{h}\bk_{2}\cdot \bq}d\bq=A_{\psi}(\bk),
\eean
which completes the proof. $\qquad \blacksquare$

\begin{rem}
The above proposition generalizes to the wave functions
from $\cS'(\IR^{d})$ (in this case the observables
need to be taken from a smaller set). In particular, the Proposition holds
for quasiperiodic delta combs and the observables from $L^{2}(\IT^{2d})$, satisfying 
$\sum_{\bk}|\hat{f}(\bk)|<\infty$.
\end{rem}

The above proposition together with Proposition \ref{evol} provides the proof that the algebraic
quantization introduced in Section \ref{catmap} coincides with a canonical one, derived in
Section \ref{CanQuant} of Appendix \ref{A}. Indeed, denoting by $U$ the 
propagator obtained by means of the canonical quantization and by $\cU$ the $^*$-automorphism constructed 
in the algebraic approach, we get (using symplecticity of $F$) what follows
\bean
\la \psi_{0}, \cU T_{\bk} \psi_{0}\ra &=& \la \psi_{0}, T_{F^{-1}\bk} \psi_{0}\ra
=\int_{R^{2d}}W_{\psi_{0}}(\bx)e^{2\pi i \frac{F^{-1}\bk \wedge \bx}{h}}d\bx \\
&=&\int_{R^{2d}}W_{\psi_{0}}(F^{-1}\bx)e^{2\pi i \frac{\bk \wedge \bx}{h}}d\bx 
= \int_{R^{2d}}W_{\psi_{1}}(\bx)e^{2\pi i \frac{\bk \wedge \bx}{h}}d\bx
=\la \psi_{1},  T_{\bk} \psi_{1}\ra \\
&=&\la U\psi_{0}, T_{\bk} U \psi_{0}\ra 
= \la \psi_{0}, U^{*}T_{\bk} U \psi_{0}\ra.
\eean

Thus the canonical quantum propagator $U$ implements the algebraic $^*$-automorphism $\cU$
on the space of  quasiperiodic delta waves.

%%%%%%%%%%%%%%%%%    BIBLIOGRAPHY    %%%%%%%%%%%%%%%%%%%%%%%%%%%

\newpage
\markright{  \rm \normalsize Bibliography}

\end{document}

%% file: ThLech.tex
\usepackage{amsmath} 
\usepackage{amssymb}
\usepackage{euscript}
\usepackage{amsfonts}
 \usepackage{graphicx}

\newcommand{\nwc}{\newcommand}

\nwc{\nwt}{\newtheorem}

\newtheorem{lem}{Lemma}[chapter]
\newtheorem{thm}[lem]{Theorem}
\newtheorem{defin}[lem]{Definition}
\newtheorem{cor}[lem]{Corollary}
\newtheorem{prop}[lem]{Proposition}
\newtheorem{rem}[lem]{Remark}

\nwc{\barr}{\begin{array}}
\nwc{\bal}{\begin{align}}
\nwc{\bequ}{\begin{equation}}
\nwc{\ben}{\begin{equation*}}
\nwc{\bea}{\begin{eqnarray}}
\nwc{\beq}{\begin{eqnarray}}
\nwc{\bean}{\begin{eqnarray*}}
\nwc{\beqn}{\begin{eqnarray*}}
\nwc{\beqast}{\begin{eqnarray*}}

\nwc{\earr}{\end{array}}
\nwc{\eal}{\end{align}}
\nwc{\eequ}{\end{equation}}
\nwc{\een}{\end{equation*}}
\nwc{\eea}{\end{eqnarray}}
\nwc{\eeq}{\end{eqnarray}}
\nwc{\eean}{\end{eqnarray*}}
\nwc{\eeqn}{\end{eqnarray*}}
\nwc{\eeqast}{\end{eqnarray*}}

\nwc{\bldm}{\boldmath} %old-tex
\nwc{\ubm}{\unboldmath} %old-tex
\nwc{\mf}{\mathbf} %Latex (as in \bf not tilted math letters)
\nwc{\blds}{\boldsymbol} %Latex 
\nwc{\ml}{\mathcal} %Latex

\nwc{\al}{\alpha}
\nwc{\vep}{\varepsilon}
\nwc{\ep}{\epsilon}
\nwc{\veps}{\varepsilon}
\nwc{\eps}{\epsilon}
\nwc{\vrho}{\varrho}
\nwc{\orho}{\bar\varrho}
\nwc{\vpsi}{\varpsi}
\nwc{\lamb}{\lambda_\varepsilon}
\nwc{\lam}{\lambda}
\nwc{\del}{\delta}
\nwc{\tht}{\theta}
\nwc{\om}{\omega}

\nwc{\btht}{\blds{\theta}}
\nwc{\bxi}{\blds{\xi}}
\nwc{\bmal}{\blds{\al}}
\nwc{\bmbe}{\blds{\beta}}

\nwc{\la}{\langle}    %like '<' - bra
\nwc{\ra}{\rangle}    %like '>' - cat
\nwc{\lp}{\left(}     %adjustable (Big) '('
\nwc{\rp}{\right)}    %adjustable (Big) ')'

\nwc{\cc}{\overline}   %complex conjugacy

\nwc{\IA}{\mathbb{A}} %algebraic
\nwc{\IB}{\mathbb{B}} %ball
\nwc{\IC}{\mathbb{C}} %complex
\nwc{\ID}{\mathbb{D}} %Dedekind
\nwc{\IE}{\mathbb{E}} %Euklides
\nwc{\IF}{\mathbb{F}} %finite field
\nwc{\IG}{\mathbb{G}} %Gauss
\nwc{\IH}{\mathbb{H}} %Hilbert\N-subgroup
\nwc{\IN}{\mathbb{N}} %natural
\nwc{\IP}{\mathbb{P}} %prime
\nwc{\IQ}{\mathbb{Q}} %rational
\nwc{\IR}{\mathbb{R}} %real
\nwc{\IS}{\mathbb{S}} %real 
\nwc{\IT}{\mathbb{T}} %torus
\nwc{\IZ}{\mathbb{Z}} %integers

\nwc{\ba}{\blds{a}}
\nwc{\bb}{\blds{b}}
\nwc{\bc}{\blds{c}}
\nwc{\bd}{\blds{d}}
\nwc{\be}{\blds{e}}
\nwc{\fb}{\blds{f}} %\bf is already defined
\nwc{\bg}{\blds{g}}
\nwc{\bh}{\blds{h}}
\nwc{\bi}{\blds{i}}
\nwc{\bj}{\blds{j}}
\nwc{\bk}{\blds{k}}
\nwc{\bl}{\blds{l}}
\nwc{\bm}{\blds{m}}
\nwc{\bn}{\blds{n}}
\nwc{\bo}{\blds{o}}
\nwc{\bp}{\blds{p}}
\nwc{\bq}{\blds{q}}
\nwc{\br}{\blds{r}}
\nwc{\bs}{\blds{s}}
\nwc{\bt}{\blds{t}}
\nwc{\bu}{\blds{u}}
\nwc{\bv}{\blds{v}}
\nwc{\bw}{\blds{w}}
\nwc{\bx}{\blds{x}}
\nwc{\by}{\blds{y}}
\nwc{\bz}{\blds{z}}

\nwc{\bA}{\blds{A}}
\nwc{\bB}{\blds{B}}
\nwc{\bC}{\blds{C}}
\nwc{\bD}{\blds{D}}
\nwc{\bE}{\blds{E}}
\nwc{\bF}{\blds{F}}
\nwc{\bG}{\blds{G}}
\nwc{\bH}{\blds{H}}
\nwc{\bI}{\blds{I}}
\nwc{\bJ}{\blds{J}}
\nwc{\bK}{\blds{K}}
\nwc{\bL}{\blds{L}}
\nwc{\bM}{\blds{M}}
\nwc{\bN}{\blds{N}}
\nwc{\bO}{\blds{O}}
\nwc{\bP}{\blds{P}}
\nwc{\bQ}{\blds{Q}}
\nwc{\bR}{\blds{R}}
\nwc{\bS}{\blds{S}}
\nwc{\bT}{\blds{T}}
\nwc{\bU}{\blds{U}}
\nwc{\bV}{\blds{V}}
\nwc{\bW}{\blds{W}}
\nwc{\bX}{\blds{X}}
\nwc{\bY}{\blds{Y}}
\nwc{\bZ}{\blds{Z}}

\nwc{\va}{{\bf a}}
\nwc{\vb}{{\bf b}}
\nwc{\vc}{{\bf c}}
\nwc{\vd}{{\bf d}}
\nwc{\ve}{{\bf e}}
\nwc{\vf}{{\bf f}}
\nwc{\vg}{{\bf g}}
\nwc{\vh}{{\bf h}}
\nwc{\vi}{{\bf i}}
\nwc{\vj}{{\bf j}}
\nwc{\vk}{{\bf k}}
\nwc{\vl}{{\bf l}}
\nwc{\vm}{{\bf m}}
\nwc{\vn}{{\bf n}}
\nwc{\vo}{{\it o}}
\nwc{\vp}{{\bf p}}
\nwc{\vq}{{\bf q}}
\nwc{\vr}{{\bf r}}
\nwc{\vs}{{\bf s}}
\nwc{\vt}{{\bf t}}
\nwc{\vu}{{\bf u}}
\nwc{\vv}{{\bf v}}
\nwc{\vw}{{\bf w}}
\nwc{\vx}{{\bf x}}
\nwc{\vy}{{\bf y}}
\nwc{\vz}{{\bf z}}

\nwc{\vA}{{\bf A}}
\nwc{\vB}{{\bf B}}
\nwc{\vC}{{\bf C}}
\nwc{\vD}{{\bf D}}
\nwc{\vE}{{\bf E}}
\nwc{\vF}{{\bf F}}
\nwc{\vG}{{\bf G}}
\nwc{\vH}{{\bf H}}
\nwc{\vI}{{\bf I}}
\nwc{\vJ}{{\bf J}}
\nwc{\vK}{{\bf K}}
\nwc{\vL}{{\bf L}}
\nwc{\vM}{{\bf M}}
\nwc{\vN}{{\bf N}}
\nwc{\vO}{{\it O}}
\nwc{\vP}{{\bf P}}
\nwc{\vQ}{{\bf Q}}
\nwc{\vR}{{\bf R}}
\nwc{\vS}{{\bf S}}
\nwc{\vT}{{\bf T}}
\nwc{\vU}{{\bf U}}
\nwc{\vV}{{\bf V}}
\nwc{\vW}{{\bf W}}
\nwc{\vX}{{\bf X}}
\nwc{\vY}{{\bf Y}}
\nwc{\vZ}{{\bf Z}}

\nwc{\cA}{\ml{A}}
\nwc{\cB}{\ml{B}}
\nwc{\cC}{\ml{C}}
\nwc{\cD}{\ml{D}}
\nwc{\cE}{\ml{E}}
\nwc{\cF}{\ml{F}}
\nwc{\cG}{\ml{G}}
\nwc{\cH}{\ml{H}}
\nwc{\cI}{\ml{I}}
\nwc{\cJ}{\ml{J}}
\nwc{\cK}{\ml{K}}
\nwc{\cL}{\ml{L}}
\nwc{\cM}{\ml{M}}
\nwc{\cN}{\ml{N}}
\nwc{\cO}{\ml{O}}
\nwc{\cP}{\ml{P}}
\nwc{\cQ}{\ml{Q}}
\nwc{\cR}{\ml{R}}
\nwc{\cS}{\ml{S}}
\nwc{\cT}{\ml{T}}
\nwc{\cU}{\ml{U}}
\nwc{\cV}{\ml{V}}
\nwc{\cW}{\ml{W}}
\nwc{\cX}{\ml{X}}
\nwc{\cY}{\ml{Y}}
\nwc{\cZ}{\ml{Z}}

\nwc{\nn}{\nonumber}
\nwc{\veta}{\mbox{\boldmath{$\eta$}}}
\nwc{\vhatk}{\hat{\bk}}
\nwc{\commentout}[1]{}
\nwc{\pdr}[2]{\frac{\partial{#1}}{\partial{#2}}}
\nwc{\odr}[2]{\frac{d{#1}}{d{#2}}}
\nwc{\summ}[2]{\sum_{#1=#2}^{\infty}}
\nwc{\summm}[3]{\sum_{#1=#2}^{#3}}
\nwc{\pdd}[1]{\frac{\partial}{\partial{#1}}}
\nwc{\odd}[1]{\frac{d}{d{#1}}}
\nwc{\m}{\mbox}
\nwc{\re}{\hbox{Re}}
\nwc{\mt}{\bar{t}}
\nwc{\ou}{\bar u}
\nwc{\noi}{\noindent}
\nwc{\non}{\nonumber}
\nwc{\half}{\frac{1}{2}}
\nwc{\third}{\frac{1}{3}}
\nwc{\pa}{\partial}
\nwc{\uP}{{\em \bf Proof: }}
\nwc{\uT}{\underline{Theorem:}}
\nwc{\epto}{{\xrightarrow{\ep\to 0}}}
\nwc{\nto}{{\xrightarrow{n\to \infty}}}
\nwc{\Tr}{\operatorname{Tr}}